\documentclass[
 amsmath,amssymb,
 aps,
prb,
floatfix,
]{revtex4-2}

\usepackage{multirow}
\usepackage{graphicx}

\begin{document}

\title{Localized States in Local Isomorphism Classes of Pentagonal Quasicrystals }

\author{M.\"O. Oktel}
\email{oktel@bilkent.edu.tr}
\affiliation{Department of Physics, Bilkent University, Ankara, 06800, TURKEY}

\date{\today}

\begin{abstract}
A family of pentagonal quasicrystals can be defined by projecting a section of the five-dimensional cubic lattice to two dimensions. A single parameter, the sum of intercepts $\Gamma=\sum_j \gamma_j$,  describes this family by defining the cut in the five-dimensional space. Each value of $0\le \Gamma \le \frac{1}{2}$ defines a unique local isomorphism class for these quasicrystals, with $\Gamma=0$ giving the Penrose lattice. Except for a few special values of $\Gamma$, these lattices lack simple inflation-deflation rules making it hard to count how frequently a given local configuration is repeated. We consider the vertex-tight-binding model on these quasicrystals and investigate the strictly localized states (LS) for all values of $\Gamma$. We count the frequency of localized states both by numerical exact diagonalization on lattices of $~10^5$ sites and by identifying localized state types and calculating their perpendicular space images. While the imbalance between the number of sites forming the two sublattices of the bipartite quasicrystal grows monotonically with $\Gamma$, we find that the localized state fraction first decreases and then increases as the distance from the Penrose lattice grows. The highest LS fraction of $~10.17\%$ is attained at $\Gamma=0.5$ while the minimum is $~4.5\%$ at $\Gamma \simeq 0.12$. The LS on the even sublattice are generally concentrated near sites with high symmetry, while the LS on the odd sublattice are more uniformly distributed. The odd sublattice has a higher LS fraction, having almost three times the LS frequency of the even sublattice at $\Gamma=0.5$. We identify 20 LS types on the even sublattice, and their total frequency agrees well with the numerical exact diagonalization result for all values of $\Gamma$. For the odd sublattice, we identify 45 LS types. However, their total frequency remains significantly below the numerical calculation, indicating the possibility of more independent LS types. 
\end{abstract}

\maketitle

\section{Introduction}
It has been almost 40 years since the discovery of quasicrystals\cite{she84,lev86,soc86}, yet there is no complete theory for describing elementary excitations in them.
The high degree of symmetry of quasicrystals does not lend itself to a simple description like Bloch's theorem to constrain electronic wavefunctions.
Recent experimental success in constructing synthetic quasicrystals in electronic\cite{col17}, atomic\cite{vie19}, or photonic systems\cite{var13} promises precise measurements in highly controlled settings. 
Consequently, there is a resurgence of interest in the quasicrystalline state beyond structural description.  

The electronic states in one-dimensional quasicrystal models, particularly in the Fibonacci chain \cite{jag21}, are relatively well understood. The spectrum is singularly continuous, and eigenstates can be localized, extended, or critically self-similar. In higher dimensions, another possibility is strictly localized states (LS) states that have exactly zero density beyond a finite region of the lattice. These states were first identified in the Penrose lattice (PL) \cite{koh86} after numerical calculations \cite{oda86,cho85} have shown that almost $10\%$ of the states are degenerate at zero energy. These LS have since been found in other quasicrystal lattices \cite{rsc95}, such as the Ammann-Beenker lattice\cite{okt21,kog20}, and closely related modes have been identified in photonic quasicrystals\cite{wan06,lin18} . For bipartite lattices, they appear at zero energy, which forms a massively degenerate manifold \cite{sutbip86}. If the Fermi energy is close to the LS manifold, interaction effects may become prominent as in flat band physics \cite{laf18}. Furthermore, these zero modes are robust with respect to perturbations; thus, their presence may be probed in less than ideal experimental conditions.

A state is strictly localized because it interferes destructively with itself at every part of the boundary of its domain. Such interference can be ensured in a bipartite tight-binding model if the two sublattices have different numbers of sites.
However, both the PL and the Ammann-Beenker lattice have an equal number of sites in their two sublattices but still have LS. A recent paper \cite{day20} argued that a local sublattice imbalance could be defined for the PL by separating the lattice into domains that favor one sublattice over the other. This sublattice imbalance was shown to account for the full LS frequency\cite{ara88,kts17} of $f_{LS}=81-50\tau$ where $\tau=(1+\sqrt{5})/2$ is the golden ratio. Furthermore, it was suggested that the robustness of the LS may be inherited from the five-dimensional cubic lattice, of which PL can be constructed by the cut-project method.

The nature of the LS in quasicrystals, particularly their connection to the sublattice imbalance and the cut-project parent lattice, needs to be understood better. To this end, we consider a family of pentagonal quasicrystals closely related to the PL \cite{bru81}. These quasicrystals are all projected from the five-dimensional cubic lattice, with the same acceptance window shape, and differ only by a single parameter $\Gamma$, which shifts the cut \cite{pav87,zob90,lin17} . Each value of $0\le \Gamma \le 1/2$ defines a unique local isomorphism (LI) class for the pentagonal quasicrystal, with $\Gamma=0$ giving the Penrose LI(PLI) class. All LI classes have the same basic rhombus tiles and support for their Fourier transform. However, one can find a local configuration that appears uniquely for each value of $\Gamma$. Hence there are regions in two quasicrystals with different values of $0\le\Gamma\le\frac{1}{2}$ that cannot be mapped to each other by translations or rotations. All these lattices are bipartite, but the two sublattices have a different number of sites for $\Gamma \ne 0$. The sublattice imbalance increases with $\Gamma$.

Previous numerical work has shown that LI classes other than the PLI support LS \cite{rsc95}, yet LS fraction has not been counted for LI classes. Most calculations of LS fraction rely on the inflation-deflation scaling symmetries, and there is no simple scaling symmetry for most general LI \cite{pav87}. As first shown by de Bruijn \cite{bru81b} the standard deflation substitution rule creates a lattice with a new intercept vector $\vec{\gamma}$. The new intercepts obey $\gamma'_n=\gamma_{n-1}+\gamma_n+\gamma_{n+1}$, thus the LI class parameter is multiplied by three $\Gamma'=3 \Gamma$. The PL with $\Gamma=0$ is mapped onto itself, while for an irrational value of $\Gamma$ infinitely many deflations are required to get back to the original LI class. There is no known substitution rule which provides a scaling symmetry for counting local environments. 

We have recently developed a method for calculating the LS fractions from perpendicular space images without deflation \cite{mok20,okt21}. In this work, we first calculate the LS fraction for LI classes by direct numerical calculation on finite lattices containing $\sim 10^5$ sites. We then identify LS types and count their frequency from their perpendicular space images. 

Both approaches show that the total LS fraction is non-monotonic with $\Gamma$. Although the sublattice imbalance increases with $\Gamma$, the LS fraction at first decreases from the PL value. After reaching a minimum value around $\Gamma\sim 0.1$, the LS fraction begins to increase and attains its highest value at $\Gamma=0.5$, which is the farthest point away from the PL. We calculate the local density of states (LDOS) at zero energy and observe that the domain structure where only one sublattice has LS in a given domain breaks down as soon as one moves away from the PLI.   The LDOS on the even lattice is concentrated around the high coordination number sites, while the LDOS on the odd lattice is much more uniformly distributed throughout the lattice. We identify 20 LS types for the even sublattice, and their frequency calculated from perpendicular space images matches closely with the numerical calculation. We identify 45 LS types for the odd sublattice, which can account for only $\sim90\%$ of the numerically observed frequency near $\Gamma=0.5$. Our results show that the domain structure and the LS types cannot be inherited from the topological properties of the parent five-dimensional lattice, and the relation between the local sublattice imbalance and LS presence is not straightforward \cite{day20}. It is also striking that other structural properties such as hyperuniformity, or restorability do not seem to have an effect on the LS frequency\cite{lin17}. 

In the next section, we summarize the basic properties of LI classes of pentagonal quasicrystals generated by the cut-project method. Numerical results on large lattices and LDOS calculations are given in section \ref{sec:Numerical}. We discuss the most prominent LS types in section \ref{sec:LS}, while the remaining LS types are given in the appendix. Our conclusions are given in section \ref{sec:Conclusion}.

\section{Cut and project definition of pentagonal quasicrystals}
\label{sec:ABL}

\begin{figure}[!htb]
    \centering
    \includegraphics[trim=8mm 8mm 8mm 8mm,clip,width=0.48\textwidth]{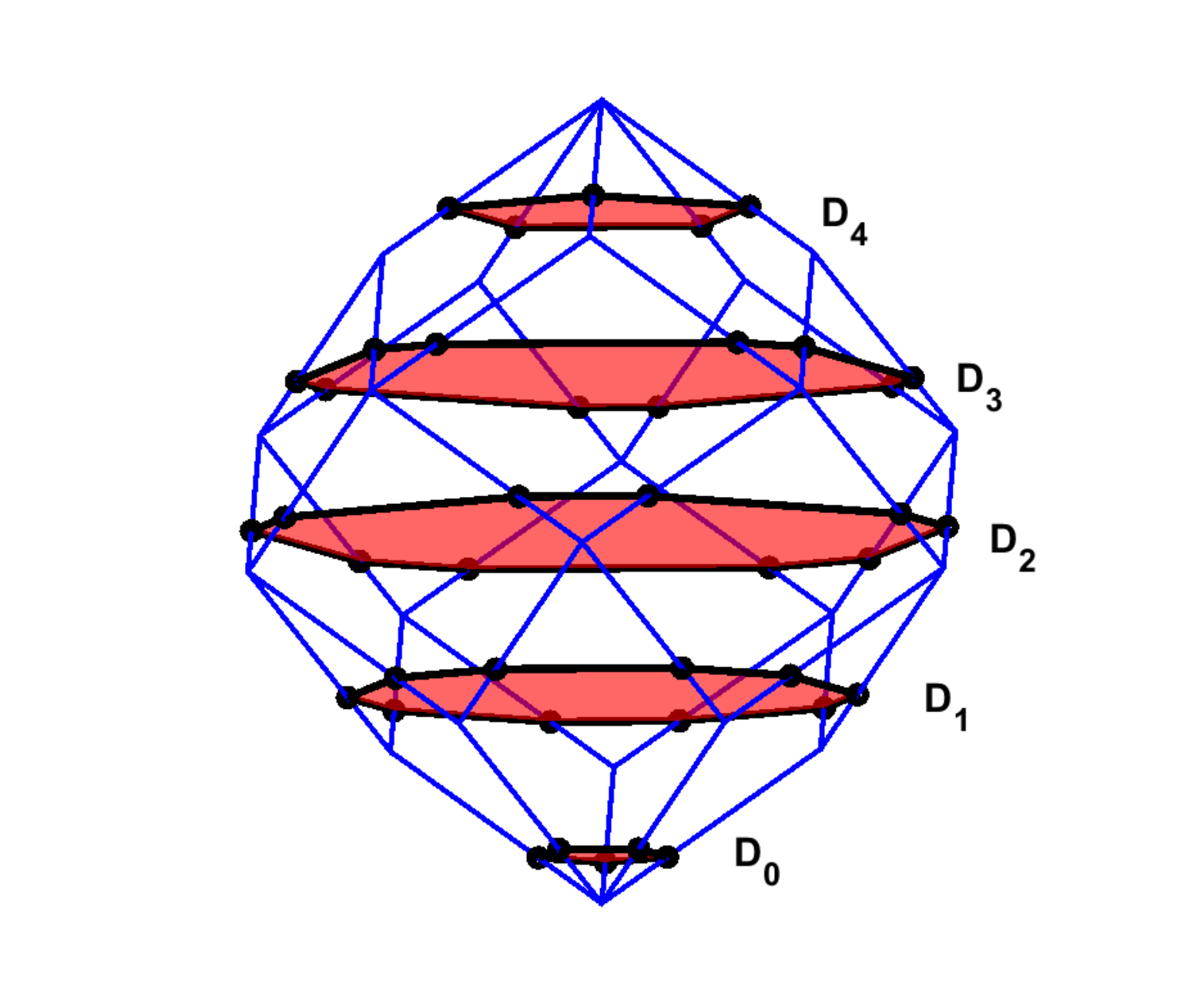}
    \caption{The acceptance window for projection in the space spanned by $\vec{c}_0,\vec{c_3},\vec{c}_4$ is a rhombic dodecahedron. The perpendicular space projections of the points forming the quasicrystal fill the five polygons $D_0,..., D_4$ densely and uniformly. The distance of $D_0$ from the lower tip, $\Gamma$, defines the LI class of the quasicrystal.}
    \label{fig:ThreeDCut}
\end{figure}

\begin{figure}[!htb]
    \includegraphics[clip,width=0.32\textwidth]{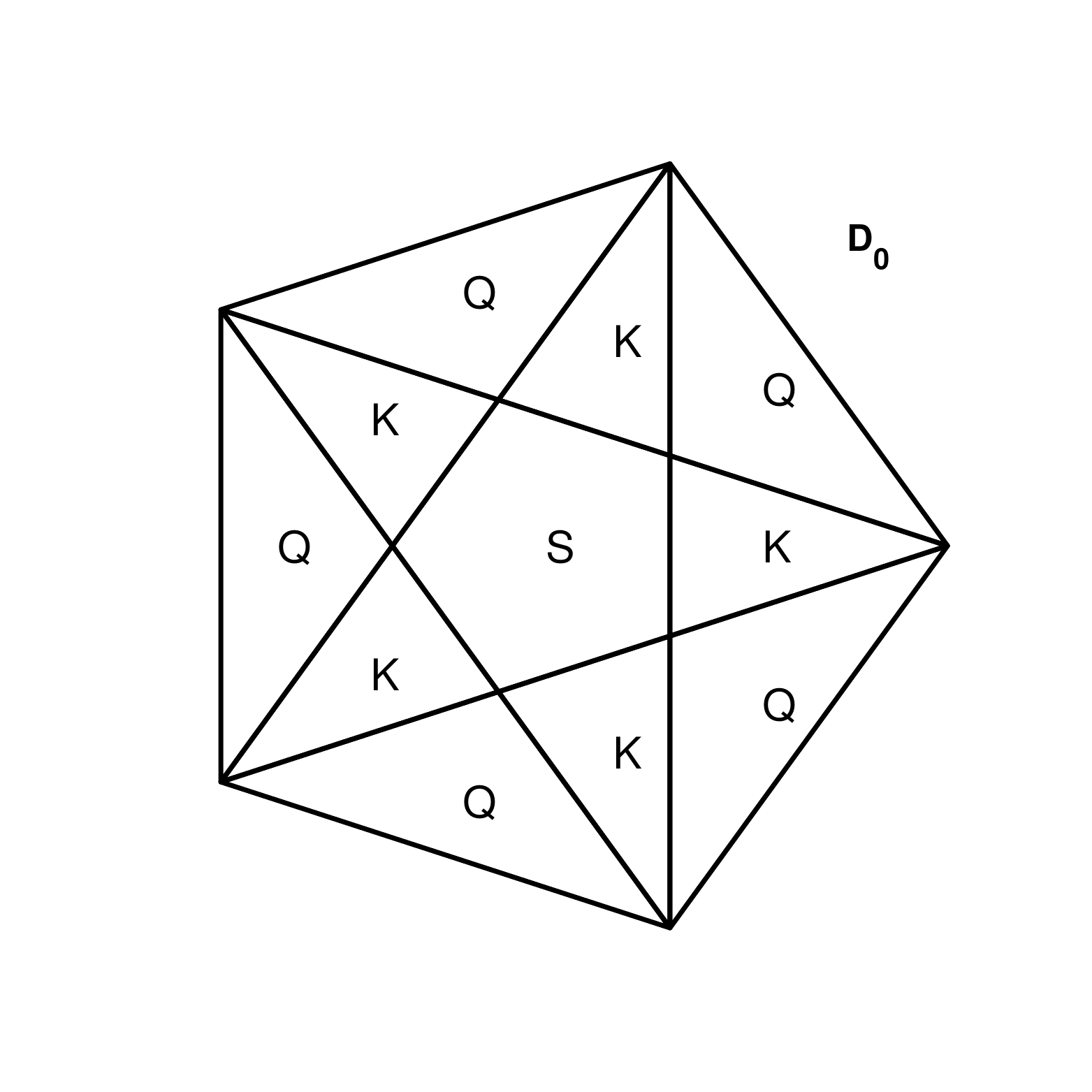} 
    \includegraphics[clip,width=0.32\textwidth]{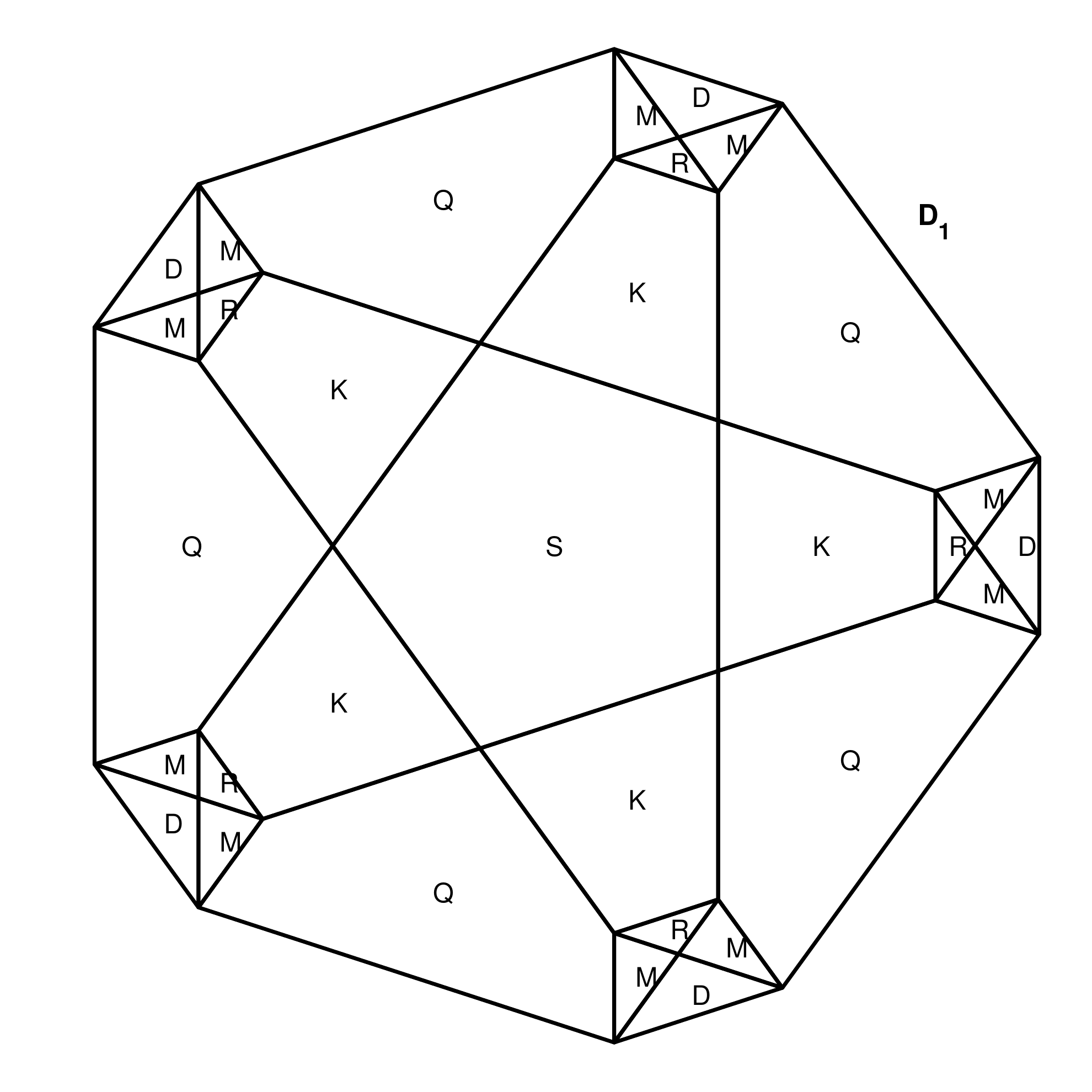} 
    \includegraphics[clip,width=0.32\textwidth]{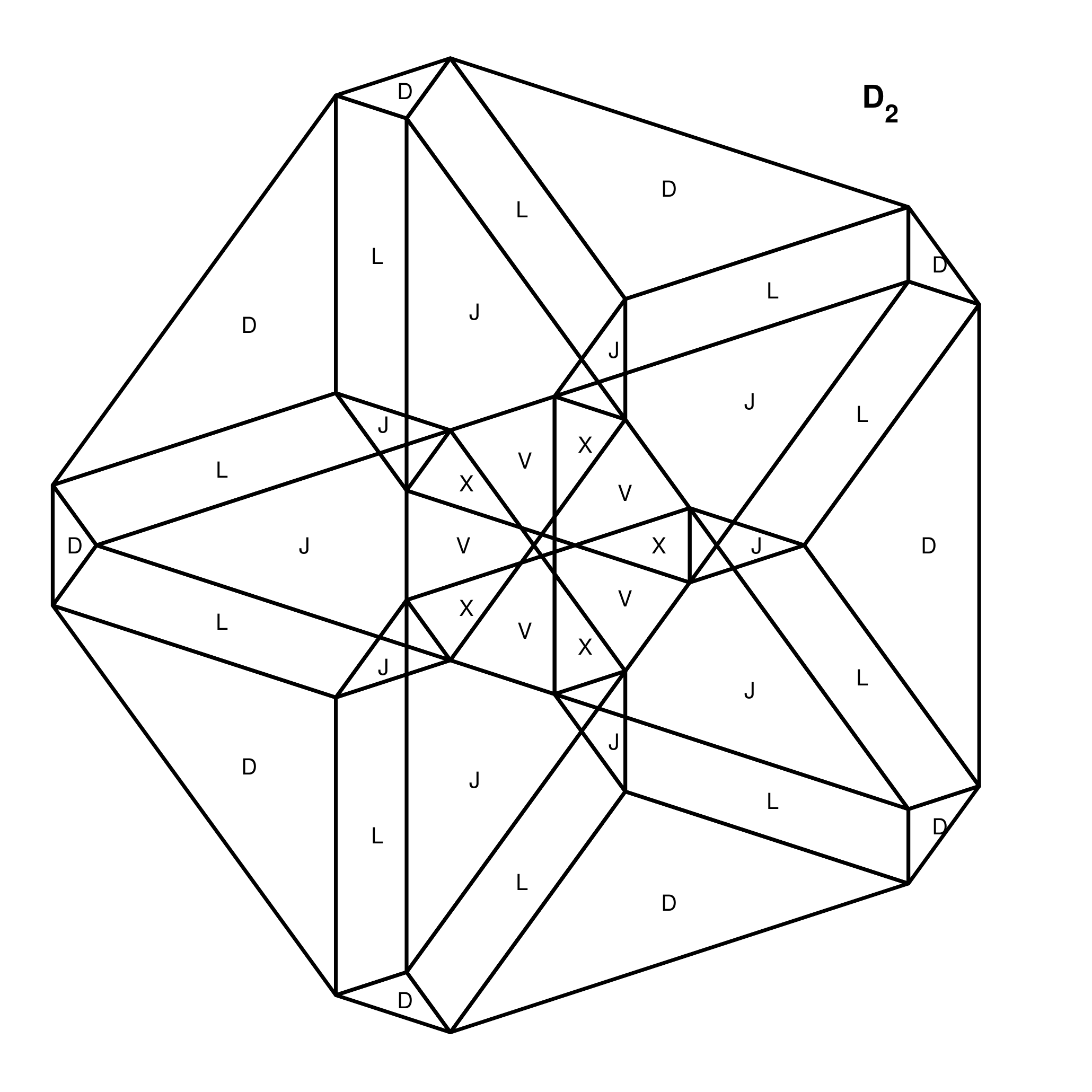} 
    \includegraphics[clip,width=0.32\textwidth]{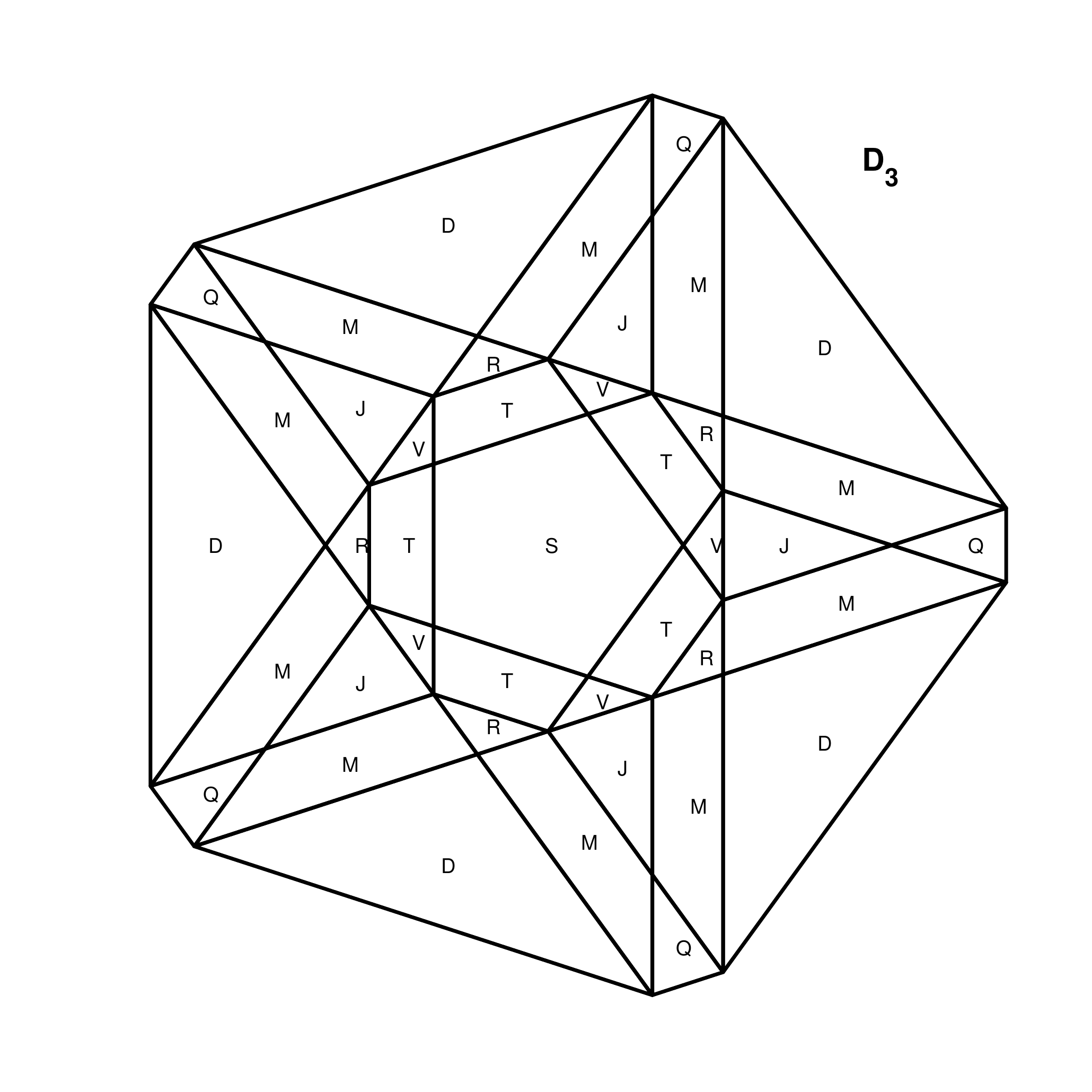}
    \includegraphics[clip,width=0.32\textwidth]{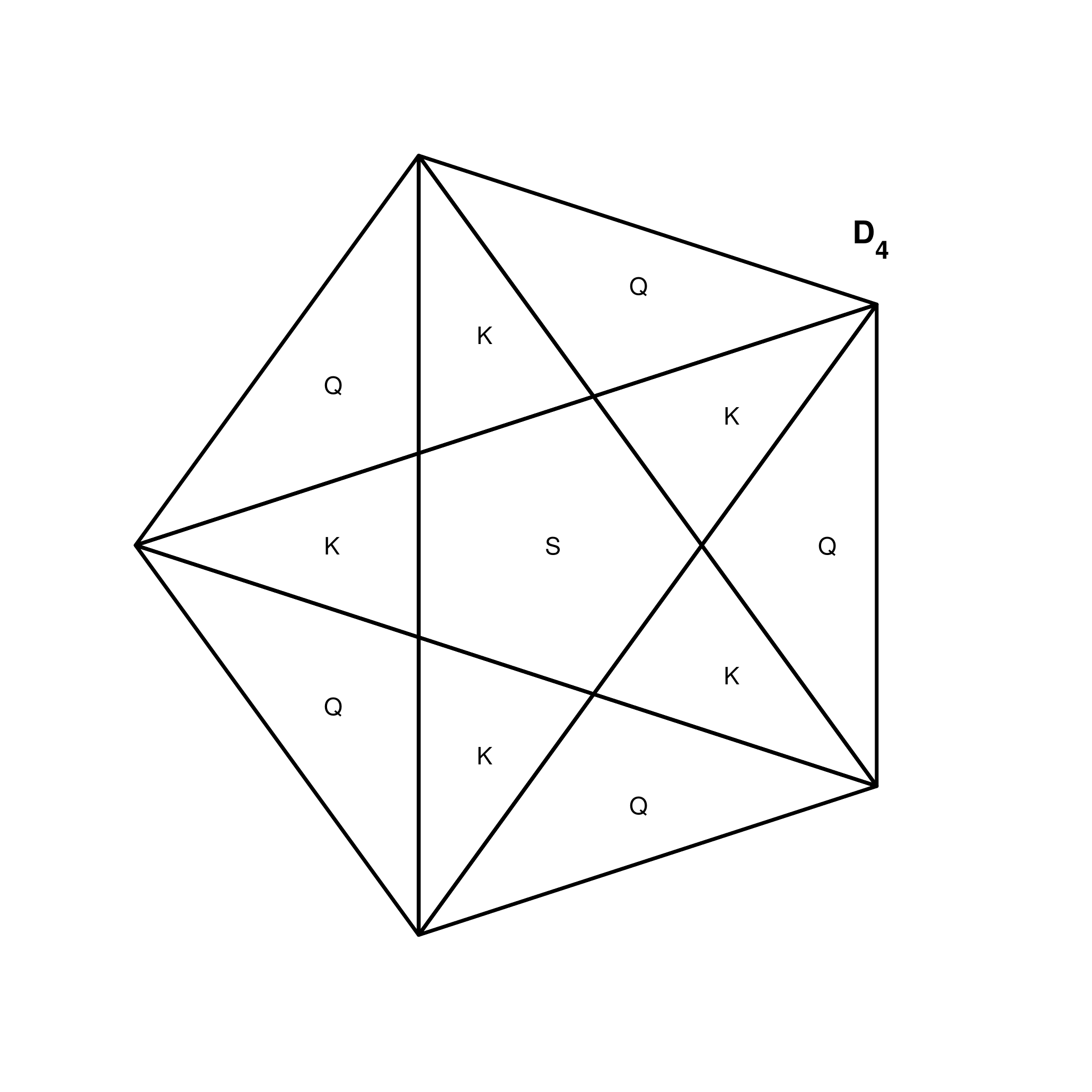} 
    
    \caption{   The five perpendicular space polygons $D_0,...,D_4$ at $\Gamma=0.2$. The points inside $D_0,D_2,$ and $D_4$ form the even sublattice, while $D_1$ and $D_3$ form the odd sublattice. The polygons are partitioned into regions which belong to a particular vertex type. An animation of the evolution of the five polygons and corresponding changes in the lattice can be found in the supplementary material.\cite{supplement}}
    \label{fig:PerpSpaceDecagons}
\end{figure}
\begin{figure}[!htb]
    \centering
    \includegraphics[clip,width=0.48\textwidth]{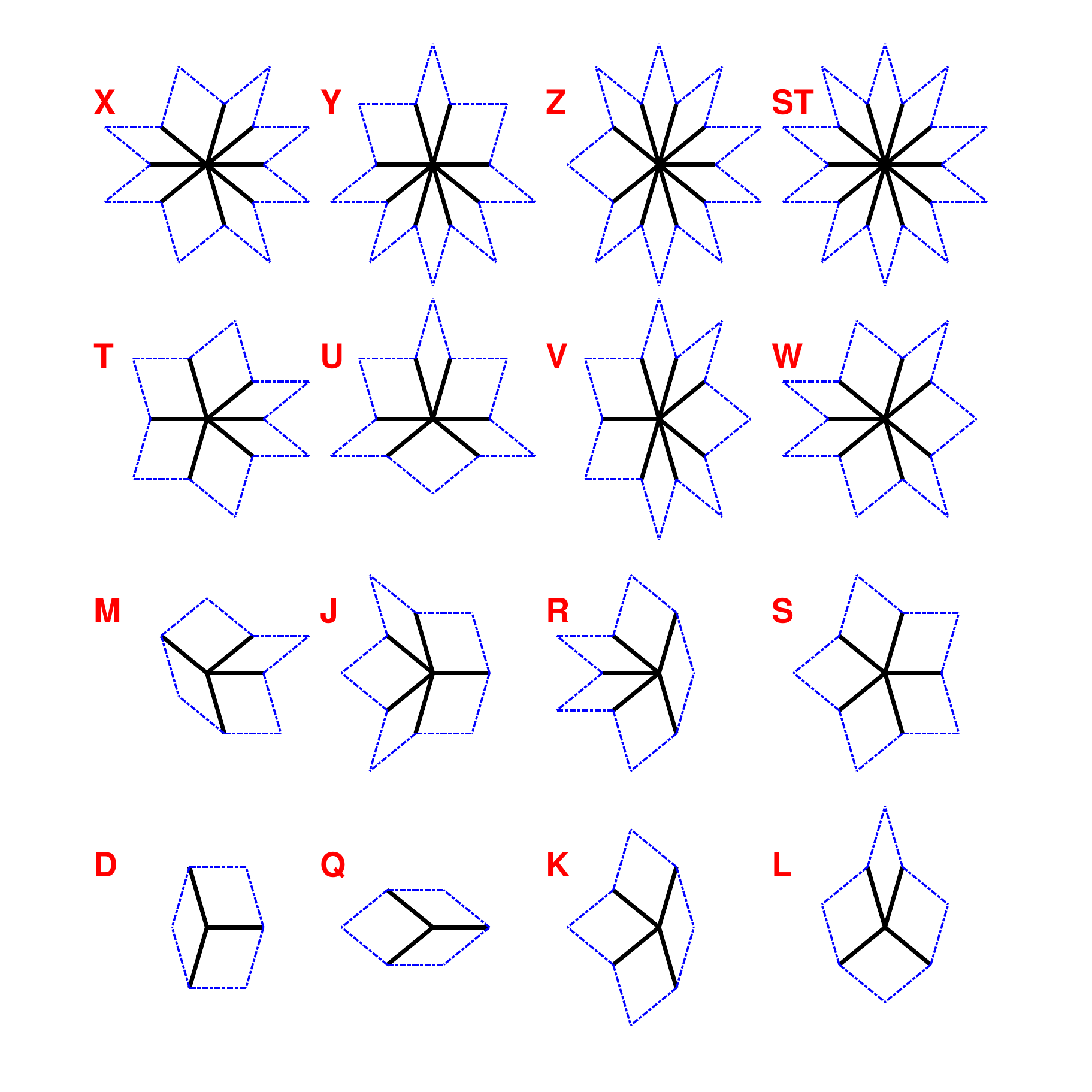}
    \caption{The sixteen possible vertex types for the LI classes considered here. Each vertex can have between three and ten neighbors. We follow the nomenclature of Ref.\cite{zob90}. }
    \label{fig:VertexTypes}
\end{figure}

\begin{figure}[!htb]
    \centering
    \includegraphics[clip,width=0.48\textwidth]{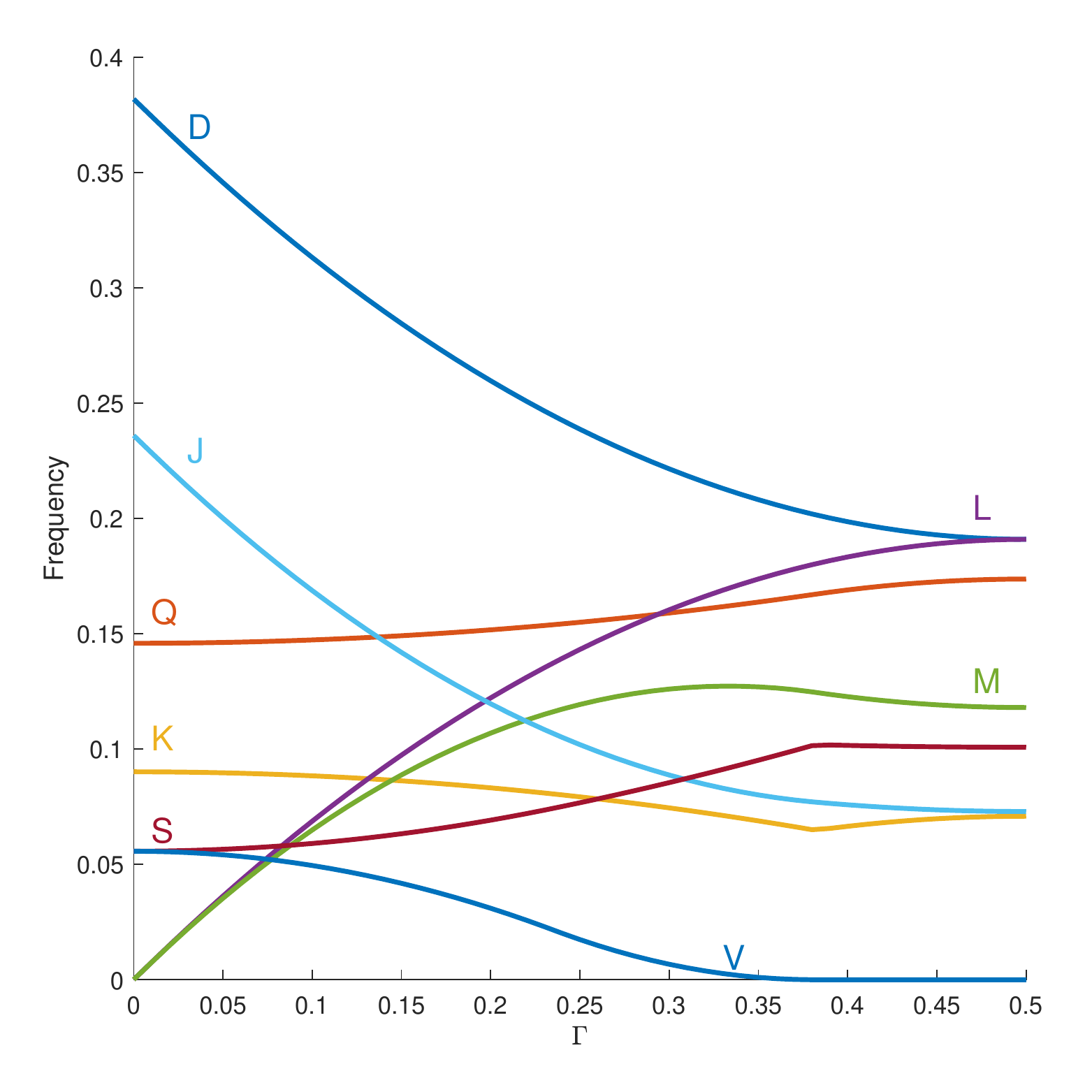}
    \includegraphics[clip,width=0.48\textwidth]{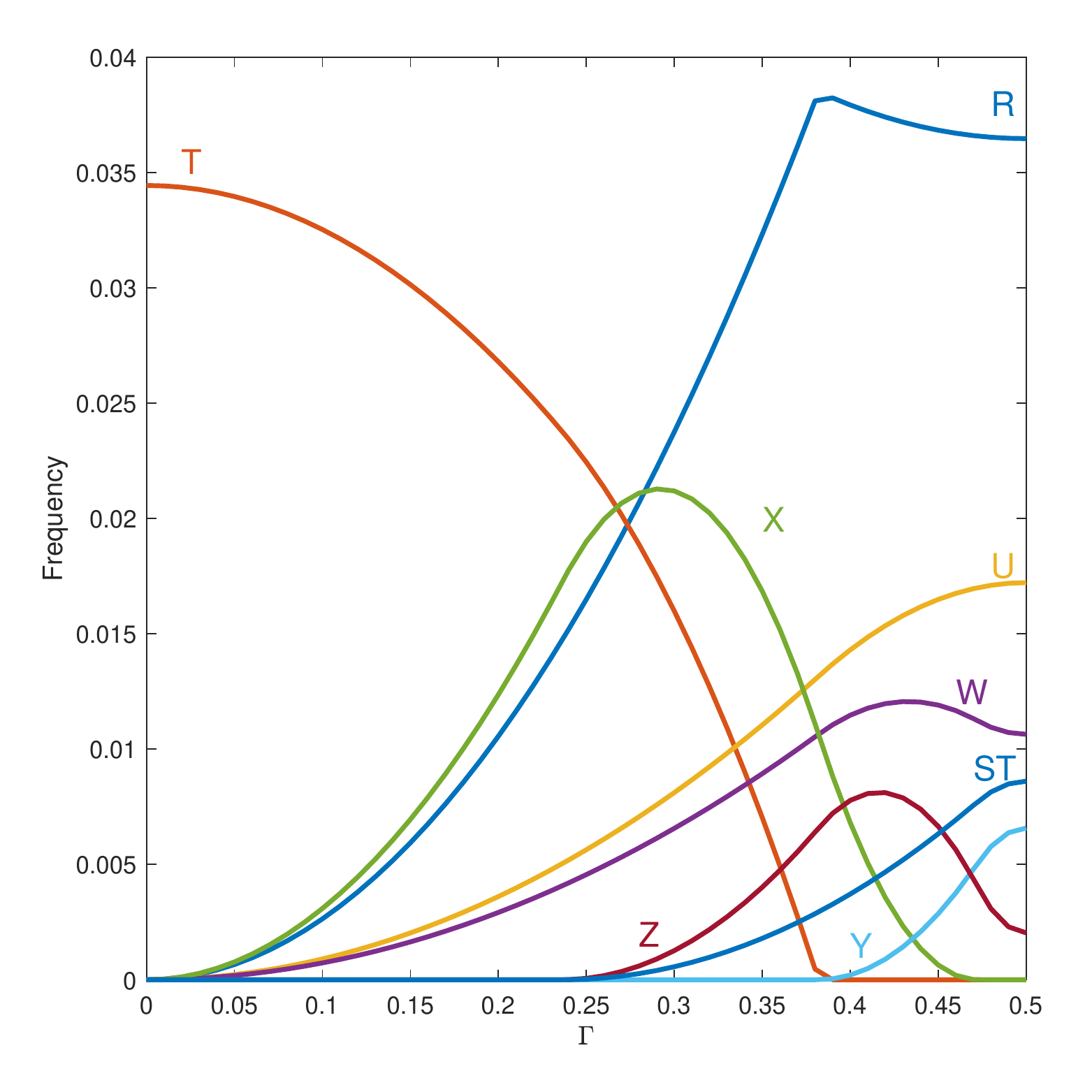}
    \caption{The frequencies of the vertex types identified in \ref{fig:VertexTypes} as a function of $\Gamma$. The figure is split into two to show vertex frequencies that remain below $5\%$ clearly. The results agree with \cite{zob90}. Out of the 16 vertex types in Fig.\ref{fig:VertexTypes}, seven (L,U,W,X,Z,ST,Y) appear only on the even sublattice and 2 (R,M) appear only on the odd sublattice. Seven vertex types (Q,K,D,J,V,T,S) appear on both. }
    \label{fig:VertexFrequencies}
\end{figure}

\begin{figure}[!htb]
    \centering
    \includegraphics[clip,width=0.48\textwidth]{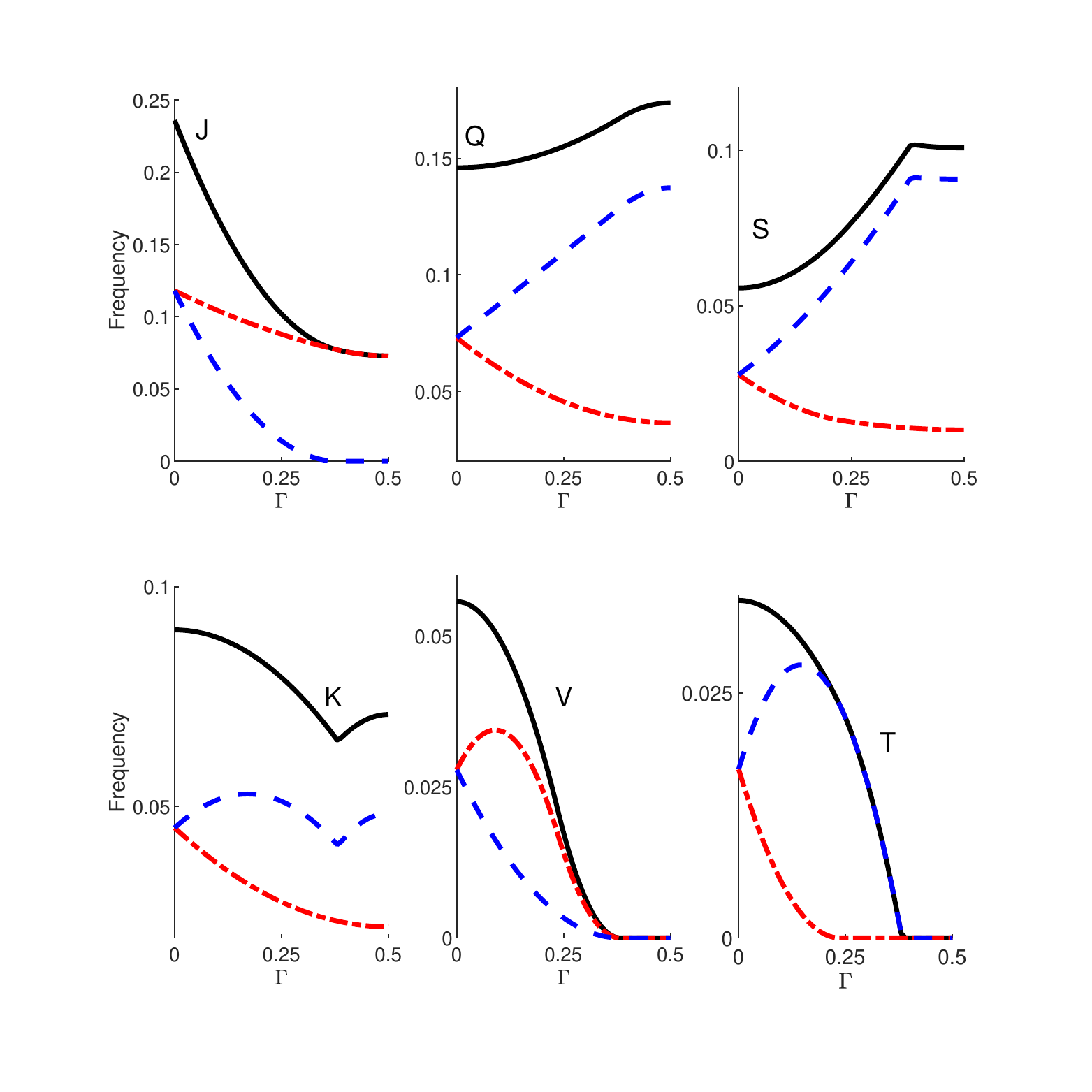}
    \includegraphics[clip,width=0.48\textwidth]{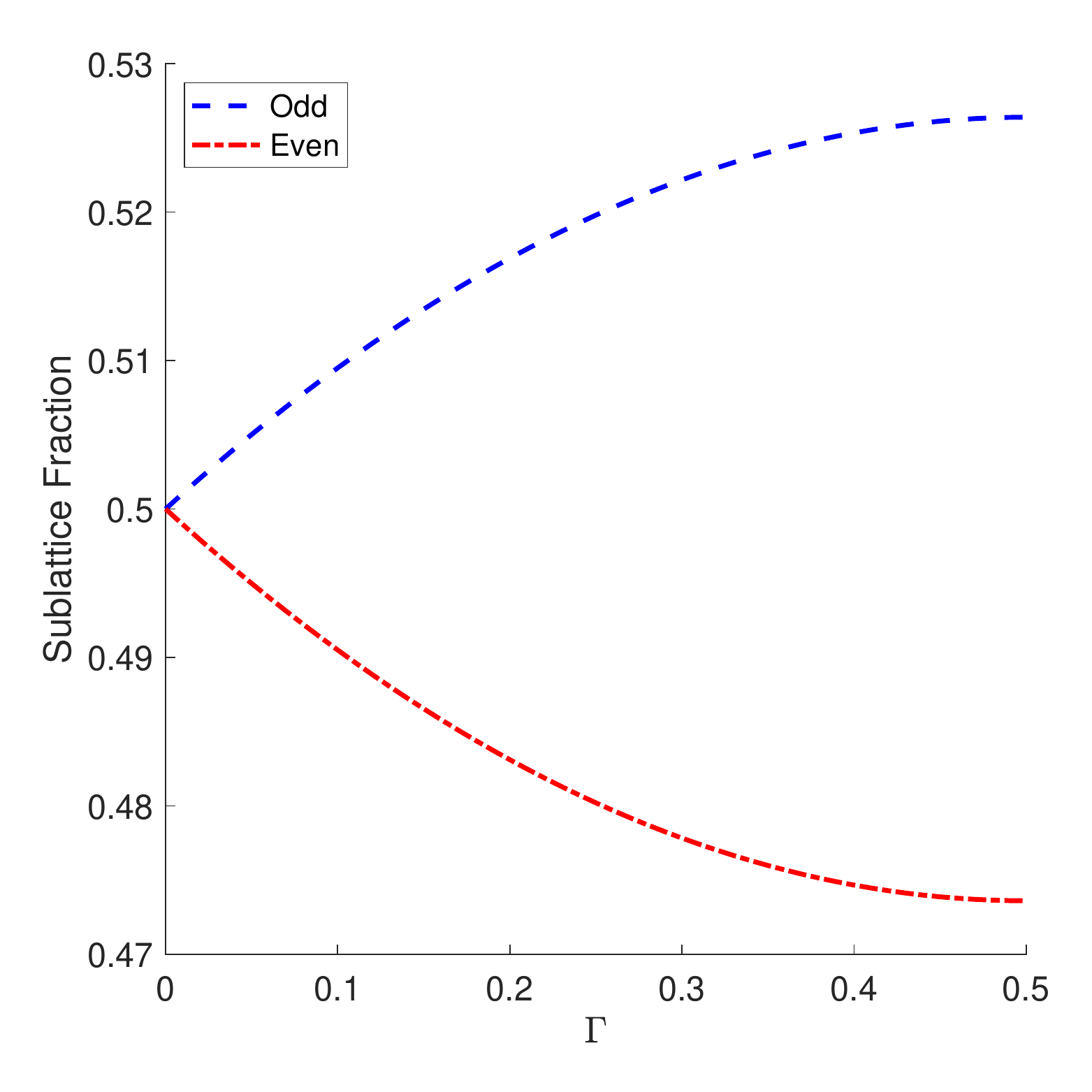}
    \caption{(a) Seven vertex types appear in both sublattices. While the fraction of D vertices is equal in both sublattices for any value of $\Gamma$, the other six are unevenly distributed as shown in the figure. Odd sublattice frequency is shown with blue dashed lines and even sublattice frequency is shown with red dash-dot line.
    (b) The fraction of vertices in the odd and even sublattices as a function of $\Gamma$. The imbalance between the sublattices grows monotonically with $\Gamma$. Even at $\Gamma=0.5$, the asymmetry remains small, with odd sublattice having less than $53\%$ of the total. }
    \label{fig:SublatticeImbalance}
\end{figure}

In this section, we give a projective definition of pentagonal quasicrystals, including the PL. Our notation and calculations follow Ref.\cite{bru81} and Ref.\cite{zob90}, but are presented here for completeness. We begin by considering the five-dimensional real space, spanned by the orthonormal set $\hat{u}_n$,  with $n=0,...,4$
\begin{equation}
    \vec{x}=\sum_n x_n \hat{u}_n.
\end{equation}
We partition this space into unit cubes 
\begin{equation}
\label{eq:OpenCube}
    k_m-1<x_m<k_m
\end{equation}
 where  $k_m$ are integers. Defining $\zeta=e^{i \frac{2 \pi}{5}}$ the following five vectors also form a basis for the five dimensional space
\begin{equation}
\label{eq:C vectors}
\vec{c}_0 =\sum_m \hat{u}_i,\quad
\vec{c}_1=\sum_m Re(\zeta^m) \hat{u}_m ,\quad
\vec{c}_2=\sum_m Im(\zeta^m) \hat{u}_m ,\quad
\vec{c}_3=\sum_m Re(\zeta^{2m}) \hat{u}_m, \quad
\vec{c}_4=\sum_m Im(\zeta^{2m}) \hat{u}_m.
\end{equation}
This set of vectors are orthogonal but not normalized, with $\hat{c}_0 \cdot \hat{c}_0=5$ and $\hat{c}_m \cdot \hat{c}_m=5/2$ for $m=1,2,3,4$.

The two-dimensional quasicrystal will be formed by first taking a slice through the five-dimensional cubic lattice and then projecting the points inside this slice. This projection window, and consequently the two-dimensional quasicrystal, is completely described by choosing a five-dimensional vector $\vec{\gamma}=\sum_m \gamma_m \hat{u}_m$. This intercept vector allows us to define a plane with the following three equations,
\begin{eqnarray}
\label{eq:Projection constraints}
(\vec{x}-\vec{\gamma})\cdot \vec{c}_0 &=& 0 \\ \nonumber
(\vec{x}-\vec{\gamma})\cdot \vec{c}_3 &=& 0 \\ \nonumber
(\vec{x}-\vec{\gamma})\cdot \vec{c}_4 &=& 0. 
\end{eqnarray}

A five-dimensional cubic lattice $\vec{R}_5=\sum_m k_m \hat{u}_m$ vertex is projected into two dimensions only if there is a point in its open unit cube defined in Eq.\ref{eq:OpenCube} which satisfies Eq.(\ref{eq:Projection constraints}). Intercept vector $\vec{\gamma}$ is chosen such that the projected lattice is not singular \cite{bru81}.  

Any five-dimensional point lying on the plane defined by Eq.(\ref{eq:Projection constraints}) can be expressed as
\begin{equation}
    (\vec{x}-\vec{\gamma})= \alpha_1 \vec{c}_1 + \alpha_2 \vec{c}_2,
\end{equation}
where $\alpha_1,\alpha_2$ are arbitrary reals. One can check if all five conditions in Eq.\ref{eq:OpenCube} are satisfied to see if a point in the five-dimensional cubic lattice should be projected to form a vertex of the quasicrystal. It is also instructive to ask what geometric constraints  Eq.\ref{eq:OpenCube} defines for the three dimensions which are orthogonal to the projection plane. This acceptance window in the perpendicular space forms a rhombic icosahedron, as shown in Fig.\ref{fig:ThreeDCut}. 
However, the perpendicular space coordinates of all the five-dimensional cubic lattice points in the cut do not fill up the three-dimensional volume of this shape. As $\sum k_m$ is an integer, these points are localized on five planes perpendicular to the $\vec{c}_0$ direction. Thus, the perpendicular space can be thought of five polygons, which we refer to as $D_0,..D_4$. These polygons have either five or ten sides. 
The shape of a perpendicular space polygon is determined by the projection of  $\vec{\gamma}$ onto $\vec{c}_0$, and the index $\sum k_m$, and can be obtained by projecting $\left(\vec{x}-\vec{\gamma}\right)$ on to $\vec{c}_3,\vec{c}_4$. 

The parameter $\Gamma=\vec{c}_0\cdot\vec{\gamma}=\sum_n \gamma_n$ controls not only the shape of the perpendicular space polygons, but it defines the LI class of the quasicrystal. If two quasicrystals which have unequal intercept vectors $\vec{\gamma}$ and $\vec{\gamma}'$ they are different in real space. However, if $\Gamma=\sum_n \gamma_n$ and $\Gamma'=\sum_n \gamma'_n$ are equal, they are locally isomorphic,i.e., any finite size section of the first crystal can be found in the second crystal. As each component of the intercept vector, $\gamma_n$, is defined up to an integer, unique LI classes are confined to $0\le \Gamma \le 1$. Furthermore, inversion maps $\Gamma$ to $1-\Gamma$, thus unique LI classes are  obtained only in the interval $0\le \Gamma \le \frac{1}{2}$.  

As $\Gamma$ is varied between zero and $1/2$, the polygons $D_0$ to $D_4$ move up through the acceptance window, with $D_0$ exactly $\Gamma$ away from the lower tip of the icosahedron. Because of the mapping between $\Gamma$ and $1-\Gamma$ it is possible say the polygons are moving down on the $\vec{c}_0$ axis, but this choice is equivalent to introducing a $D_5$ polygon at the top and relabeling the polygons. At $\Gamma=0$, $D_0$ is reduced to a point and $D_1$ to $D_4$ become pentagons. This is exactly the perpendicular space structure of the PL. Thus $\Gamma=0$ corresponds to the PLI, and as $\Gamma$ increases, we can say that the lattice becomes less similar to the PL.

At any value of $\Gamma$, the lattice is formed by the same tiles, the thin and the thick rhombuses. Consequently, all the lattices we consider are bipartite. This bipartite structure is reflected in the perpendicular space as well. A point with perpendicular space image in the decagon $D_m$ can only have neighbors in the polygons $D_{m-1}$ or $D_{m+1}$. We refer to the points in $D_0,D_2,D_4$ as the even sublattice, and $D_1, D_3$ as the odd sublattice. For the PLI $D_0$ is reduced to a point, furthermore inversion maps $D_1$ to $D_4$ and $D_2$ to $D_3$. Thus, there is complete symmetry between the odd and even sublattices, the fraction of vertices in one sublattice is exactly $50\%$. However, this symmetry is broken as soon as $\Gamma\ne 0$, and there are more vertices in the odd sublattice. The sublattice fractions are shown in Fig.\ref{fig:SublatticeImbalance}. The imbalance between the sublattices increases monotonically with $\Gamma$, reaching the maximum value at $\Gamma=0.5$ where  $52.6\%$ of the vertices are in the odd sublattice. 

The definitions of the set in Eq.\ref{eq:C vectors} show that any point in the quasicrystal can have neighbors only in one of the ten directions. We can define the star vectors in real space as $\hat{e}_m = Re(\zeta^m) \hat{i} + Im(\zeta^m) \hat{j}$ and in perpendicular space as  $\hat{\tilde{e}}_m = Re(\zeta^{2m}) \hat{i}_\perp + Im(\zeta^{2m}) \hat{j}_\perp$ . A point with perpendicular space position $\vec{r}_\perp$ in the decagon $D_n$ has a neighbor in the $\hat{e}_m$ direction in real space only if $\vec{r}_\perp+\hat{\tilde{e}}_m$ is in the decagon $D_{n+1}$. Similarly, a neighbor in the $-\hat{e}_m$ direction is possible only if $\vec{r}_\perp-\hat{\tilde{e}}_m$ is in $D_{n-1}$. This simple construction allows one to obtain the local structure in real space in terms of the perpendicular space position. 

We can classify all vertices based on their nearest neighbor configurations. Any vertex can have between 3 and 10 edges connecting them to nearest neighbors. In total, there are 16 configurations (up to rotations) for the nearest neighbors, as shown in Fig.\ref{fig:VertexTypes}. Only seven of these appear in the PL. The polygons $D_n$ in perpendicular space can be split into regions by superimposing five shifted copies of $D_{n-1}$ and five shifted copies of $D_{n+1}$. Each one of these regions would contain the perpendicular space images of vertices with the same nearest neighbor configuration. The partition of the polygons for $\Gamma=0.2$ is given in Fig.\ref{fig:PerpSpaceDecagons}. Furthermore, the mapping from five dimensions to the perpendicular space is linear, and the image is dense within the polygons. Thus the area of each region is proportional to the frequency of that local configuration appearing in the infinite lattice. Using the areas of the perpendicular space regions, we calculated the frequency of each vertex type, as shown in Fig.\ref{fig:VertexFrequencies}. 

As the above construction provides all the structural information necessary for the construction of the lattice we can now define the vertex tight binding model. We consider a single Wannier state at each vertex $|\vec{R}\rangle$ and uniform tunneling amplitude over each bond,
\begin{equation}
\label{eq: Hamiltonian}
    {\cal H}=-\sum_{<ij>} |\vec{R}_i\rangle \langle \vec{R}_j |.
\end{equation}
The eigenstates of this Hamiltonian can be extended, critical, or localized. In this paper, we are considering a particular subset of the localized eigenstates, which have zero amplitude beyond a finite region of the lattice. Such strictly localized states  appear at zero energy for bipartite lattices. Furthermore, by using the bipartite property, we can split the LS manifold into two parts, LS on the even and odd sublattices. Consider a LS that has components on both sublattices
\begin{equation}
    |\Psi_{1}\rangle=\sum_{Odd} \psi_{\vec{R}_j} |\vec{R}_j\rangle + \sum_{Even} \psi_{\vec{R}_j} |\vec{R}_j\rangle 
\end{equation} which satisfies ${\cal H} |\Psi_1\rangle=0$. For a bipartite lattice, regardless of the coupling between the sublattices or the symmetry between the two sublattices the following wavefunction
\begin{equation}
    |\Psi_{2}\rangle=\sum_{Odd} \psi_{\vec{R}_j} |\vec{R}_j\rangle - \sum_{Even} \psi_{\vec{R}_j} |\vec{R}_j\rangle 
\end{equation} also satisfies ${\cal H} |\Psi_2\rangle=0$. The sum and difference of these two states will form the LS which remain in only one sublattice.
For the PLI, the two sublattices can be mapped to each other by inversion, so the LS properties are the same for both sublattices. Even when the symmetry between the two sublattices is broken the lattice remains bipartite so LS can be chosen to lie in only one sublattice. However the behavior of LS in the two sublattices will be markedly different. In the next section, we describe our numerical method for counting the LS fraction in both sublattices for LI classes beyond the PL.

\section{Localized state fraction and zero energy local density of states}
\label{sec:Numerical}

\begin{figure}[!htb]
    \centering
    \includegraphics[clip,width=0.48\textwidth]{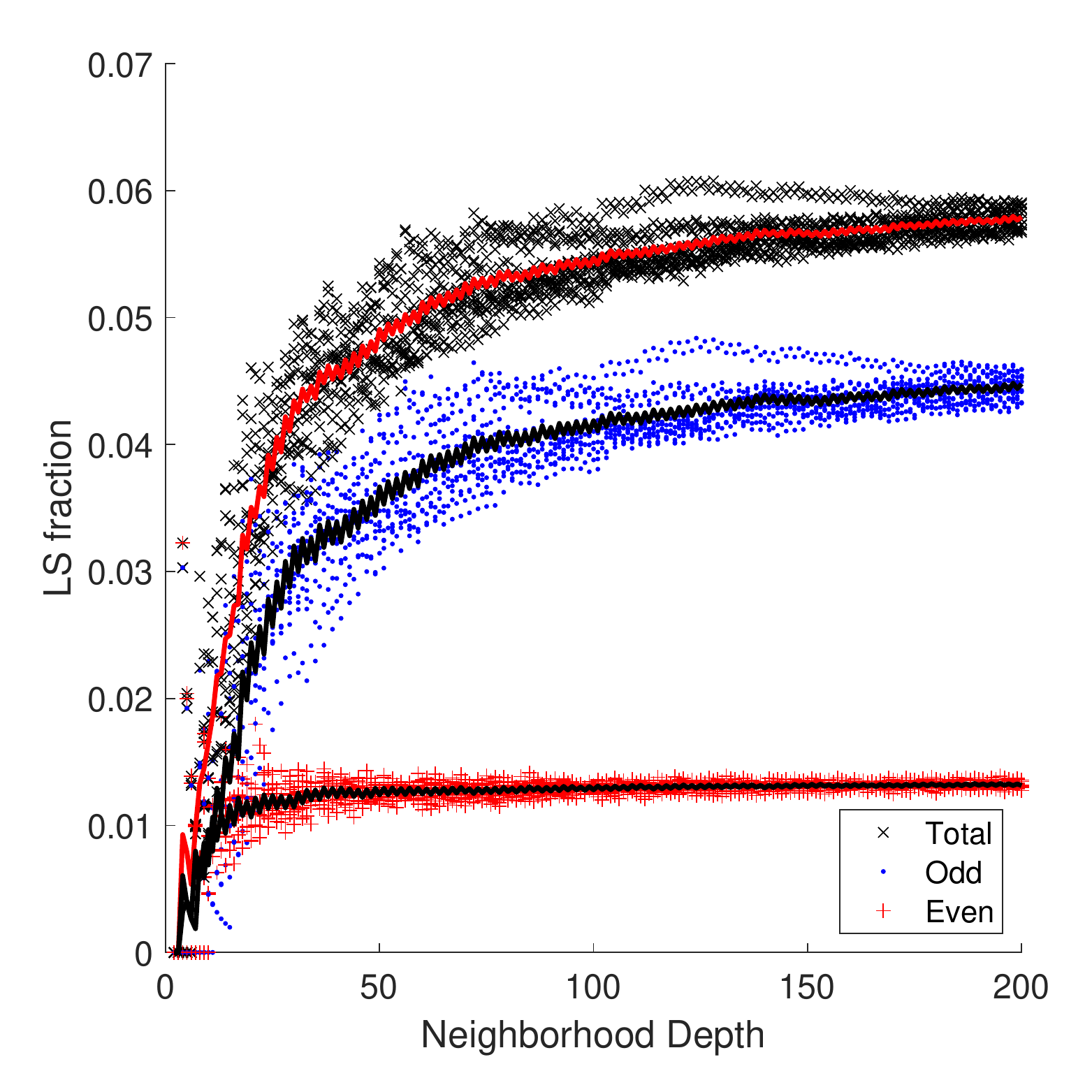}
    \caption{The numerically calculated LS fractions at $\Gamma=0.2$, as a function of neighborhood depth. Data for ten different initial points are shown with individual markers, while their average is indicated with a different color. While the LS fraction depends on the choice of the initial point for small neighborhoods, results converge for large lattices. The odd sublattice has more significant variation compared to the even sublattice. The largest lattices at depth 200 have close to 100,000 sites. }
    \label{fig:NumericalTestPoints}
\end{figure}
\begin{figure}[!htb]
    \centering
    \includegraphics[clip,width=0.48\textwidth]{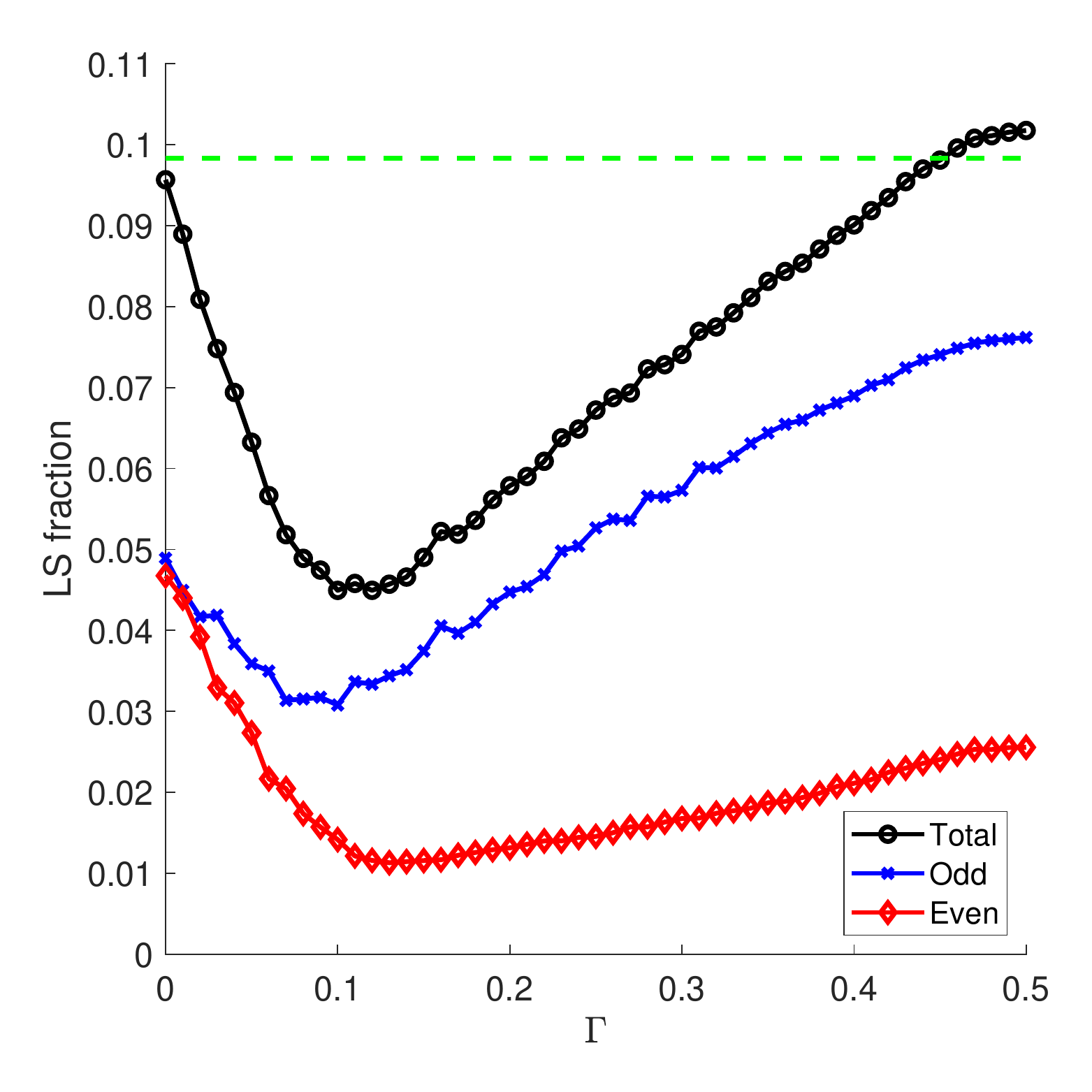}
    \caption{Numerically obtained LS fractions as a function of $\Gamma$. The fraction on the odd and even sublattices and the total fraction are displayed. The result at $\Gamma=0$ agrees with the analytical expression for the PL $81-50\tau$ shown by the dotted line. The highest LS fraction is at $\Gamma=0.5$, where more than $10\%$ of the states are LS. The odd sublattice has more LS than the even sublattice for all $\Gamma$ except $\Gamma=0$. }
    \label{fig:NumericalLSFractions}
\end{figure}

\begin{figure}[!htb]
    \centering
    \includegraphics[clip,width=0.34\textwidth]{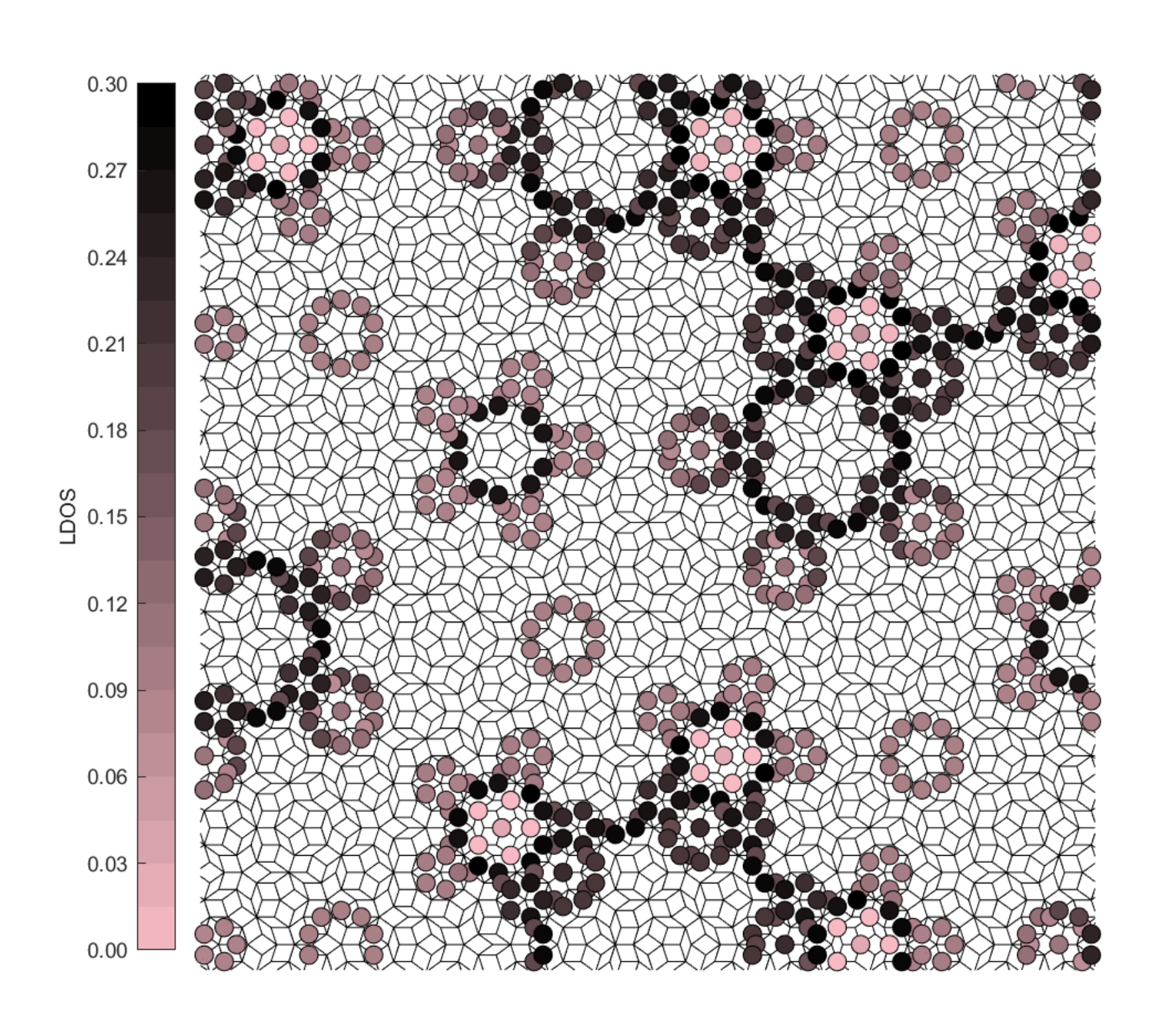}
    \includegraphics[clip,width=0.31\textwidth]{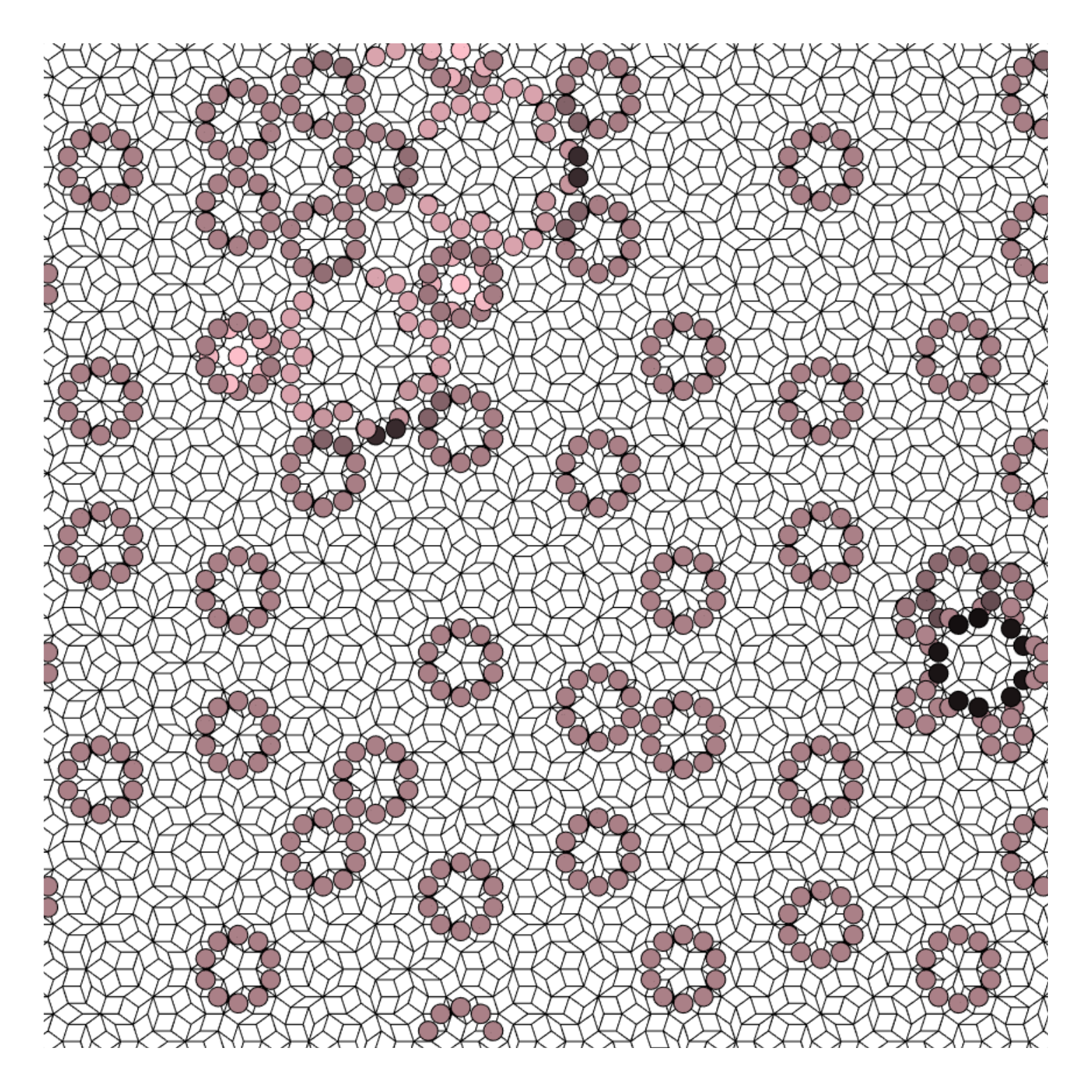}
    \includegraphics[clip,width=0.31\textwidth]{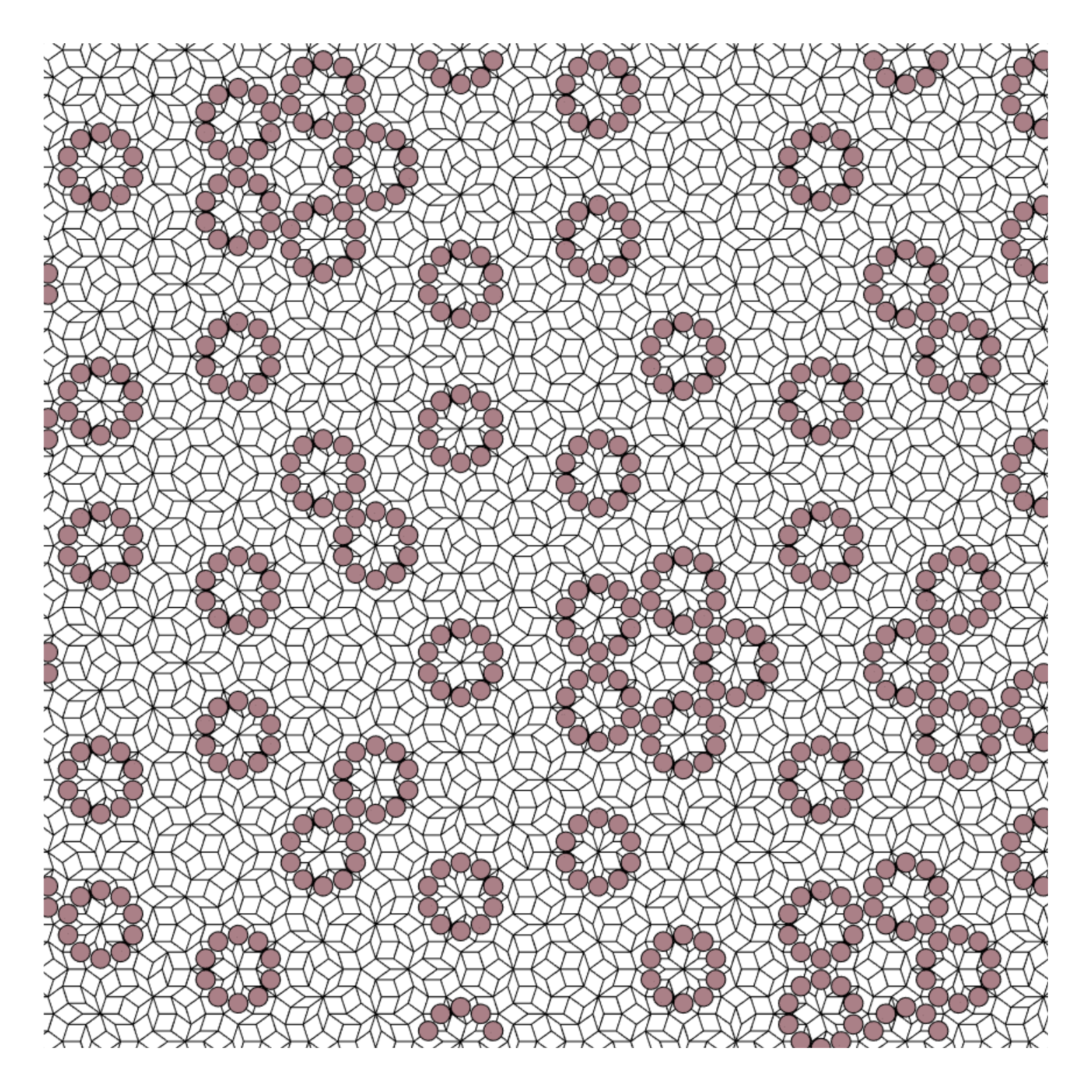}
    \caption{LDOS for the even sublattice for three $\Gamma$ values, $\Gamma=0,0.1,0.2$. For the PLI $\Gamma=0$, one can notice the forbidden regions and "strings" separating them. The LS density quickly falls as $\Gamma$ increases, and the LDOS forms isolated regions centered around highly connected vertices. An animation of the evolution of the LDOS is given in the supplementary material.}
    \label{fig:EvenLDOS01}
\end{figure}

\begin{figure}[!htb]
    \centering
    \includegraphics[clip,width=0.31\textwidth]{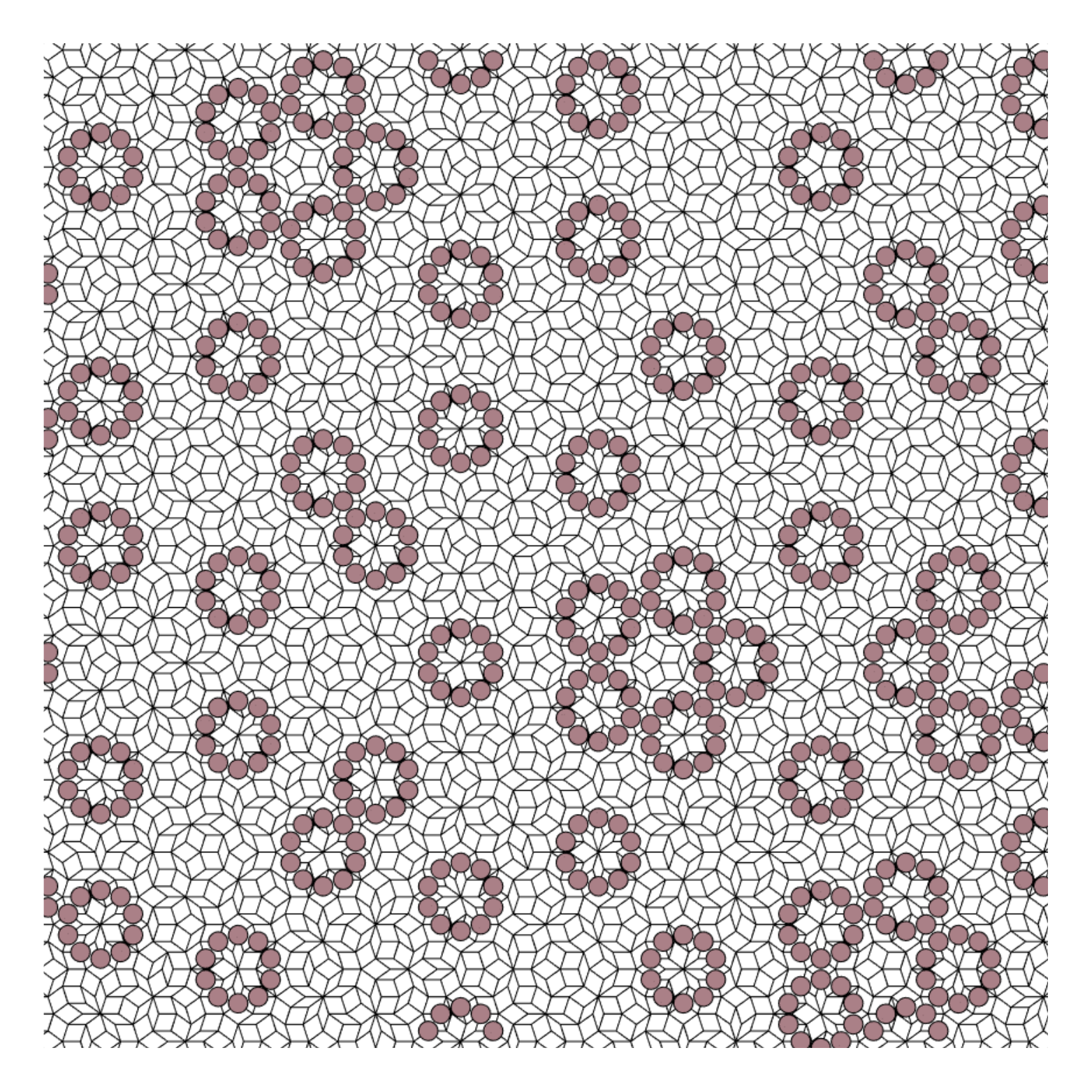}
    \includegraphics[clip,width=0.31\textwidth]{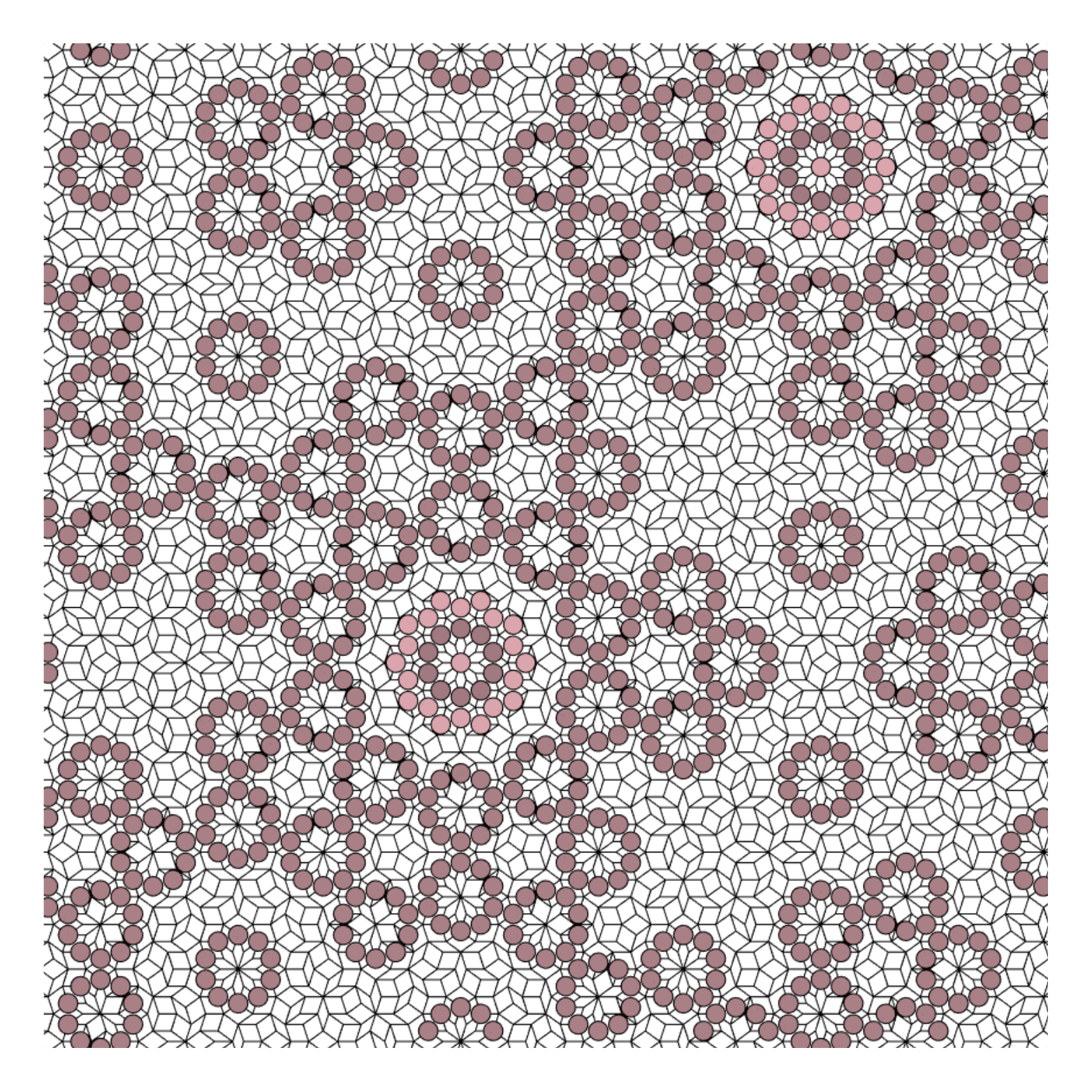}
    \includegraphics[clip,width=0.31\textwidth]{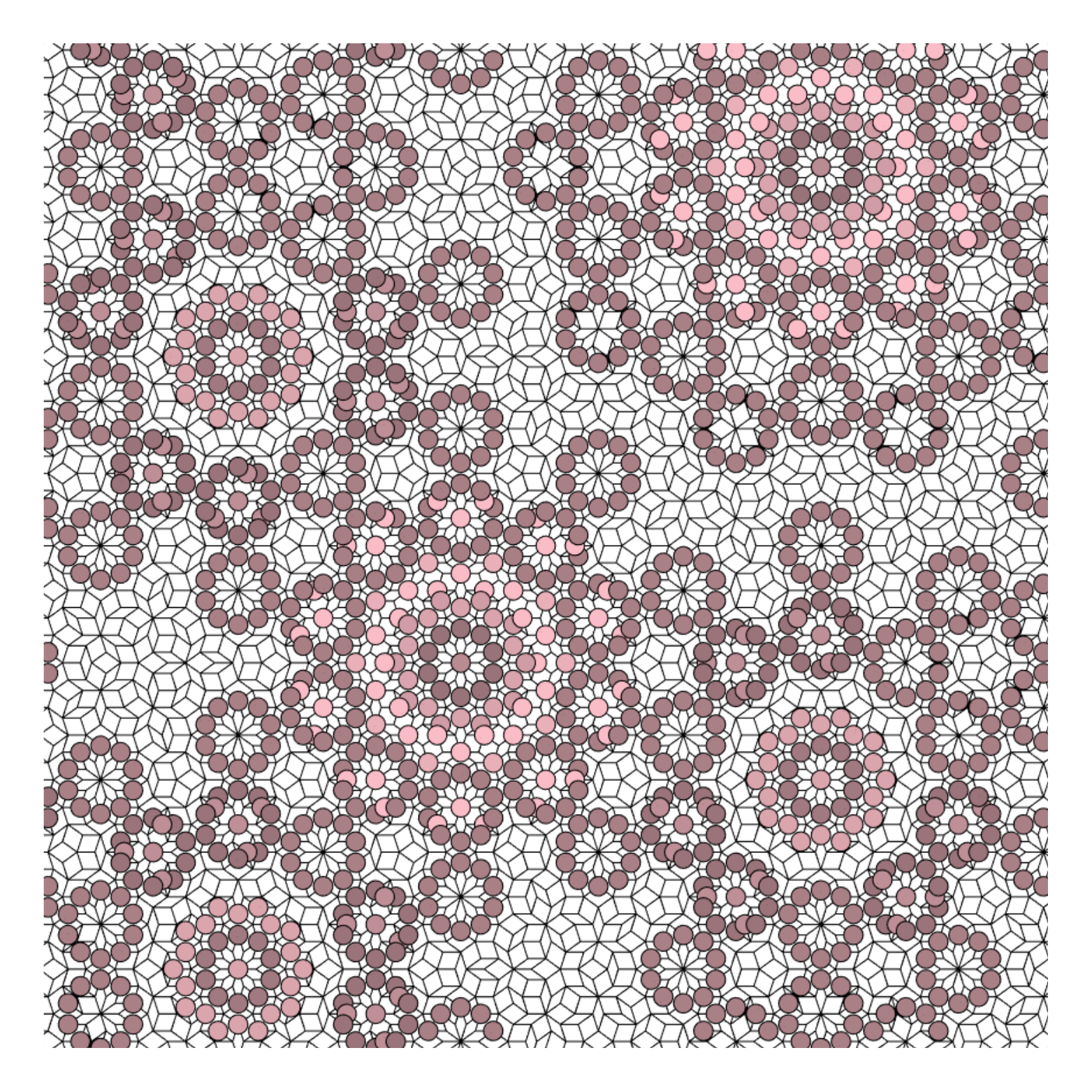}
    \caption{LDOS in the even sublattice for $\Gamma=0.3,0.4,0.5$. LS fraction increases; however, LDOS mostly stays in isolated regions with high symmetry.  }
    \label{fig:EvenLDOS02}
\end{figure}

\begin{figure}[!htb]
    \centering
    \includegraphics[clip,width=0.47\textwidth]{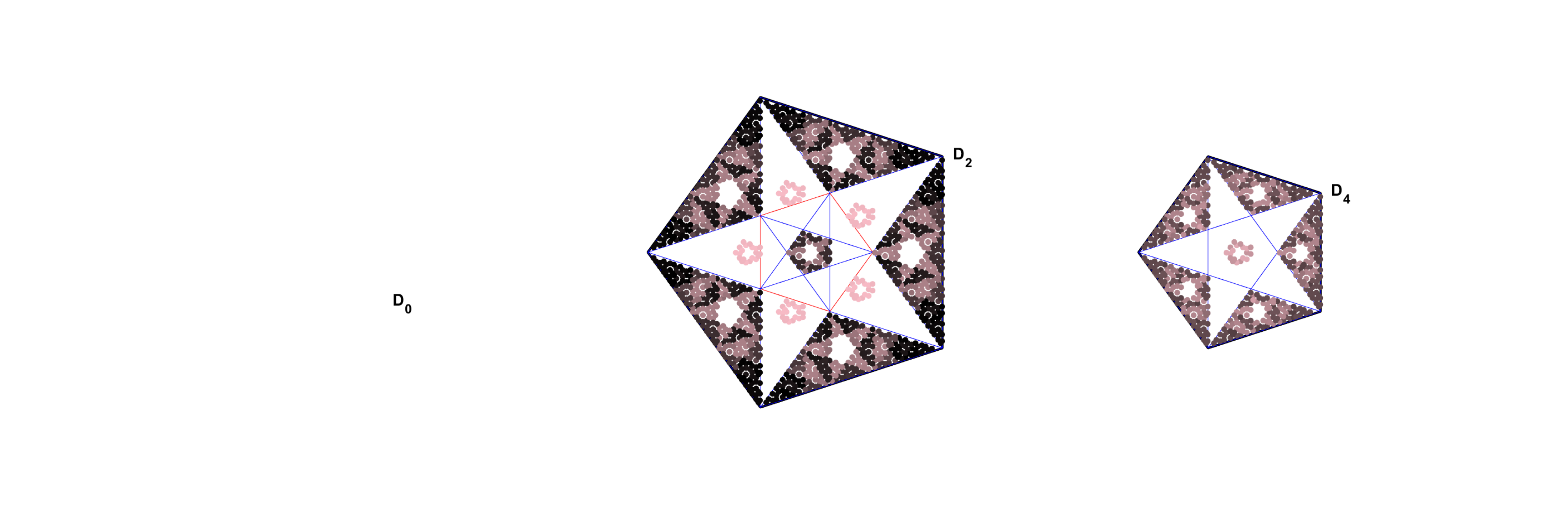}
    \includegraphics[clip,width=0.47\textwidth]{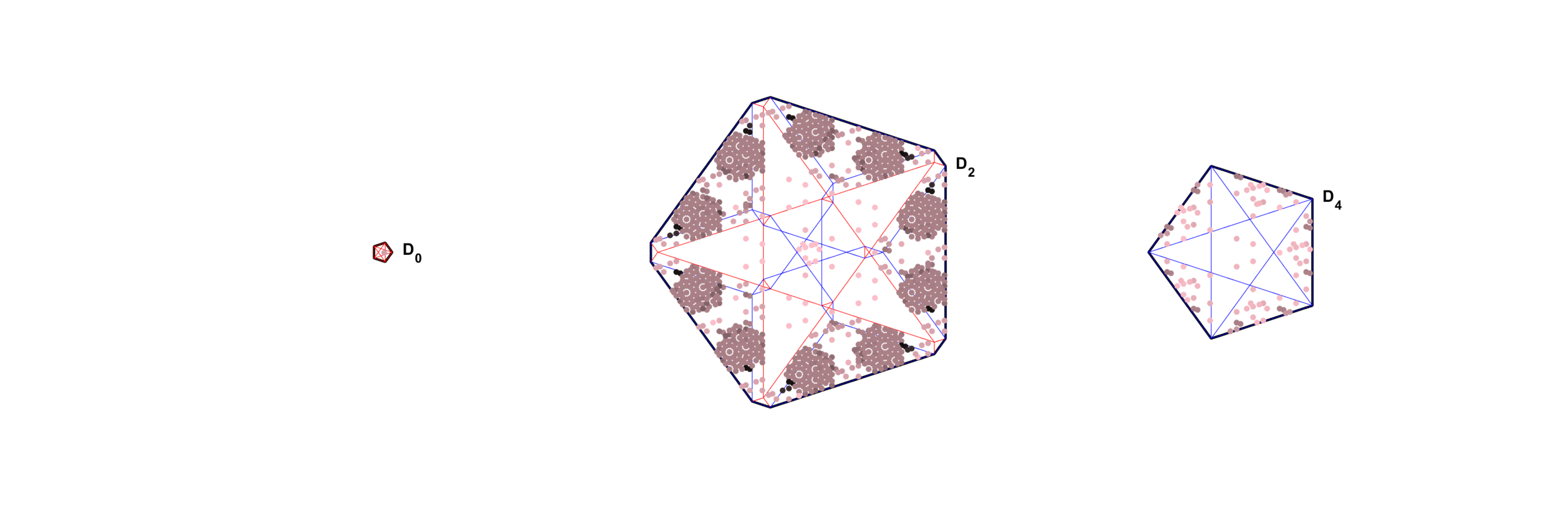}
    \includegraphics[clip,width=0.47\textwidth]{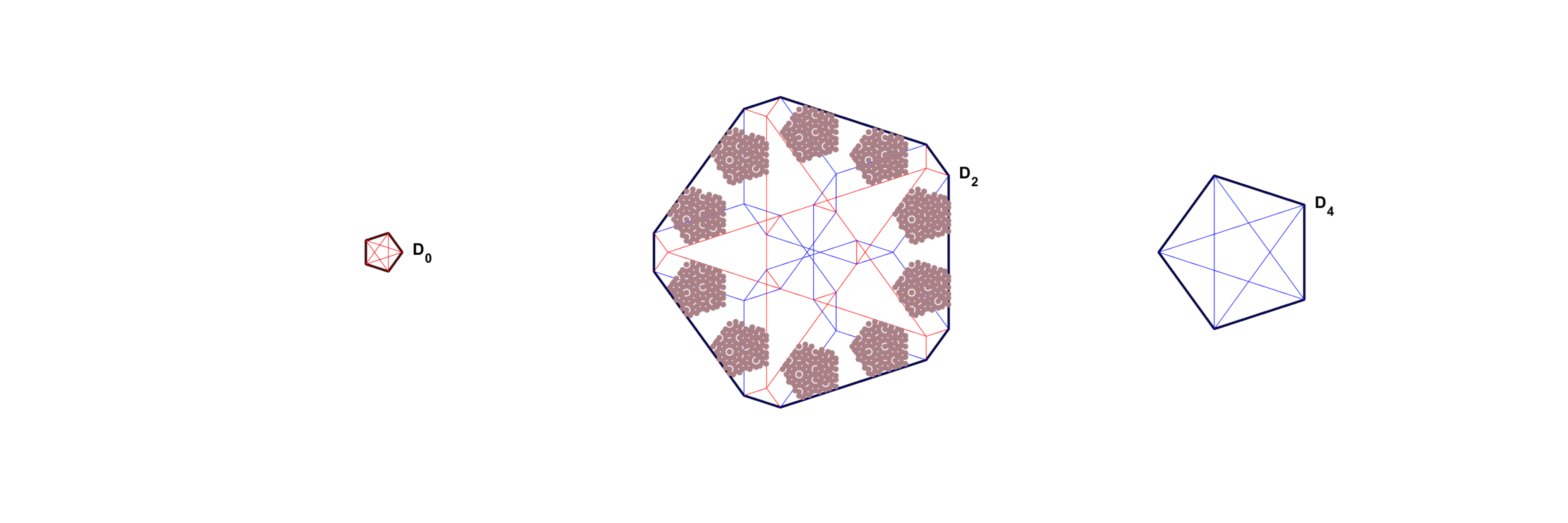}
    \includegraphics[clip,width=0.47\textwidth]{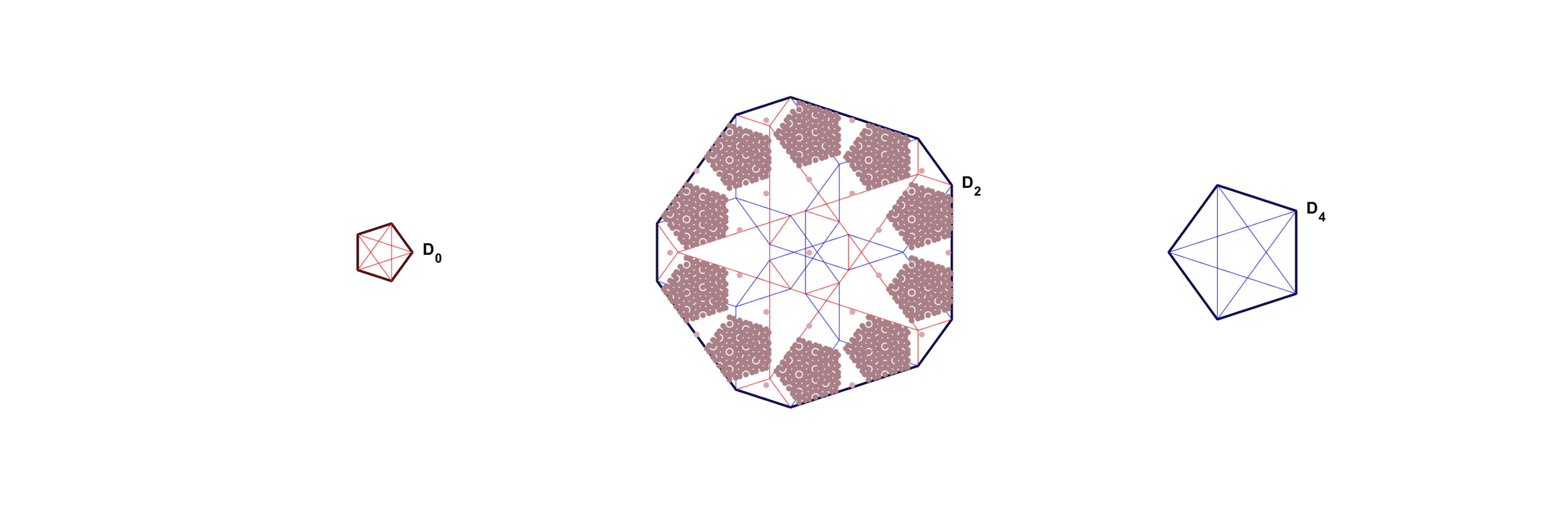}
    \includegraphics[clip,width=0.47\textwidth]{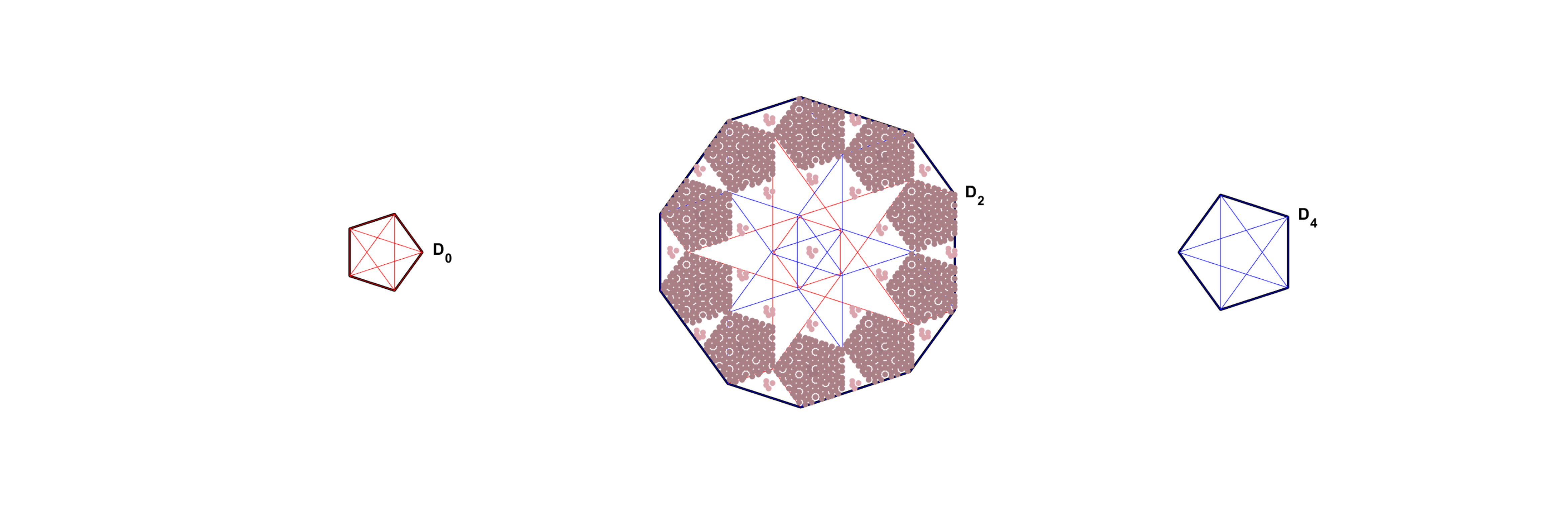}
    \includegraphics[clip,width=0.47\textwidth]{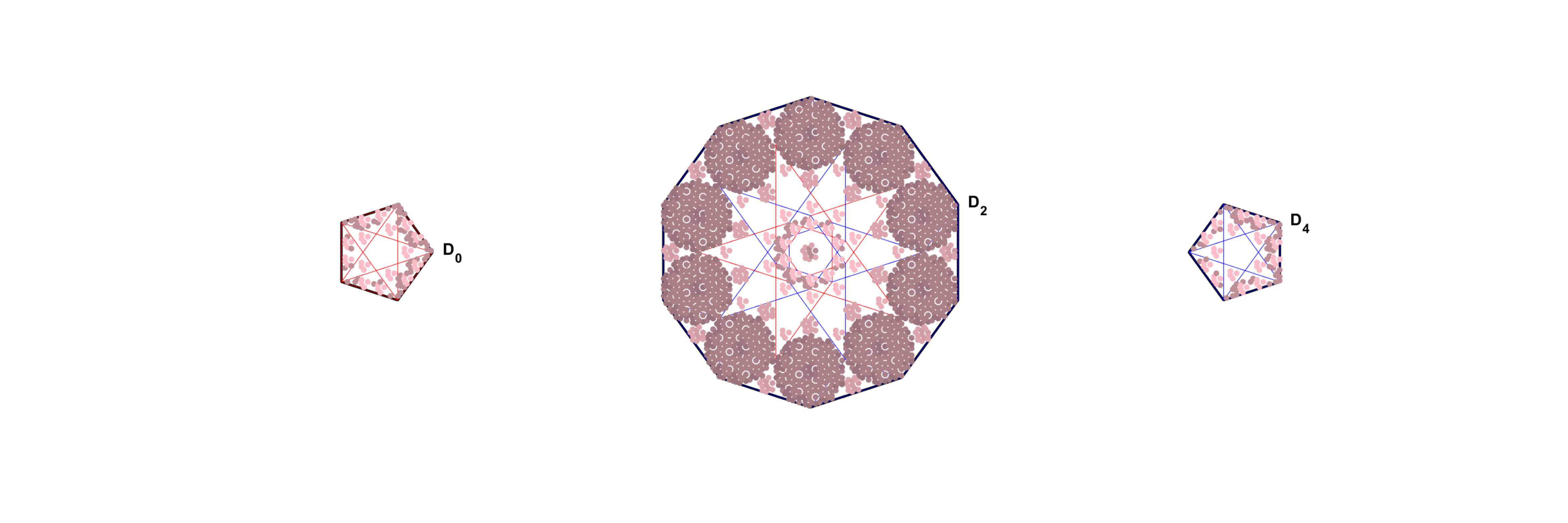}
    \caption{LDOS of the even sublattice in the perpendicular space for $\Gamma=0.0,...,0.5$.   }
    \label{fig:EvenLDOSPerp}
\end{figure}

\begin{figure}[!htb]
    \centering
    \includegraphics[clip,width=0.34\textwidth]{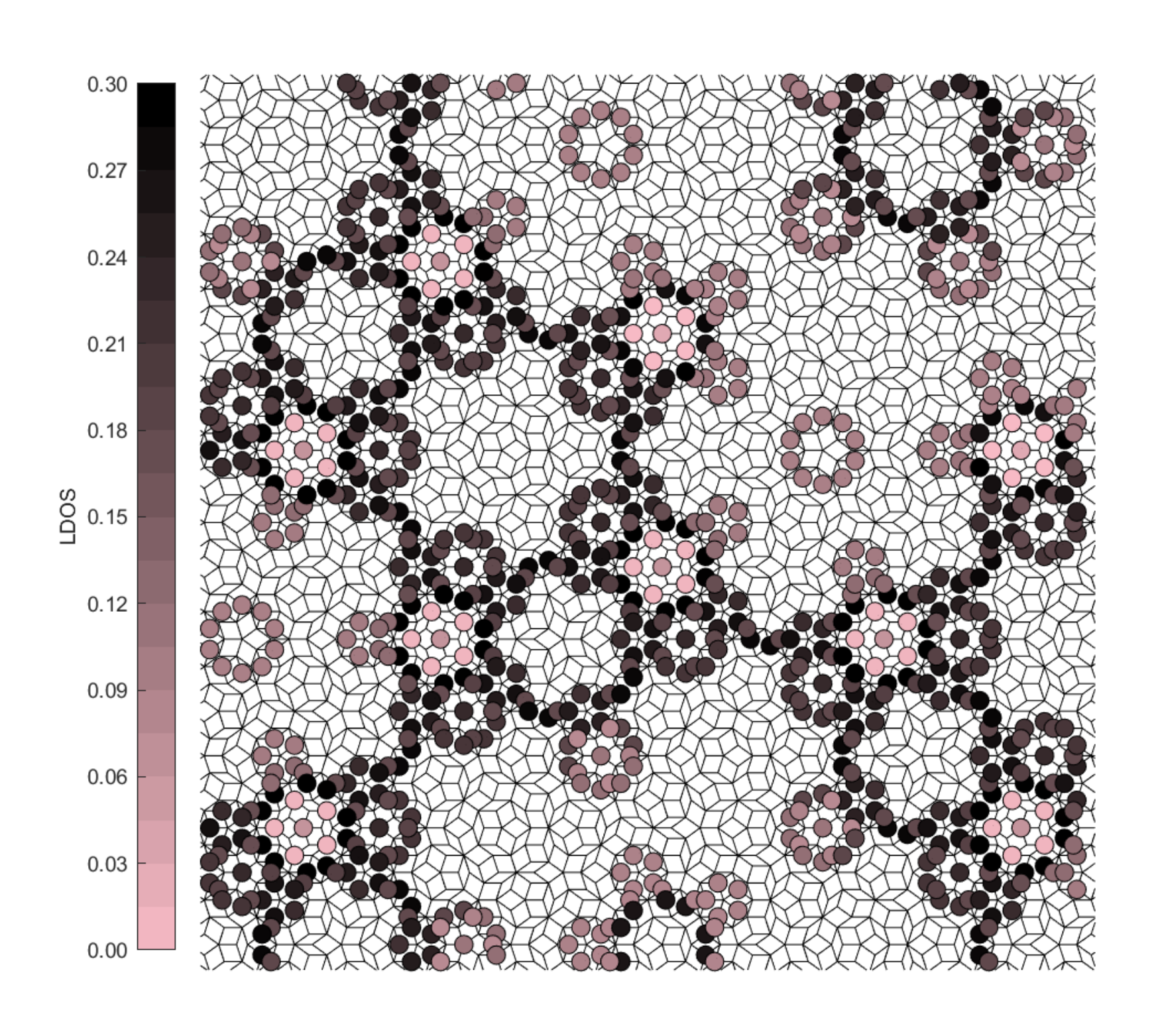}
    \includegraphics[clip,width=0.31\textwidth]{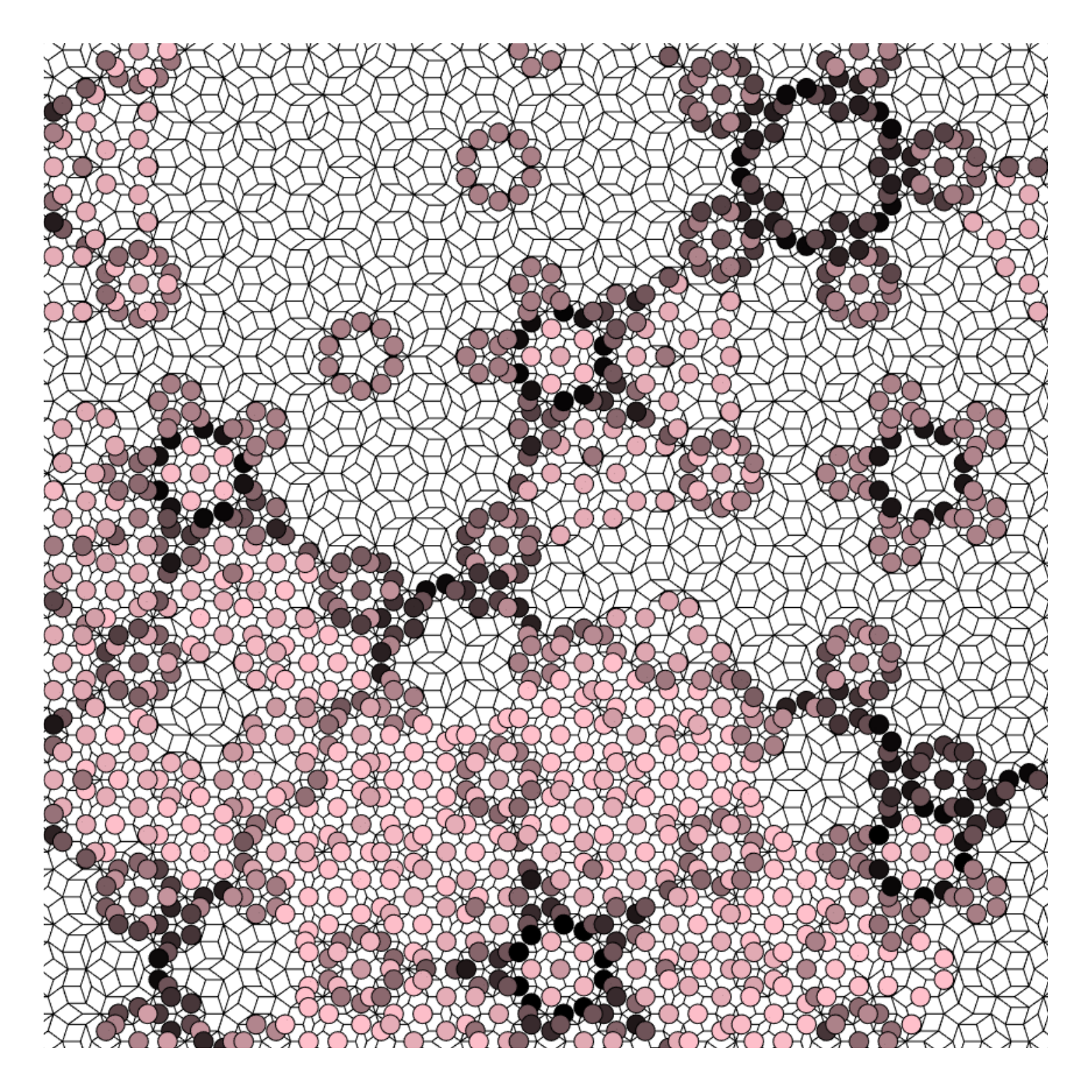}
    \includegraphics[clip,width=0.31\textwidth]{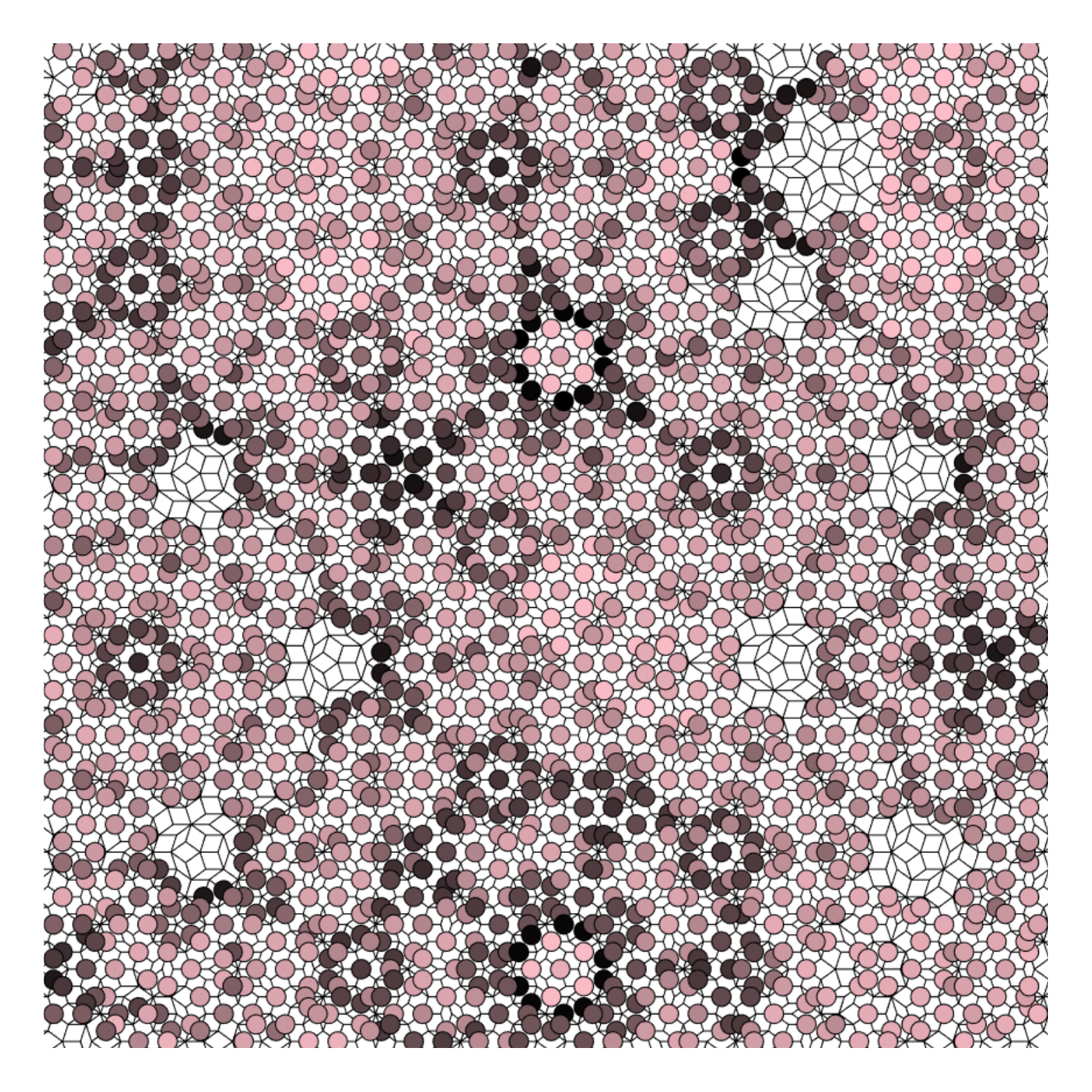}
    \caption{ LDOS on the odd sublattice for $\Gamma=0,0.1,0.2$. For the PLI $\Gamma=0$, one can notice the separation of the lattice into two regions in which only one sublattice has LDOS by comparing with Fig.\ref{fig:EvenLDOS01}. The LS density falls as $\Gamma$ increases, but the LDOS gets less concentrated and moves into "forbidden" regions for the odd sublattice. A more detailed evolution can be found in the supplementary material.}
    \label{fig:OddLDOS01}
\end{figure}

\begin{figure}[!htb]
    \centering
    \includegraphics[clip,width=0.31\textwidth]{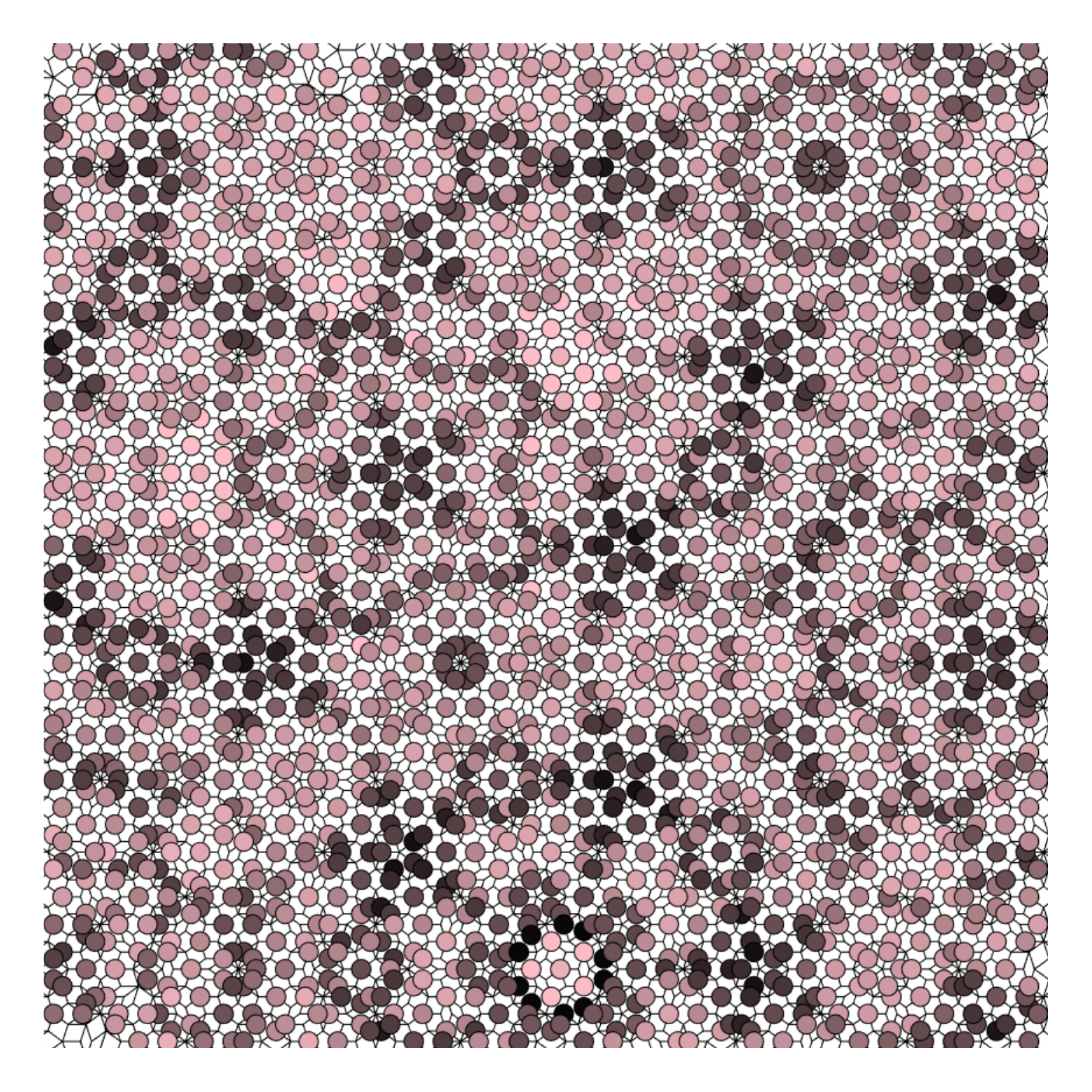}
    \includegraphics[clip,width=0.31\textwidth]{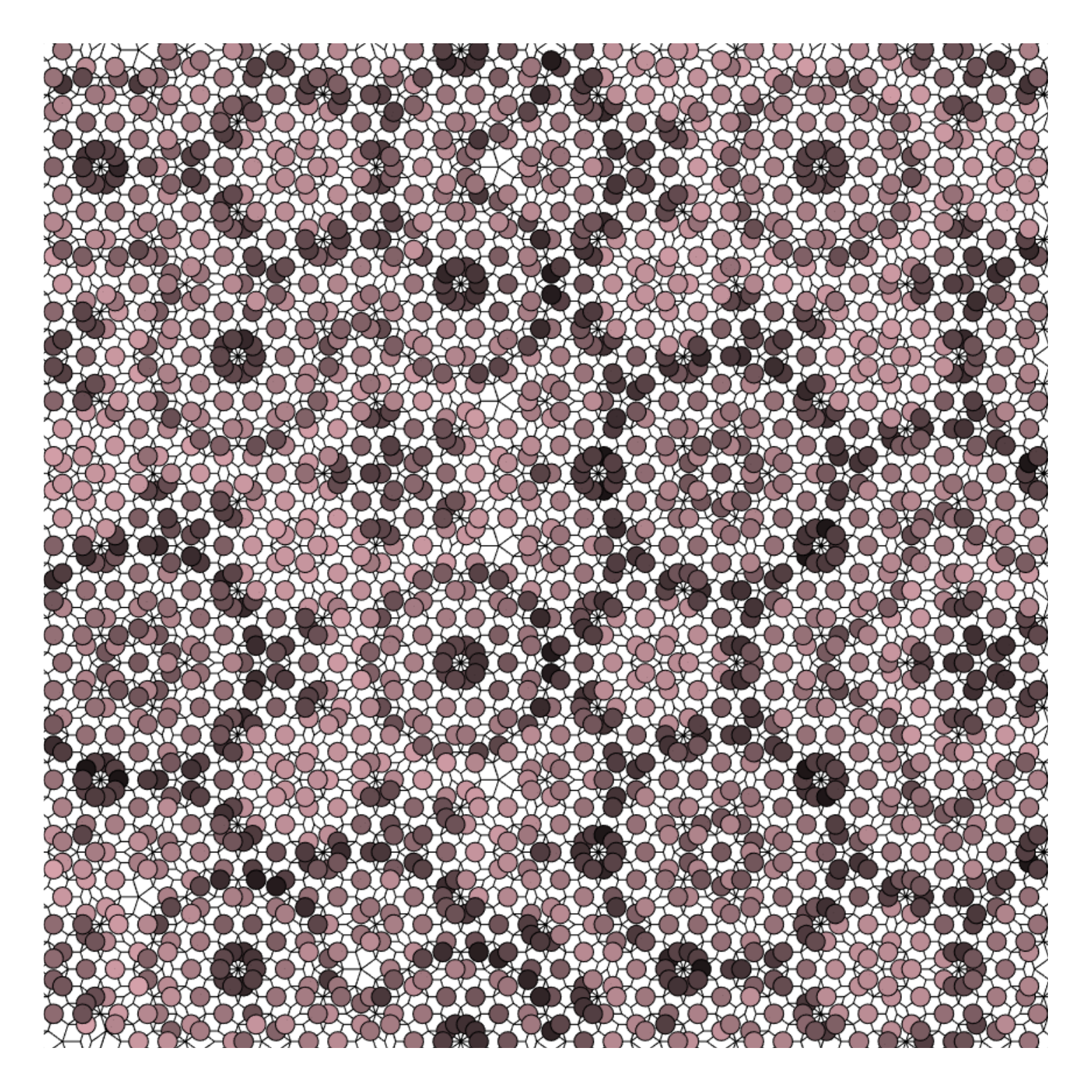}
    \includegraphics[clip,width=0.31\textwidth]{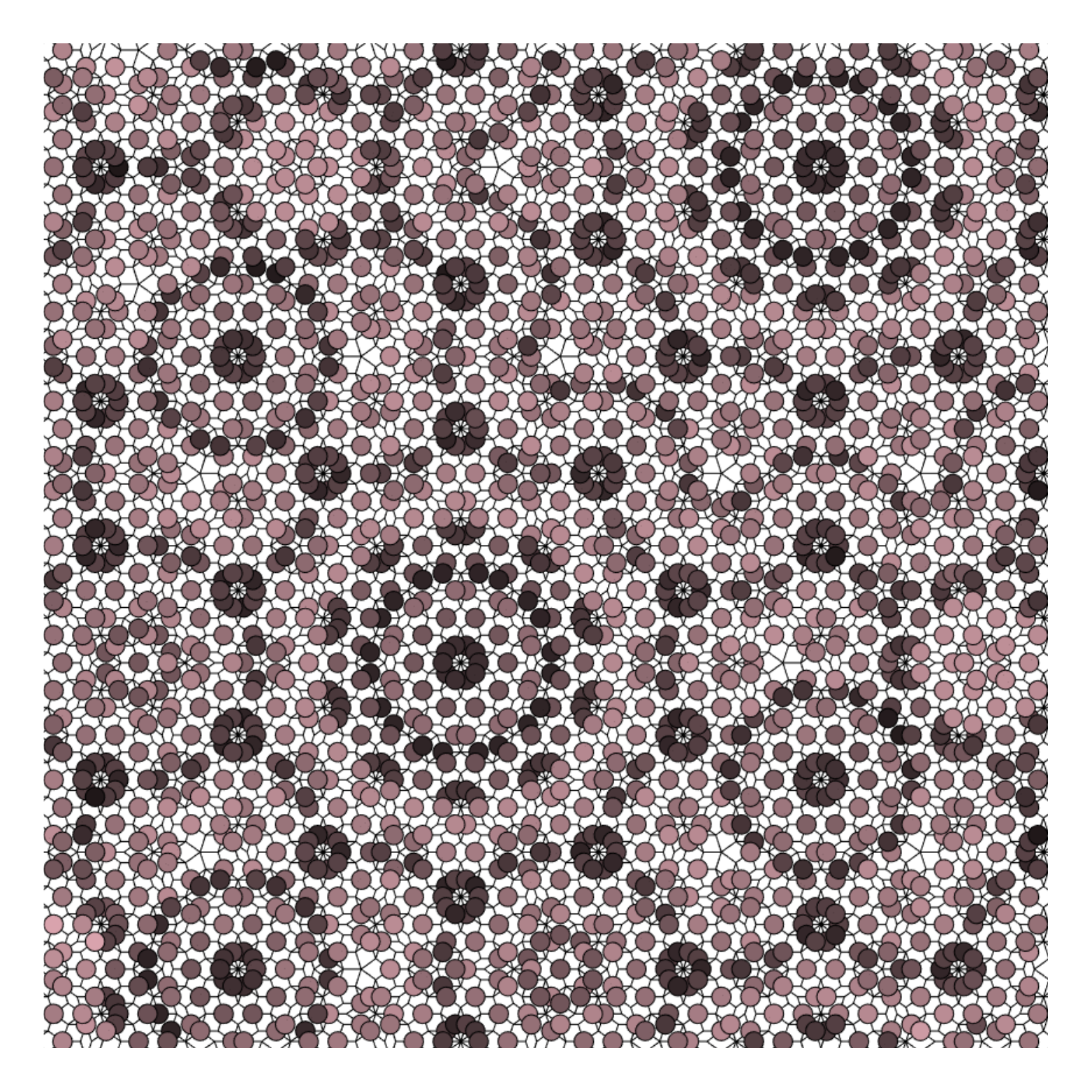}
    \caption{ LDOS on the odd sublattice for $\Gamma=0.3,0.4,0.5$.Almost all sites in the odd sublattice have some LDOS, and the LS fraction increases to its peak value at $\Gamma=0.5$. }
    \label{fig:OddLDOS02}
\end{figure}
\begin{figure}[!htb]
    \centering
    \includegraphics[clip,width=0.47\textwidth]{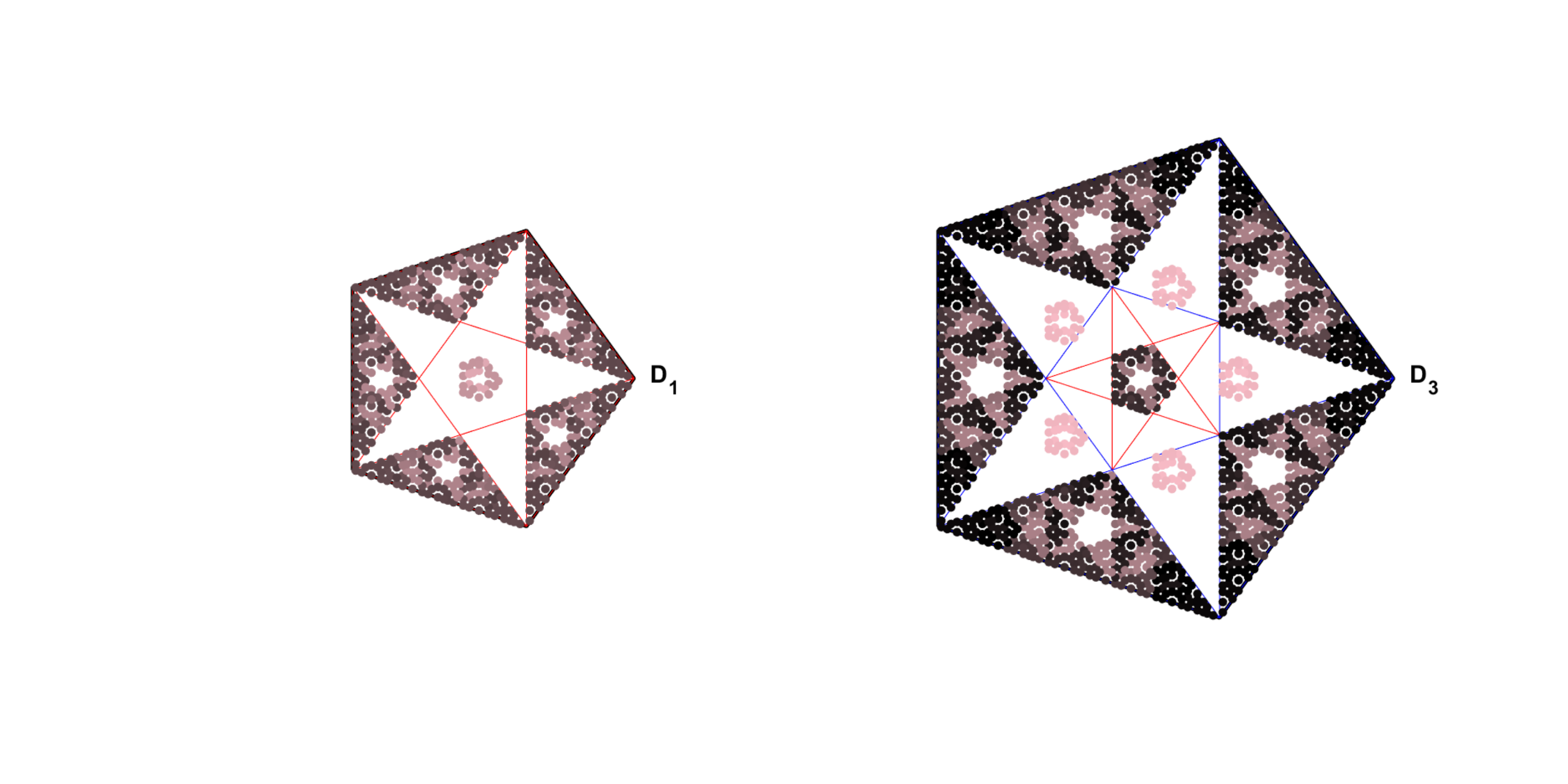}
    \includegraphics[clip,width=0.47\textwidth]{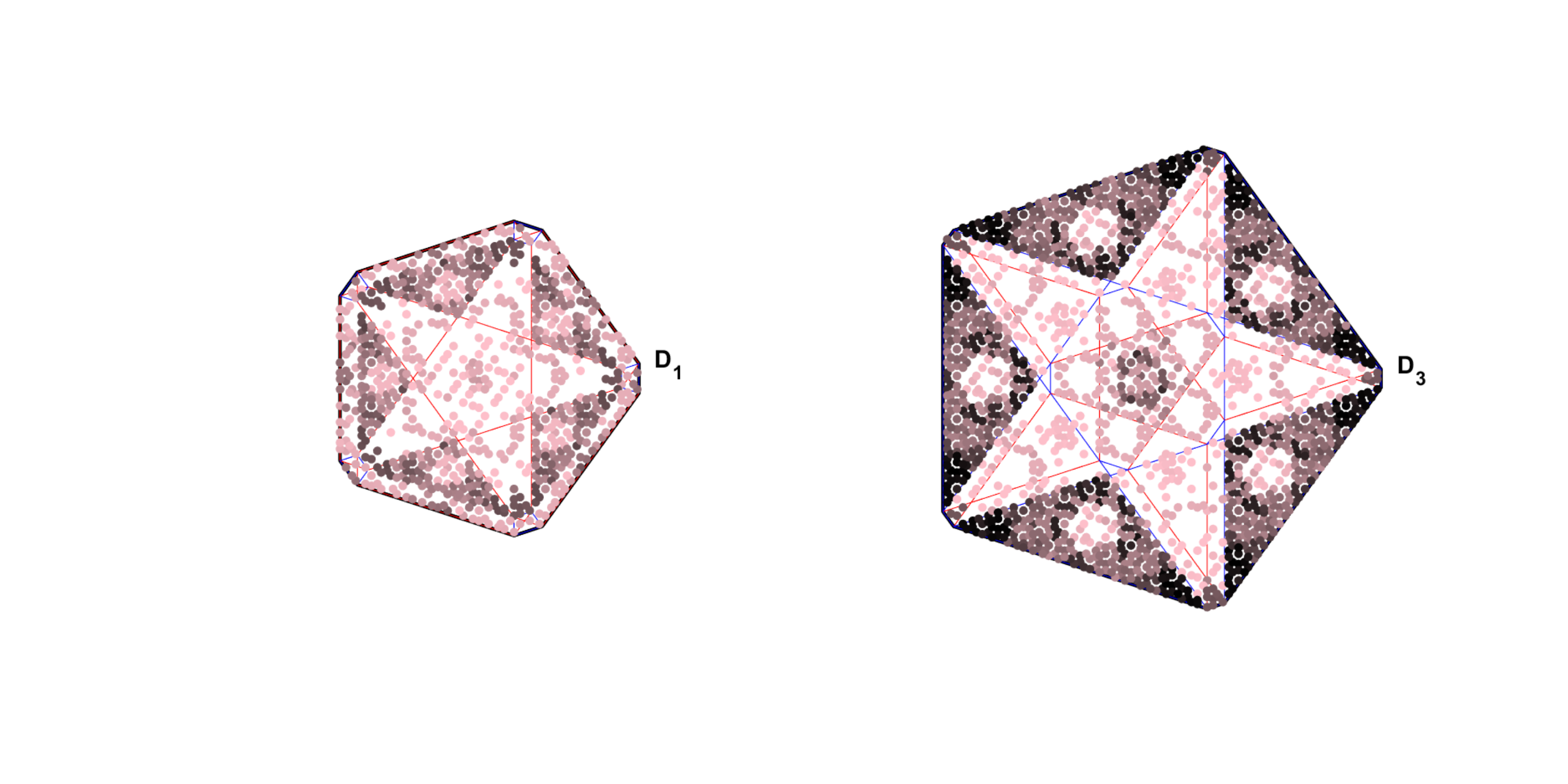}
    \includegraphics[clip,width=0.47\textwidth]{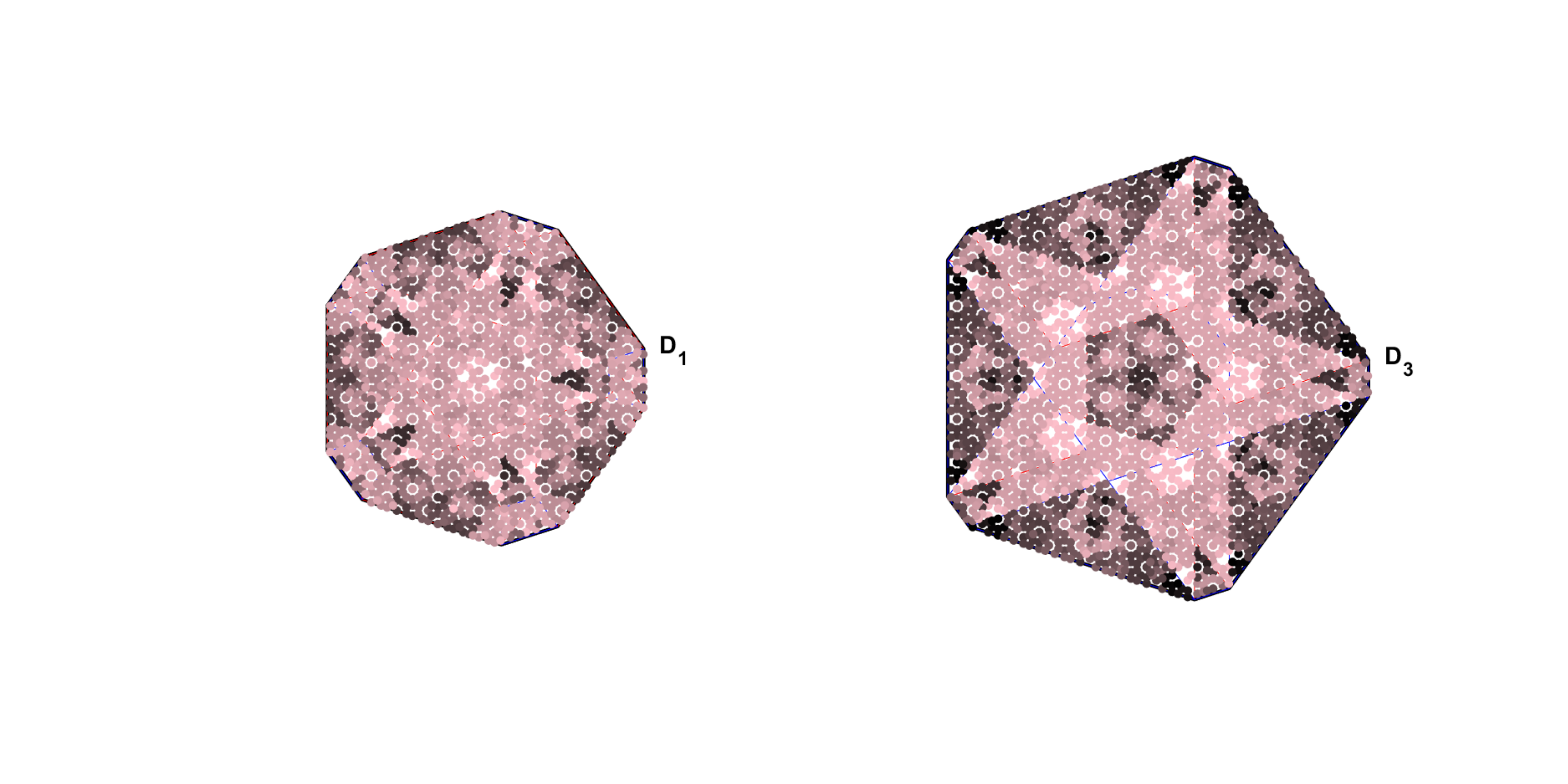}
    \includegraphics[clip,width=0.47\textwidth]{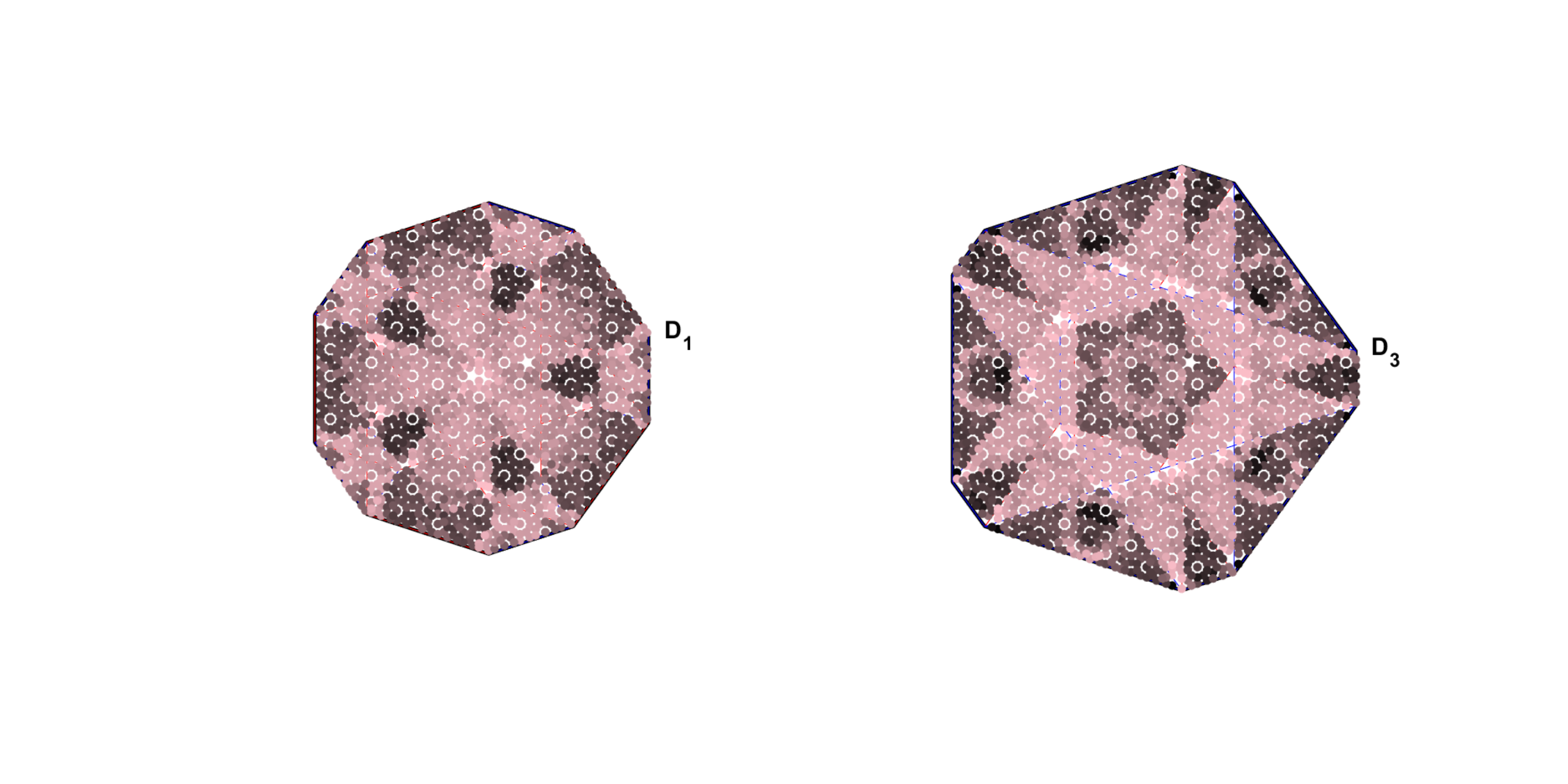}
    \includegraphics[clip,width=0.47\textwidth]{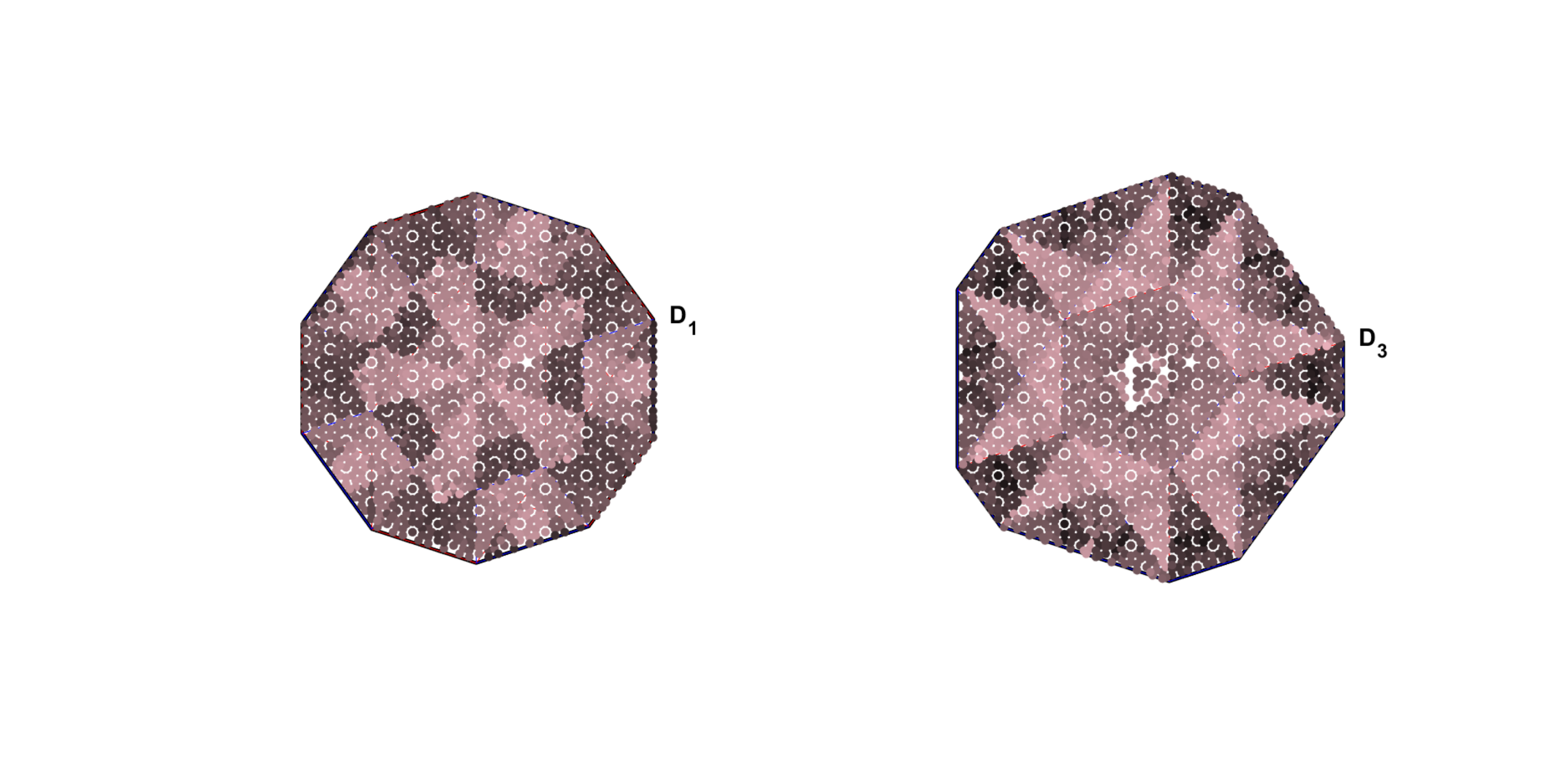}
    \includegraphics[clip,width=0.47\textwidth]{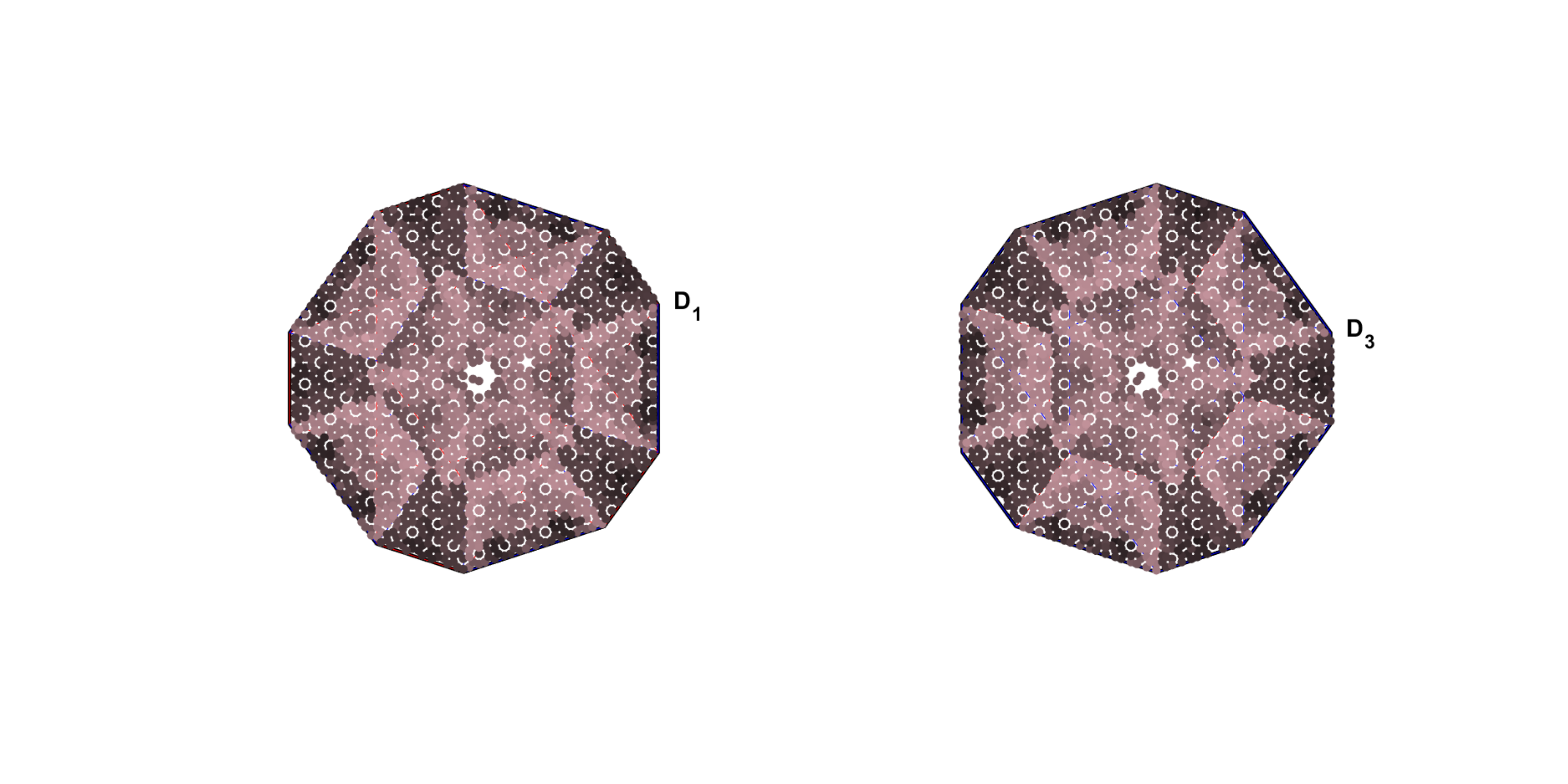}
    \caption{LDOS of the odd sublattice in the perpendicular space for $\Gamma=0.0,...,0.5$.   }
    \label{fig:OddLDOSPerp}
\end{figure}

To be able to simulate the tight-binding model on the quasicrystal, we need to translate the structural information of the quasicrystal into a tight binding matrix. The most commonly used methods for lattice generation rely on either approximants \cite{lin18} or scaling symmetry \cite{ara88}. Both approaches are not well suited to the current problem. Approximants necessarily introduce defects into the quasicrystal, which may result in spurious LS even when the unit cell size is large. While generating the lattice through deflation can quickly create lattices of large sizes, such scaling symmetry is not present for the quasicrystal LI classes considered here.

Instead, we rely on the perpendicular space to generate a finite-size lattice \cite{okt21,mok20}. We start by choosing the perpendicular space coordinates of a single point, i.e., specify a point on one of the five polygons. This point has at most ten nearest neighbors; their possible perpendicular space coordinates are obtained by adding the vectors $\pm \hat{\tilde{e}}_m$. The number of first neighbors and their perpendicular space positions are thus easily obtained from this list by checking if they reside inside the five polygons. The same process can be repeated on the first neighbors to generate the second neighbors and iterated to create a neighborhood of the initial point up to n$^{th}$ nearest neighbors. 

This generation method adds a layer of vertices, all of which are in the same sublattice.
 It is thus easy to keep track of the tight-binding Hamiltonian
\begin{equation}
\label{eq:MatrixC}
    {\cal H}=\begin{bmatrix}
0 & {\cal C} \\
{\cal C}^T & 0
\end{bmatrix},
\end{equation}
in the bipartite form. 
The ${\cal C}$ matrix connects even sublattice sites to odd sublattice sites and is composed of overlapping blocks connecting $n^{th}$ to $(n+1)^{th}$ neighbors.

As we are interested in LS, which can be chosen to be localized to a single sublattice, their number can be chosen by finding the dimension of the null space of ${\cal C}$ or ${\cal C}^T$. Although we carry out this calculation on a finite lattice, all of the vertices in the boundary belong to the same sublattice. Thus, we are assured that we find all the LS strictly inside the finite region we consider for the  other sublattice. For example, if we start with an initial point on the even sublattice and generate the neighborhood up to $10^{th}$ nearest neighbors, the last added layer will be all in the even sublattice. Any LS we find on the odd sublattice will be present in the odd sublattice of the infinite quasicrystal as well.

We count the number of LS in both sublattices as we increase the size of the neighborhood. The largest lattices we use are 200-deep neighborhoods of the initial point and contain approximately 100,000 lattice sites. The LS fraction is calculated by dividing the number of LS states by the number of lattice sites. For small lattices, there is a large deviation for the LS fraction depending on the initial perpendicular space position. However, as the lattice size grows, the LS fraction converges to a narrow band. In Fig.\ref{fig:NumericalTestPoints} we show how LS fraction changes with neighborhood depth for 10 randomly selected initial lattice points for $\Gamma=0.2$. It is worthwhile to note that the range of calculated LS fractions on the even lattice is more narrow than the odd sublattice. As we use a finite-size lattice, we expect the main source of error to be the number of LS crossing the boundary of our region. The larger variation in the odd sublattice is a reflection of the larger size of LS types in the odd sublattice.   

We use the average of the results at the largest lattice size as the numerically calculated value of the LS fraction at a given $\Gamma$. Repeating this procedure at $\Gamma$ values separated by 0.01 we obtain Fig.\ref{fig:NumericalLSFractions}. At $\Gamma=0$, we have the PL, and our numerical result is 0.001 below the analytical result $81-50\tau$. As the two sublattices are symmetric under inversion, their LS fractions must be equal. Once again, our numerical calculation finds LS fraction values within 0.001 of each other for both lattices. Based on these two observations and fluctuation of the value between neighboring $\Gamma$ values, we expect 0.001 to be a good estimate for the typical error in our numerical value. 
As $\Gamma$ starts increasing from 0, the total LS fraction steeply drops. This is somewhat unexpected, as increasing $\Gamma$ makes the number of sites on the two sublattices different, as shown in Fig.\ref{fig:SublatticeImbalance}. Thus, unlike certain bipartite flat-band models where the zero energy states are directly linked to the overall imbalance between the number of sites between sublattices, the local distribution of this imbalance seems to be a larger factor for quasicrystal LS. As $\Gamma$ increases the total LS fraction takes a minimum value of $\sim 0.048$ near $\Gamma \sim 0.12$ and then monotonically increases until $\Gamma=0.5$. The maximum LS fraction is achieved at $\Gamma=0.5$ where  $\sim 10.17\%$ of the states are LS.  

As soon as $\Gamma$ is different from zero, the symmetry between the two sublattices is broken. We find that the odd sublattice has a higher LS fraction for all values of $\Gamma$ except at the PL. Both sublattices' LS fraction first decreases and then increases with $\Gamma$, but their minima occur at different values. After $\Gamma \sim 0.1$, the odd sublattice has almost three times more LS than the even sublattice. At $\Gamma=0.5$, the odd sublattice provides $\sim 0.076$, and the even sublattice gives $\sim 0.026$  LS fraction. The maximum imbalance between the number of sites is at $\Gamma=0.5$, where the odd sublattice has $\sim 52.6 \% $ of the sites. Thus, the three-fold difference between the LS fraction of even and odd sublattices cannot be explained by the difference in their vertex numbers. The two sublattices must have a significant difference in their local connectivity.

We further explore the difference between the sublattices by defining the LDOS at zero energy.
\begin{equation}
    \rho(\vec{R}_i)=\sum_m \delta(E_m) |\langle \vec{R}_i | \Psi_m \rangle |^2,
\end{equation}
where $|\Psi_m \rangle$ are the normalized eigenstates of the tight-binding Hamiltonian Eq.\ref{eq: Hamiltonian} with energy $E_m$. Furthermore, as all LS in the zero-energy manifold can be chosen to lie on only one of the sublattices, we can investigate LDOS on the odd and even sublattices separately. In figures \ref{fig:EvenLDOS01},\ref{fig:EvenLDOS02},\ref{fig:EvenLDOSPerp},\ref{fig:OddLDOS01},\ref{fig:OddLDOS02},\ref{fig:OddLDOSPerp} we show LDOS for both sublattices both in real and perpendicular space, at different values of $\Gamma$. The figures show a region that is 20 lattice constants away from the edges so that the effects of the boundary are minimized. The LDOS for both sublattices are calculated in the neighborhood of the same initial lattice point for all values of $\Gamma$. Hence, these figures show a snapshot of how LDOS evolves with $\Gamma$ without changing other parameters.

At $\Gamma=0$, the PL splits into regions where the LDOS is non-zero only on one of the sublattices. These regions are separated by 'strings' formed by rhombi with two three-edge vertices, as first identified in Ref.\cite{ara88}. However, when $\Gamma \ne 0$, it is impossible to define such regions as LDOS can be non-zero for both sublattices in the same region of space. The behavior of the LDOS is markedly  different on the odd and even sublattices. On the even sublattice, the LDOS is mostly non-zero in disjoint regions centered around vertices with a high number of edges. This is consistent with the observation in the numerical calculation that the LS fraction in a finite region does not fluctuate significantly as the boundaries move. It also allows us to identify LS in terms of a small number of LS types in the next section, as the perpendicular space LDOS is also confined to well defined regions.
On the other hand, the LDOS on the odd sublattice gets more evenly distributed throughout the lattice as $\Gamma$ increases. At $\Gamma=0.5$, almost all the sites in the odd sublattice seem to have some overlap with zero energy LS. The effects of boundaries are observed for larger sizes compared to the even sublattice. Both the LDOS calculation and the considerable variation in the LS fraction, point to the presence of  LS that cannot be reduced to smaller size LS types. It is not clear why the behavior of the LS is so different in the two sublattices.

Before we analyze the LS manifold in terms of LS types in the next section, we want to point out that LDOS is a property that can be experimentally probed in scanning tunneling microscope experiments on synthetic surface quasicrystals as in Ref\cite{col17}. Similarly, a cold atom experiment\cite{vie19} can explore LDOS by investigating the non-dispersing part of a wave packet's evolution in an optical quasicrystal. 

\section{Localized State Types}
\label{sec:LS}

\begin{figure}[!htb]
    \centering
    \includegraphics[clip,width=0.31\textwidth]{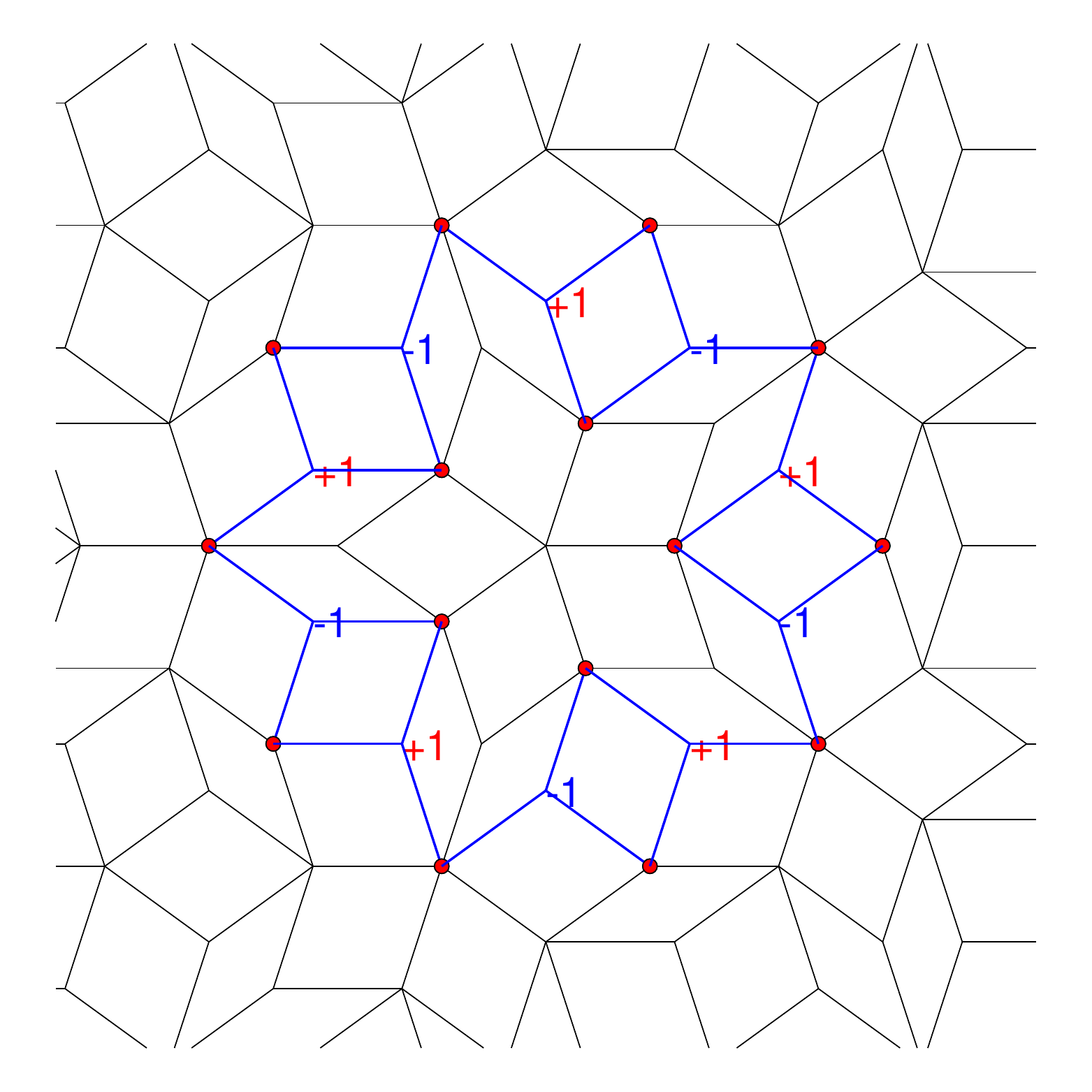}
    \includegraphics[clip,width=0.31\textwidth]{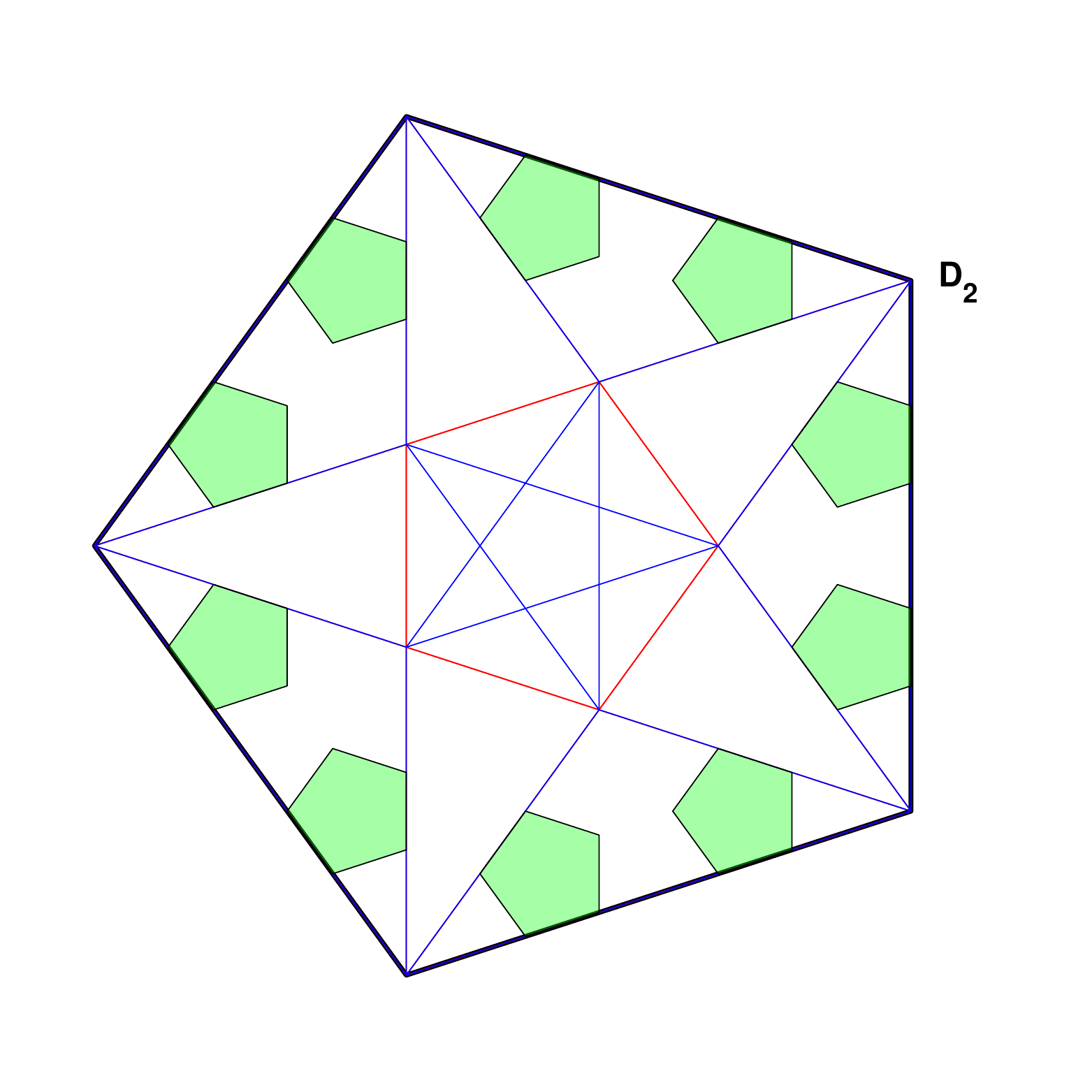}
    \includegraphics[clip,width=0.31\textwidth]{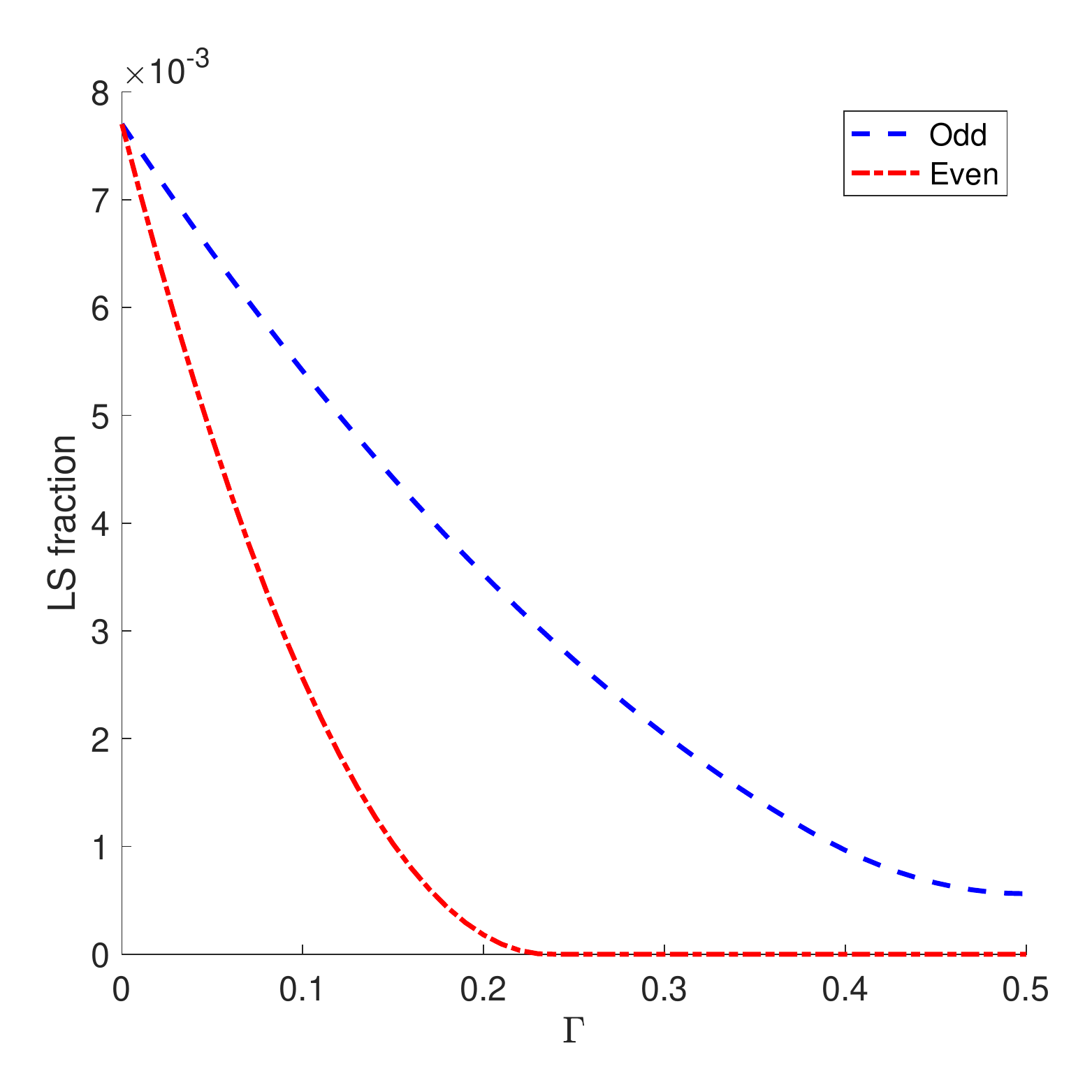}
    \caption{(a)Type-2 state has a support of 10 D vertices arranged around a central S vertex. (b) The allowed perpendicular space positions for Type-E2 state at $\Gamma=0$ form ten pentagons inside $D_2$. The area of these pentagons can be used to calculate the frequency of the type-E2 state. (c) The frequency for the type-E2 on the even and Type-O2 on the odd sublattice as a function of  $\Gamma$. After $\Gamma\simeq.22$, there is no type-E2 state in the quasicrystal. }
    \label{fig:Type2}
\end{figure}

\begin{figure}[!htb]
    \centering
    \includegraphics[clip,width=0.48\textwidth]{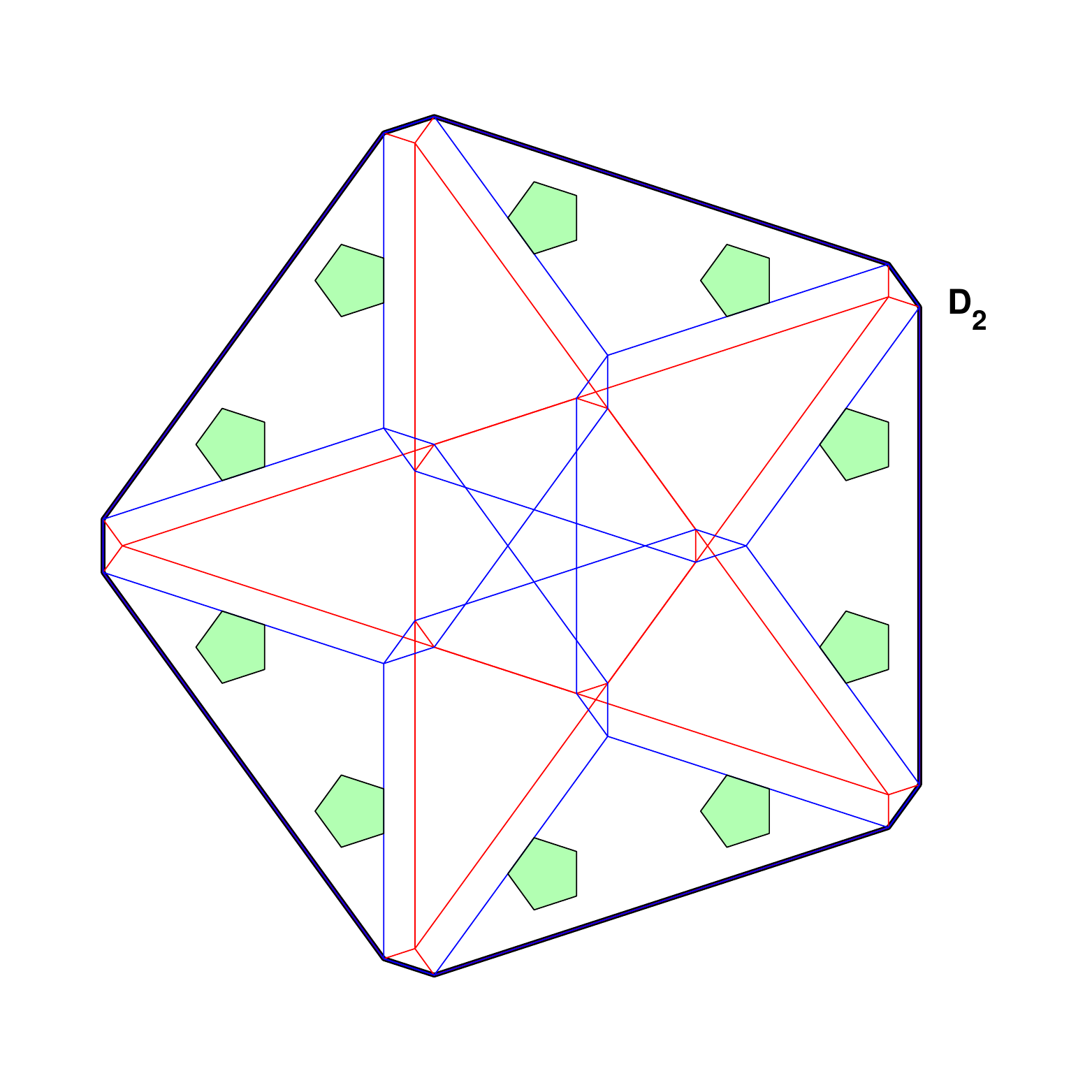}
    \includegraphics[clip,width=0.48\textwidth]{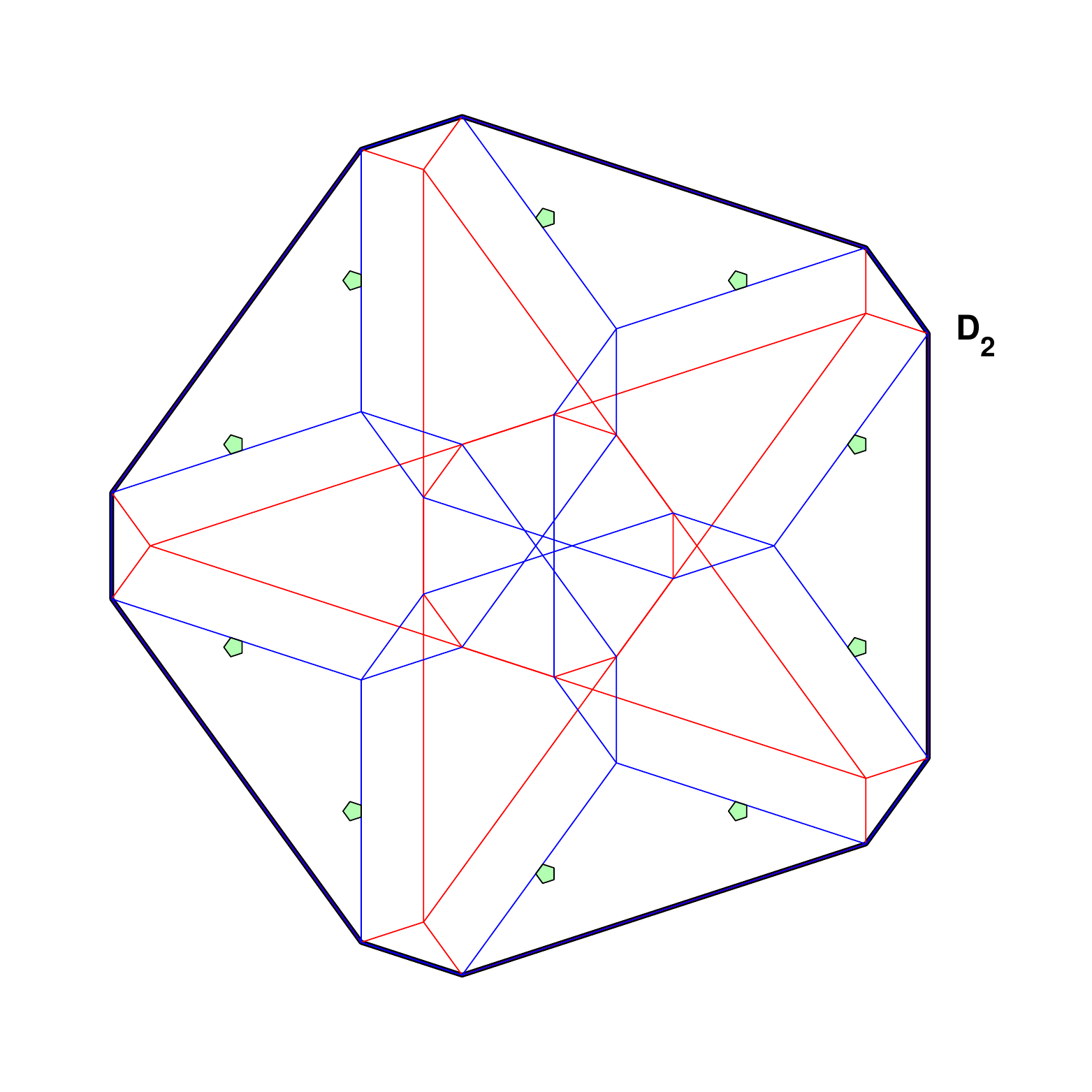}
    \caption{The allowed regions for type-E2 state at $\Gamma=0.1,0.2$ show the frequency decrease through the shrinking allowed areas. The supplementary material contains an animation of the allowed areas evolution. }
    \label{fig:Type2EvenScan}
\end{figure}

\begin{figure}[!htb]
    \centering
    \includegraphics[clip,width=0.24\textwidth]{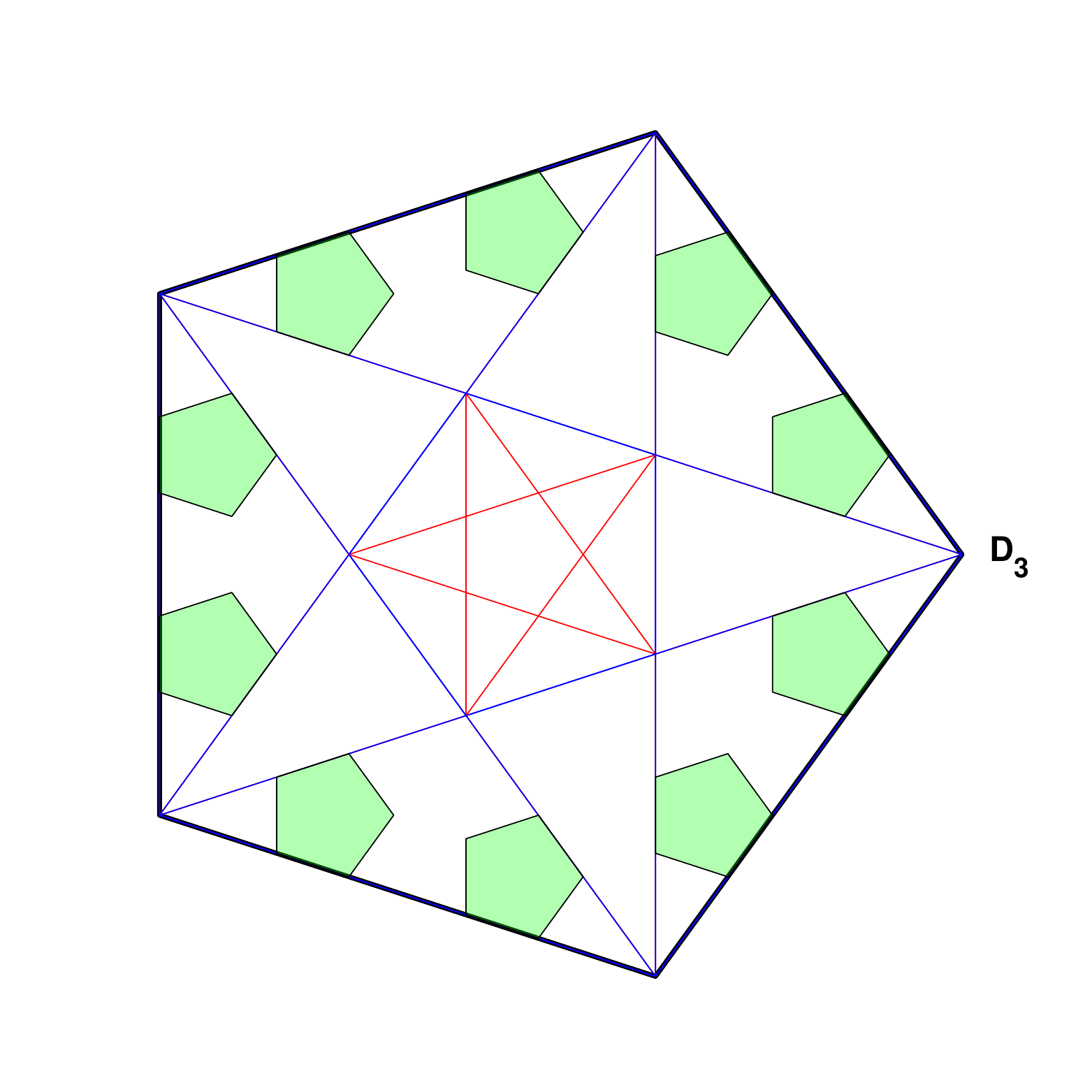}
    \includegraphics[clip,width=0.24\textwidth]{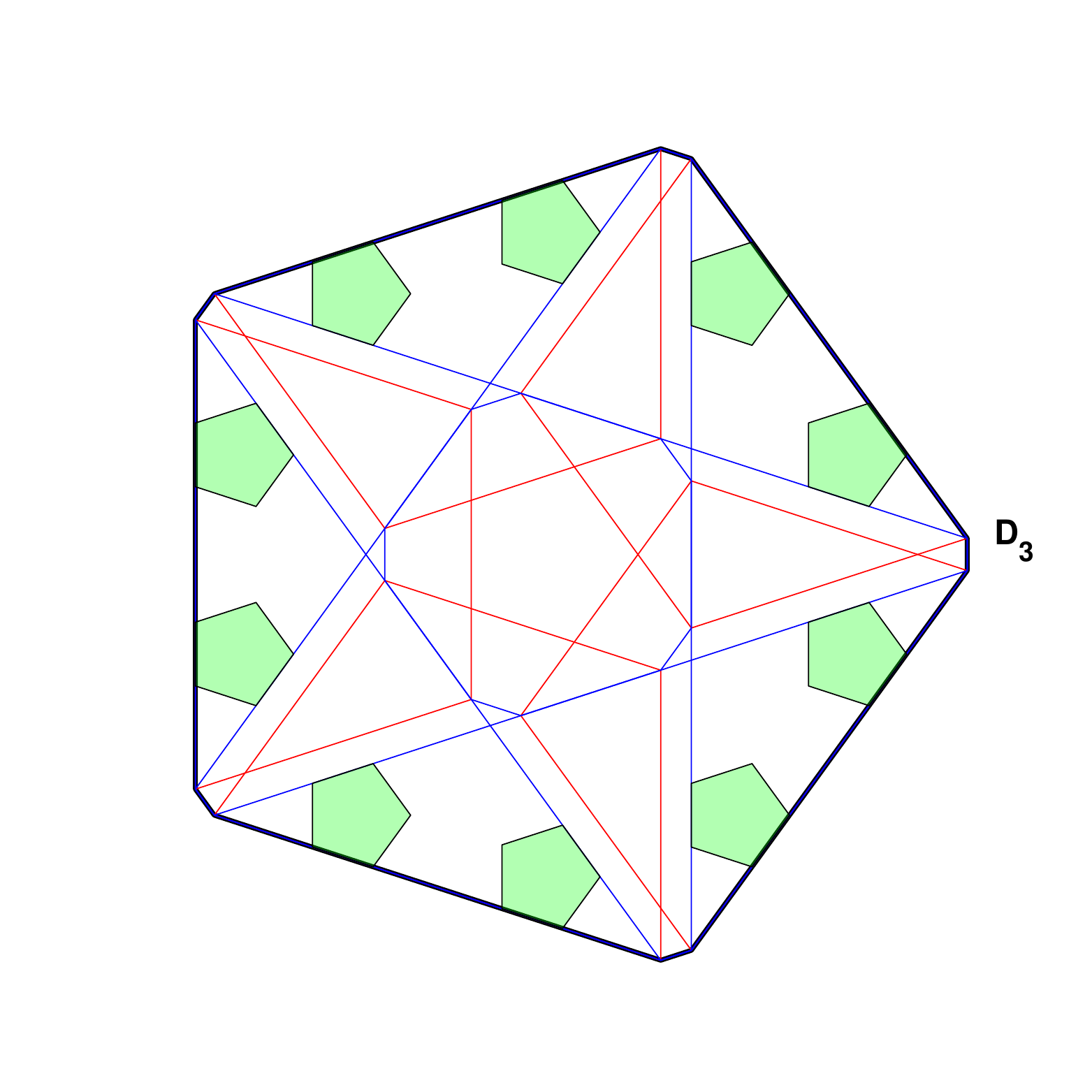}
    \includegraphics[clip,width=0.24\textwidth]{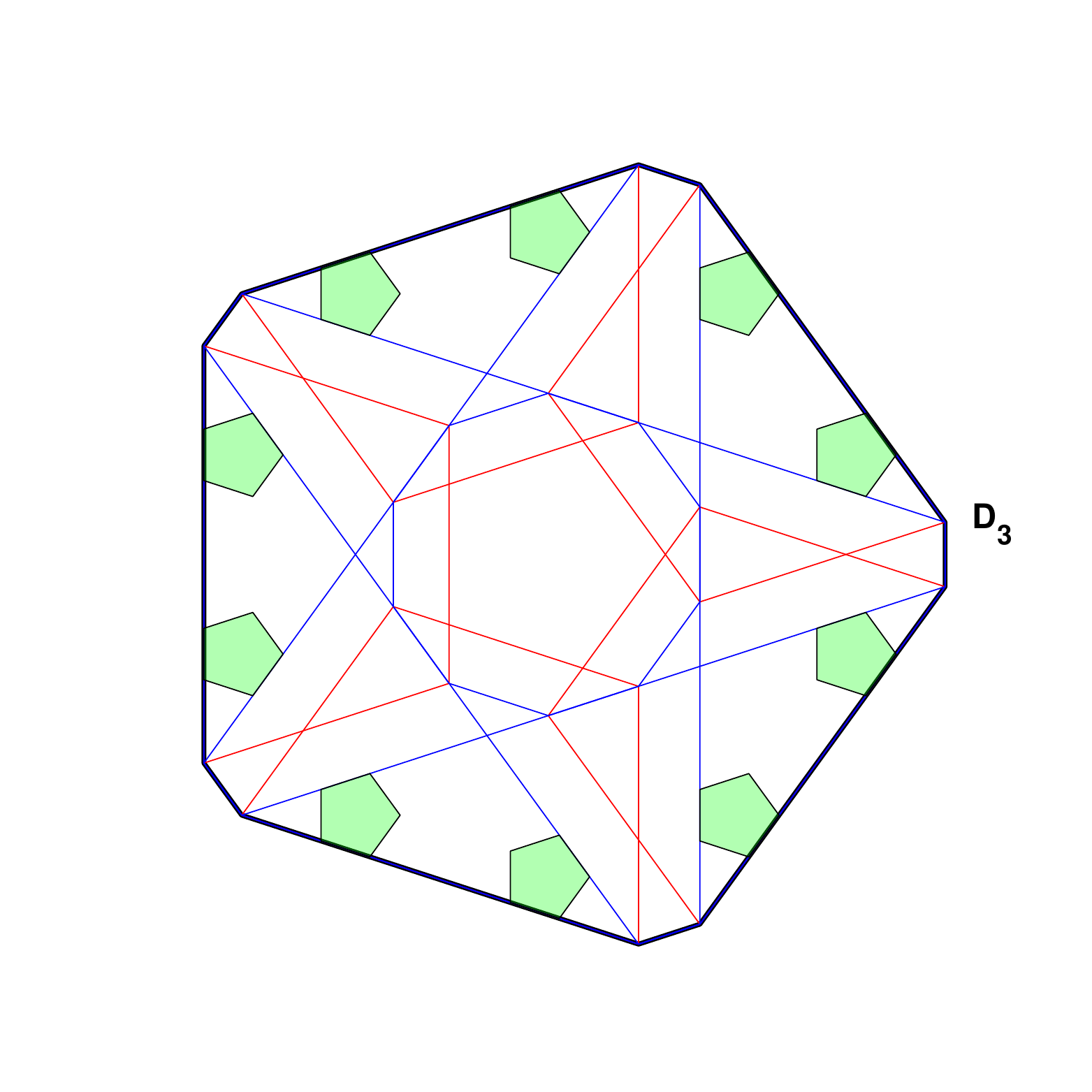}
    \includegraphics[clip,width=0.24\textwidth]{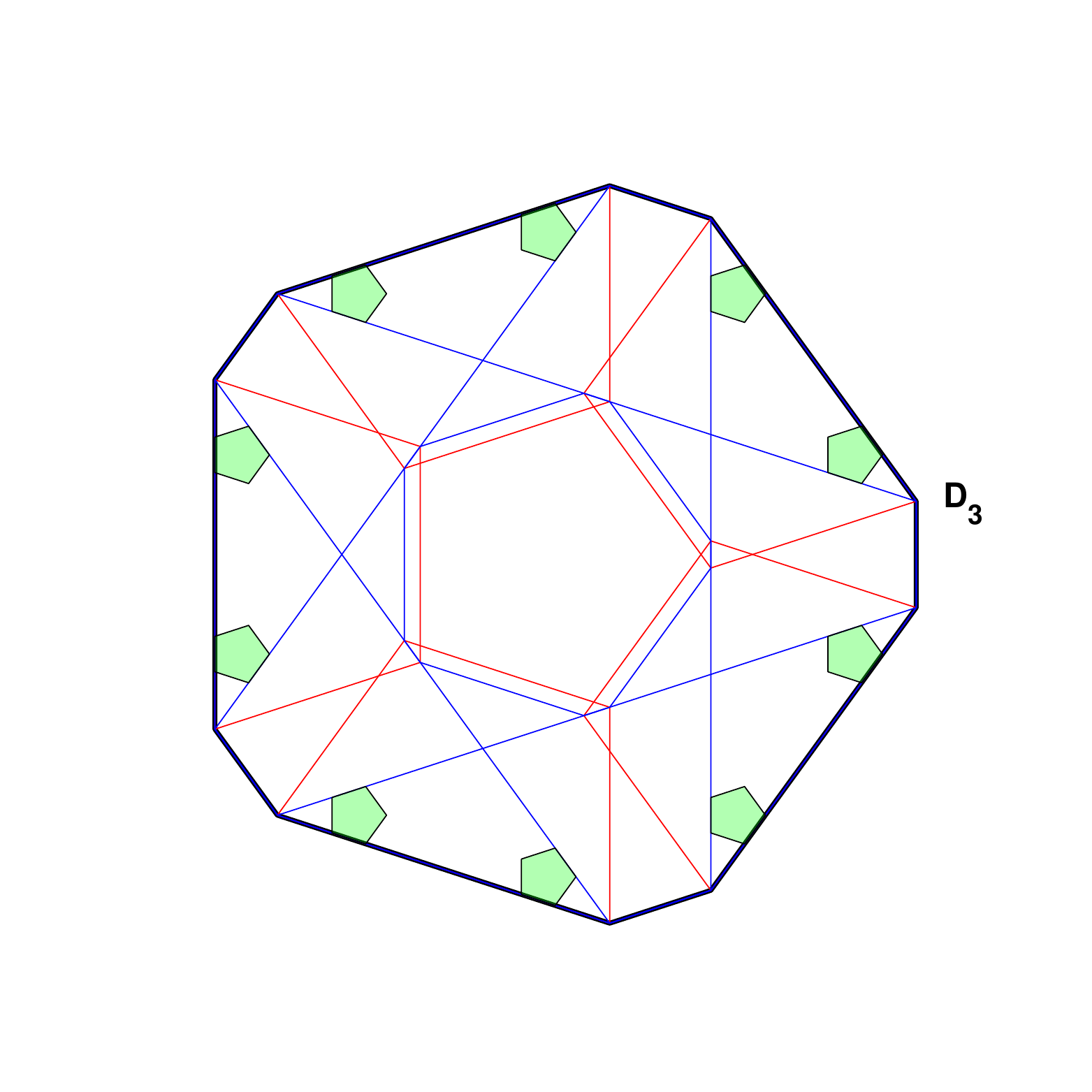}
    \includegraphics[clip,width=0.24\textwidth]{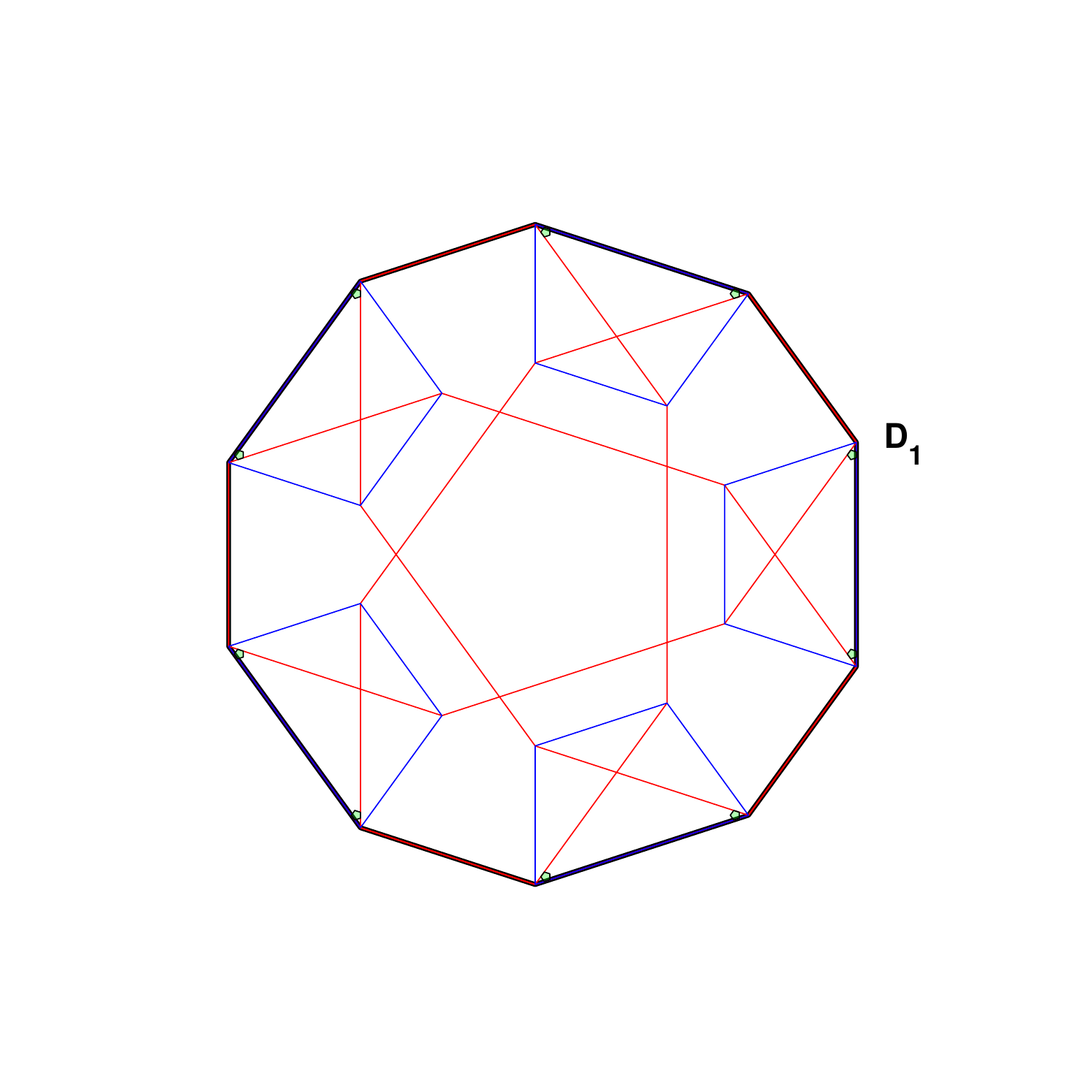}
    \includegraphics[clip,width=0.24\textwidth]{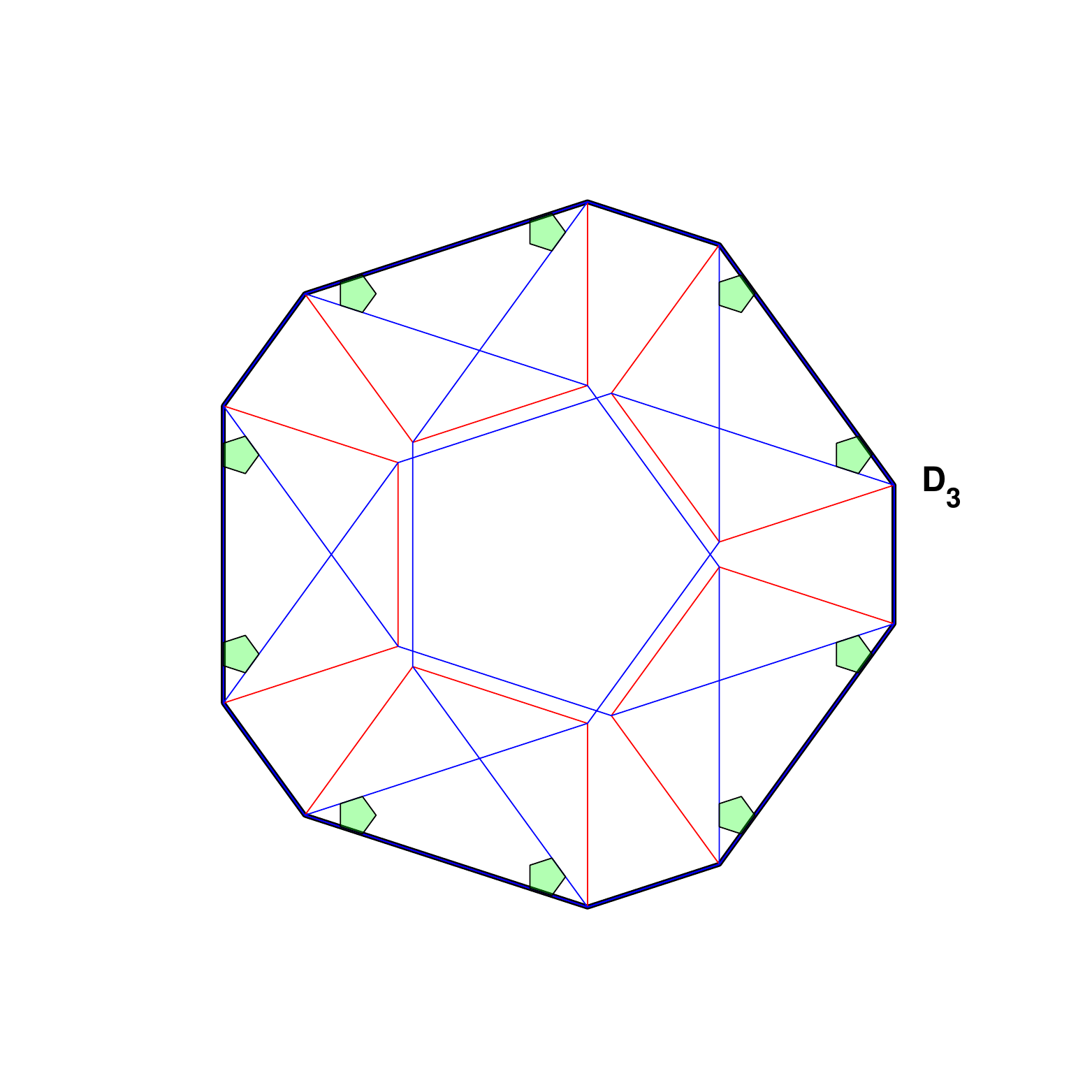}
    \includegraphics[clip,width=0.24\textwidth]{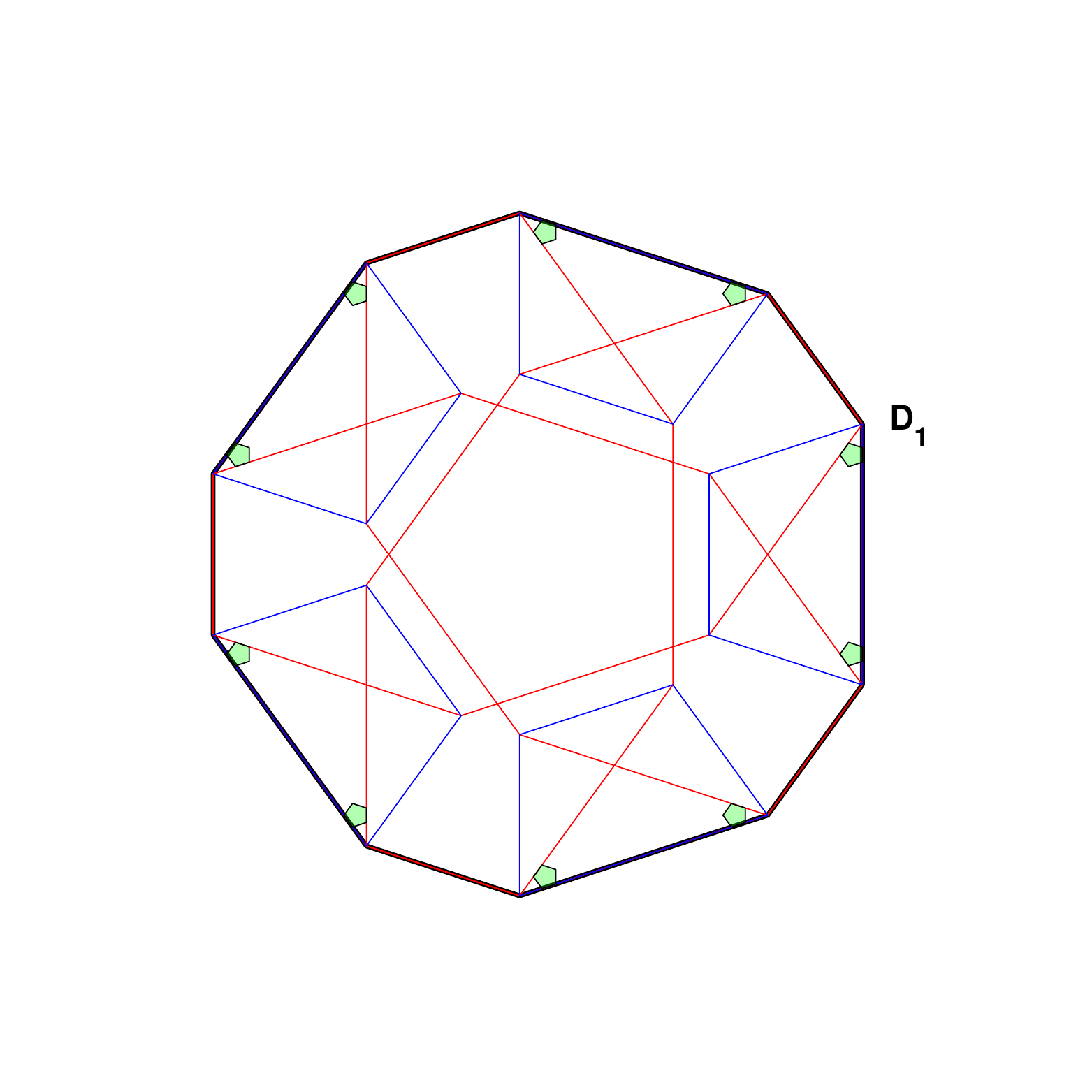}
    \includegraphics[clip,width=0.24\textwidth]{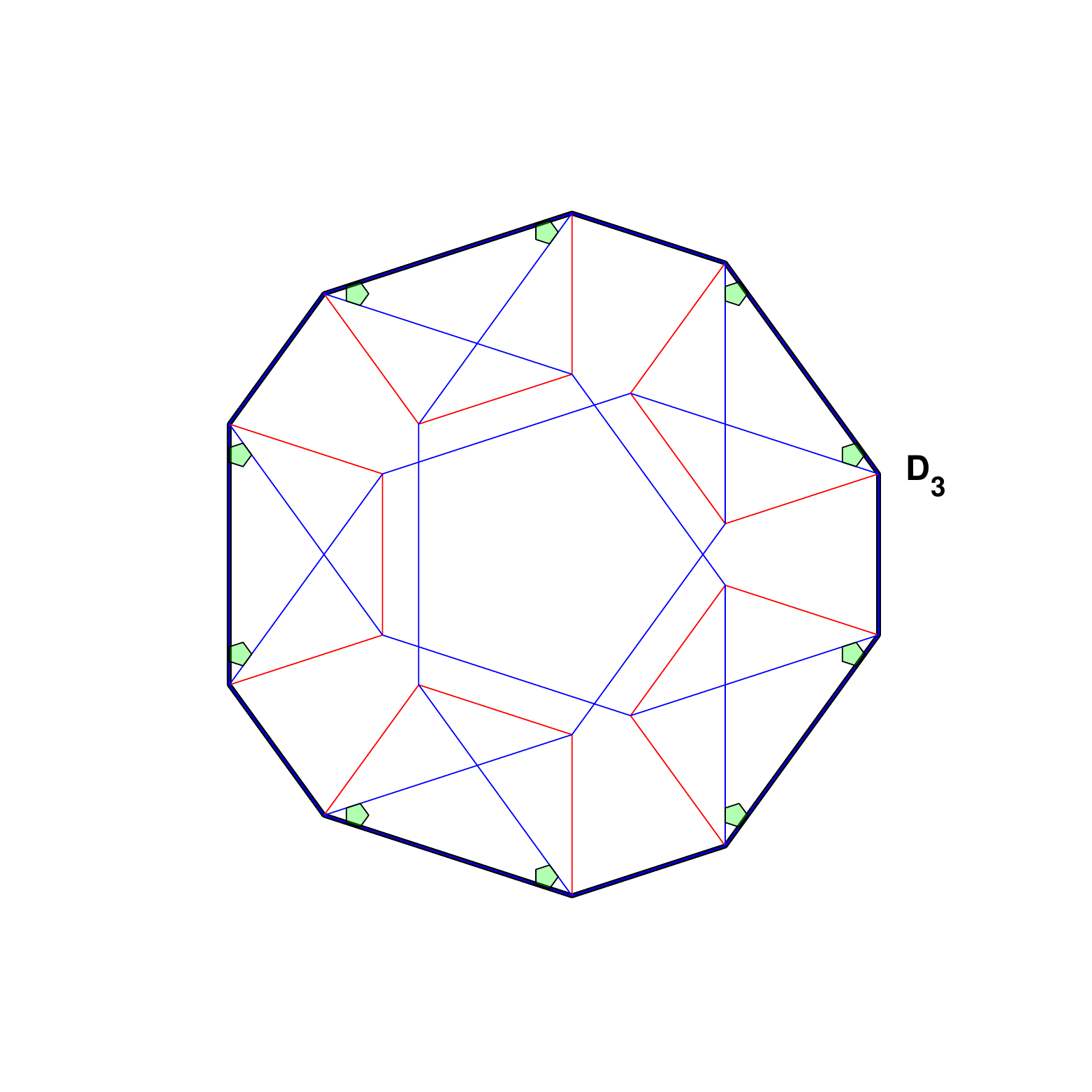}

    \caption{Allowed regions for type-O2 LS, for $\Gamma=0,0.1,0.2,0.33$ on the upper row and for $\Gamma=0.43,0.5$ in the lower row. There is an allowed type-02 state with D vertices in $D_1$ for $\Gamma=0.43,0.5$, thus both polygons are pictured. At $\Gamma=0.5$ inversion maps $D_1$ to $D_3$, so this doubling is expected. More detailed evolution is given in the supplementary material. }
    \label{fig:Type2OddScan}
\end{figure}

\begin{figure}[!htb]
    \centering
    \includegraphics[clip,width=0.24\textwidth]{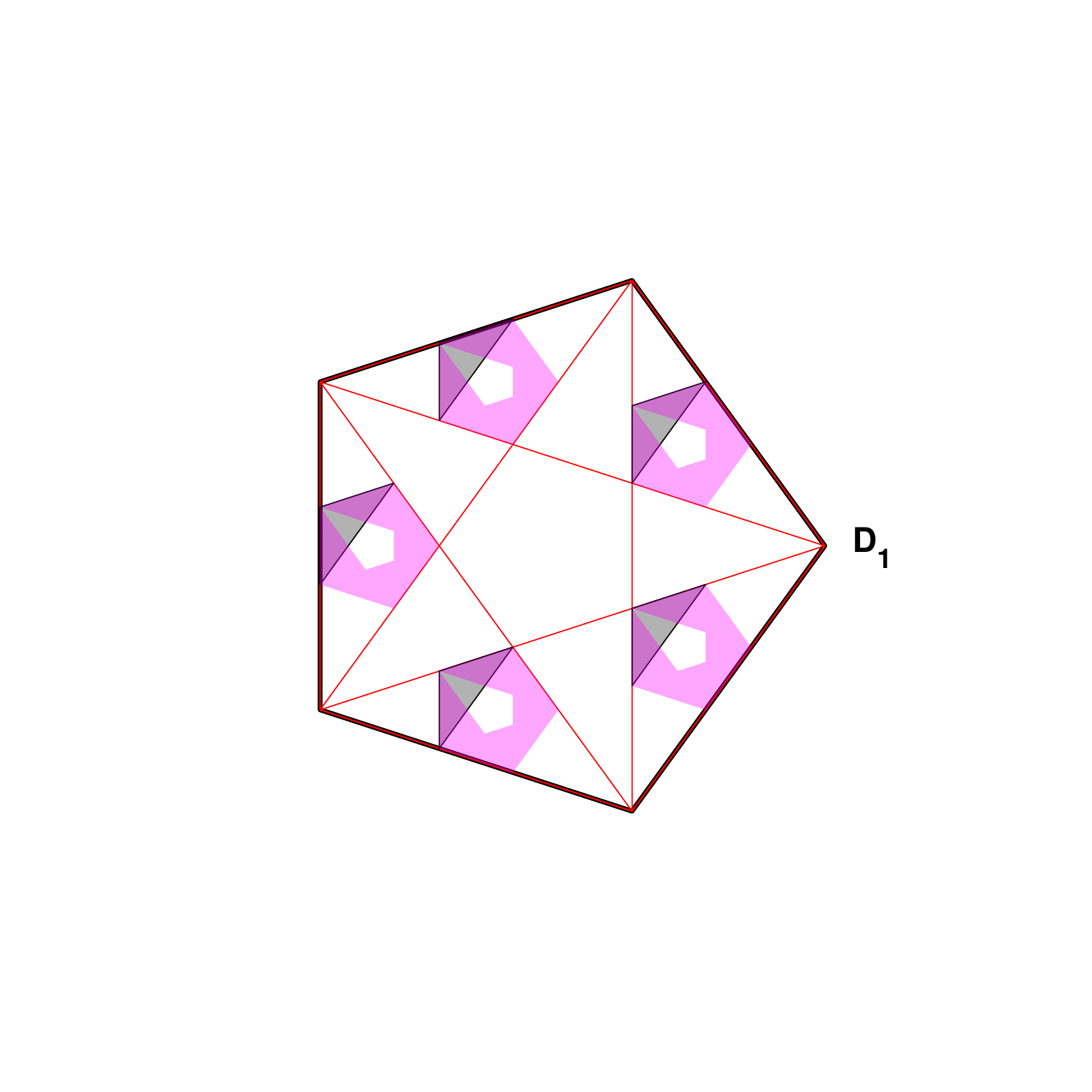}
    \includegraphics[clip,width=0.24\textwidth]{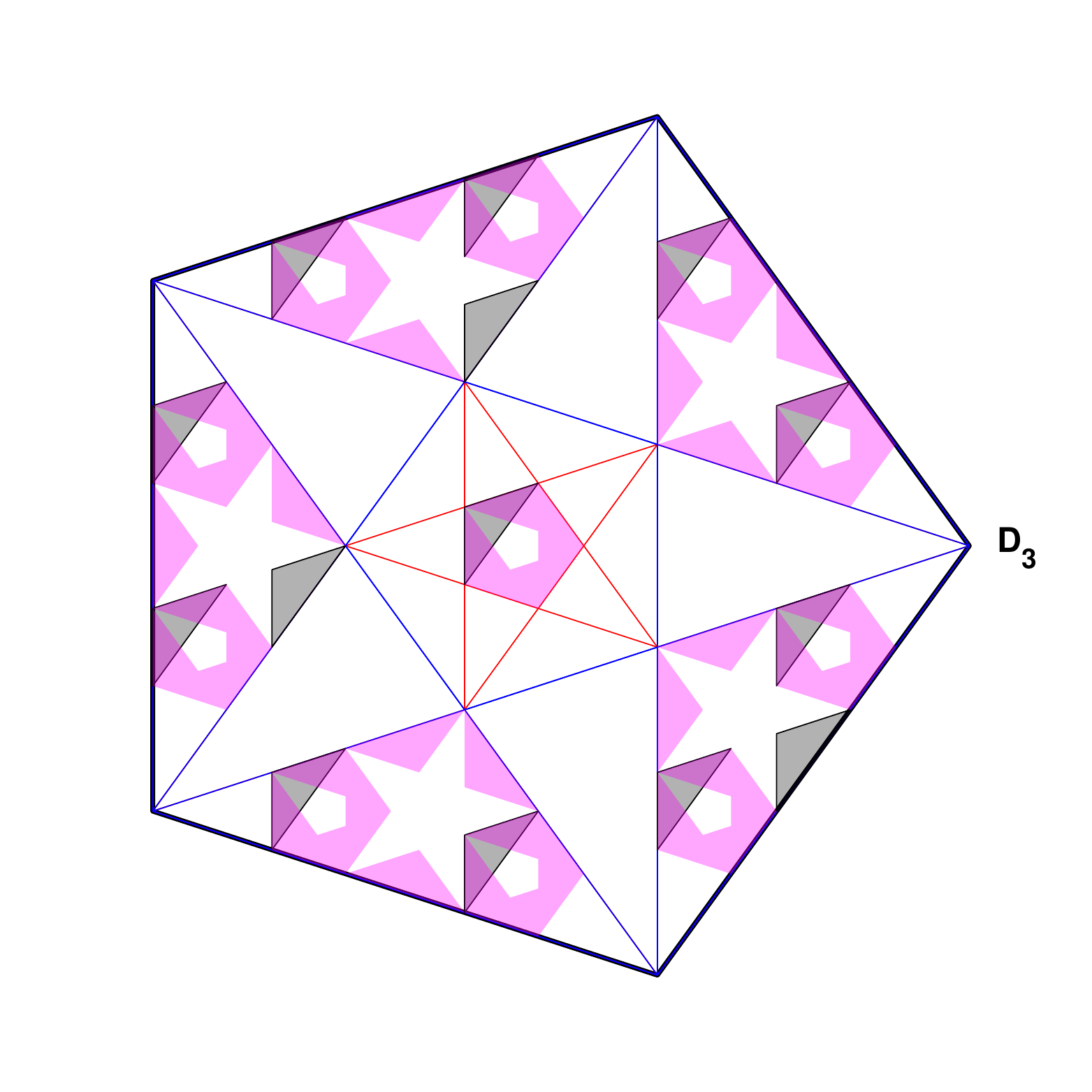}
    \includegraphics[clip,width=0.24\textwidth]{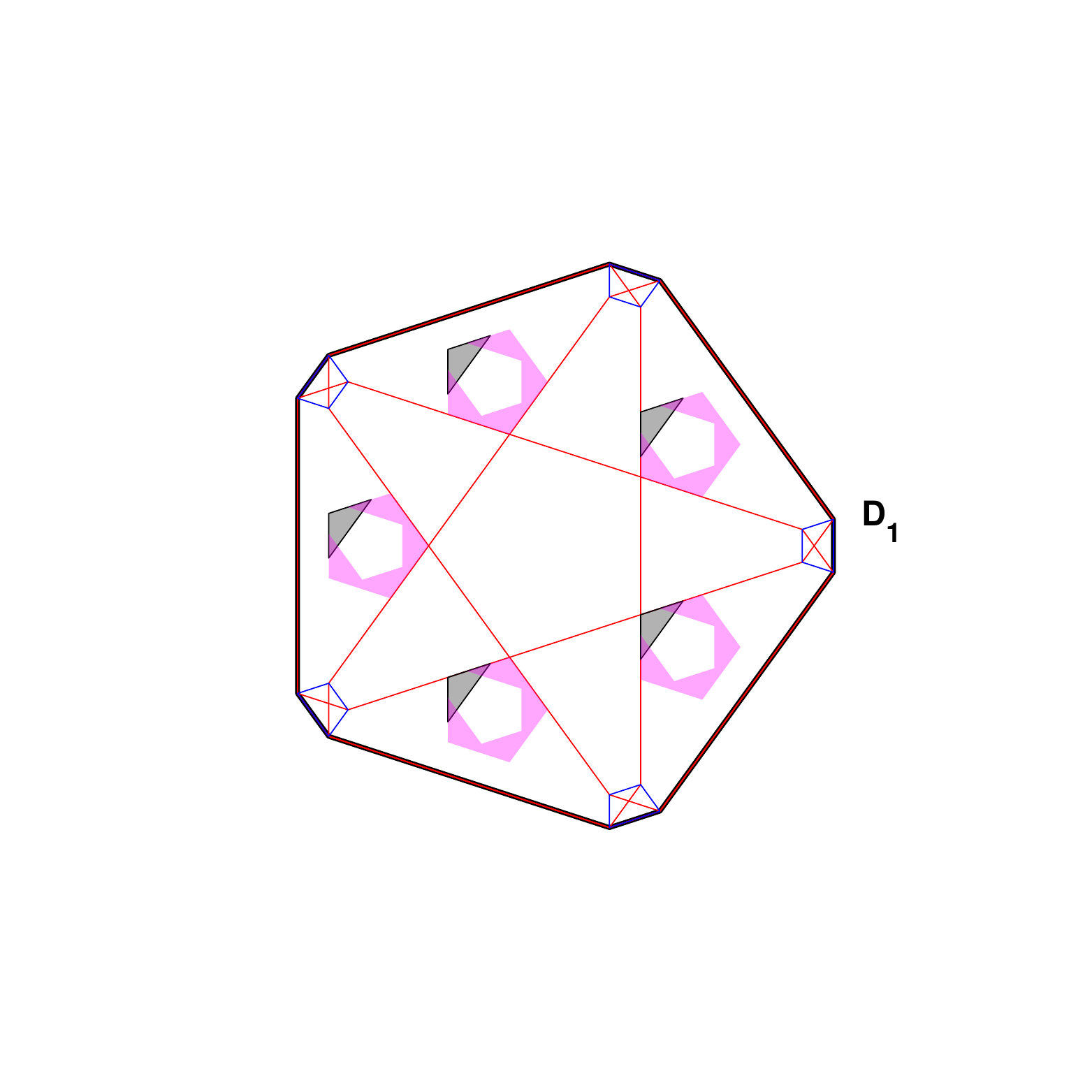}
    \includegraphics[clip,width=0.24\textwidth]{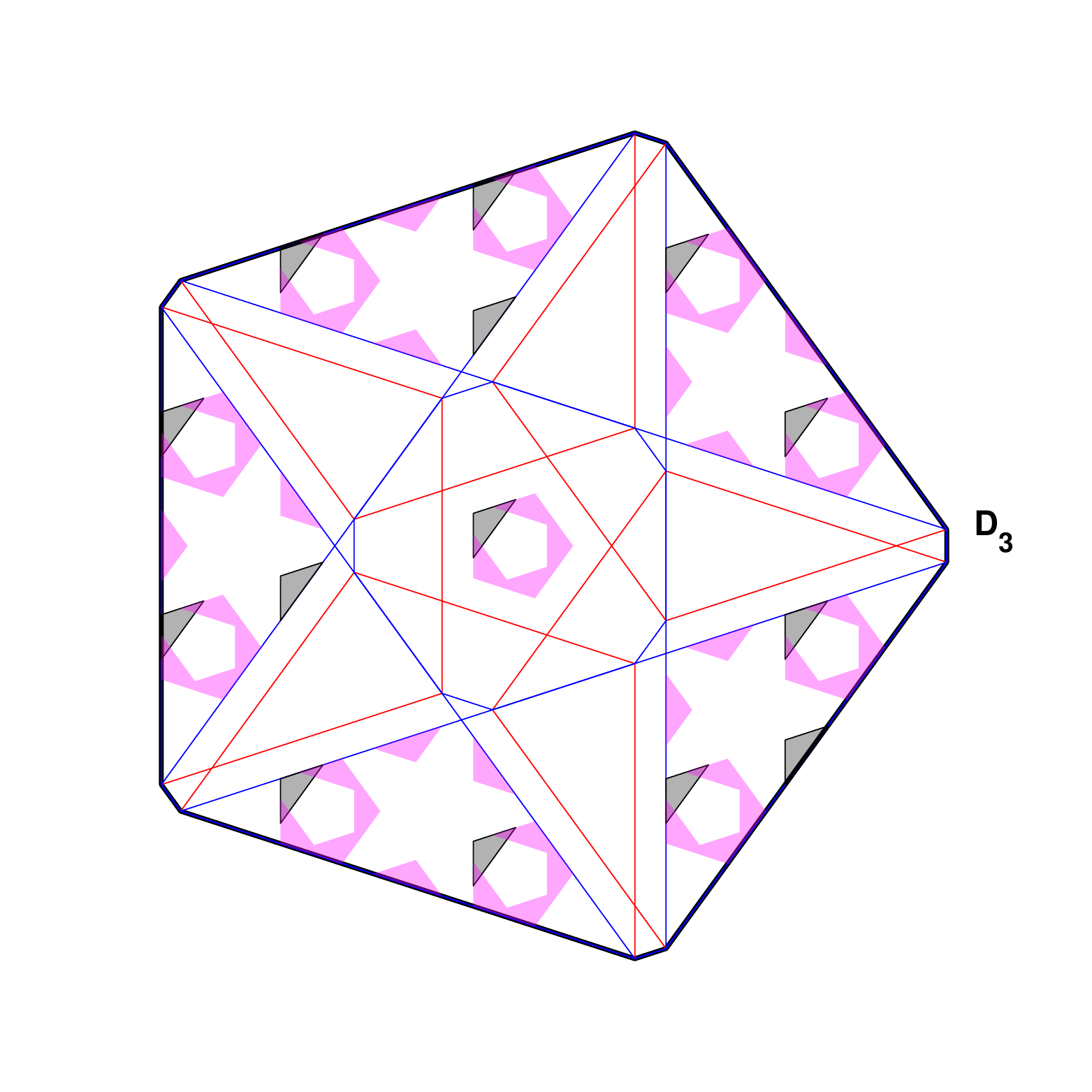}
    \includegraphics[clip,width=0.24\textwidth]{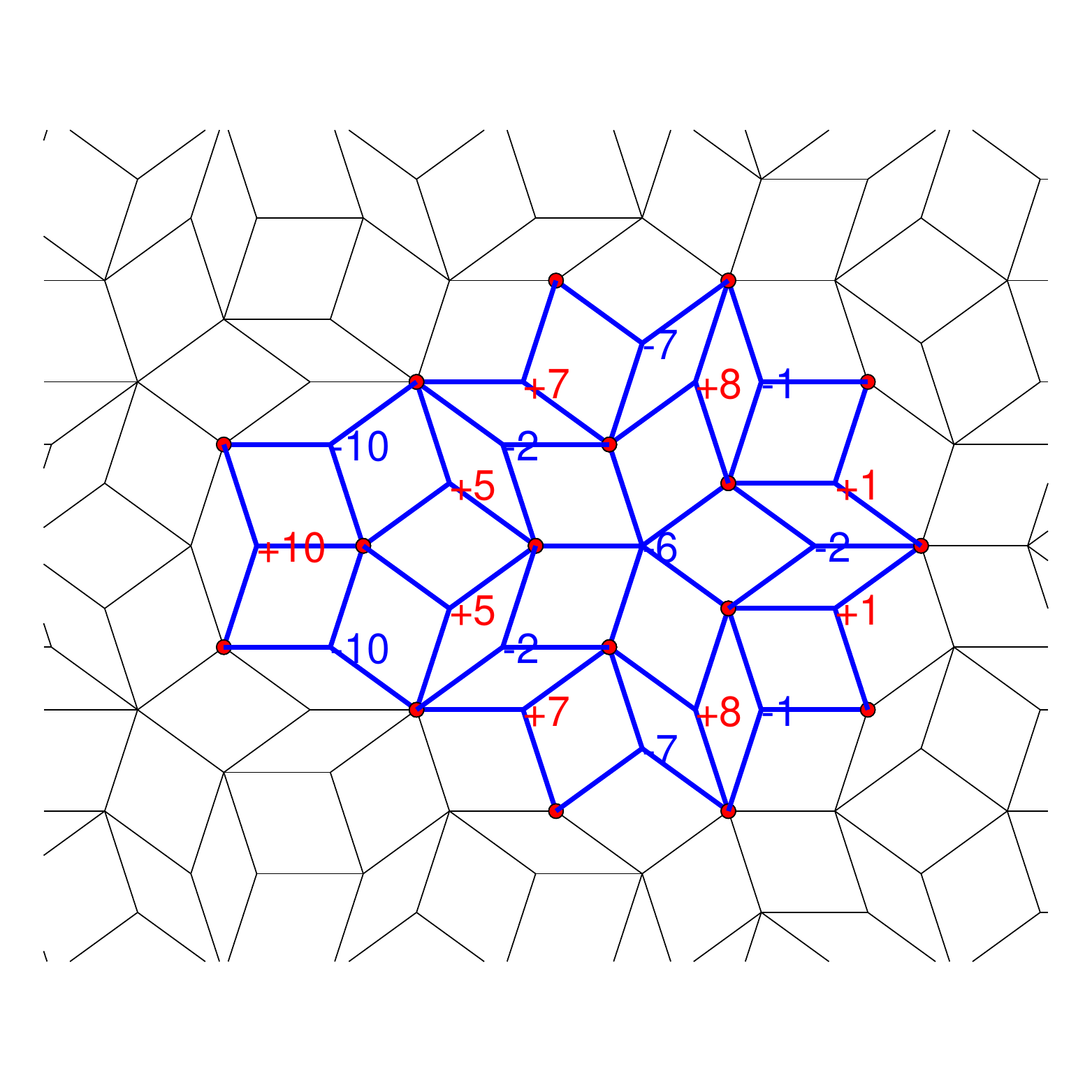}
    \includegraphics[clip,width=0.24\textwidth]{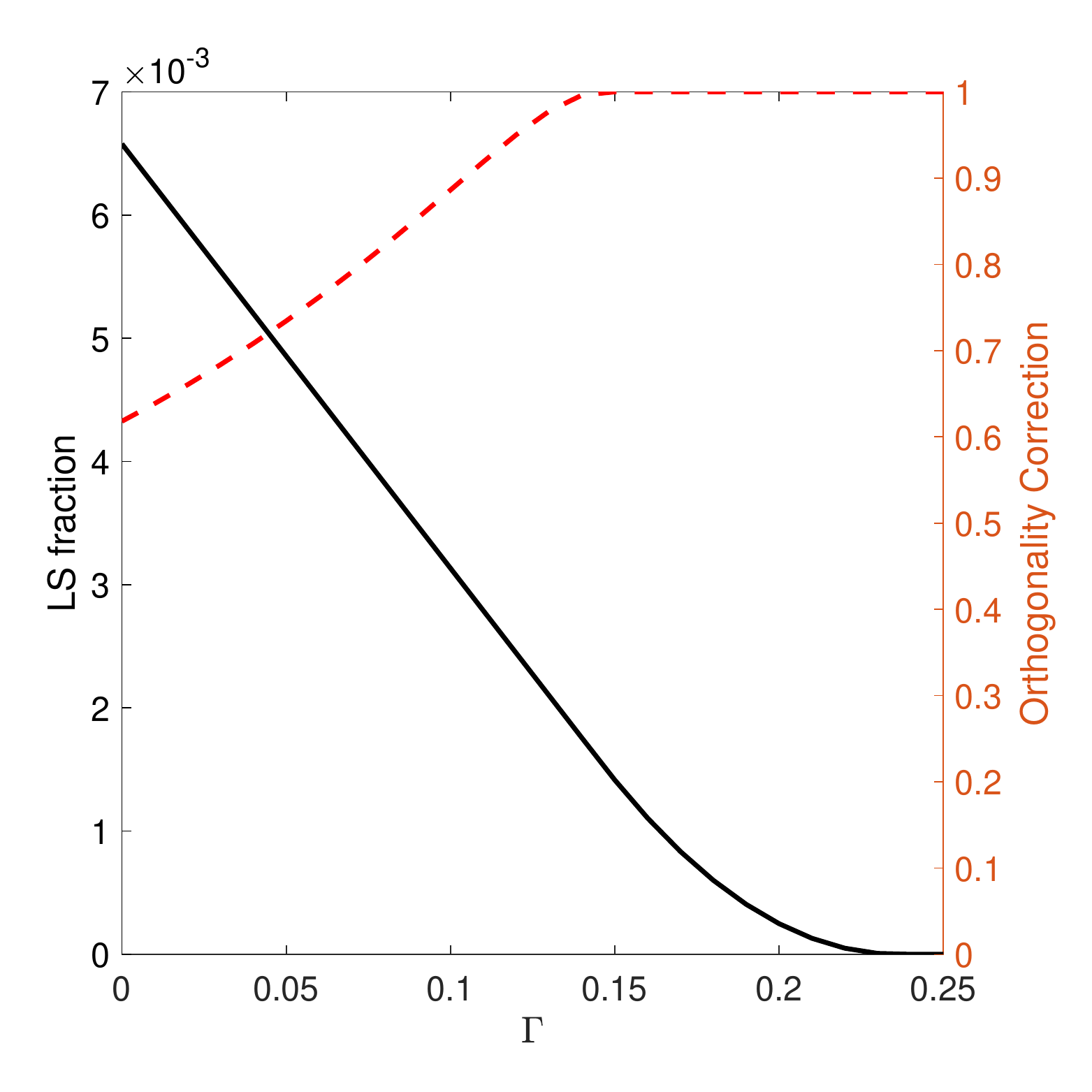}
    
    \caption{(a) Type-05 allowed regions, shown for both a single orientation and all five possible orientations, at $\Gamma=0.0$. (b) The same for $\Gamma=0.1$, showing that the allowed regions and the overlaps of rotated allowed regions change with $\Gamma$. (c) Real space configuration for the type-O5 LS. (d) LS Frequency and orthogonality correction for the type-O5 LS. Orthogonality correction is defined as the ratio of the LS frequency to the total covered area by all orientations of a given LS.  }
    \label{fig:Type5}
\end{figure}

\begin{figure}[!htb]
    \centering
    \includegraphics[clip,width=0.31\textwidth]{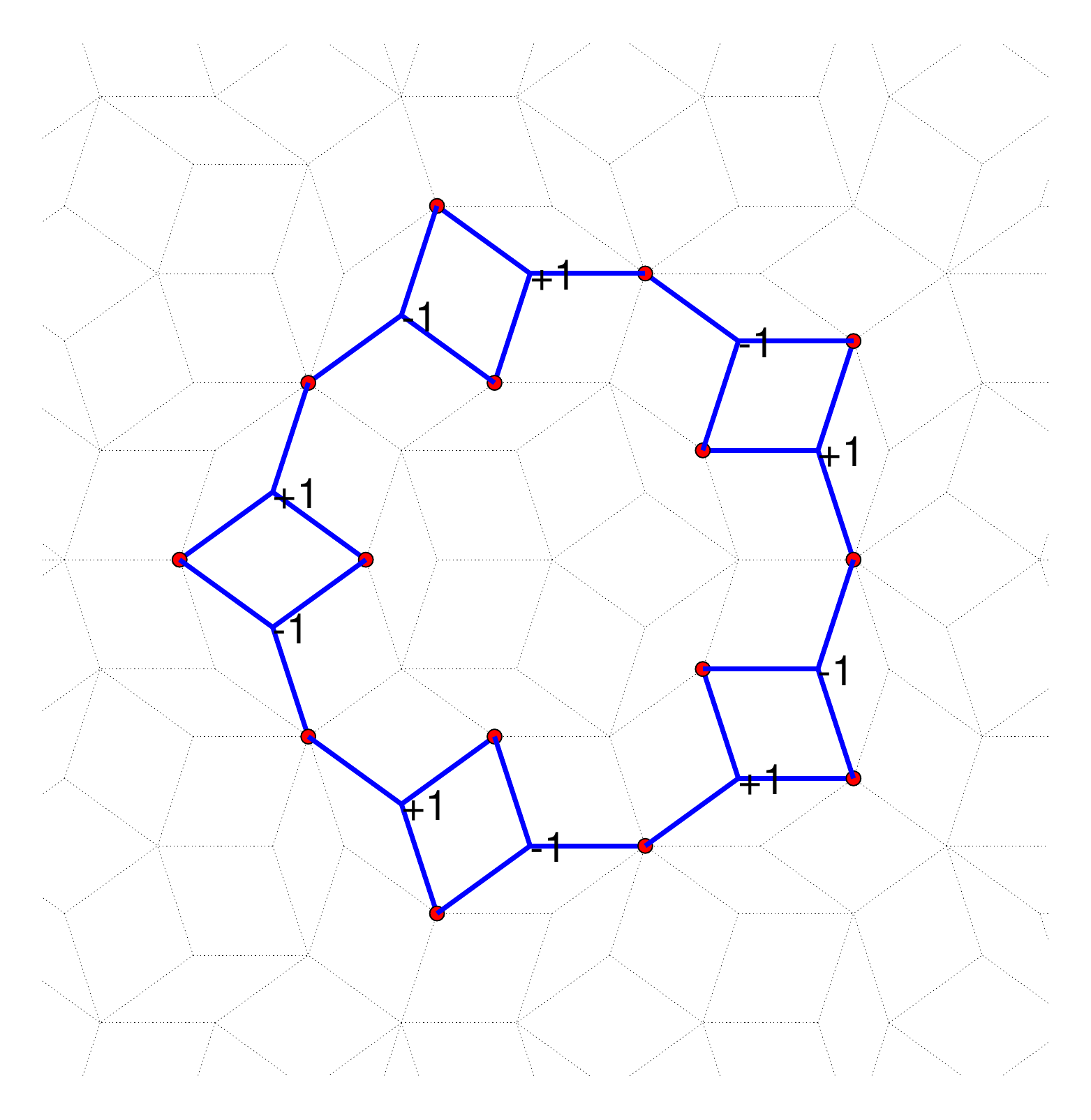}
    \includegraphics[clip,width=0.31\textwidth]{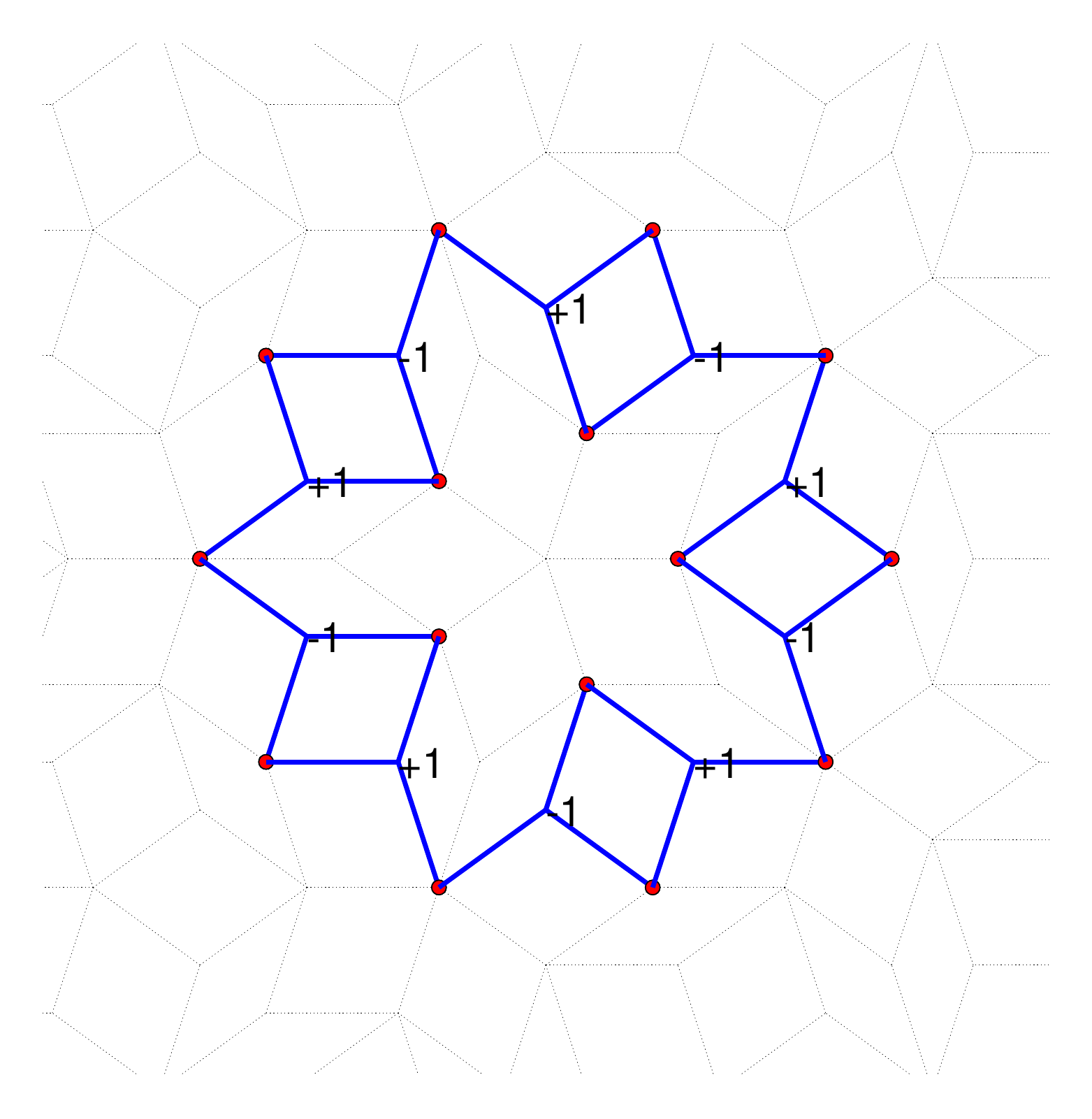}
    \includegraphics[clip,width=0.31\textwidth]{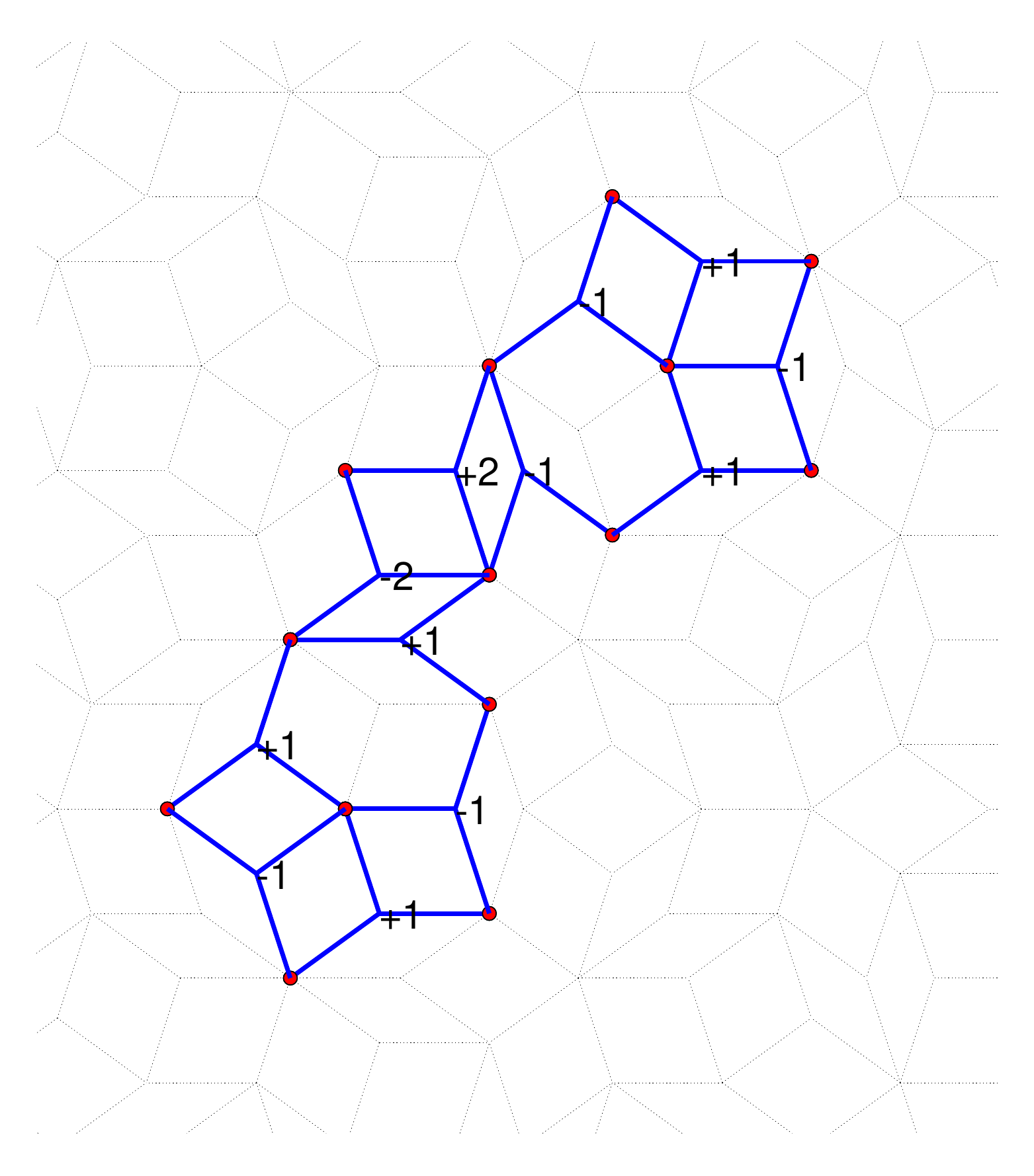}
    \includegraphics[clip,width=0.31\textwidth]{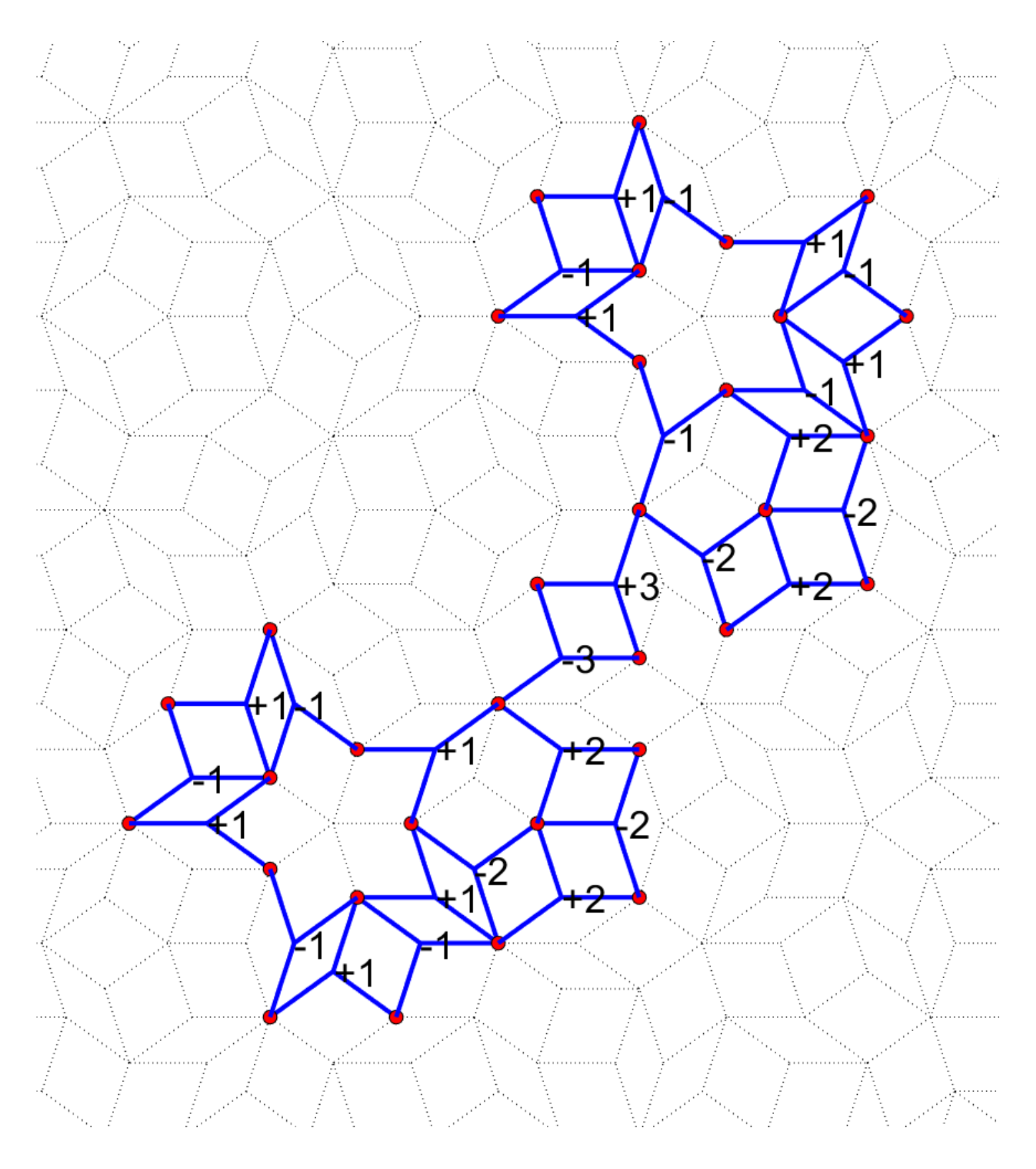}
    \includegraphics[clip,width=0.31\textwidth]{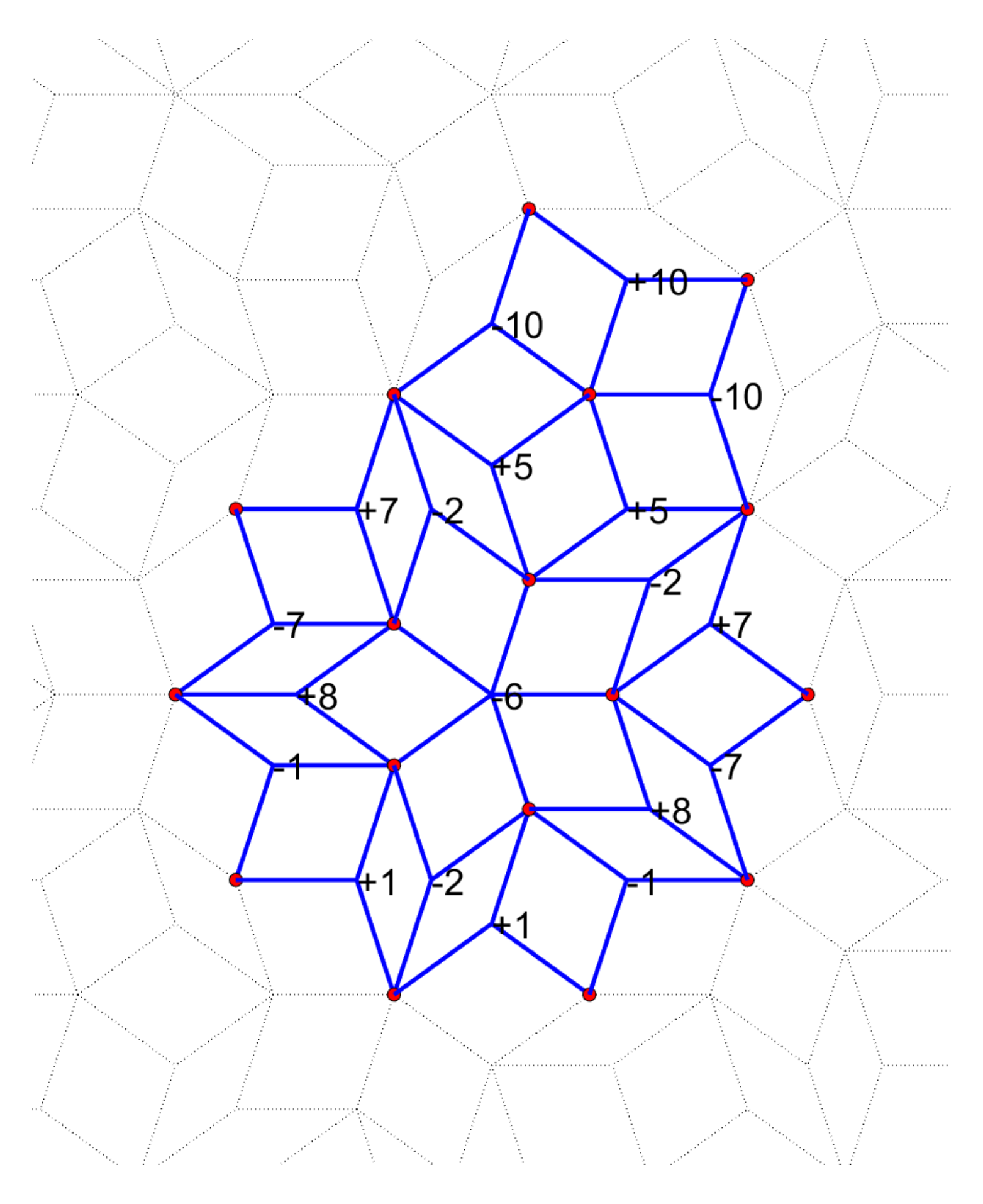}
    \includegraphics[clip,width=0.31\textwidth]{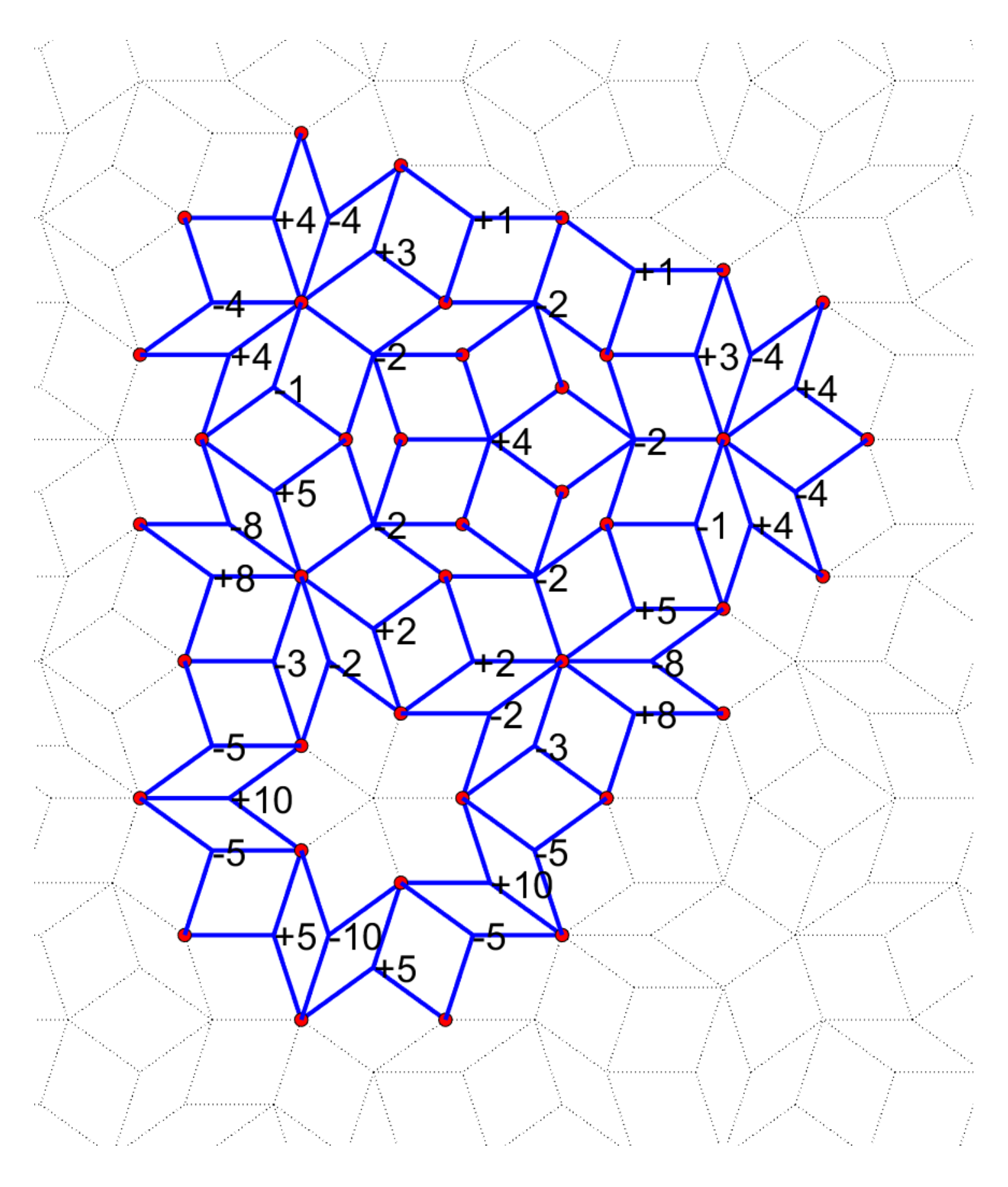}
    
    \caption{The six LS types defined on the PL. All six appear both on the even and the odd sublattice. }
    \label{fig:First6RealSpace}
\end{figure}
\begin{figure}[!htb]
    \centering
    \includegraphics[clip,width=0.48\textwidth]{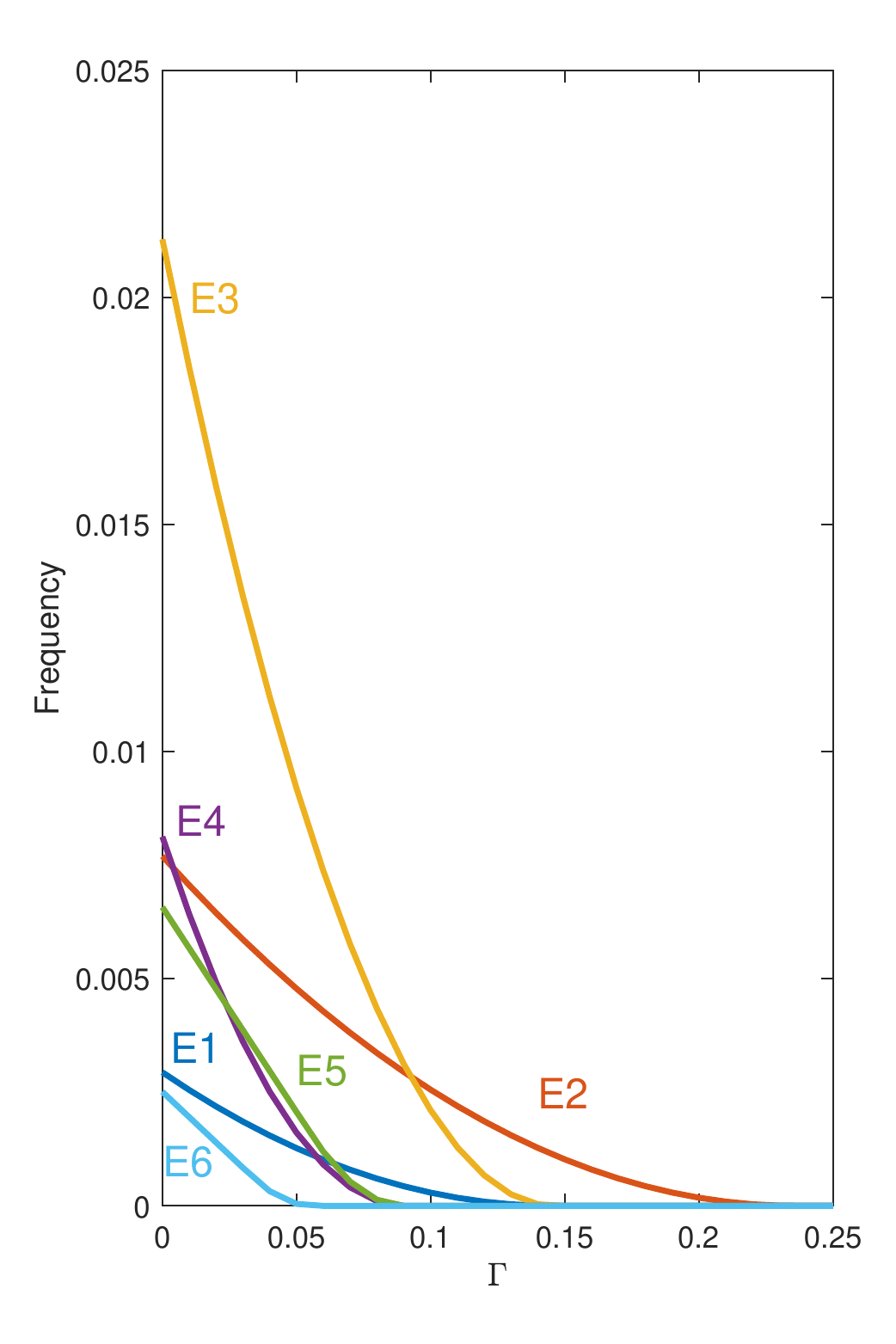}
    \includegraphics[clip,width=0.48\textwidth]{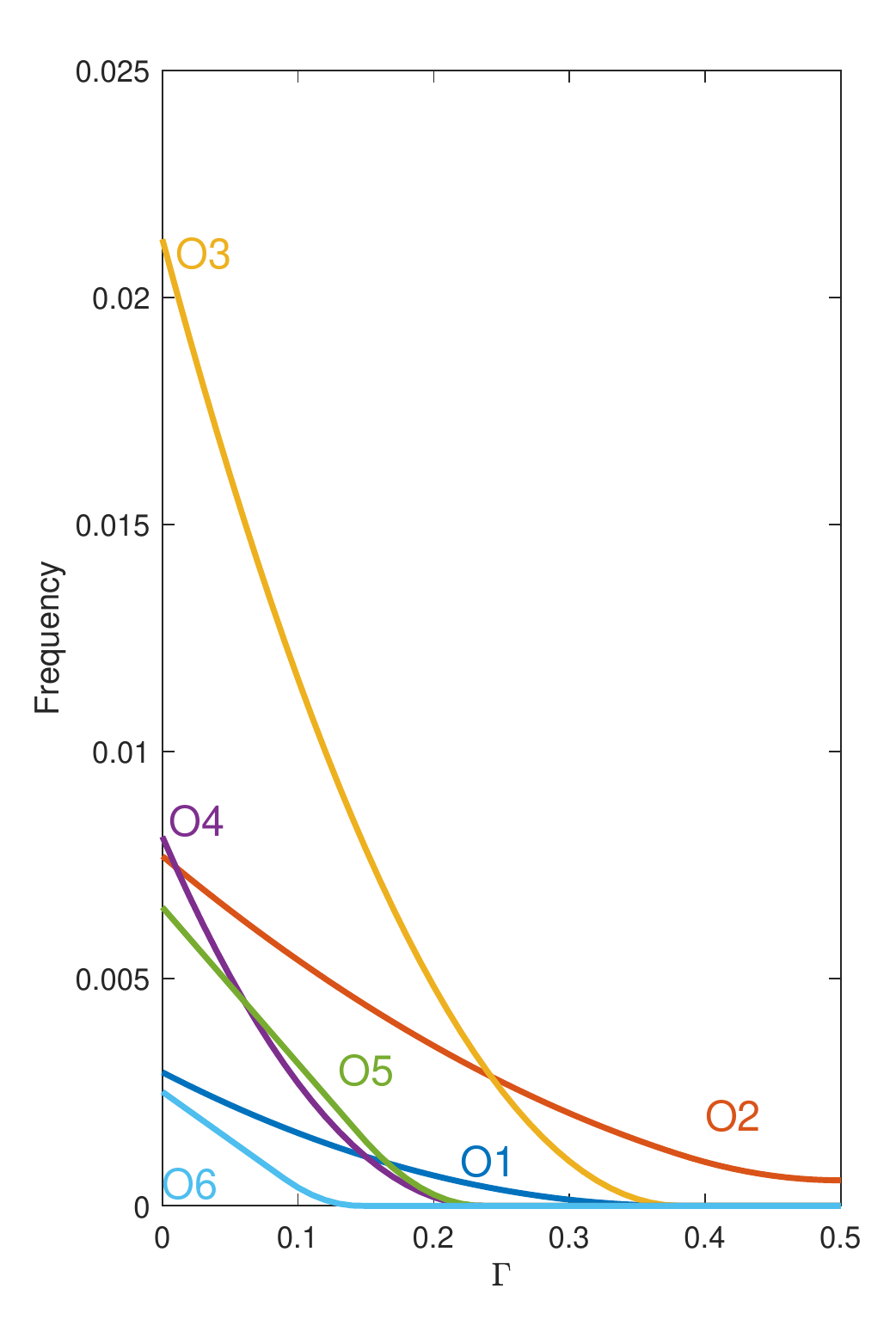}
    \caption{The frequencies of the six LS types of the PL as a function $\Gamma$ in both sublattices. Notice that their frequencies decrease faster on the even sublattice compared to the odd sublattice.}
    \label{fig:First6LSFrequencies}
\end{figure}
\begin{figure}[!htb]
    \centering
    \includegraphics[clip,width=0.31\textwidth]{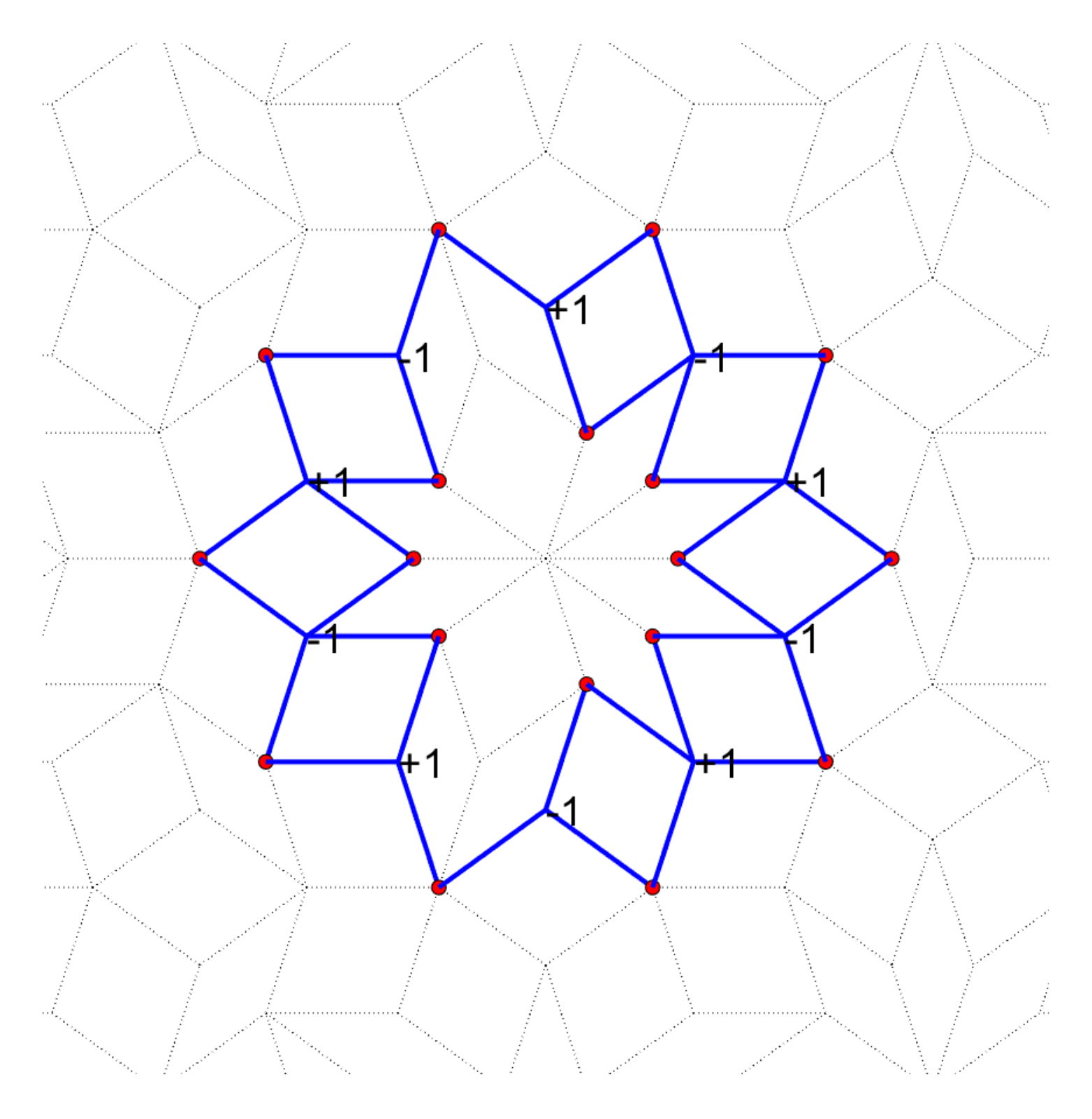}
    \includegraphics[clip,width=0.31\textwidth]{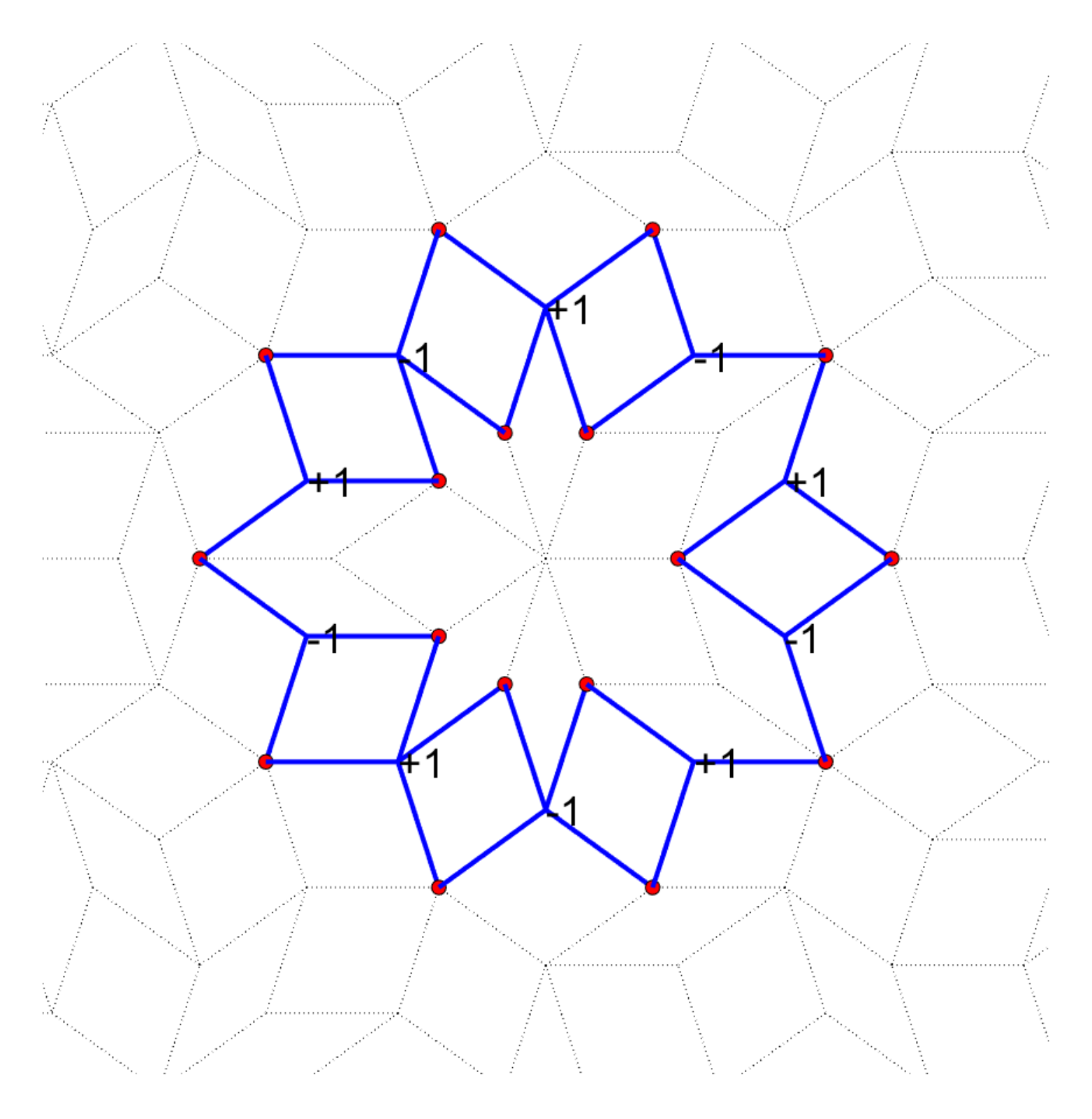}
    \includegraphics[clip,width=0.31\textwidth]{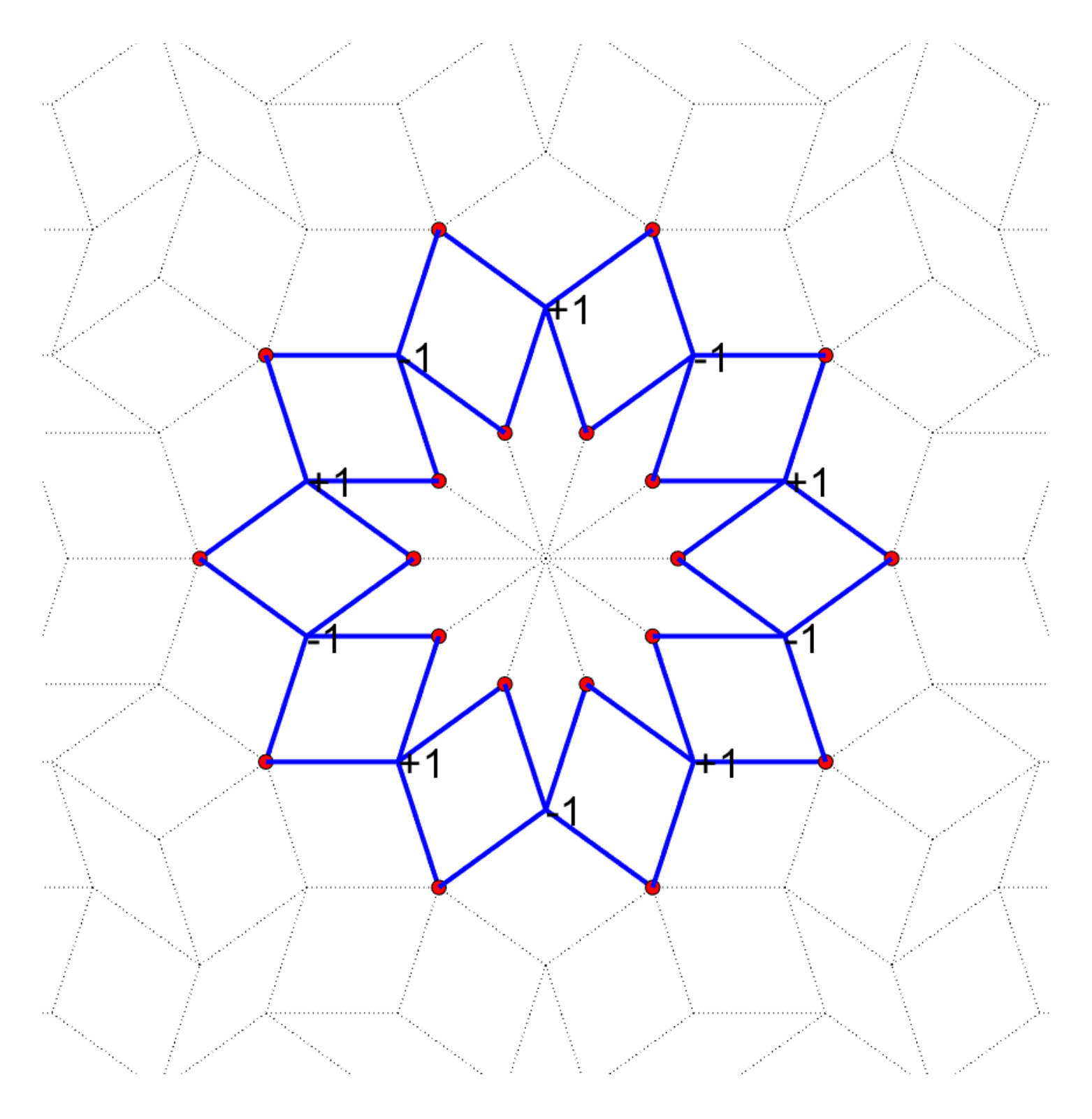}
    \includegraphics[clip,width=0.31\textwidth]{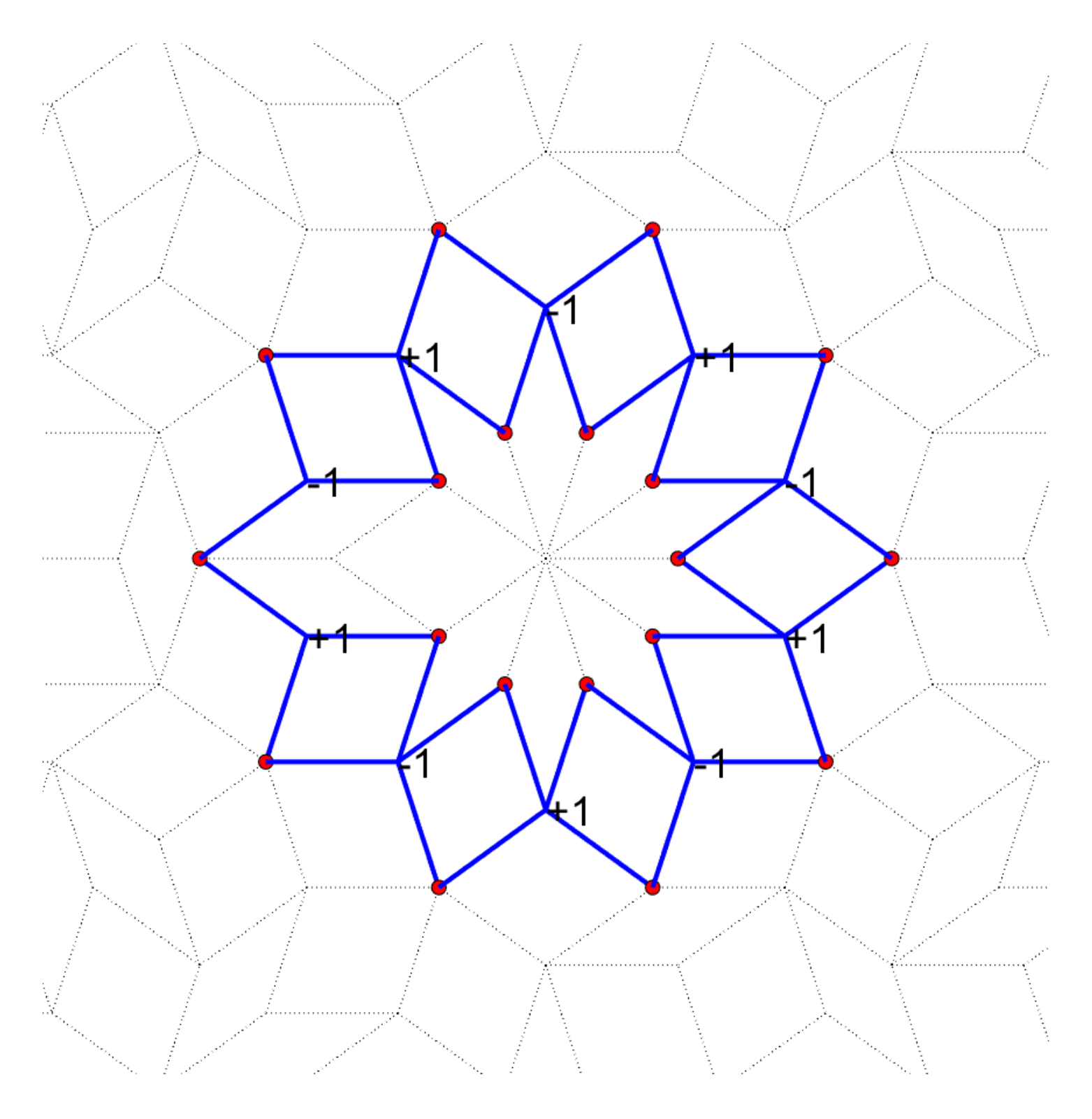}
    \includegraphics[clip,width=0.31\textwidth]{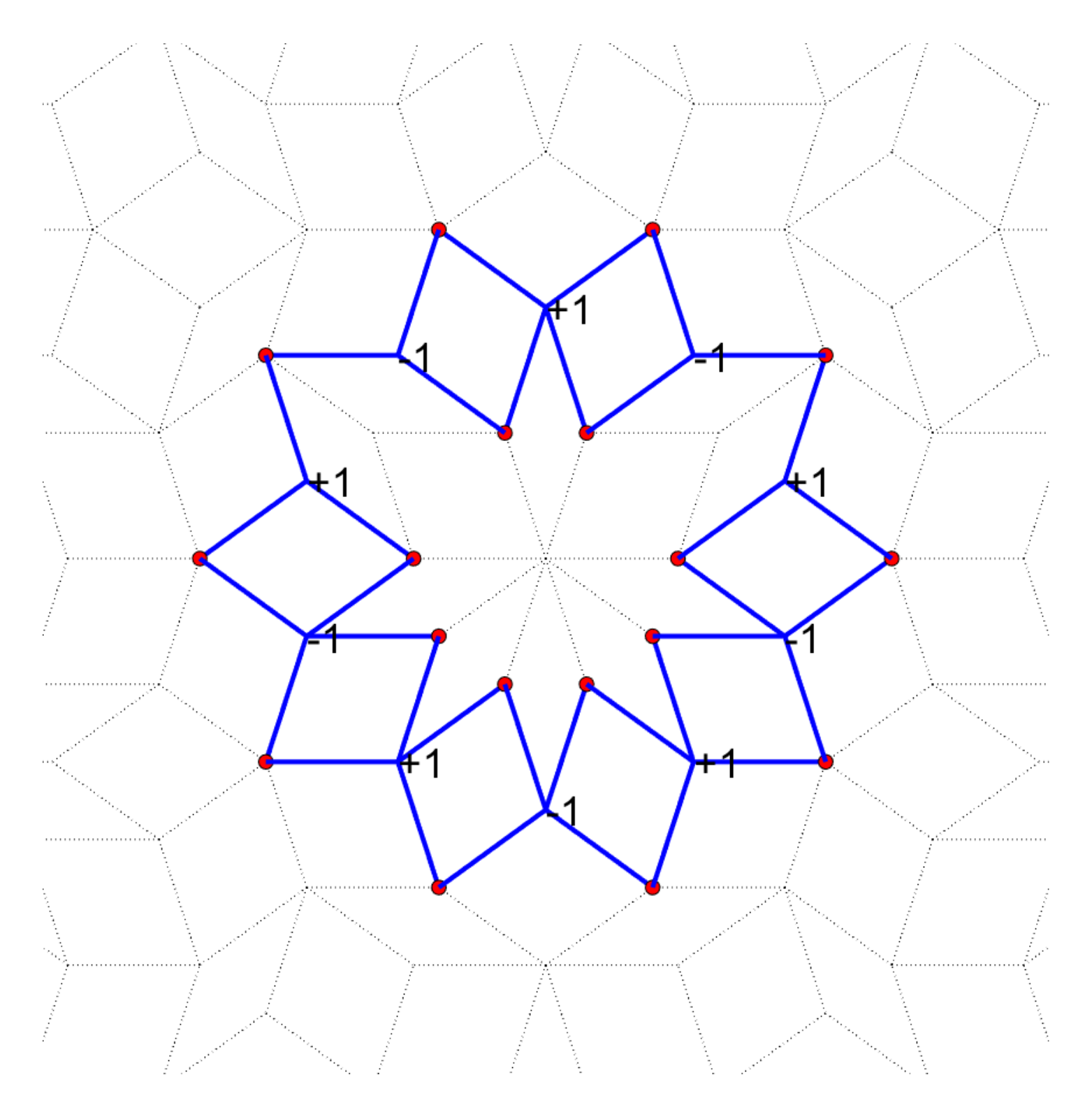}
    \includegraphics[clip,width=0.31\textwidth]{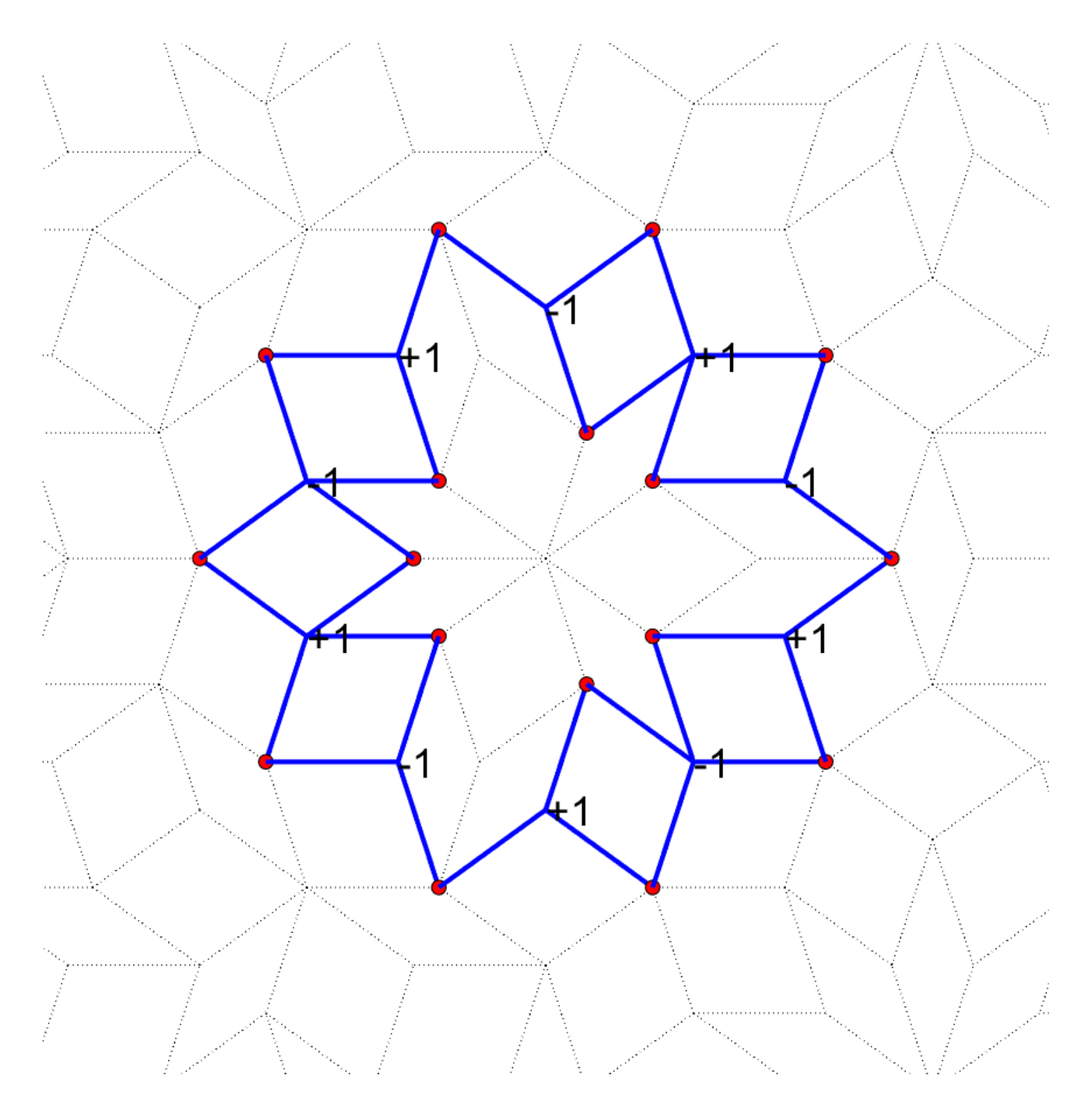}
    
    \caption{Real space structure for type-E7 to type-E12 on the even sublattice. }
    \label{fig:Next6EvenRealSpace}
\end{figure}

\begin{figure}[!htb]
    \centering
    \includegraphics[clip,width=0.48\textwidth]{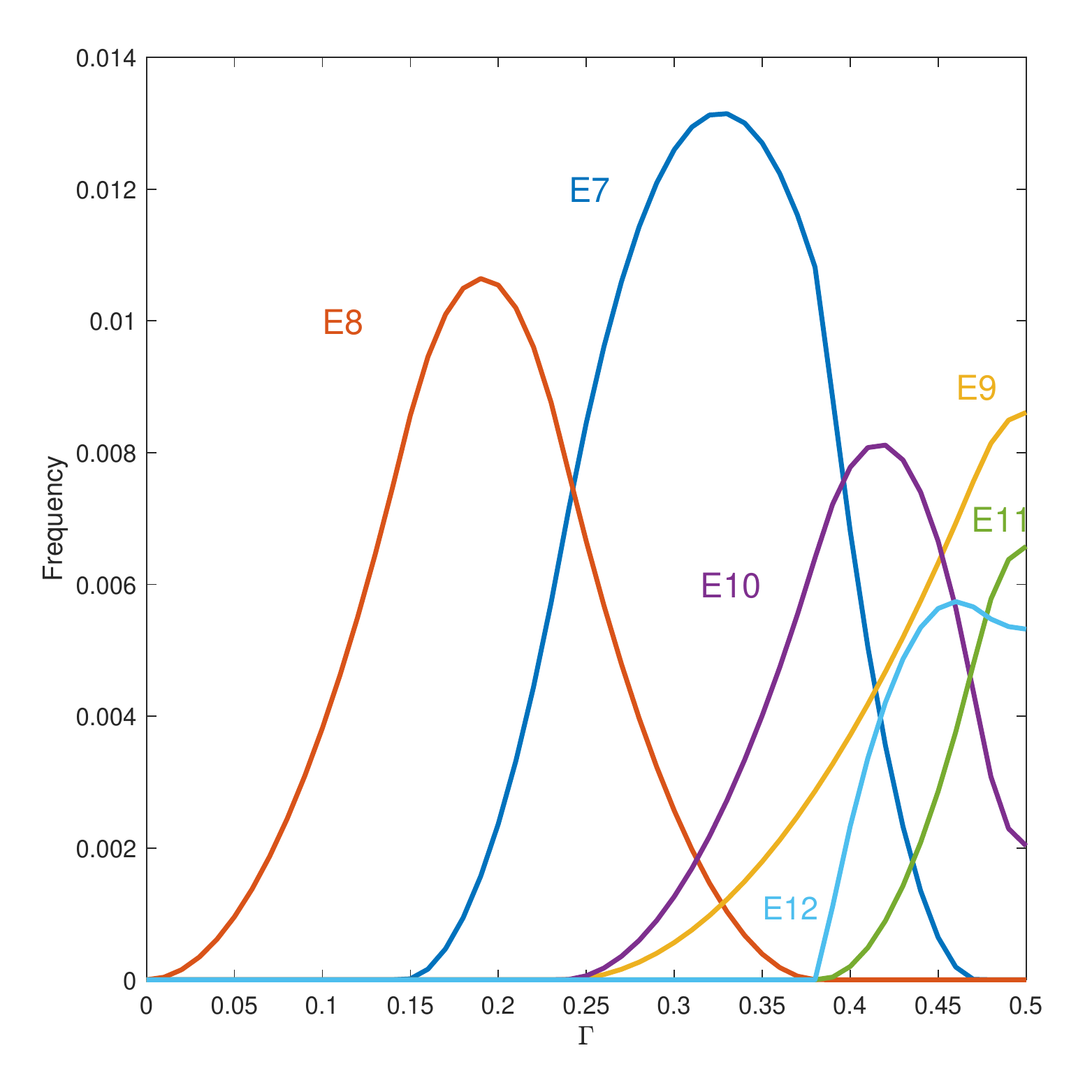}
    \caption{Frequencies of the LS types given in Fig.\ref{fig:Next6EvenRealSpace}.}
    \label{fig:NextSixEvenFrequencies}
\end{figure}

\begin{figure}[!htb]
    \centering
    \includegraphics[clip,width=0.31\textwidth]{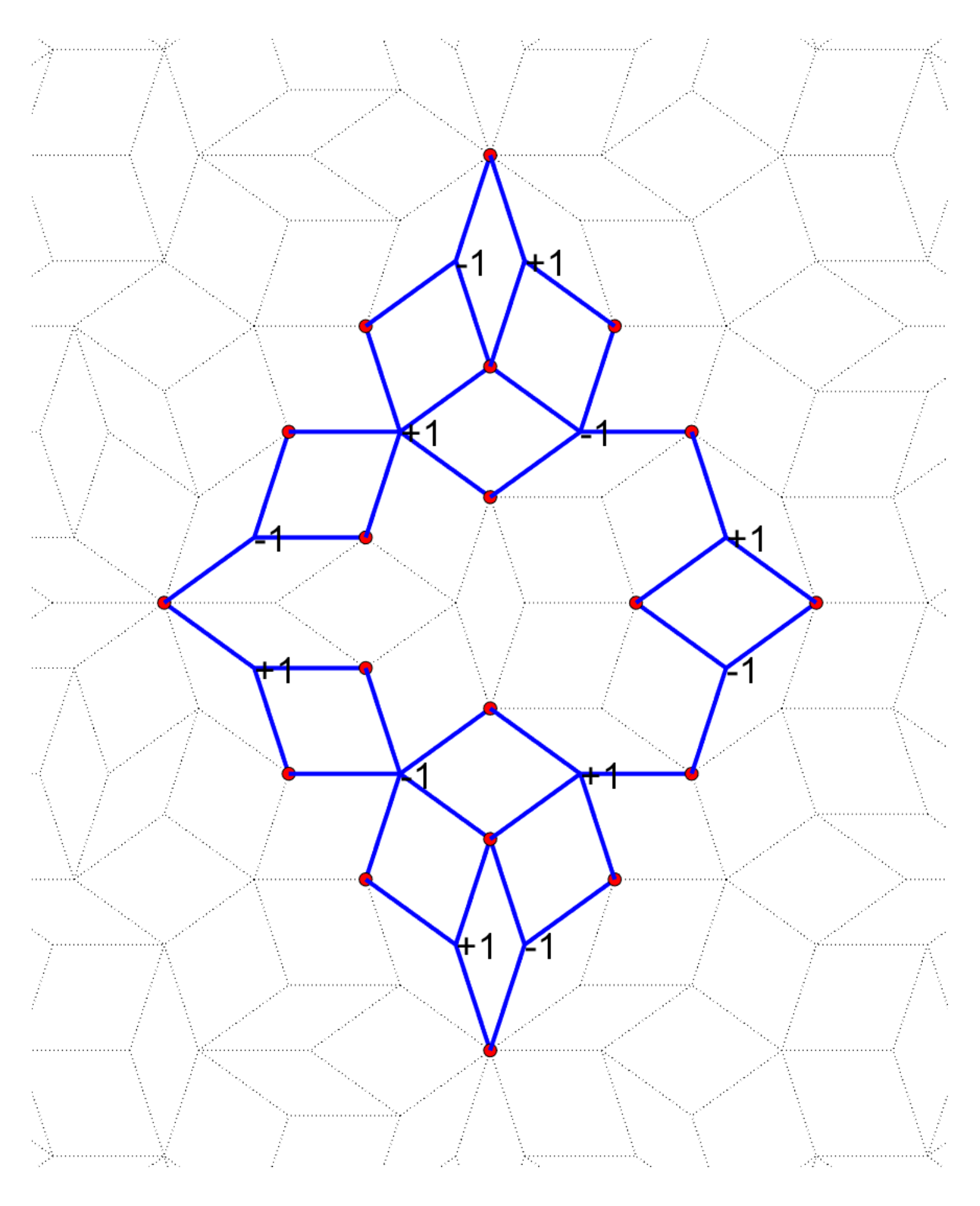}
    \includegraphics[clip,width=0.31\textwidth]{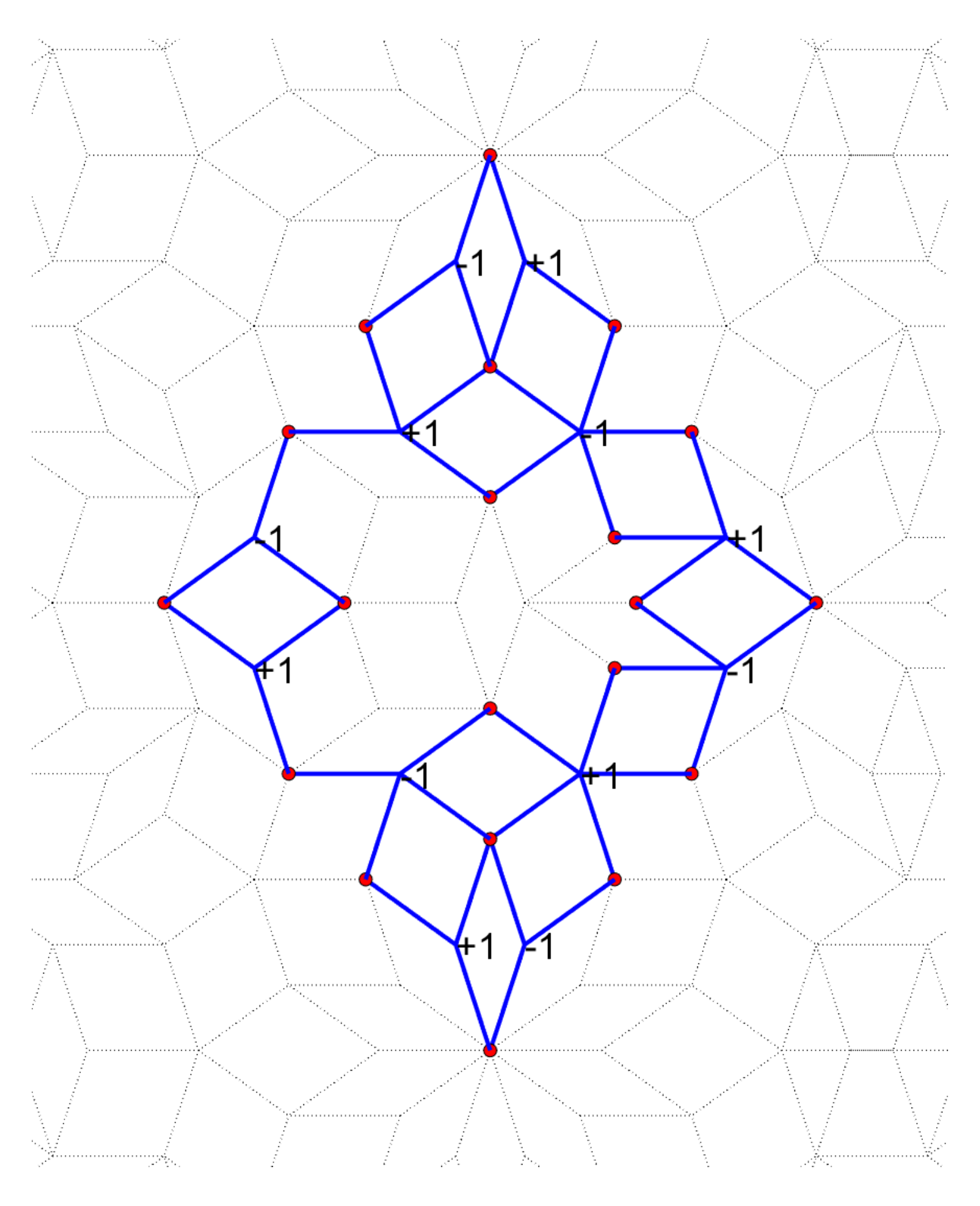}
    \includegraphics[clip,width=0.31\textwidth]{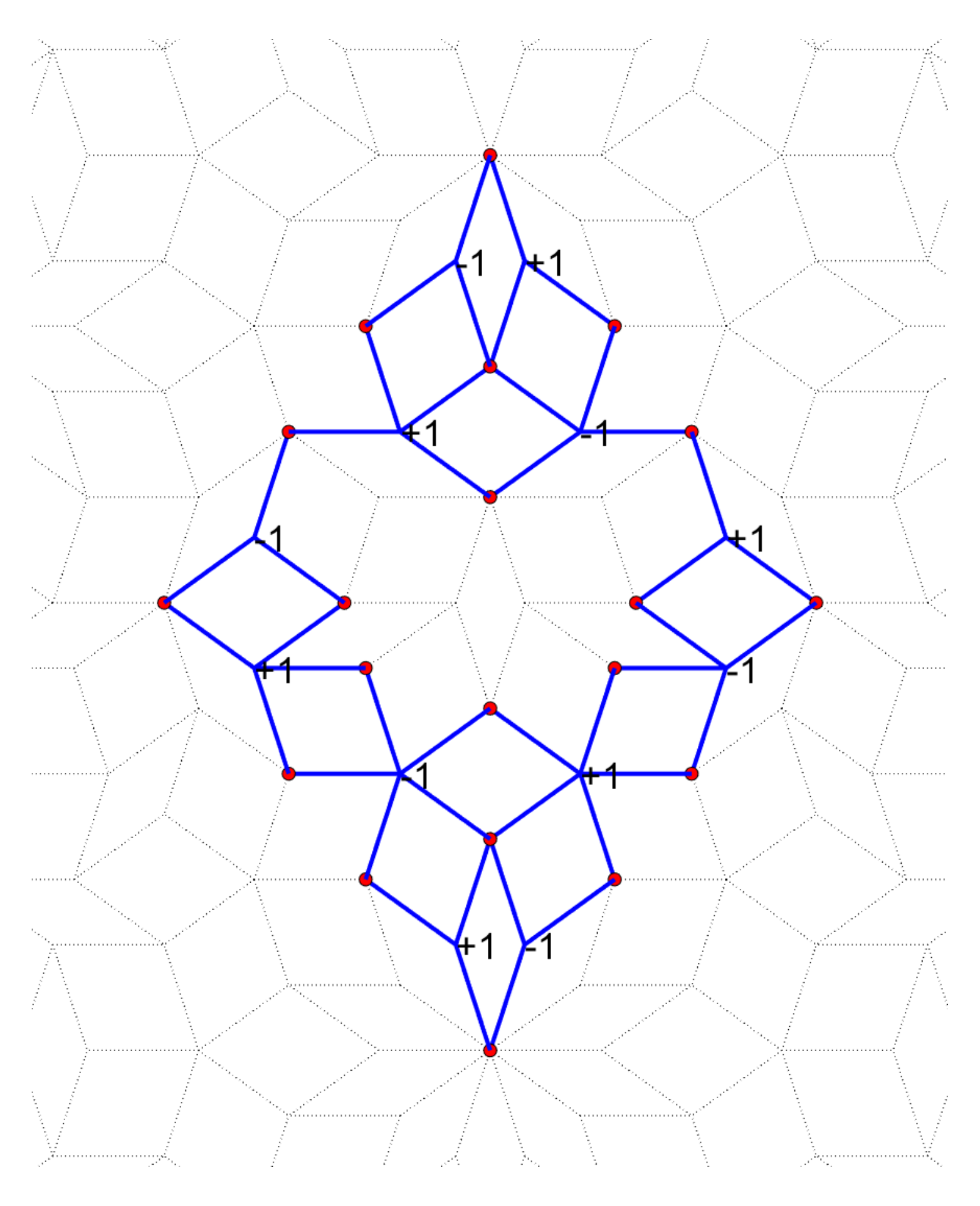}
    \includegraphics[clip,width=0.31\textwidth]{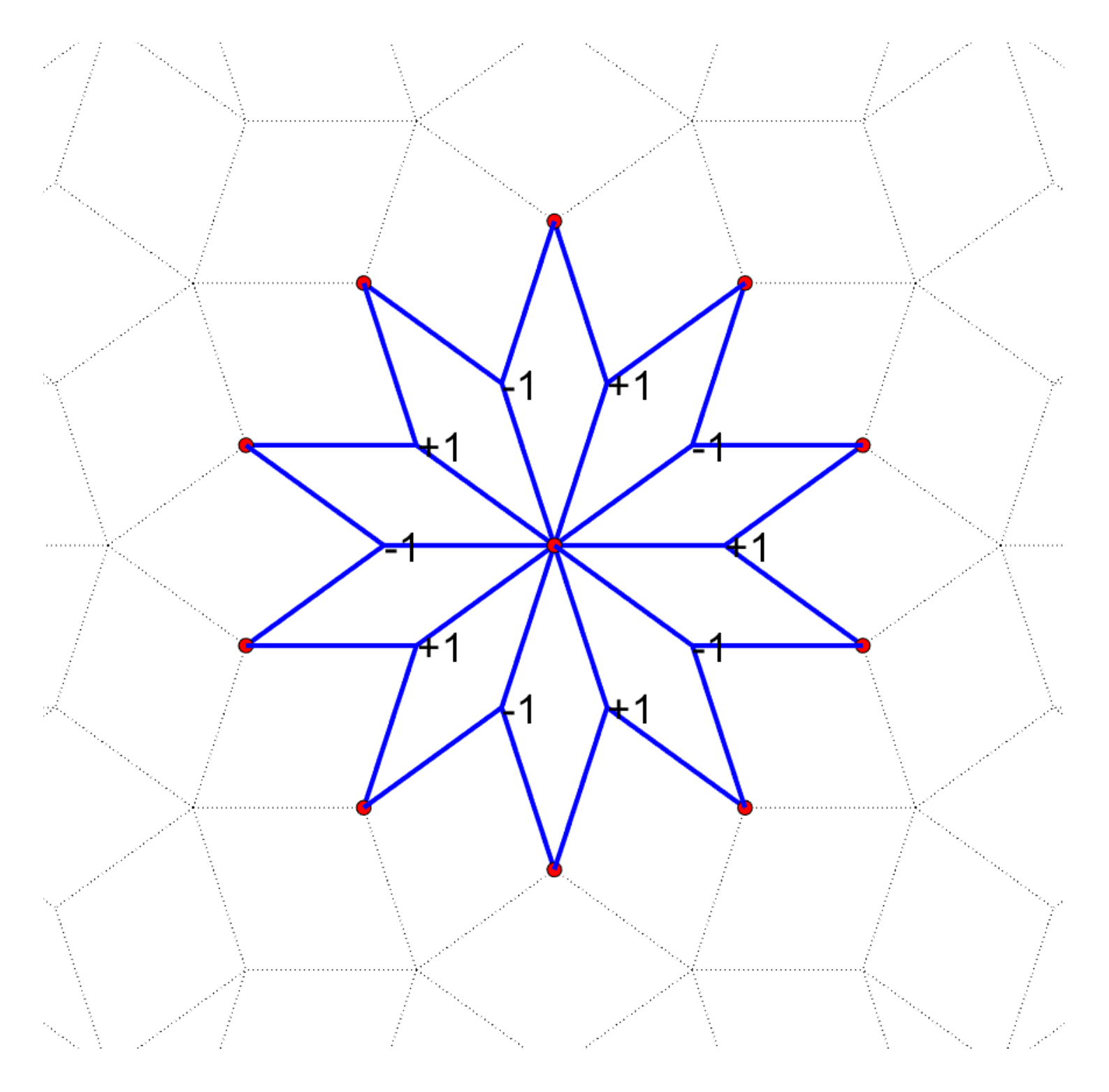}
    \includegraphics[clip,width=0.31\textwidth]{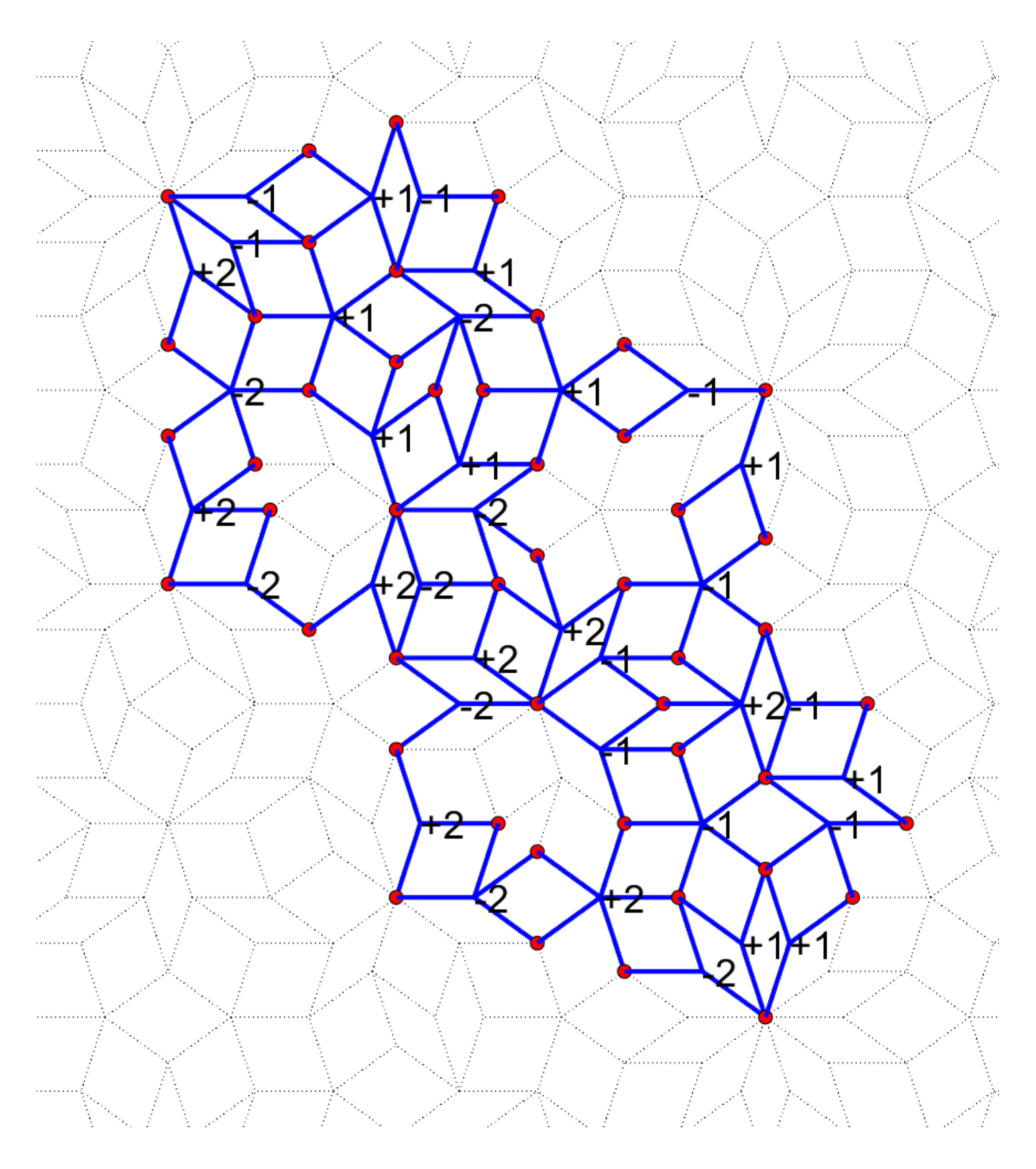}
    \includegraphics[clip,width=0.31\textwidth]{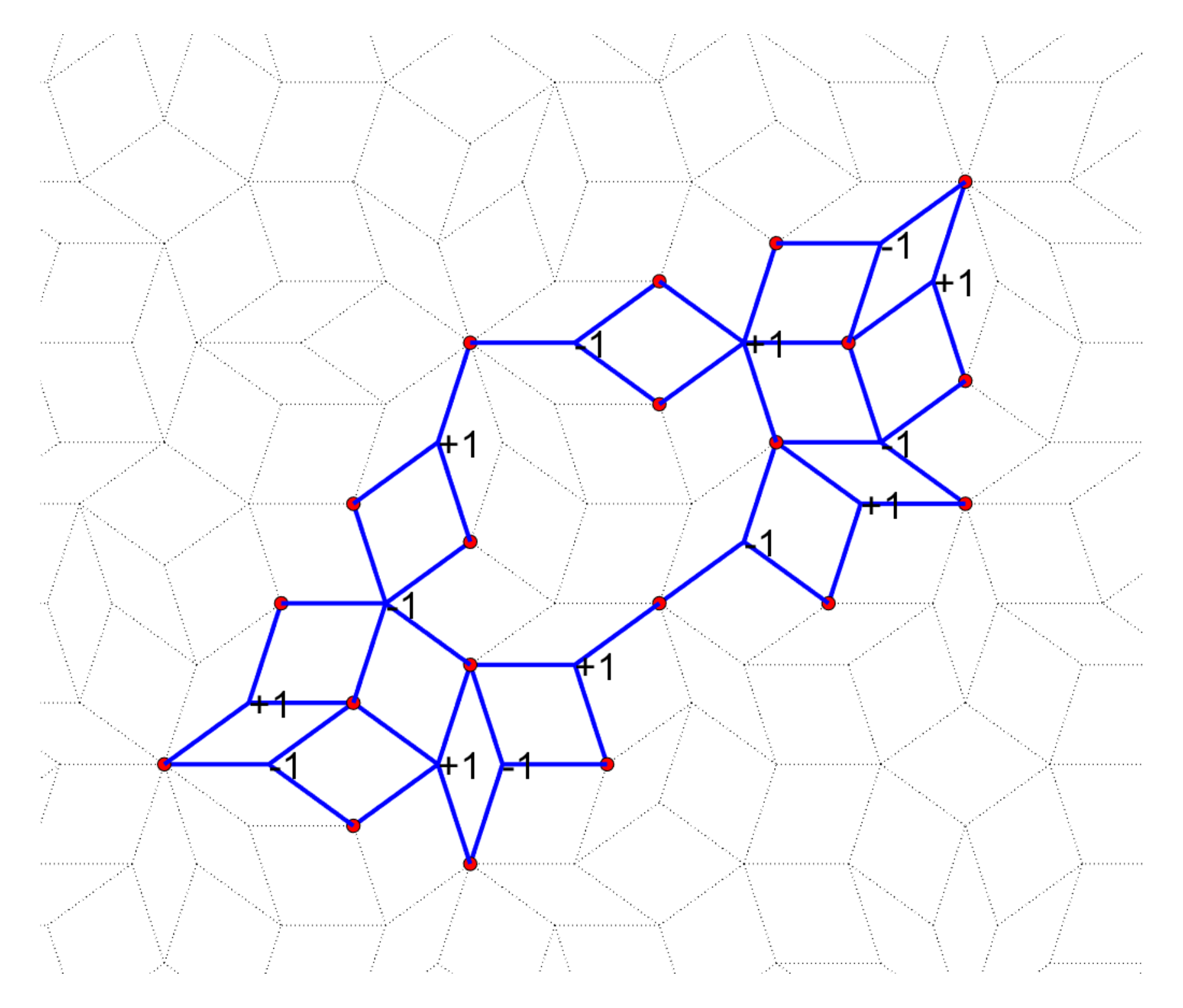}
    
    \caption{ Real space structure for type-O7 to type-O12 on the odd sublattice.}
    \label{fig:Next6OddRealSpace}
\end{figure}

\begin{figure}[!htb]
    \centering
    \includegraphics[clip,width=0.49\textwidth]{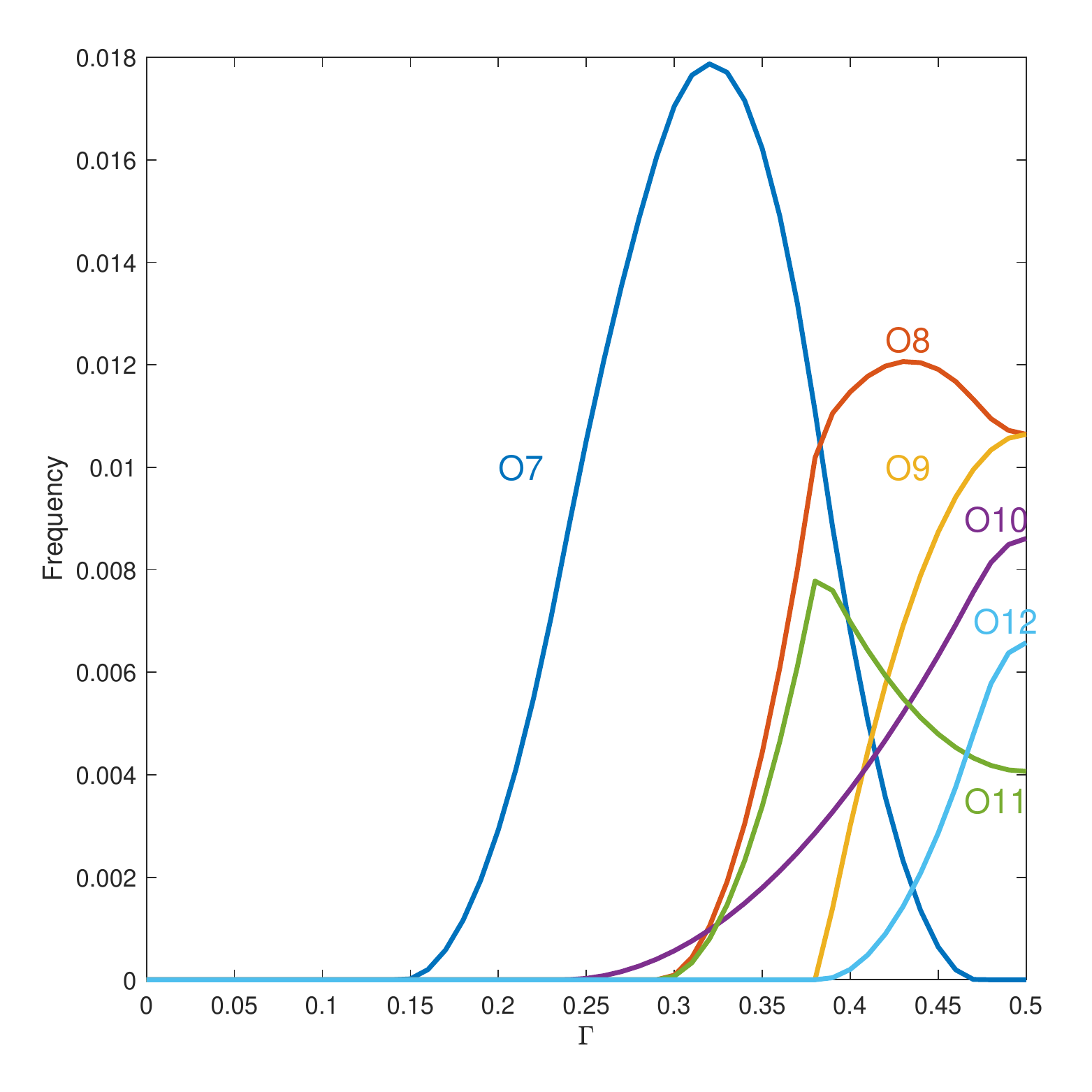}
    \caption{ Frequencies of the LS types given in Fig.\ref{fig:Next6OddRealSpace}.}
    \label{fig:NextSixOddFrequencies}
\end{figure}

The zero-energy manifold formed by LS is massively degenerate. While the numerical calculation in the previous section gives us the overall degeneracy and some information about the spatial structure through LDOS, a more in-depth understanding of LS can be obtained by defining LS types. An LS type is a localized state wavefunction defined in a finite section of the lattice. The same wavefunction, up to five-fold rotations and inversion, will appear infinitely many times throughout the lattice, thus will account for a portion of the degeneracy of the zero-energy manifold. While any linear combination of two LS types can be defined as a new LS type, the aim is to find a set of linearly independent LS types that can span most of the LS manifold. For the PL, six LS types were identified in Ref.\cite{ara88} and there is good evidence that any LS can be expressed as a linear combination of just these six LS types \cite{kts17,mok20}. See Fig. \ref{fig:First6RealSpace} for the real space structure of these LS types. However, the Amman Beenker lattice seems to require an infinite number of LS types \cite{kog20,okt21}. These LS types are arranged in generations, with the first four generations accounting for more than $99\%$ of the degeneracy. It is unclear what determines the minimum number of LS types necessary to span the LS manifold for any quasicrystal.

In the PL, the even and odd sublattices are related by inversion; thus, they have exactly the same LS fraction and LS types. When we consider the LI classes other than the PLI, we need to distinguish between LS types in the odd and even sublattice. Beyond specifying the sublattice, we define an LS type by specifying the properties of sites whose density is non-zero, i.e., its support. We keep the list of vertex types, as described in Fig.\ref{fig:VertexTypes} for each vertex in the support. We also retain the value of the wave function at each of these points. Finally, we keep a list of vectors that specify the relative positions of each vertex in the support. LS that have the same vertex type list and the same wavefunction but differ in the relative vector list are classified as the same type if five-fold rotations and inversion relate their relative vector lists.
 
For example, type-2 LS, as identified in Ref.\cite{ara88}, is now split into type-O2 on the odd sublattice and type-E2 on the even sublattice. The first panel in Fig.\ref{fig:Type2} shows the type-E2 state on the even sublattice. The support consists of 10 type-D vertices, all lying in $D_2$, and the wavefunction alternates as $\pm 1$ over the support. We count the frequency of this LS type by considering the perpendicular space images of all the points in its support. For the type-E2 LS these form ten pentagons on $D_2$ as shown in Fig. \ref{fig:Type2}. The area of one of the pentagons divided by the total area of the perpendicular space gives the frequency of type-E2 on the lattice. We have previously used this method for the Penrose and the Ammann-Beenker lattices and showed that the frequencies agree with previous work. The power of the perpendicular space counting method is that it can be applied to quasicrystals without simple scaling rules, such as the LI classes considered here. As $\Gamma$ changes, the perpendicular space polygons evolve, changing the areas for the allowed regions for the vertices in the support of the LS type. The evolution of $D_2$ and allowed areas for type-E2 are shown in Fig.\ref{fig:Type2EvenScan}, where the shrinking regions indicate the decreasing type-E2 frequency. Beyond $\Gamma \simeq 0.22$, the even sublattice has no regions supporting a type-E2 state. The situation is quite different on the odd sublattice. For $\Gamma=0$, the PLI class, type-O2 LS has support only on the perpendicular space decagon $D_3$, and as $\Gamma$ increases, the allowed regions get smaller. However, the allowed regions shrink slowly compared to the even sublattice, and never disappear. Furthermore, beyond $\Gamma\simeq0.42$, the type-02 state can exist on D type vertices in $D_1$. This is expected as inversion symmetry maps $D_1$ to $D_3$ at $\Gamma=0.5$. The evolution of the allowed regions for type-02 is shown in Fig.\ref{fig:Type2OddScan}, and the LS frequency on both sublattices as a function of $\Gamma$ is given in the last panel of Fig.\ref{fig:Type2}.

Once the real space structure of an LS type is identified, it is easy to calculate its frequency through the perpendicular space images. However, defining LS types aims to find an independent basis that can span the zero energy manifold. Even for the PL, the independence of the LS types is not easy to determine. For example, the type-1 and type-2 states of the PL are not orthogonal\cite{mok20}. However, one can still prove that they are independent as the support of every type-2 state contains at least one point which is not covered by the allowed areas of the type-1. This is the simplest method of ensuring that a newly defined LS type is independent. However, a new LS can have density only on sites that are in the support of other LS but be independent of them. For example, for the PL,  type-4 states have support on sites that can be covered by the combination of a type-1, type-2, type-3, and another (rotated) type-4 LS. Still, one can show that each type-4 state is independent. For type-5 LS, this independence is not guaranteed. Adding a type-2 and type-3 LS to a type-5 state can result in a rotated type-5 state \cite{kts17,mok20}. Only $1/\tau$ of the type-5 states are independent for the PL when other LS types are considered. For LI classes other than PLI, the independence relations are also modified. In Fig.\ref{fig:Type5} we show how the perpendicular space areas for one orientation and all five orientations of type-05 state change and the evolution of LS frequency for this LS type. 

We find that the total frequency of the initial six LS types quickly goes down as $\Gamma$ increases, see Fig.\ref{fig:First6LSFrequencies}. However, new LS types are possible as the local topology of the lattice changes with $\Gamma$. We used null space calculations on small-sized lattices to identify LS as compact as possible and calculated their perpendicular space images. We decided to include a new LS type only if its support covers a previously uncovered area in perpendicular space. While this approach is likely to miss some LS types, it ensures that the LS types we count form an independent set. Thus, we expect the sum of the frequencies of the identified LS types to be a lower bound for the total LS fraction. 

For the even sublattice, we identify 14 LS types in addition to the six present for PLI, giving a total of 20 LS types. The most prominent (highest frequency for any $\Gamma$) six of the new types, type-E7 to type-E12 are displayed in Fig.\ref{fig:Next6EvenRealSpace}, and their frequencies as a function of $\Gamma$ is given in Fig.\ref{fig:NextSixEvenFrequencies}.The remaining eight even sublattice LS types and their frequencies are given in the appendix. Most of the even sublattice LS types are concentrated around a vertex with a high number of edges, in line with the LDOS results. 

\begin{figure}[!htb]
    \centering
    \includegraphics[clip,width=0.48\textwidth]{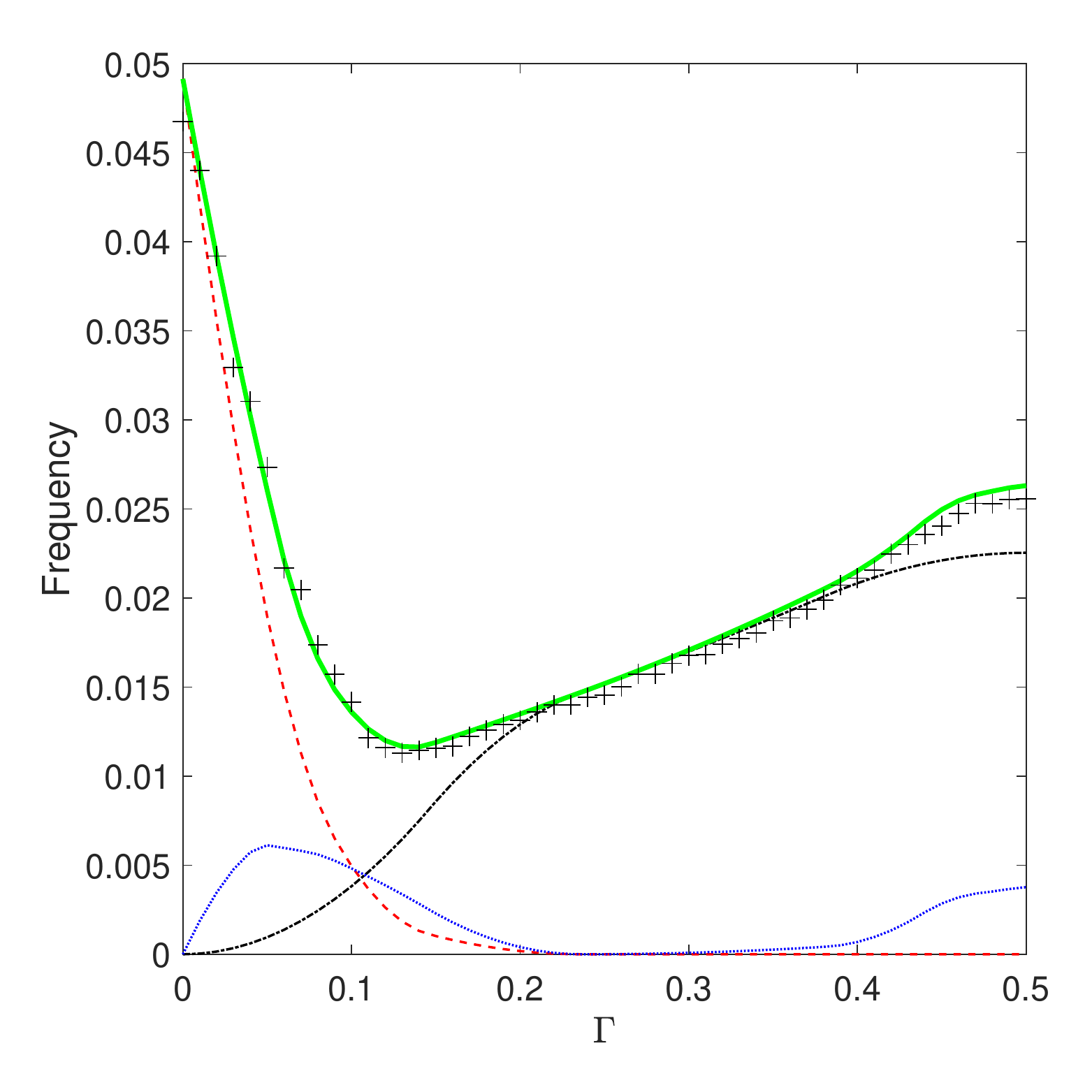}
    \caption{The total LS frequency calculated from the perpendicular space areas on the even sublattice (green line) compared with numerical results (black plusses). The total frequencies for the first six (red dotted line), the following six (black line), and the remaining eight (blue line) LS types are also shown. The agreement between the two results shows that most of the zero-energy manifold for the even sublattice can be expanded in terms of the 20 identified LS types. }
    \label{fig:EvenComparison}
\end{figure}

In Fig.\ref{fig:EvenComparison} we compare the LS fraction on the even sublattice from the numerical calculation with the total obtained for the 20 LS types. The result obtained from perpendicular space areas is within the expected error bounds of the numerical result. It is reasonable to expect that the 20 LS types given here span the zero energy manifold for the even sublattice.   

Using a similar method, we identify 45 LS types for the odd sublattice. The six most prominent odd LS types which are not present in the PL are given in Fig.\ref{fig:Next6OddRealSpace}, and their frequencies are plotted in Fig.\ref{fig:NextSixOddFrequencies}. The remaining 33 LS types and their frequencies are reported in the appendix. In contrast to the even sublattice, our numerical approach found larger LS types in the odd sublattice. Larger LS types are much less frequent, consistent with LDOS pictures obtained on the odd sublattice. Overlap of many large LS types gives an almost uniform distribution for the LDOS.  

\begin{figure}[!htb]
    \centering
    \includegraphics[clip,width=0.48\textwidth]{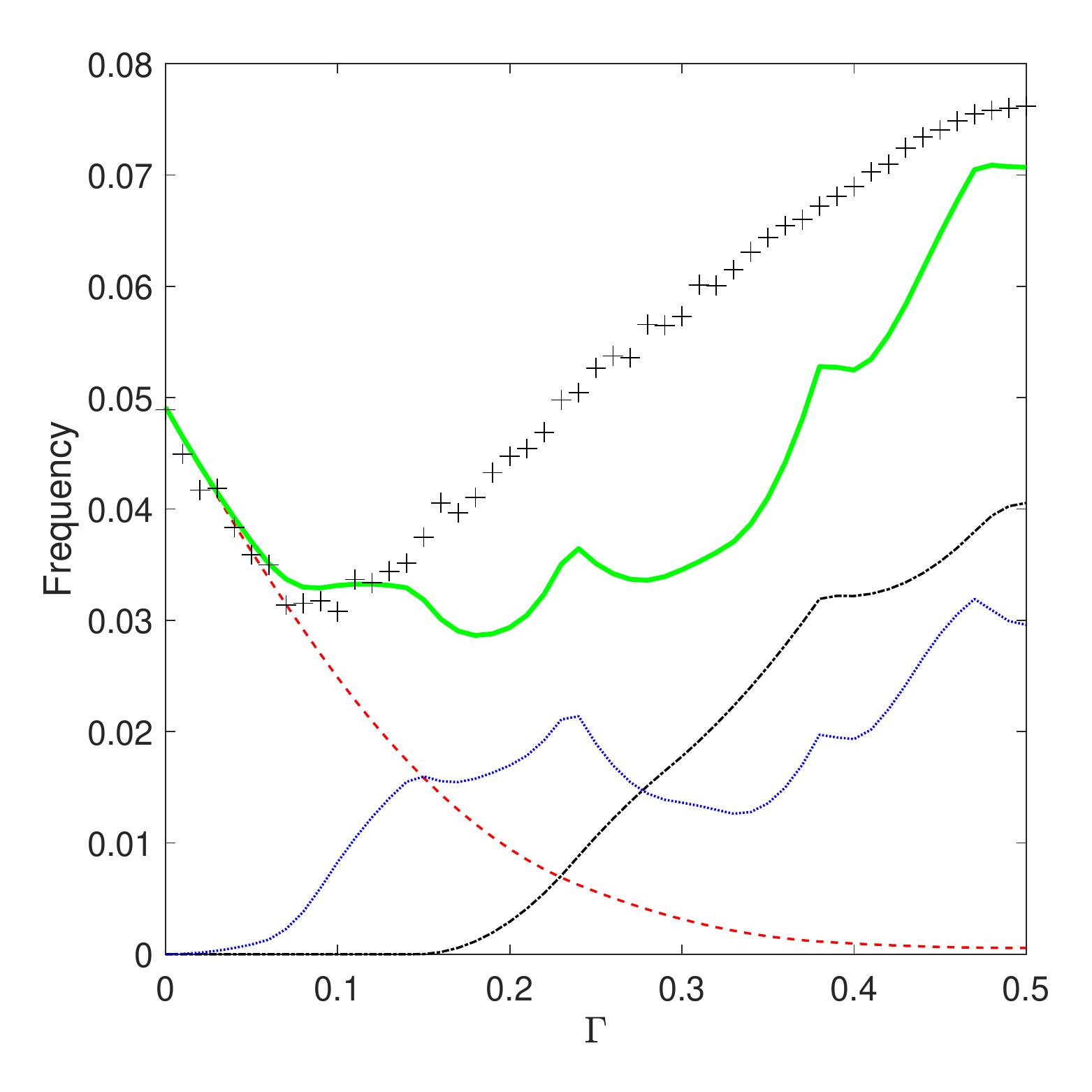}
    \caption{The total LS frequency calculated from the perpendicular space areas on the odd sublattice (green line) compared with numerical results (black plusses). The total LS frequencies for the first six (red dotted line), the following six (black line), and the remaining 33 LS types (blue line) are also shown. The 45 identified LS types are not enough to account for the numerically obtained degeneracy. }
    \label{fig:OddComparison}
\end{figure}

In Fig.\ref{fig:OddComparison}, we compare the numerically obtained LS fraction with the sum of the LS fractions of the 45 LS types we identified. While the two calculations are in good agreement for $\Gamma <0.1$, we see a significant deficit for the frequency obtained from LS types for larger $\Gamma$. There must be other LS types on the odd sublattice. Our algorithm identifies an LS type only if there is a unique site in its support that is not in the support of a previously identified LS types. It is thus quite likely that we are missing LS types that share all points in their support with other LS types but are independent of them due to their wavefunction. It is less likely, but the odd sublattice may have many large LS types that are not reducible to LS types with small support. In this case, our approach would be missing them as we use small neighborhoods to identify LS types. 

\section{Conclusion}
\label{sec:Conclusion}

We consider the vertex tight-binding model on LI classes of pentagonal quasicrystals, obtained from the five-dimensional cubic lattice by the cut and project method. These crystals share the same five-dimensional lattice and the same projection window size. However, the projection window shifts by a parameter $\Gamma$, which uniquely defines LI classes for 
$0\le \Gamma \le \frac{1}{2}$ 
with $\Gamma=0$ 
giving the PLI class \cite{pav87,zob90}.  

One common feature of tight-binding models on quasicrystals is the presence of strictly localized states. These states form almost $10\%$ of all eigenstates for the PL and can be expanded in terms of just six LS types \cite{ara88,kts17,mok20}. The frequency of LS types has been counted by relying on the scaling symmetries of the PL, and this method is not suitable for application to other LI classes which lack simple scaling symmetries. Instead, we used a recently developed method based on the perpendicular space projections to calculate the frequency of LS and compared it with direct numerical calculation on finite lattices.

We find that the total LS fraction first drops as $\Gamma$ increases from the PLI value, makes a minimum around $\Gamma \simeq 0.12$, and then monotonically increases until $\Gamma=0.5$. The highest LS fraction, above $10\%$, is obtained at this limit. This non-monotonic behavior contrasts with the sublattice imbalance, which monotonically increases with $\Gamma$. We also observe that the odd sublattice has more LS than the even sublattice everywhere except at $\Gamma=0$. The higher frequency of LS on the odd sublattice is significantly more pronounced than the number imbalance between the sublattices. At $\Gamma=0.5$, the odd sublattice has $53\%$ of the sites but more than $75\%$ of the LS. 

We further investigate the difference between the sublattices by calculating the LDOS as a function of $\Gamma$. For the PLI at $\Gamma=0$, the LDOS displays regions where only one of the two sublattices has LS, and the lattice is equally split between the two possibilities. As $\Gamma$  increases, the LDOS on the even sublattice forms mostly isolated regions concentrated around high symmetry vertices. LDOS expands throughout the lattice on the odd sublattice, becoming relatively uniform over the whole plane.  
 
We analyze the zero-energy manifold in terms of LS types. For the even sublattice, we identify 20 LS types and calculate their frequencies for all $\Gamma$. Most LS types are confined to regions that encircle a high edge number vertex in agreement with the LDOS result. The total LS fraction obtained from these 20 LS types closely matches the numerical calculation for all values of $\Gamma$. For the odd sublattice, we identified 45 LS types which generally have larger support than even sublattice LS types. Their total frequency matches the numerical calculation only within $\Gamma < 0.11$, while for larger values a significant part of the LS fraction is not captured by these LS types. 

Extending the study of LS to other LI classes makes it clear that many properties of the LS found for the PL, such as the domain structure in their LS, are not inherited from the five-dimensional cubic lattice. If there is any connection between the topological properties of the parent lattice this connection must sensitively depend on the projection window. 
Similarly, the presence of LS does not simply follow from a local imbalance between the sublattices for any LI class different from the PL \cite{day20}. The numerically calculated LS fraction is continuous and smooth as a function of $\Gamma$, which is natural when viewed as a consequence of continuously shrinking or increasing allowed areas of LS types. The behavior of the local environments seem uncorrelated with other structural measures such as hyperuniformity \cite{lin17}.      
It is unclear why the odd sublattice and the even sublattice have markedly different behavior in the LS structure and LDOS. It would also be interesting to address the robustness of the zero-energy manifold and the properties of states that are not strictly localized through perpendicular space methods.

\appendix*
\section{LS types with low frequency }

\begin{figure}[!htb]
    \centering
    \includegraphics[clip,width=0.31\textwidth]{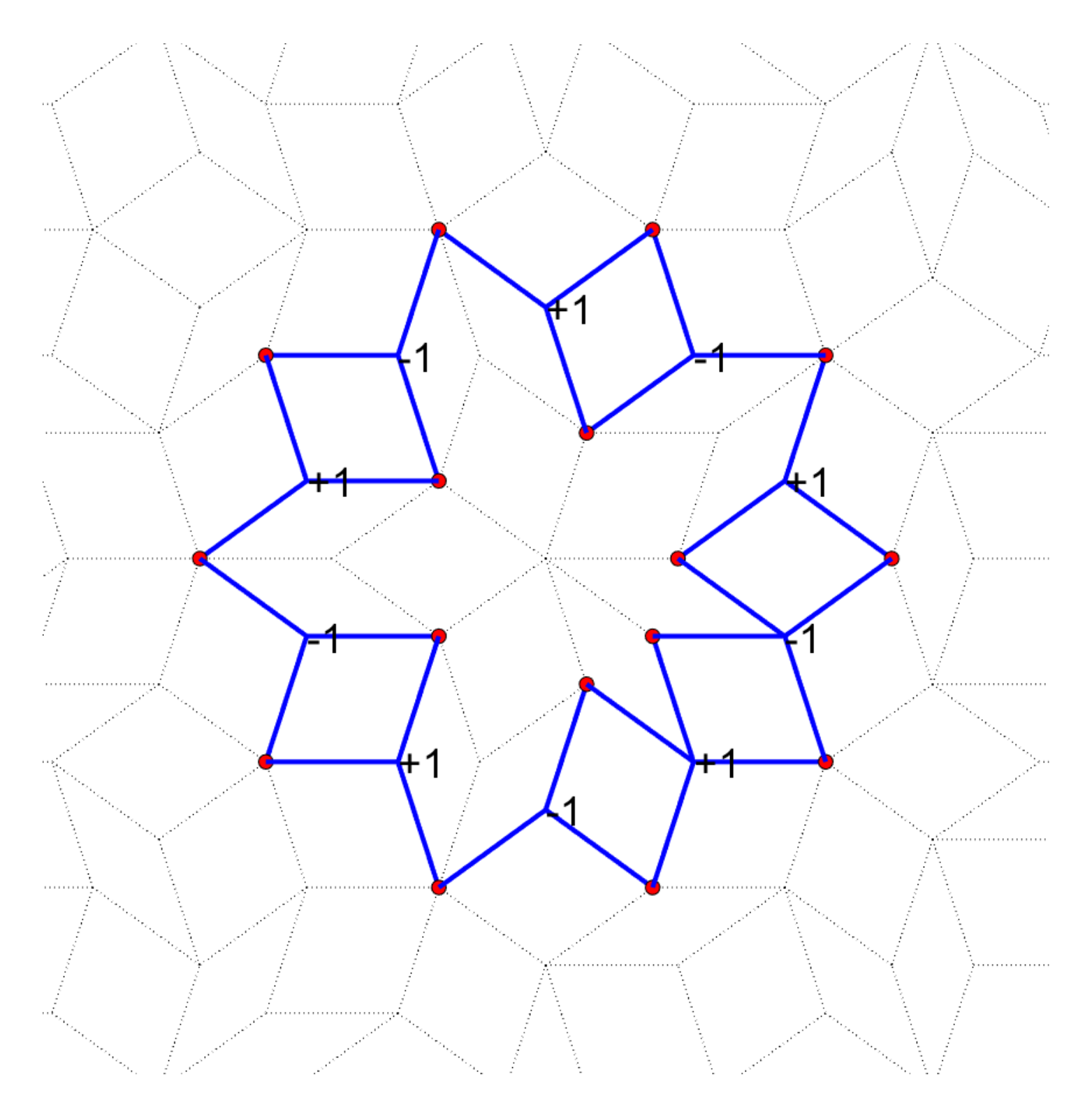}
    \includegraphics[clip,width=0.31\textwidth]{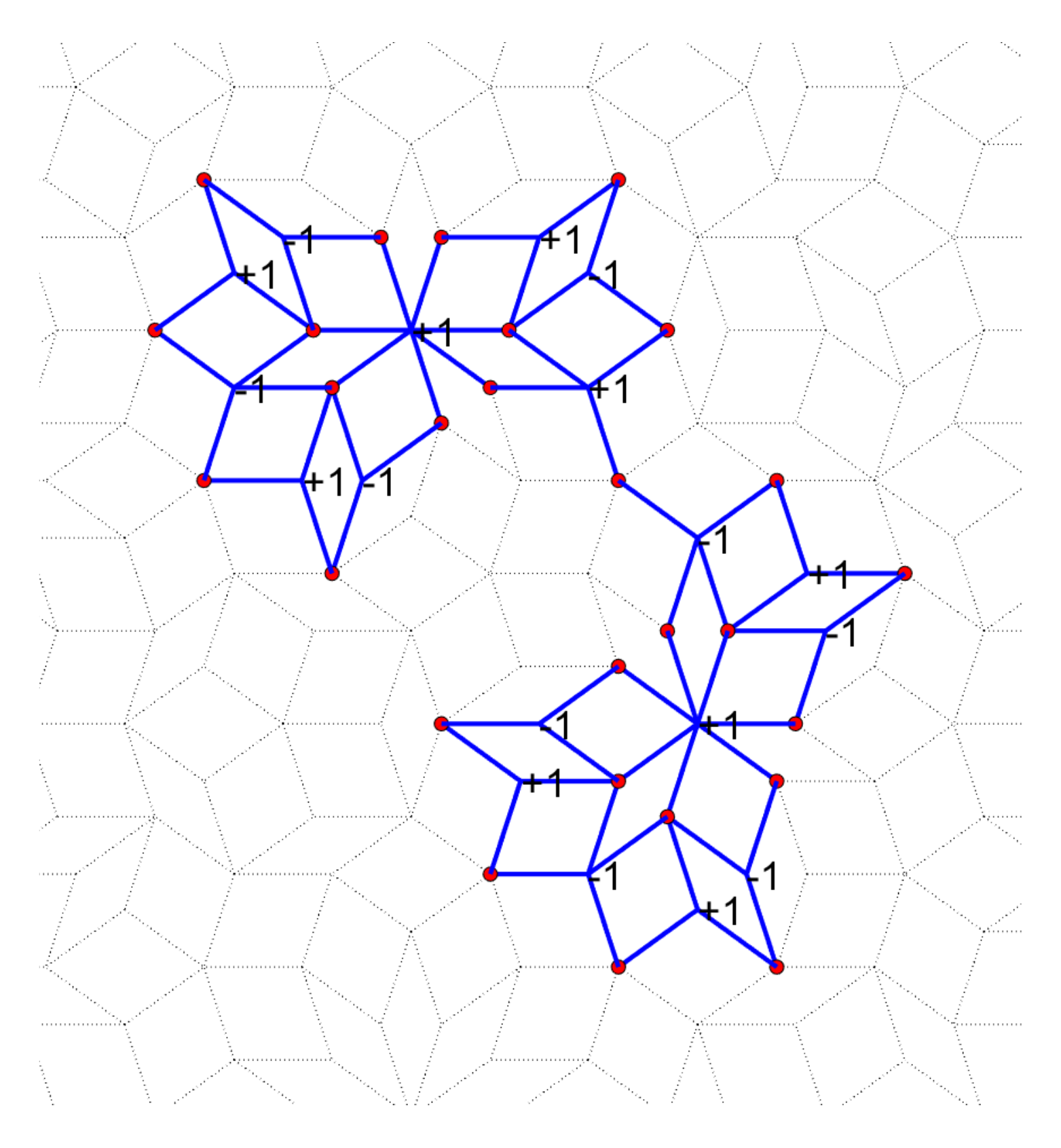}
    \includegraphics[clip,width=0.31\textwidth]{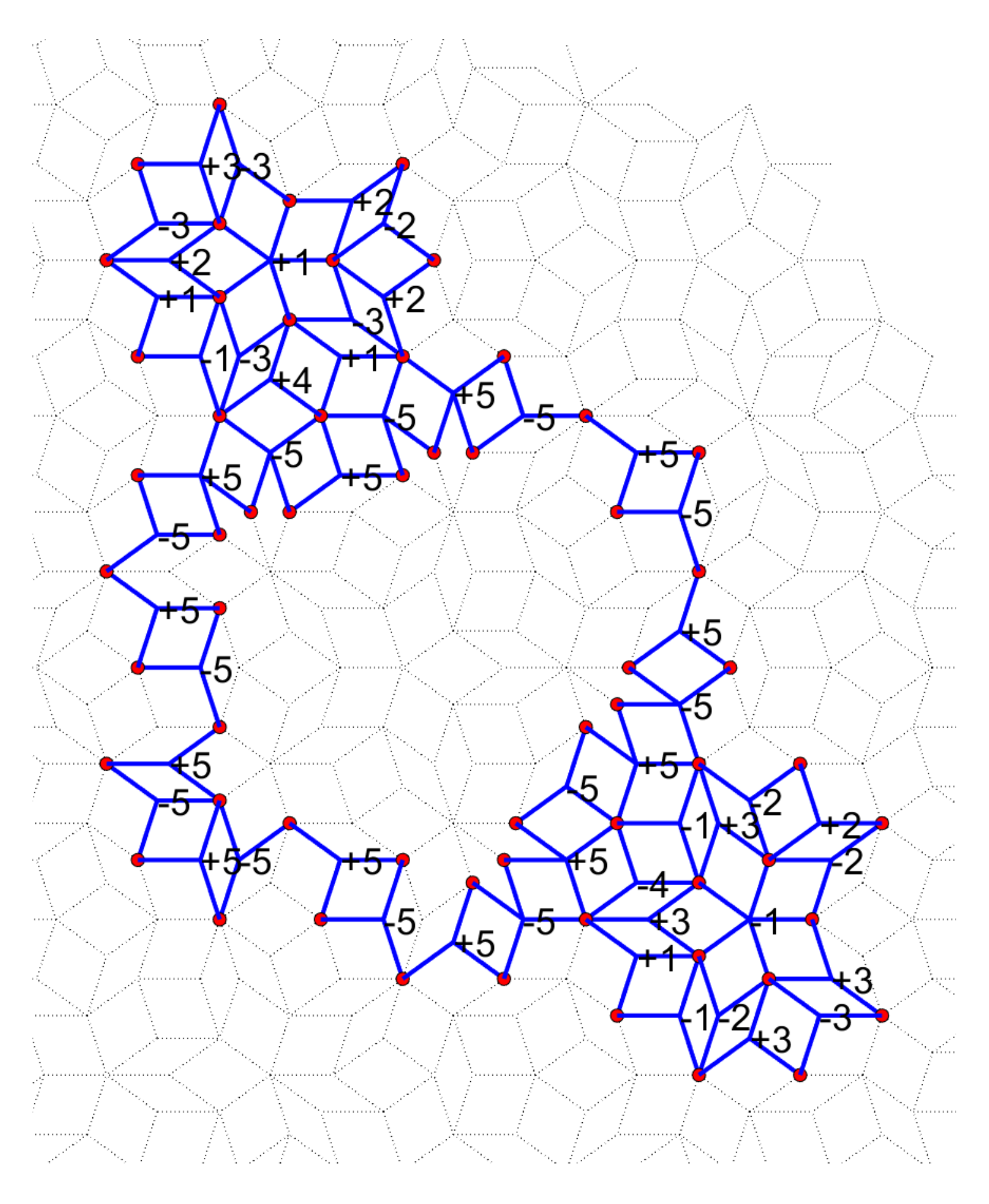}
    \includegraphics[clip,width=0.31\textwidth]{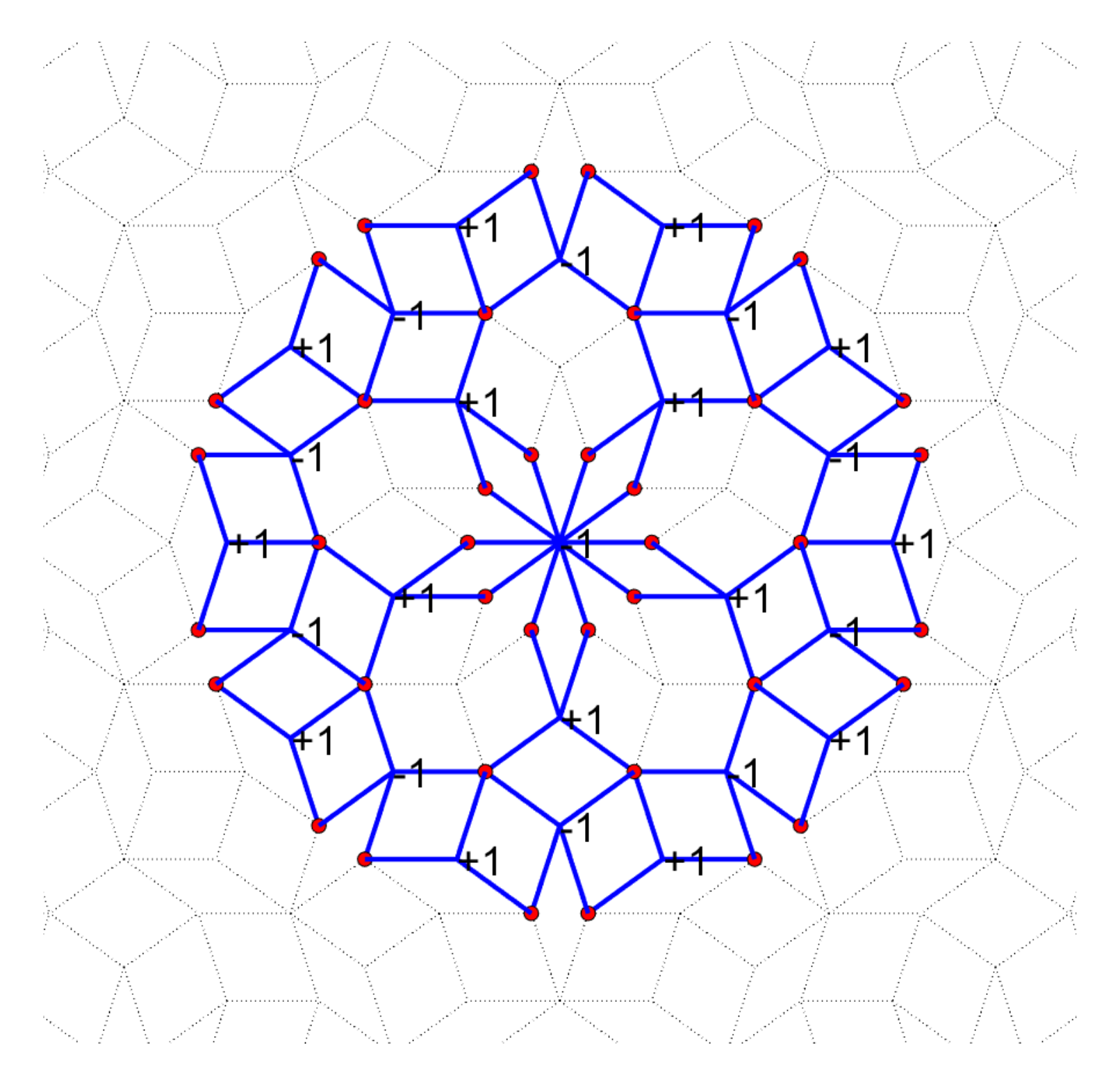}
    \includegraphics[clip,width=0.31\textwidth]{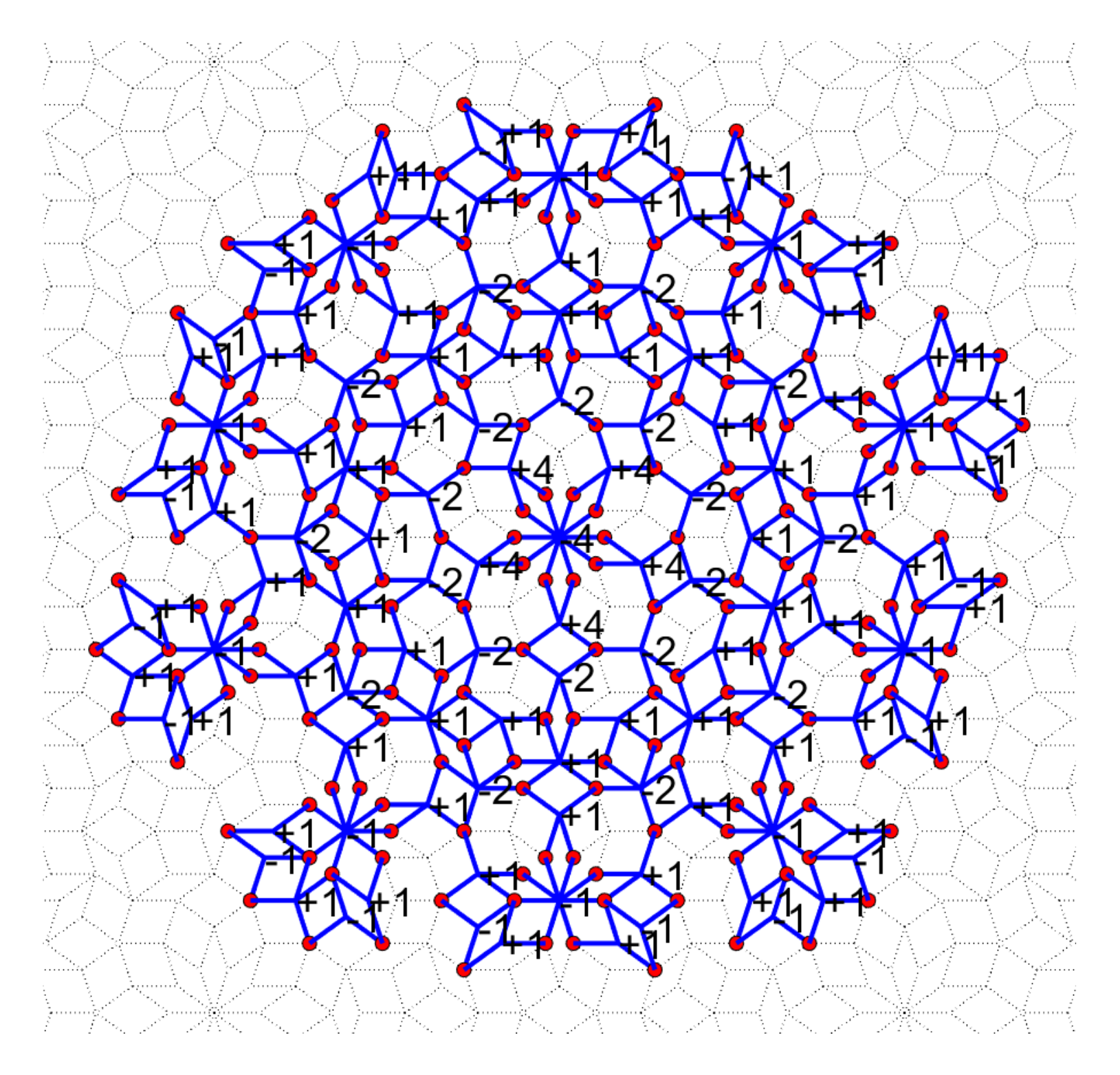}
    \includegraphics[clip,width=0.31\textwidth]{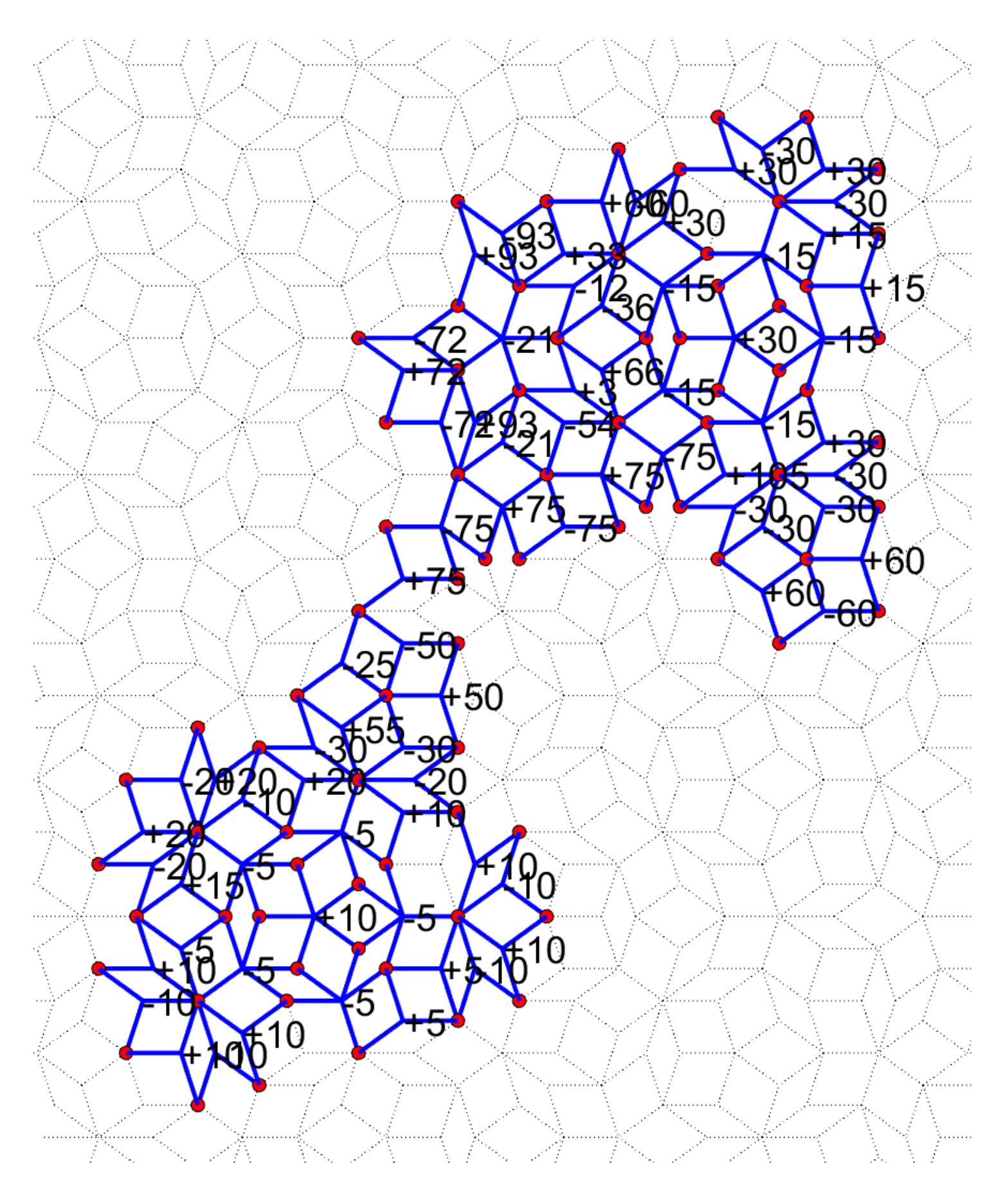}
    \includegraphics[clip,width=0.31\textwidth]{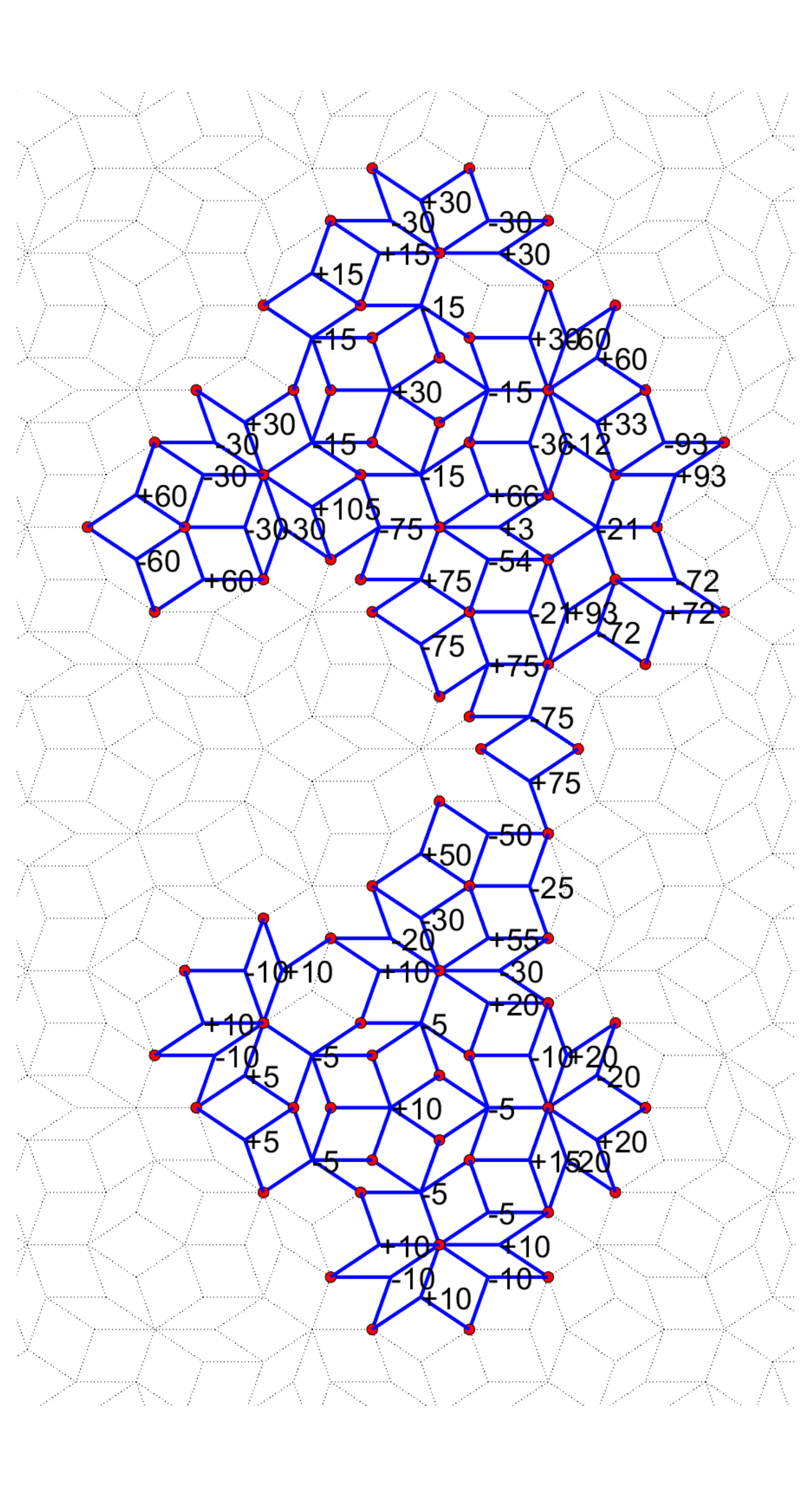}
    \includegraphics[clip,width=0.31\textwidth]{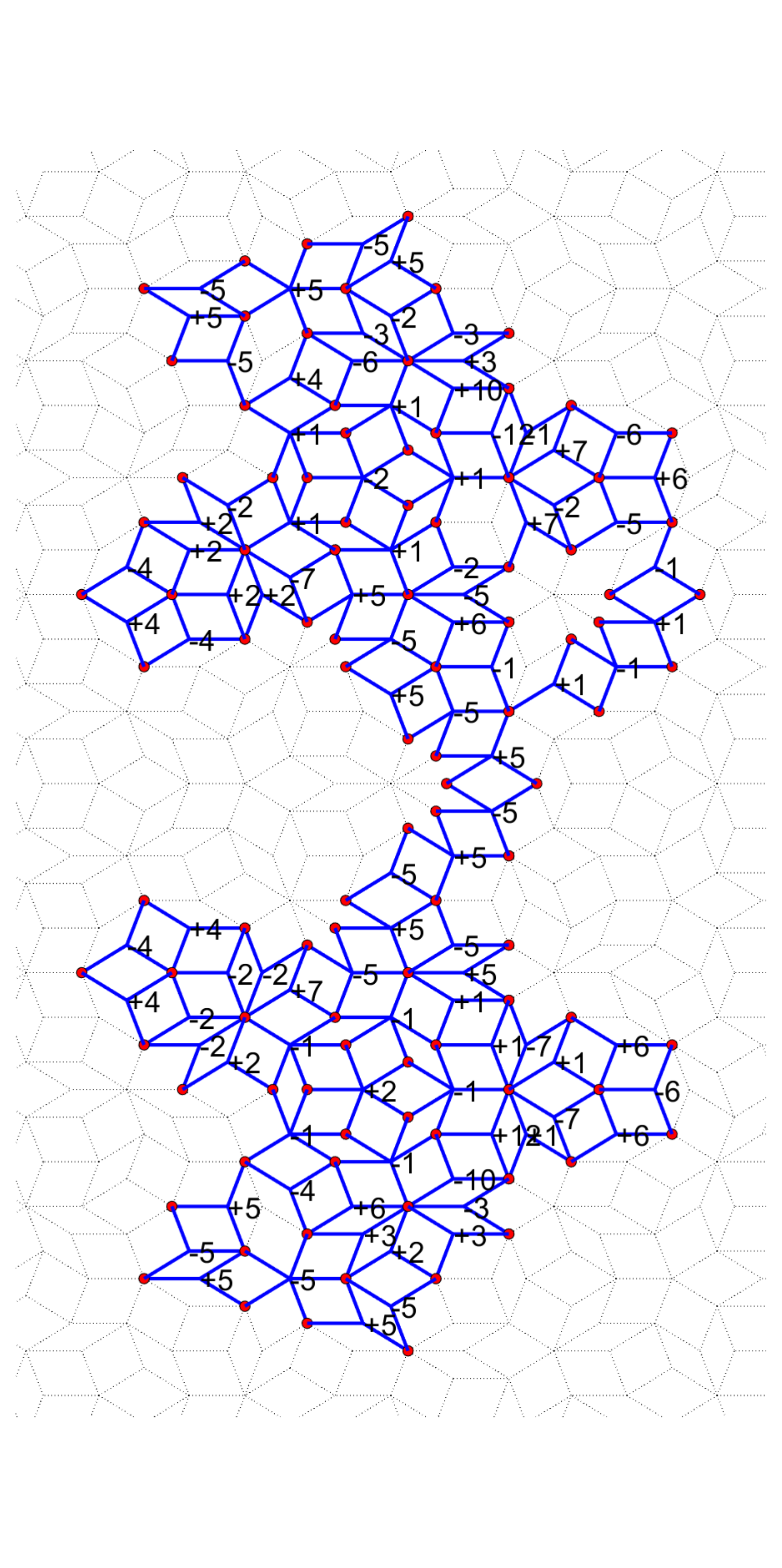}
    
    \caption{LS types type-E13 to type-E20 on the even sublattice.Type-E17 can be made more symmetrical by adding other LS types. }
    \label{fig:Last8EvenRealSpace}
\end{figure}

\begin{figure}[!htb]
    \centering
    \includegraphics[clip,width=0.48\textwidth]{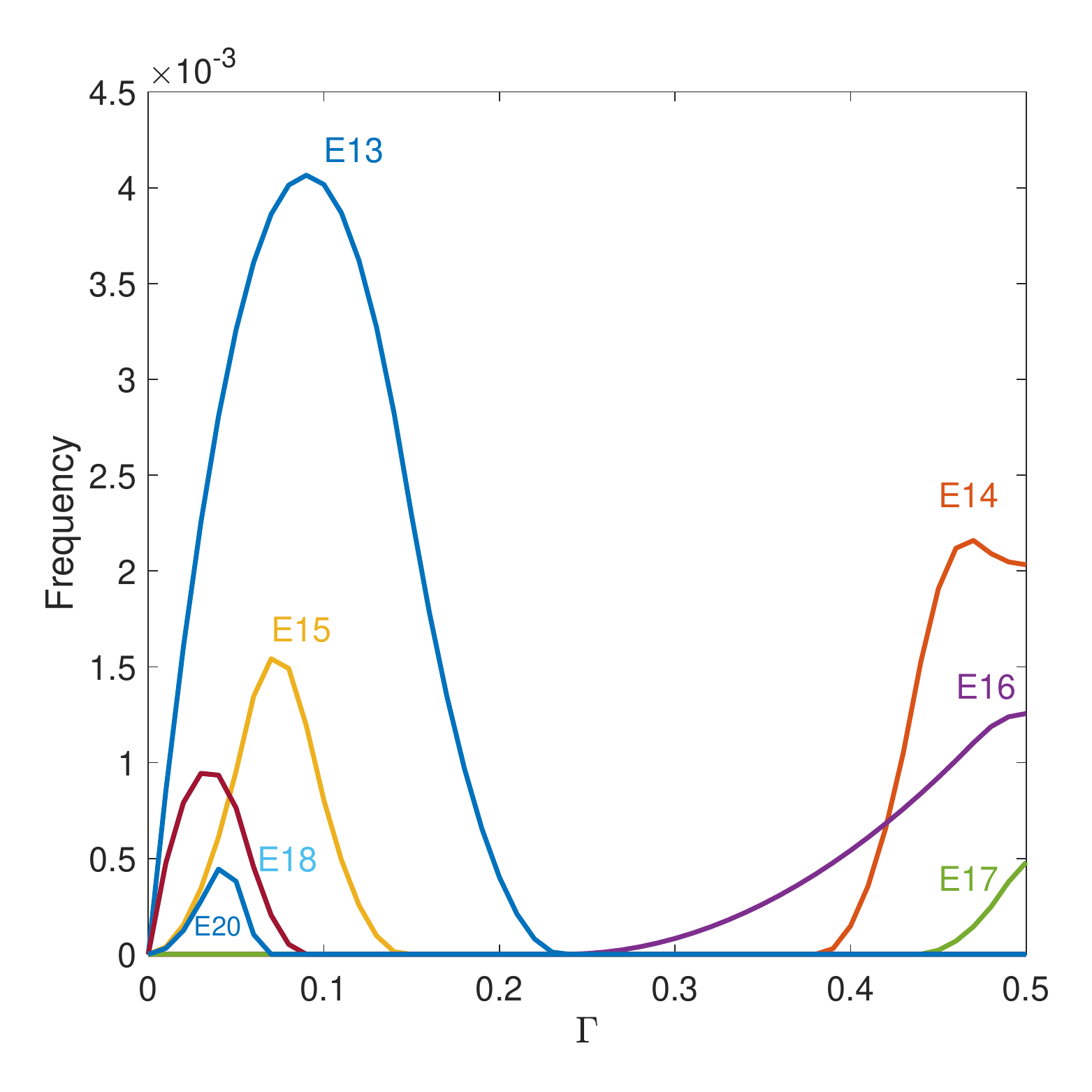}
    \caption{Frequencies of the LS types in Fig.\ref{fig:Last8EvenRealSpace}. Types E18 and E19 are related by reflection hence share the same frequency. }
    \label{fig:Last8EvenLSFrequency}
\end{figure}

\begin{figure}[!htb]
    \centering
    \includegraphics[clip,width=0.31\textwidth]{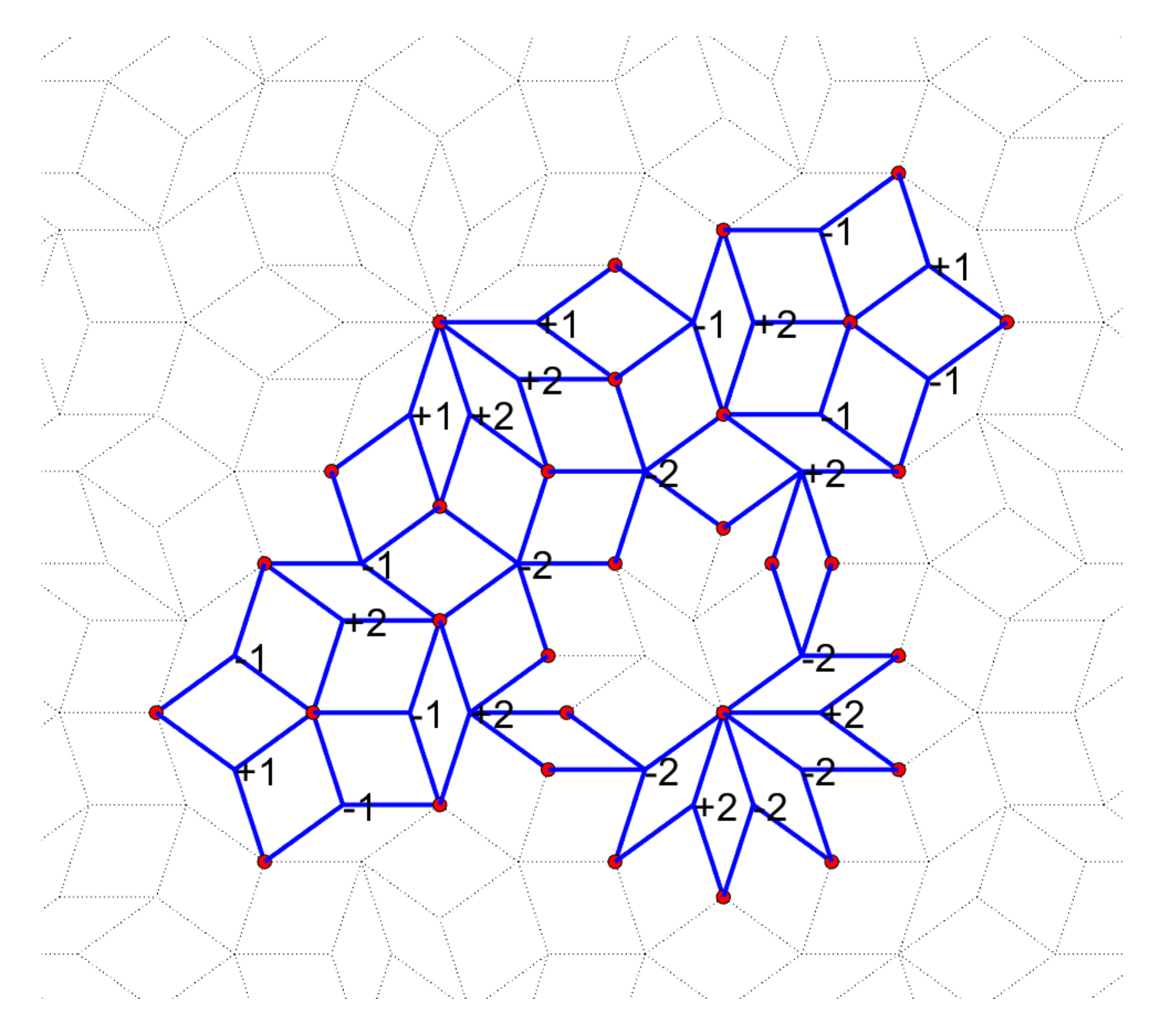}
    \includegraphics[clip,width=0.31\textwidth]{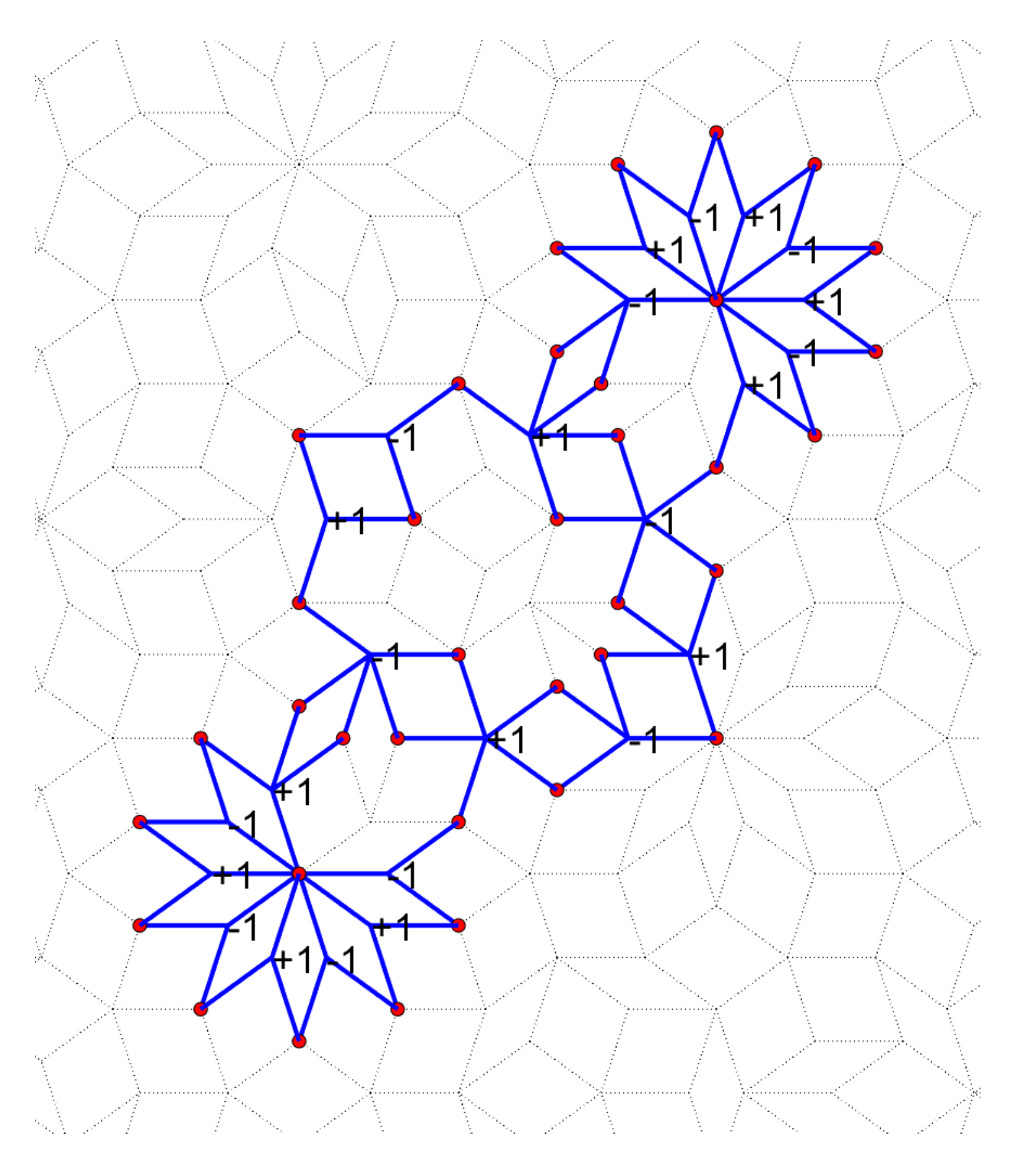}
    \includegraphics[clip,width=0.31\textwidth]{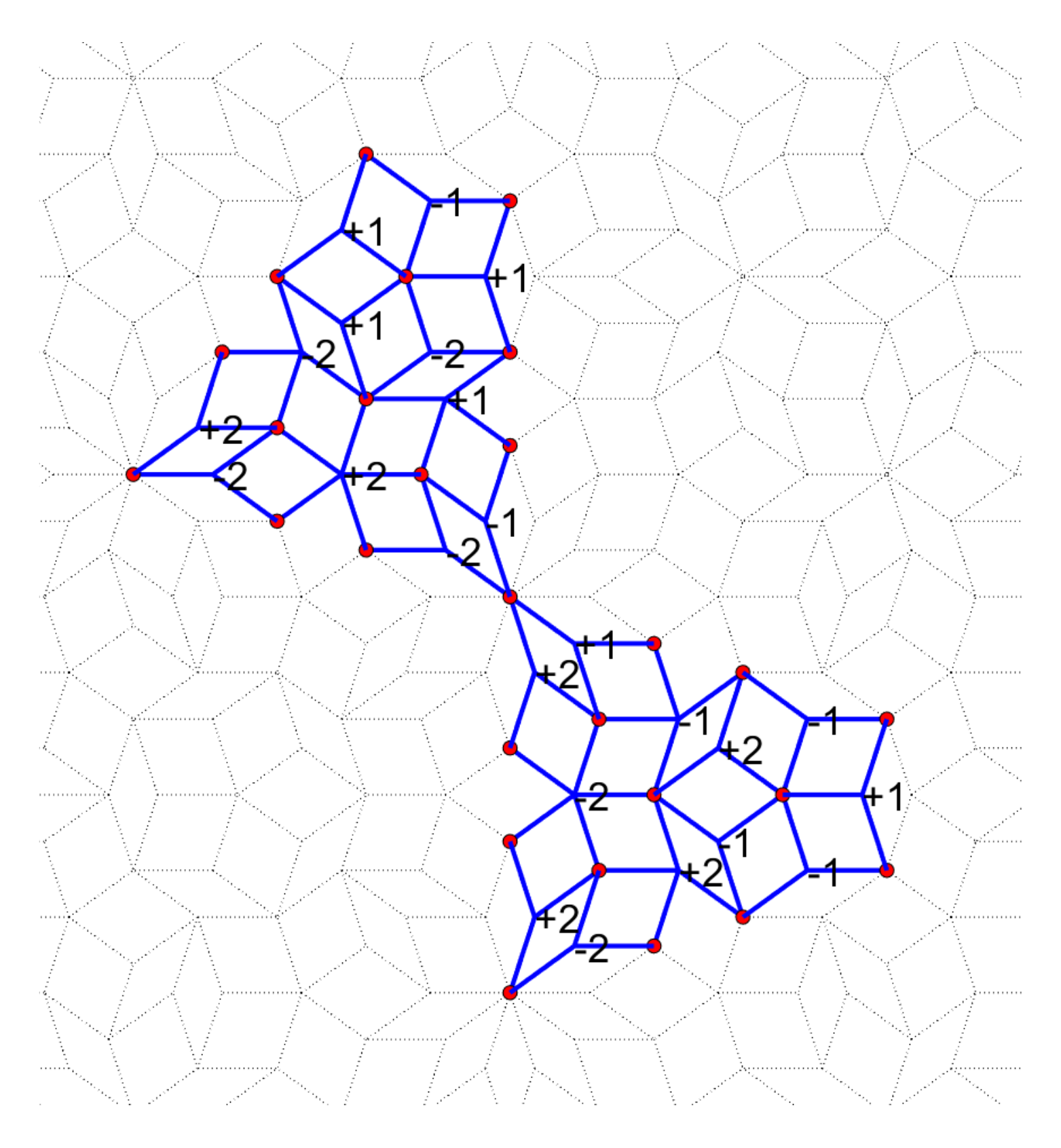}
    \includegraphics[clip,width=0.31\textwidth]{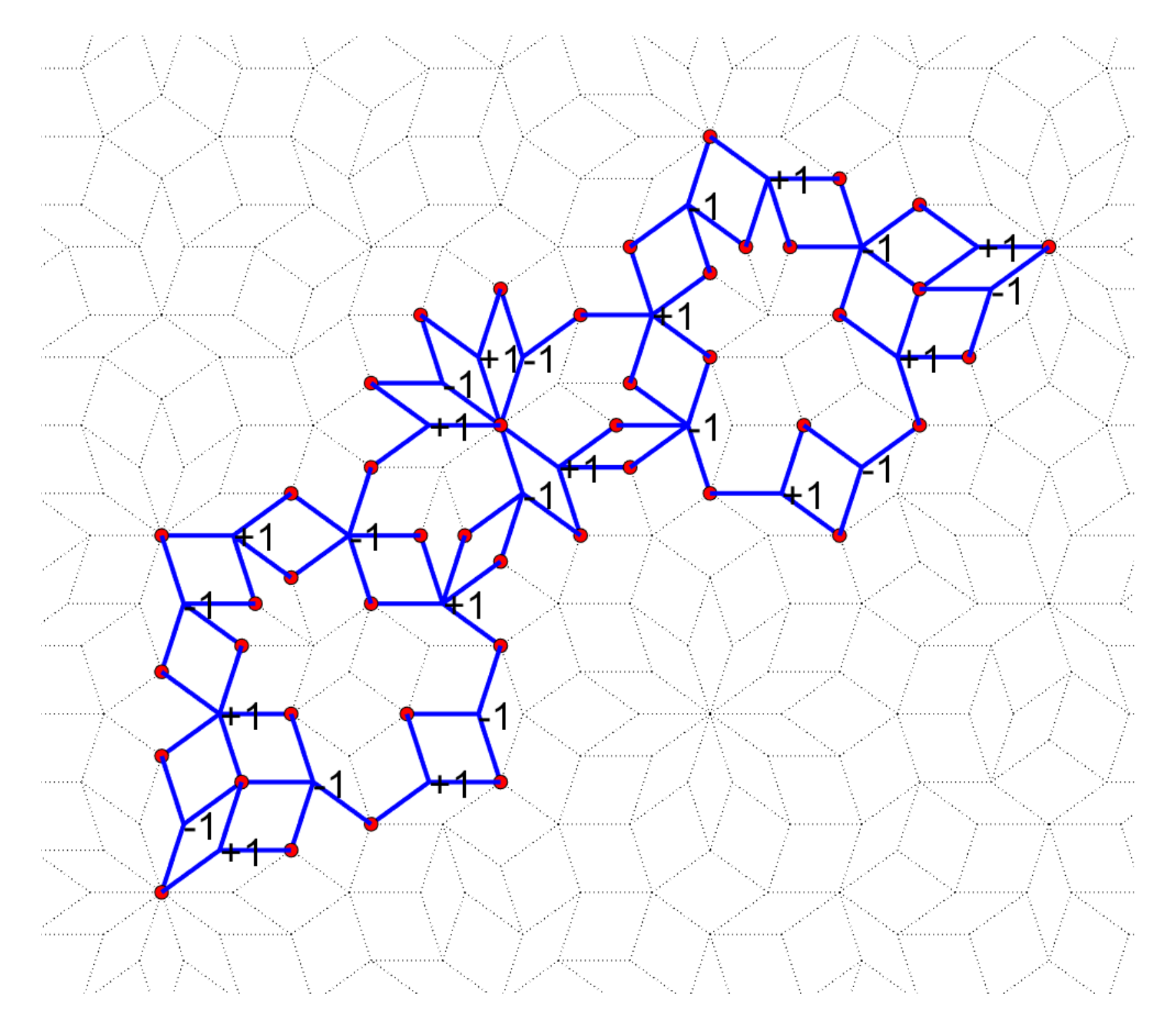}
    \includegraphics[clip,width=0.31\textwidth]{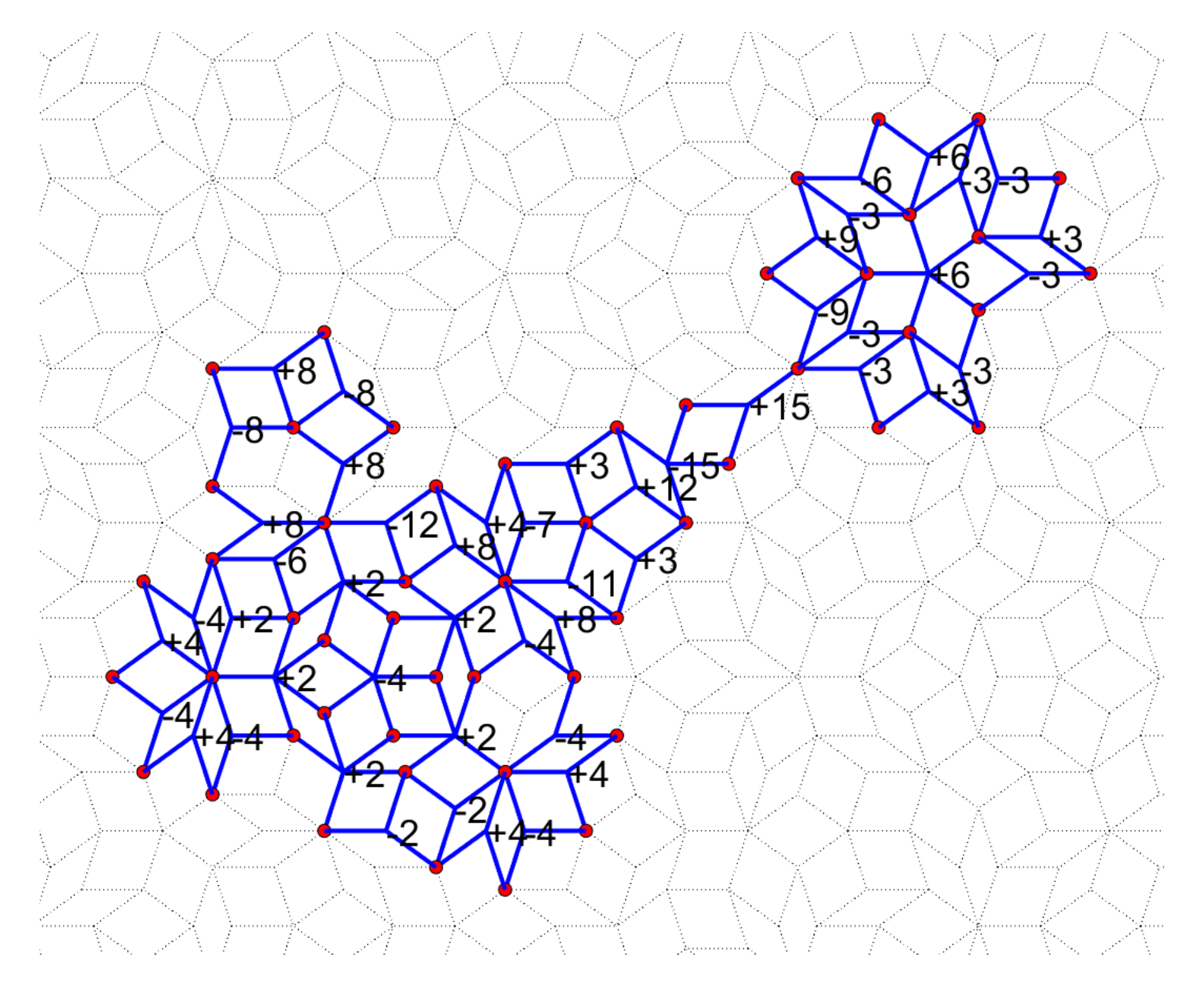}
    \includegraphics[clip,width=0.31\textwidth]{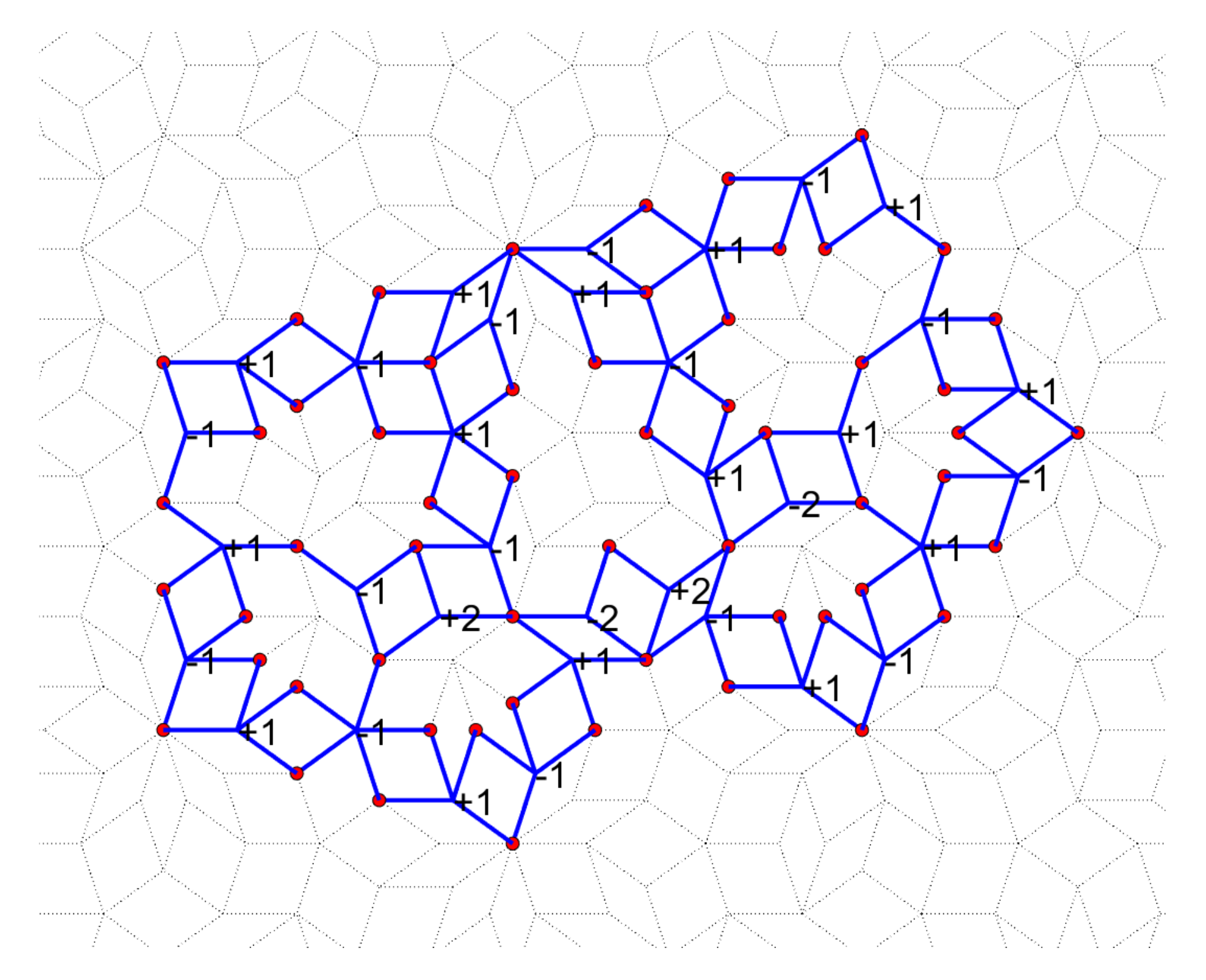}
    \includegraphics[clip,width=0.31\textwidth]{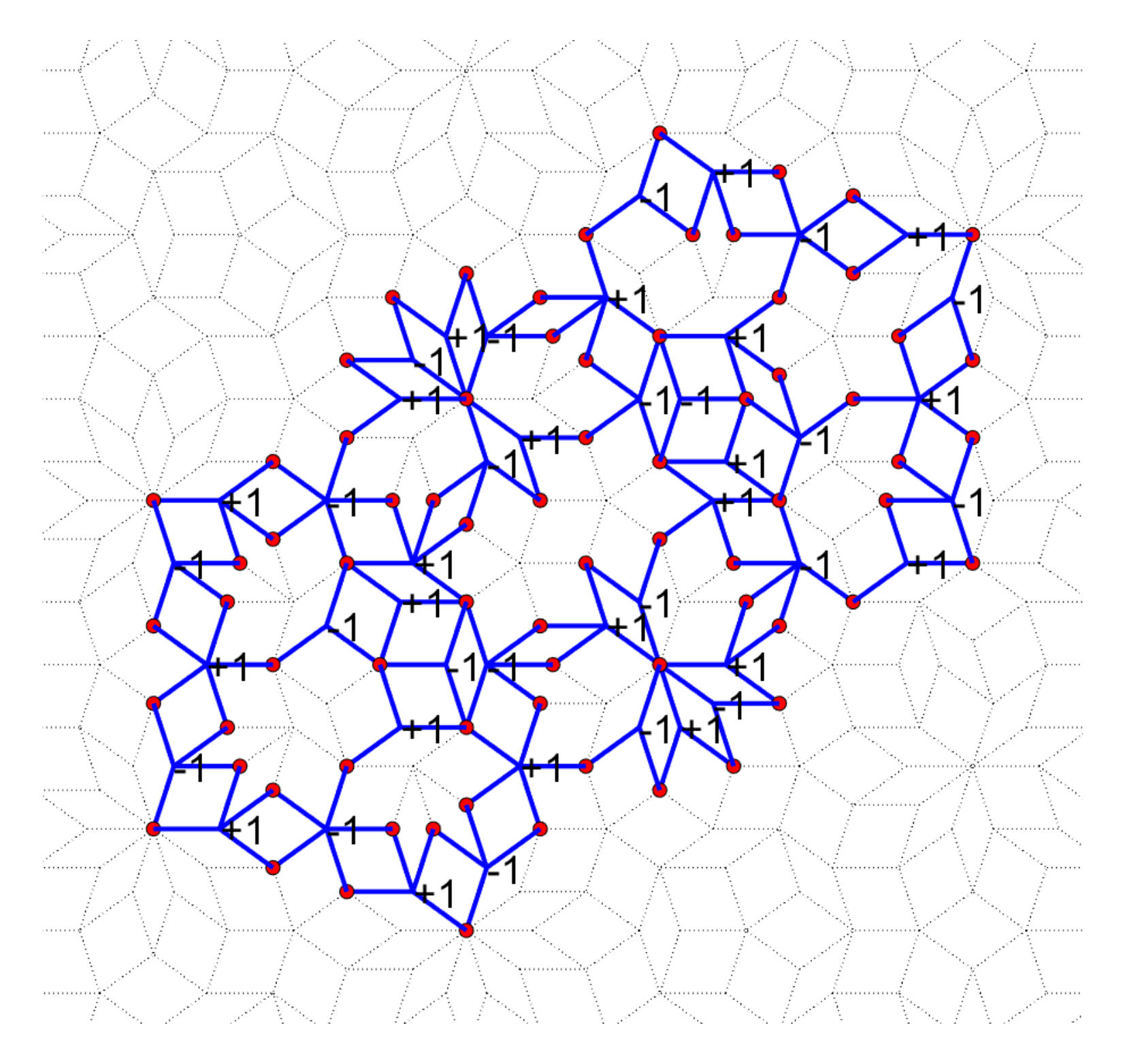}
    \includegraphics[clip,width=0.31\textwidth]{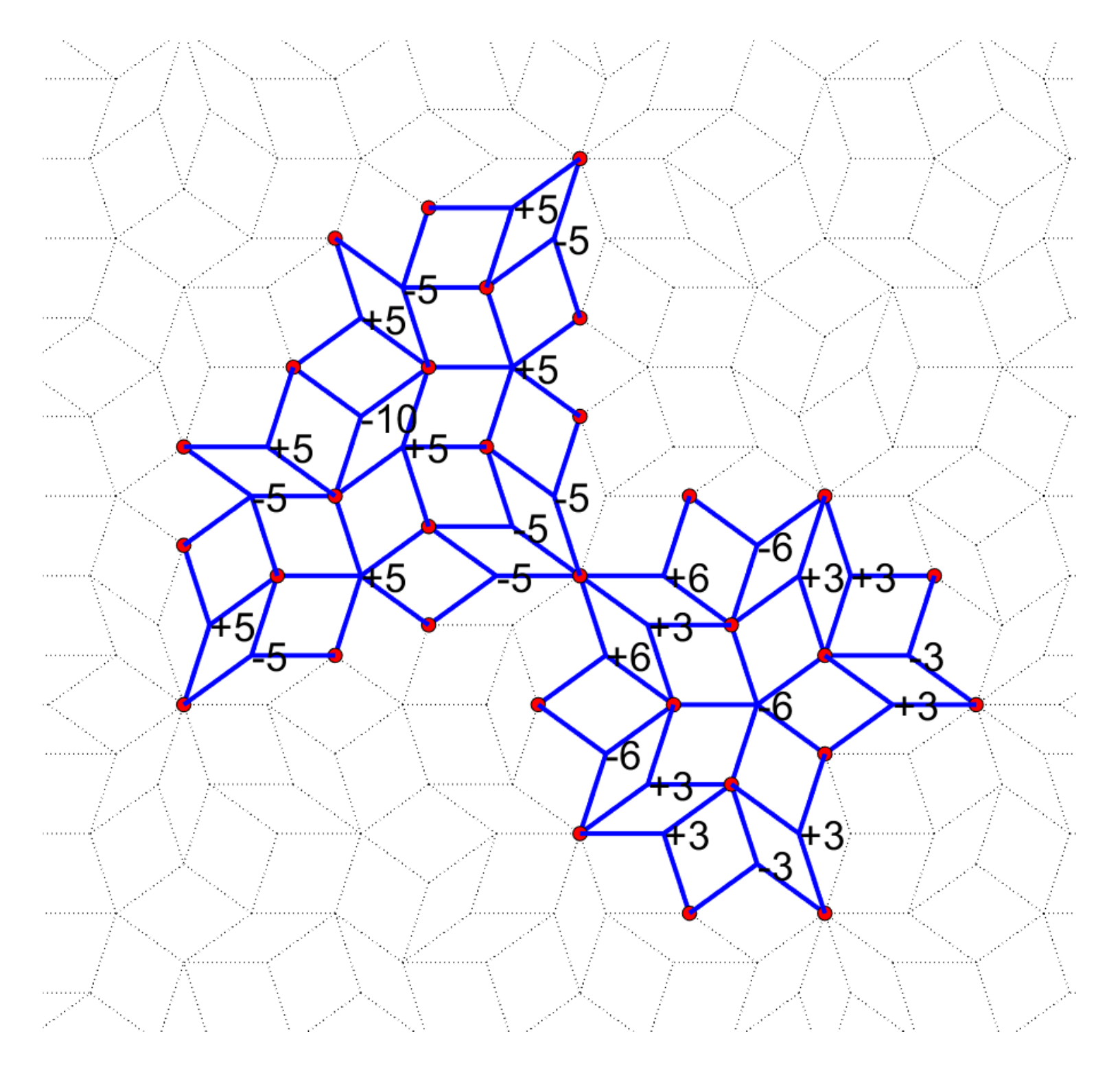}
    \includegraphics[clip,width=0.31\textwidth]{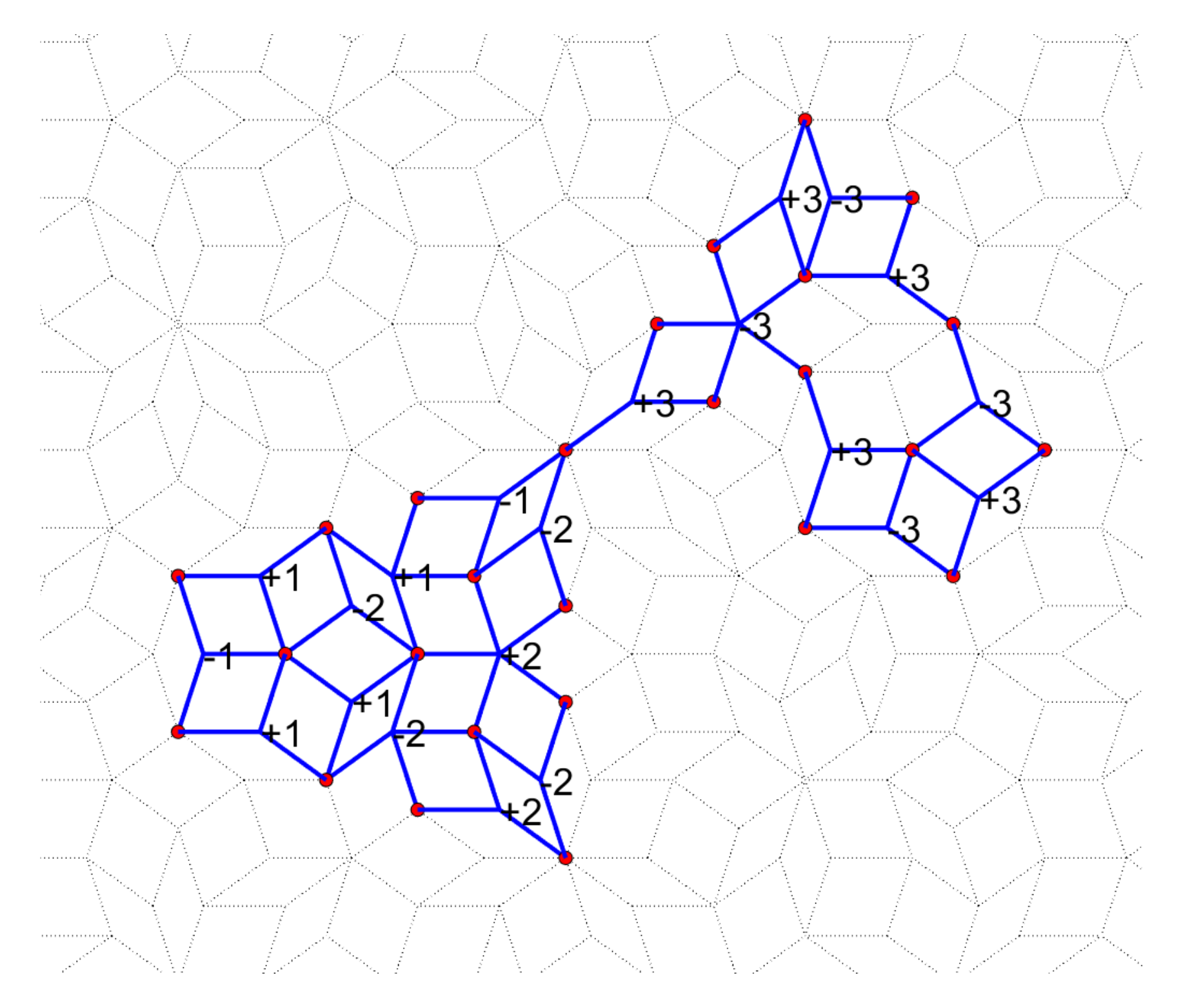}
    \caption{Types O13 to O21 on the odd sublattice. }
    \label{fig:LS1321OddRealSpace}
\end{figure}

\begin{figure}[!htb]
    \centering
    \includegraphics[clip,width=0.31\textwidth]{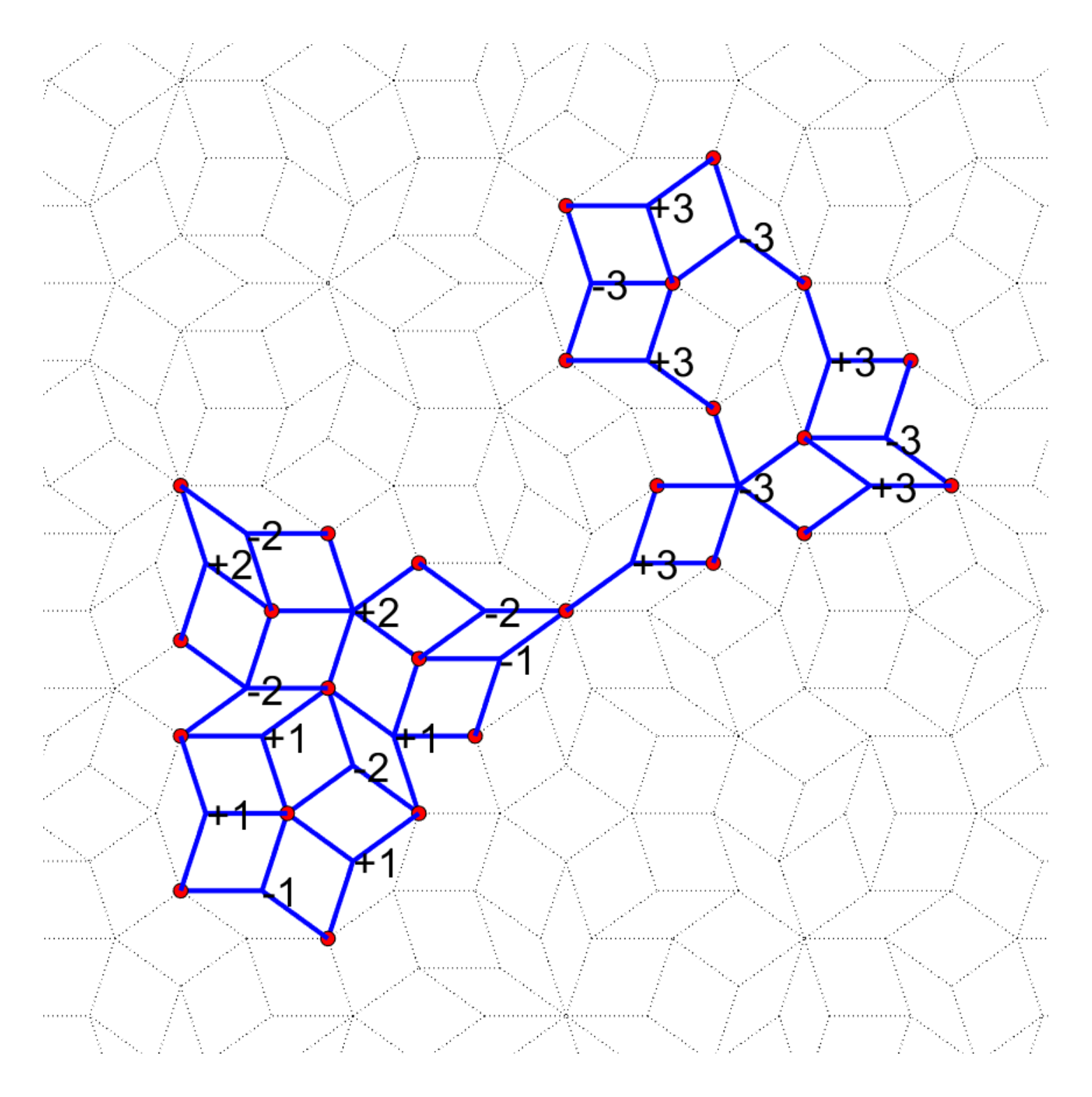}
    \includegraphics[clip,width=0.31\textwidth]{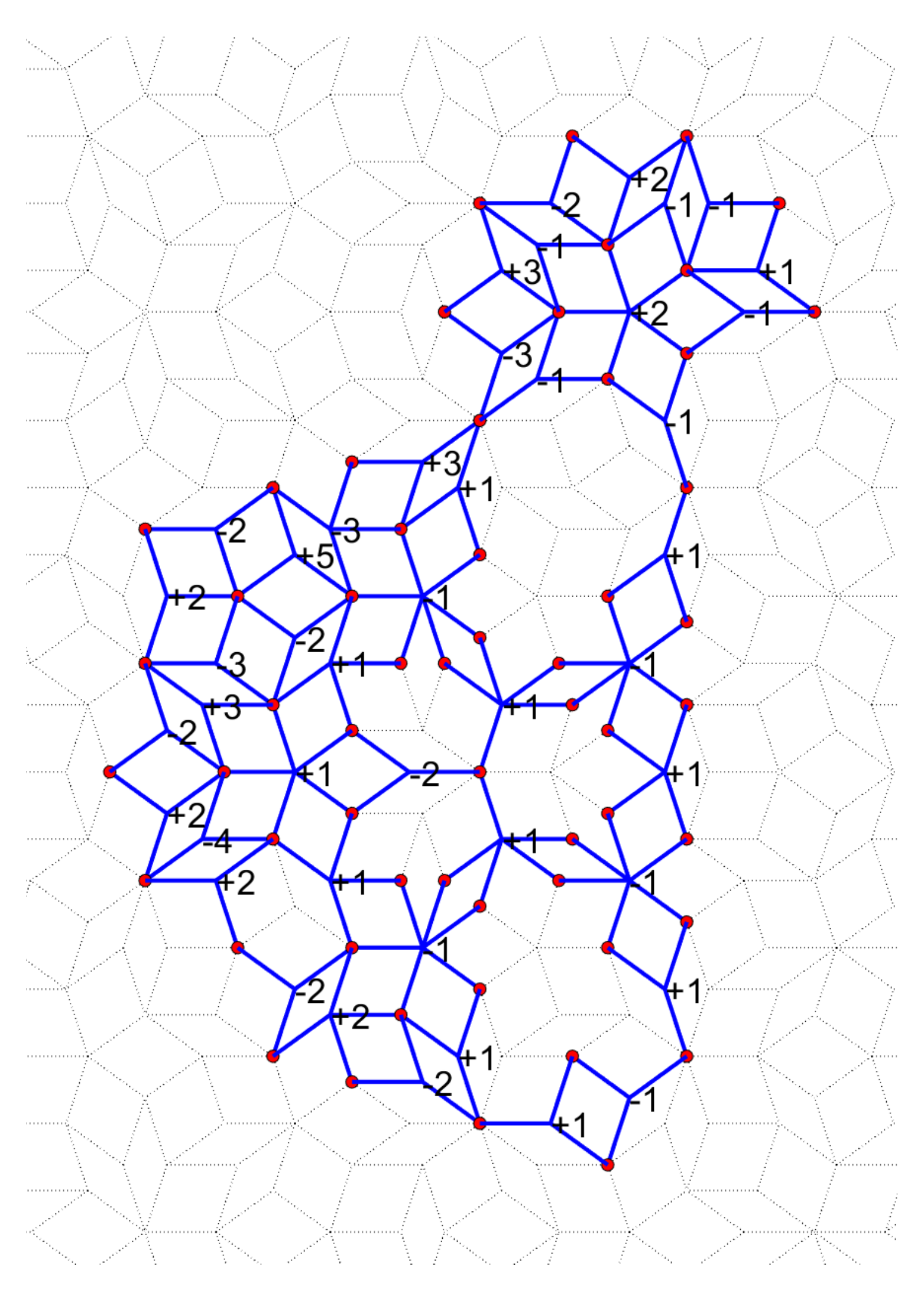}
    \includegraphics[clip,width=0.31\textwidth]{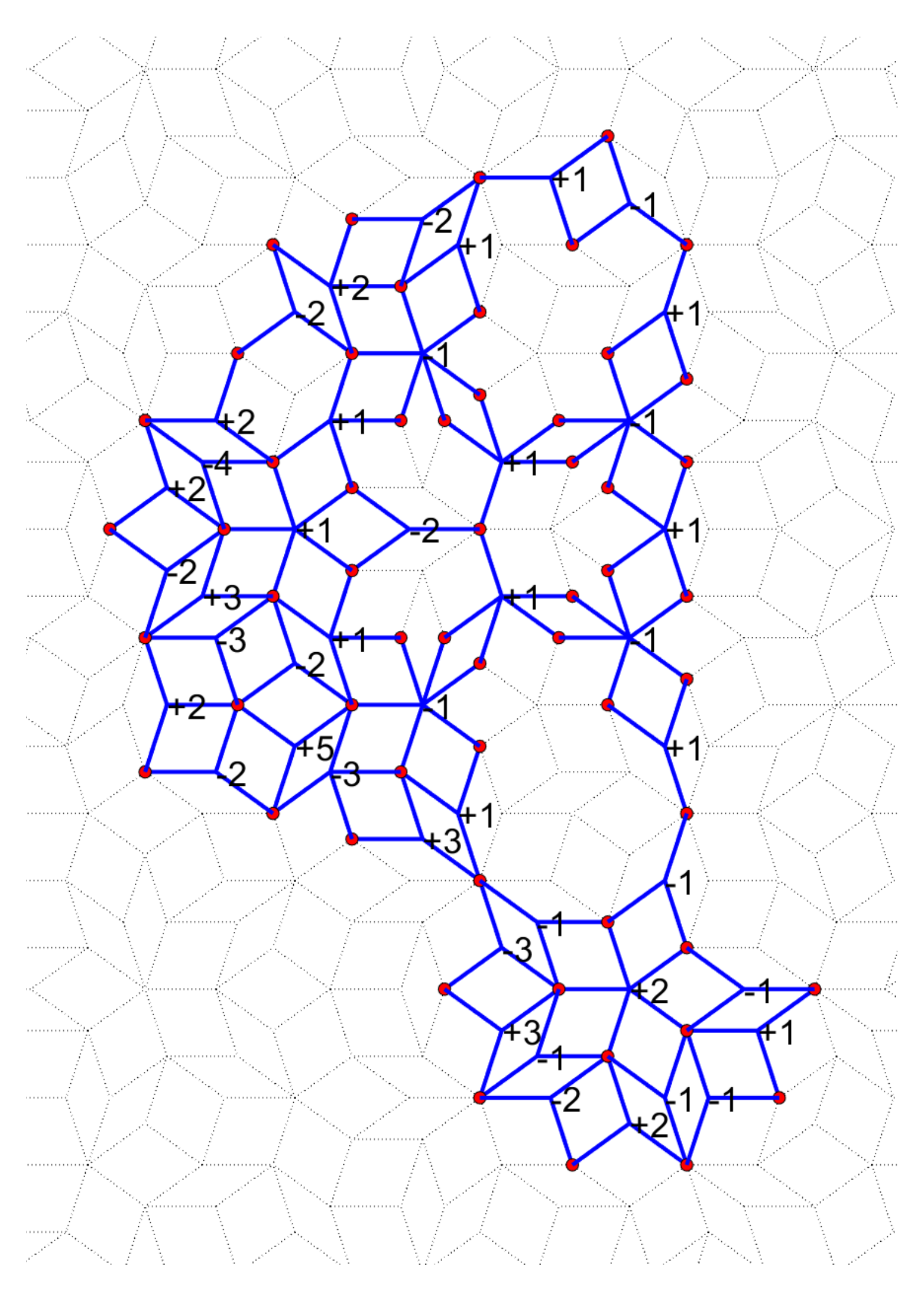}    
    \includegraphics[clip,width=0.31\textwidth]{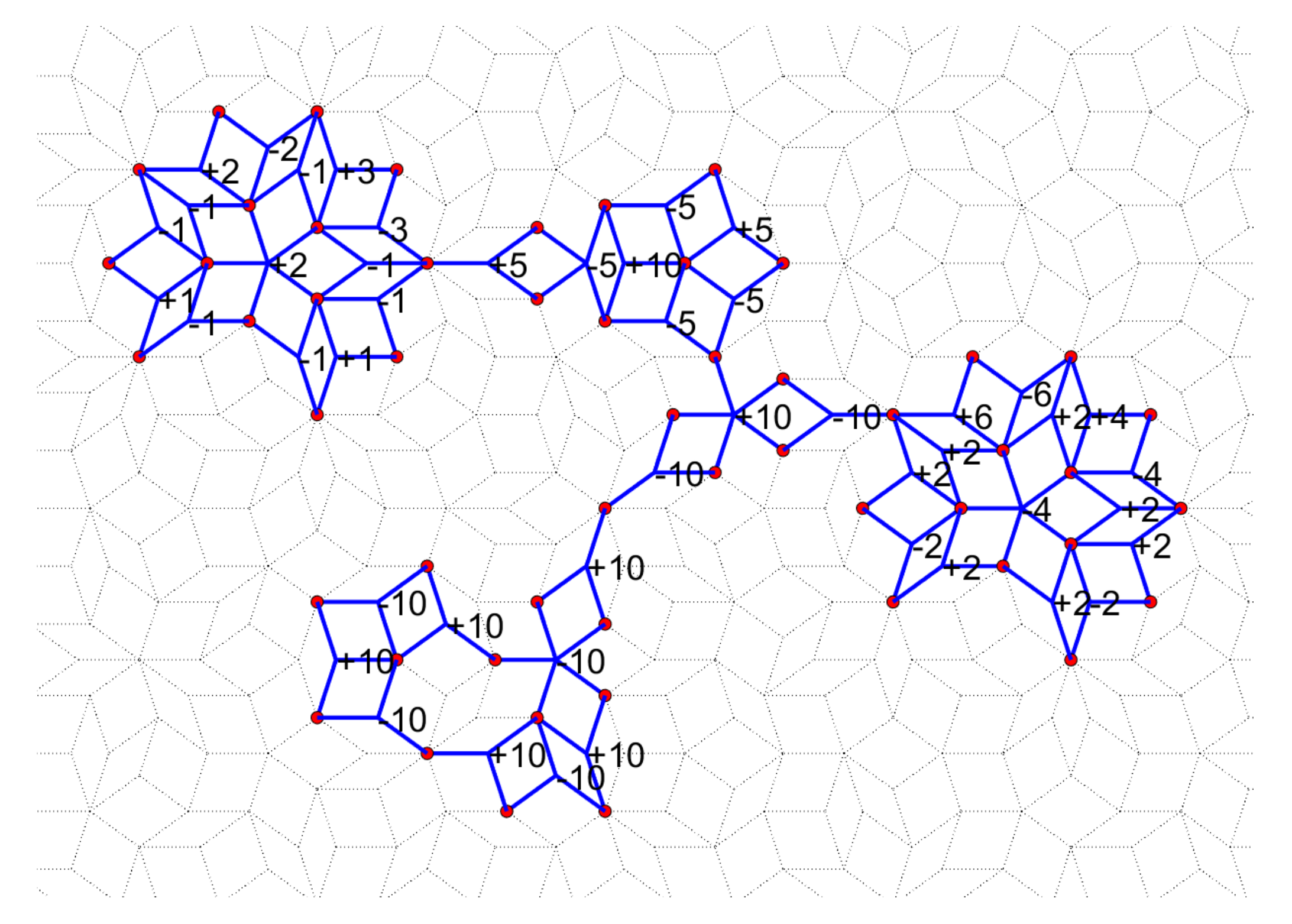}
    \includegraphics[clip,width=0.31\textwidth]{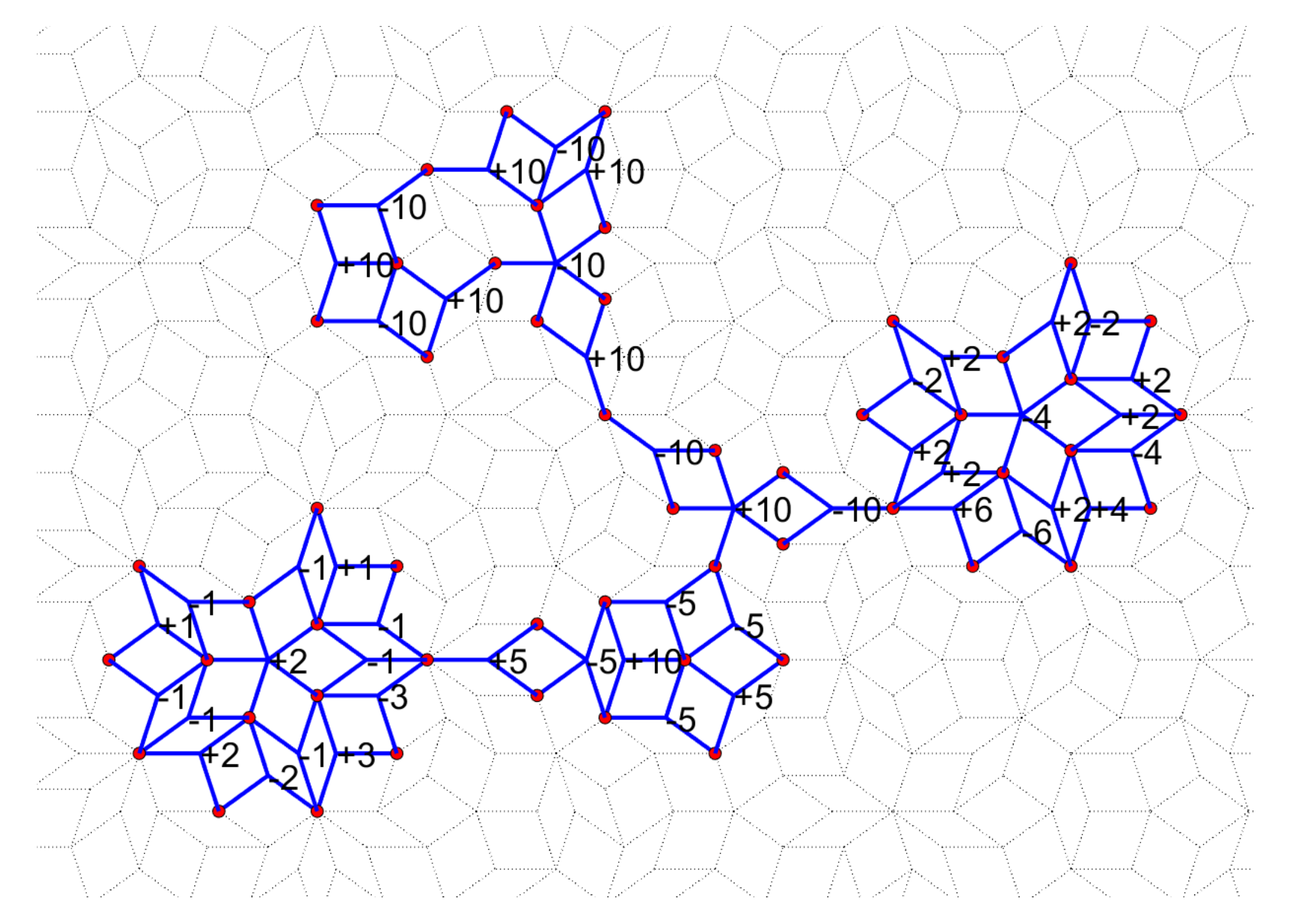}
    \includegraphics[clip,width=0.31\textwidth]{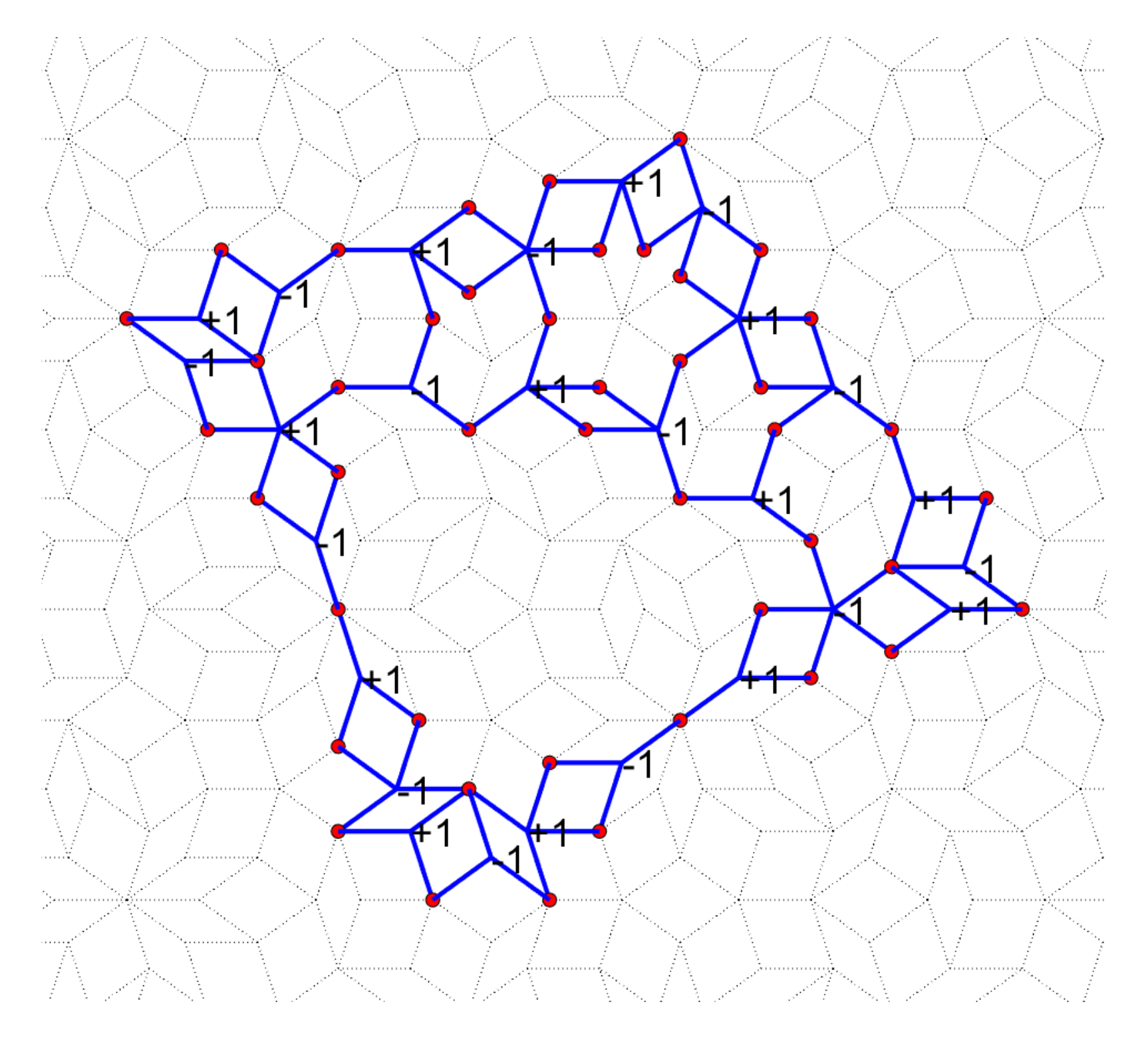}
    \includegraphics[clip,width=0.31\textwidth]{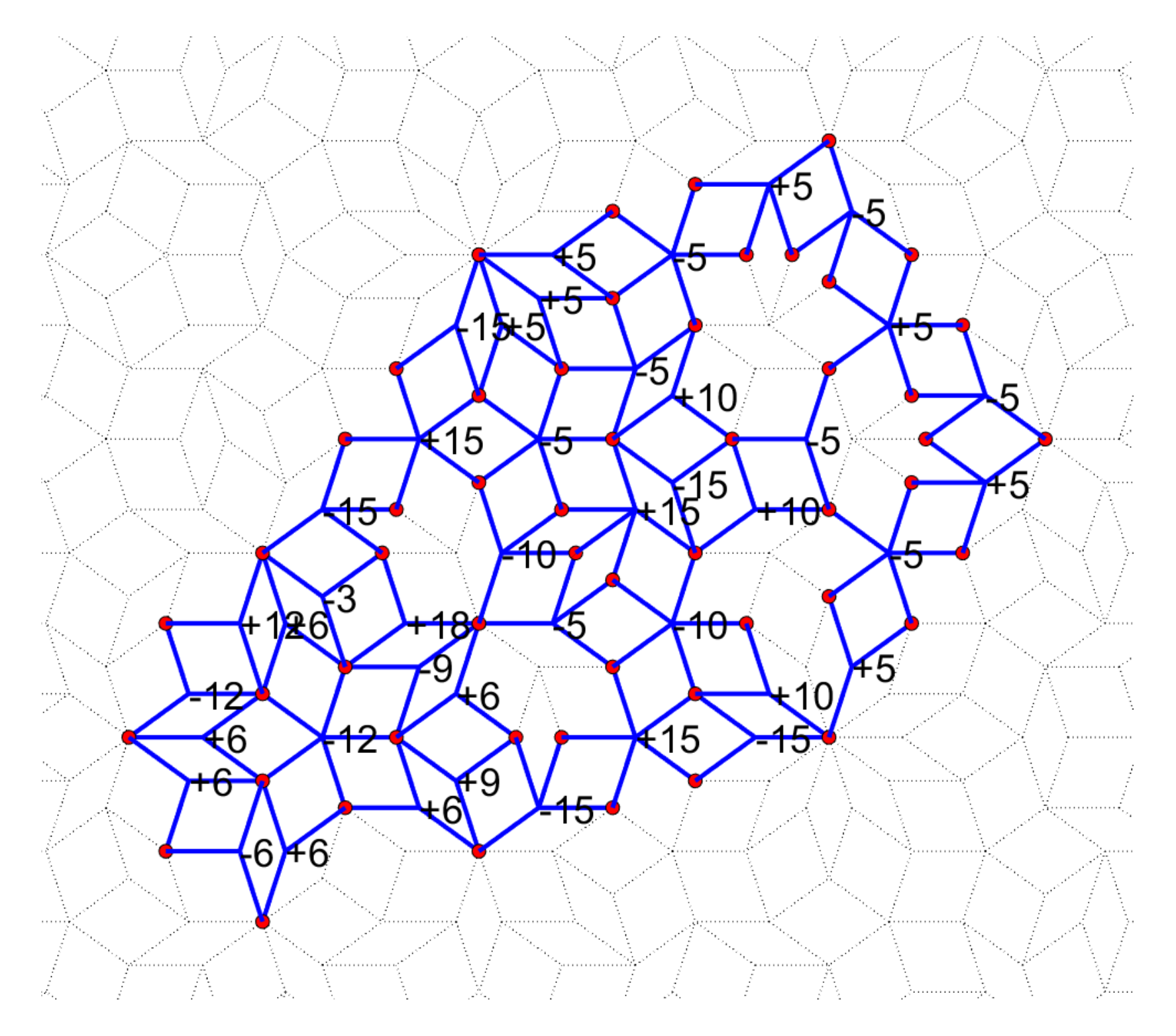}
    \includegraphics[clip,width=0.31\textwidth]{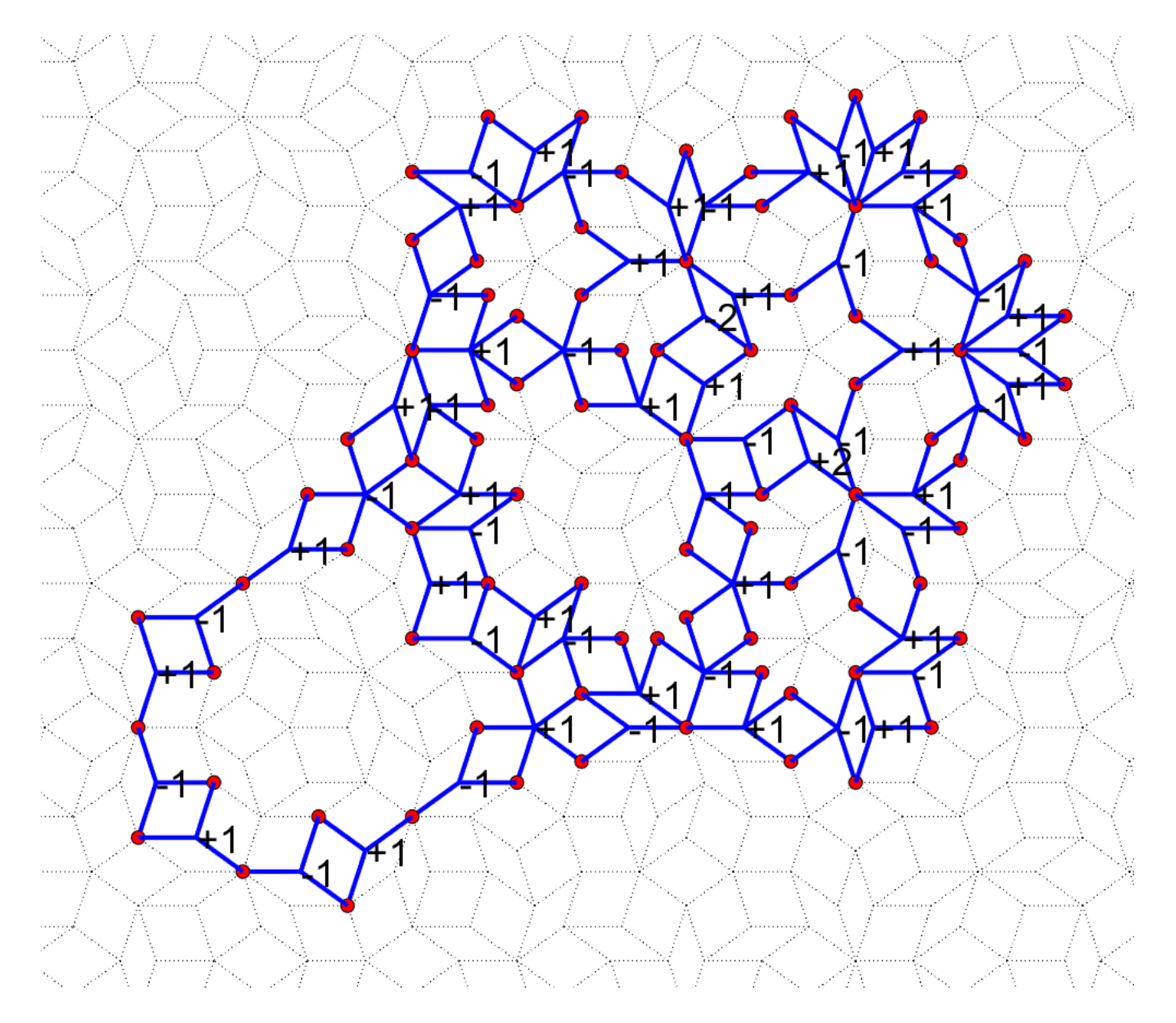}
    \includegraphics[clip,width=0.31\textwidth]{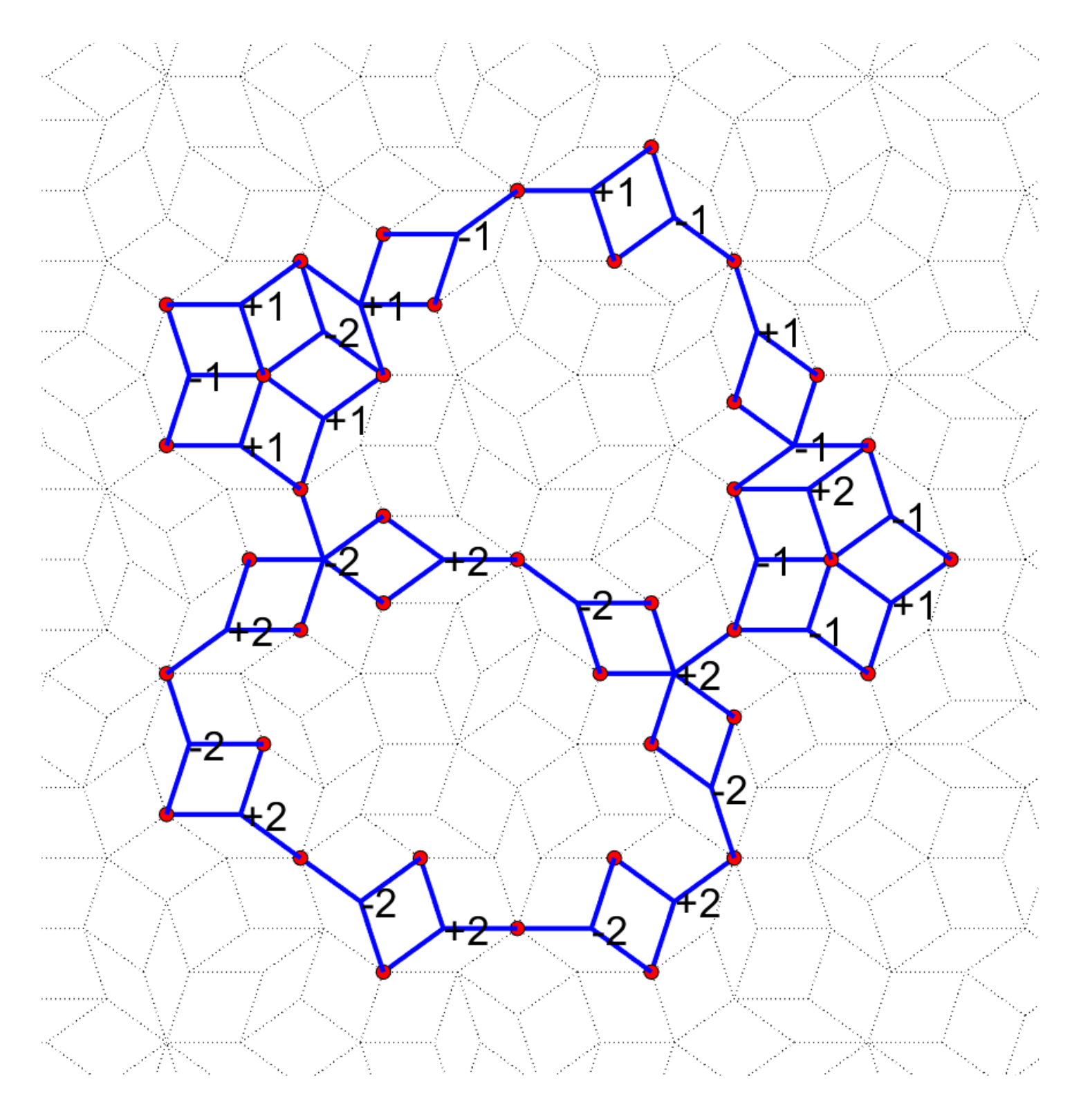}
    \caption{ Types O22 to O30 on the odd sublattice.}
    \label{fig:LS2230OddRealSpace}
\end{figure}

\begin{figure}[!htb]
    \centering
    \includegraphics[clip,width=0.31\textwidth]{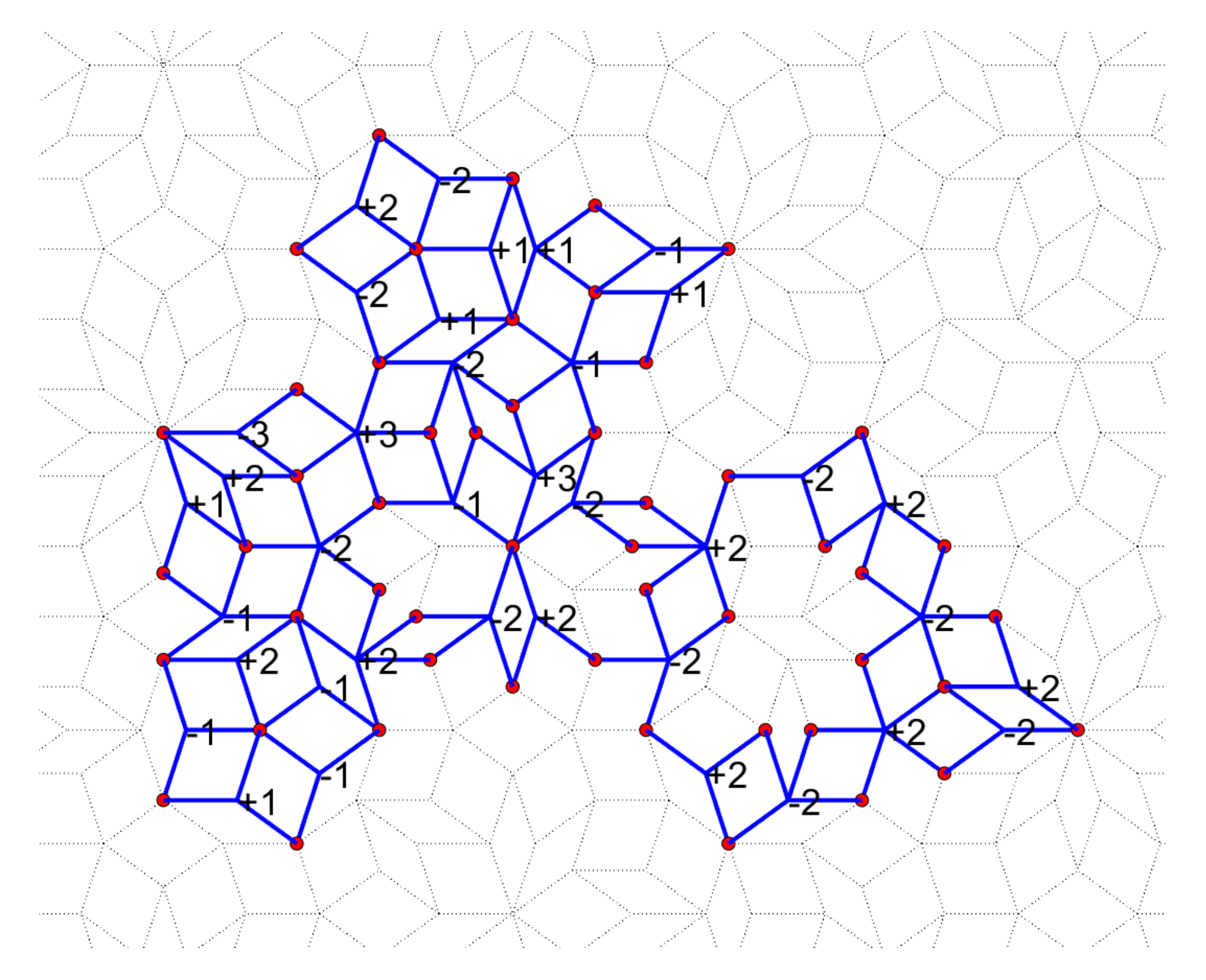}
    \includegraphics[clip,width=0.31\textwidth]{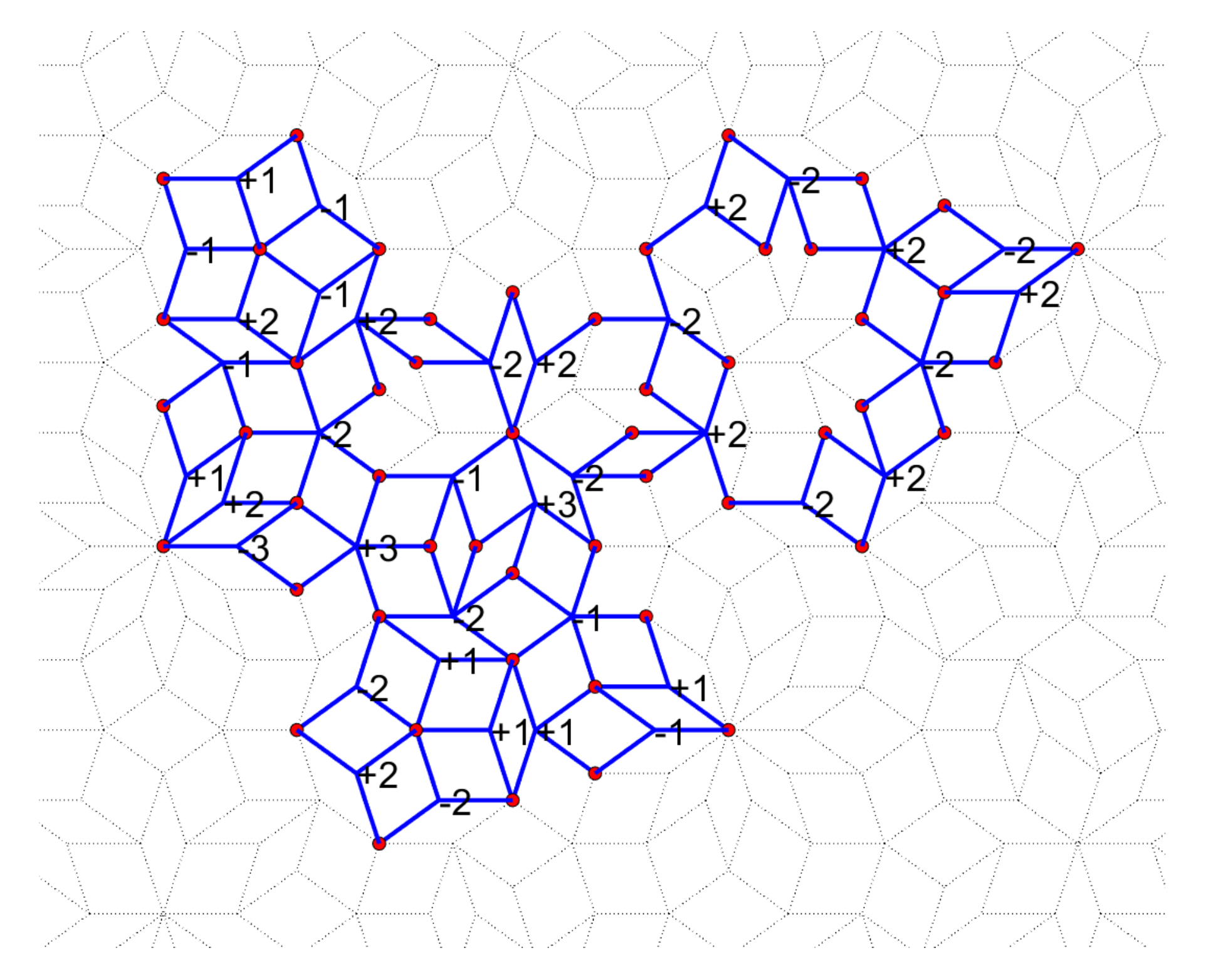}
    \includegraphics[clip,width=0.31\textwidth]{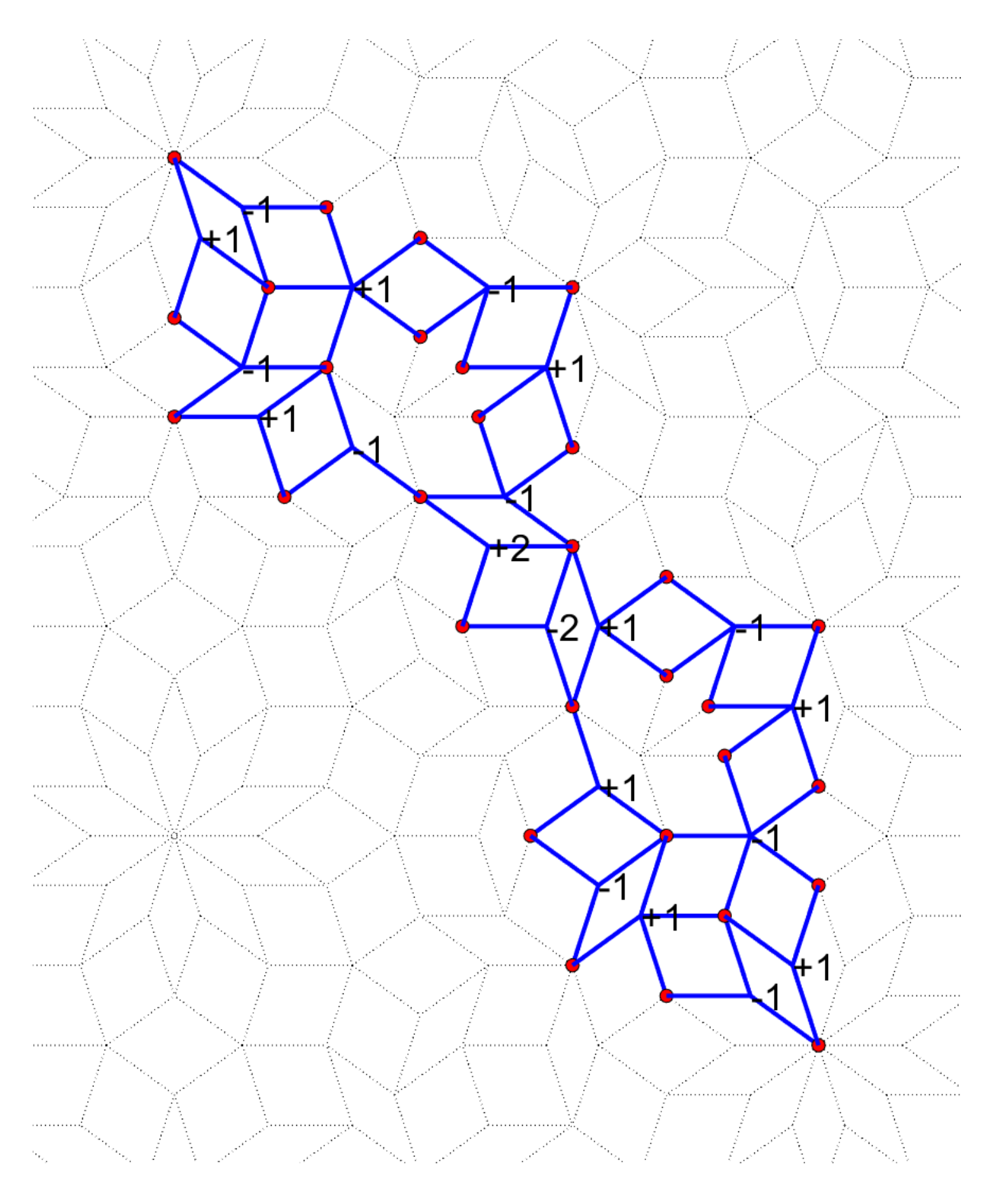}    
    \includegraphics[clip,width=0.31\textwidth]{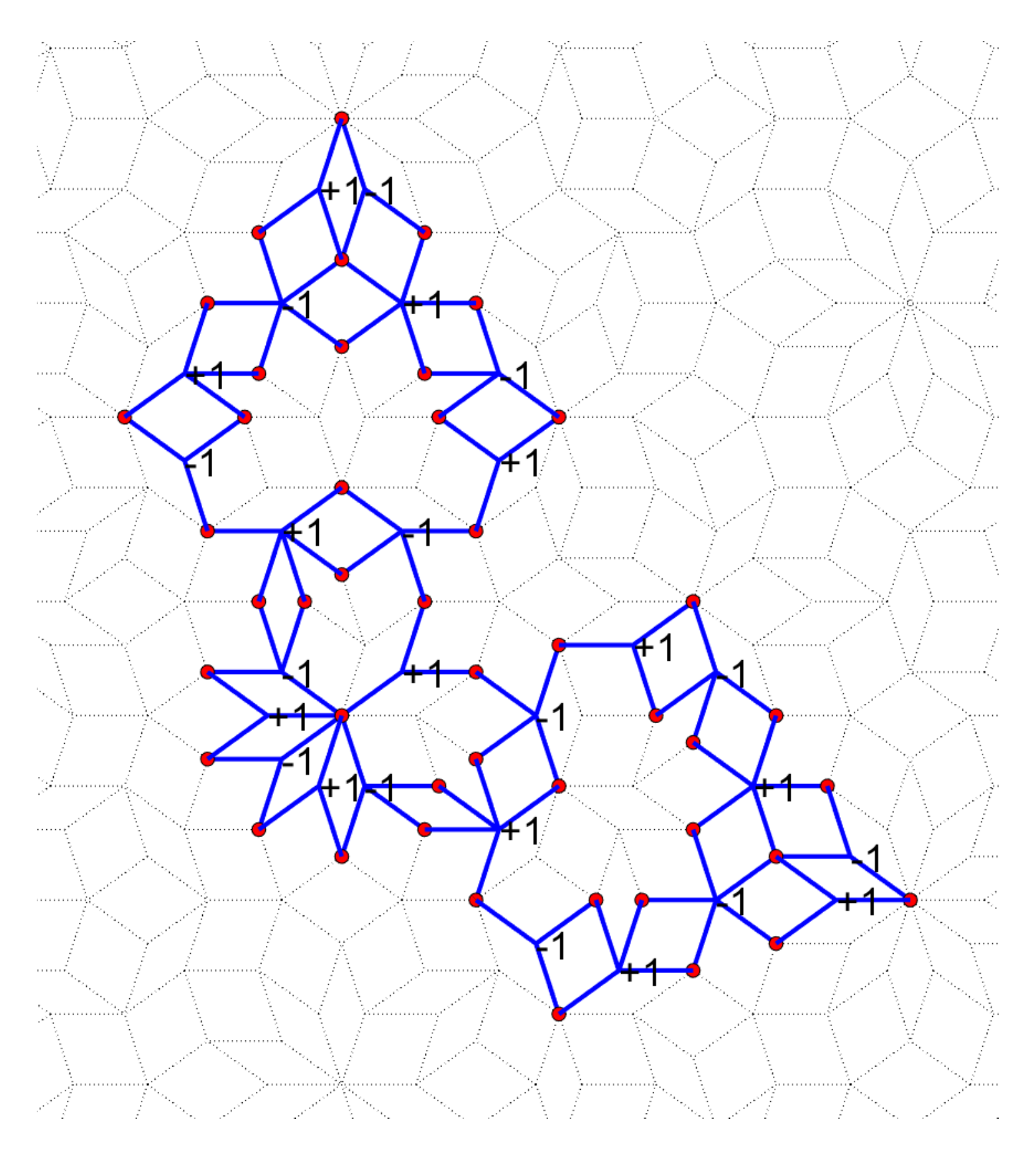}
    \includegraphics[clip,width=0.31\textwidth]{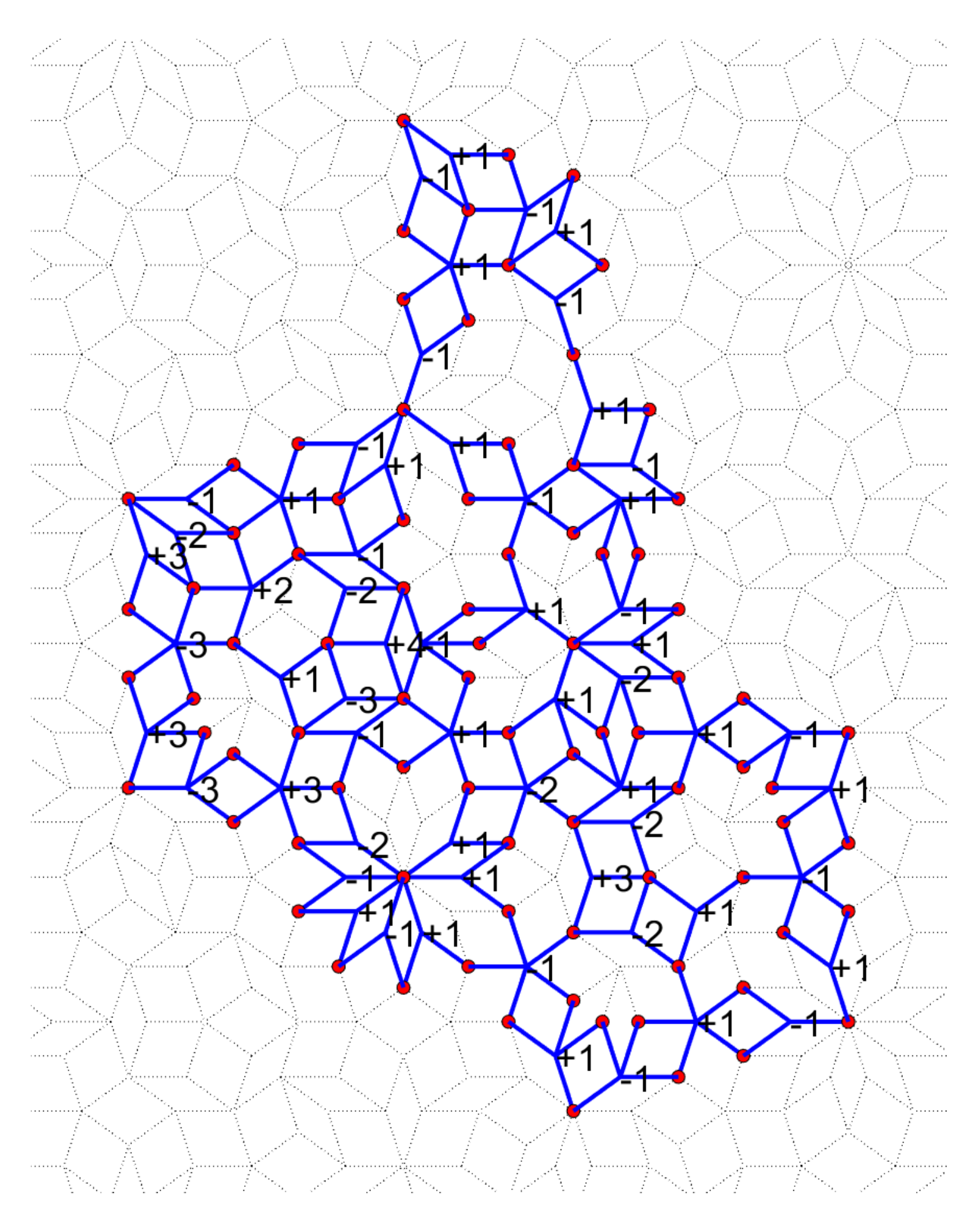}
    \includegraphics[clip,width=0.31\textwidth]{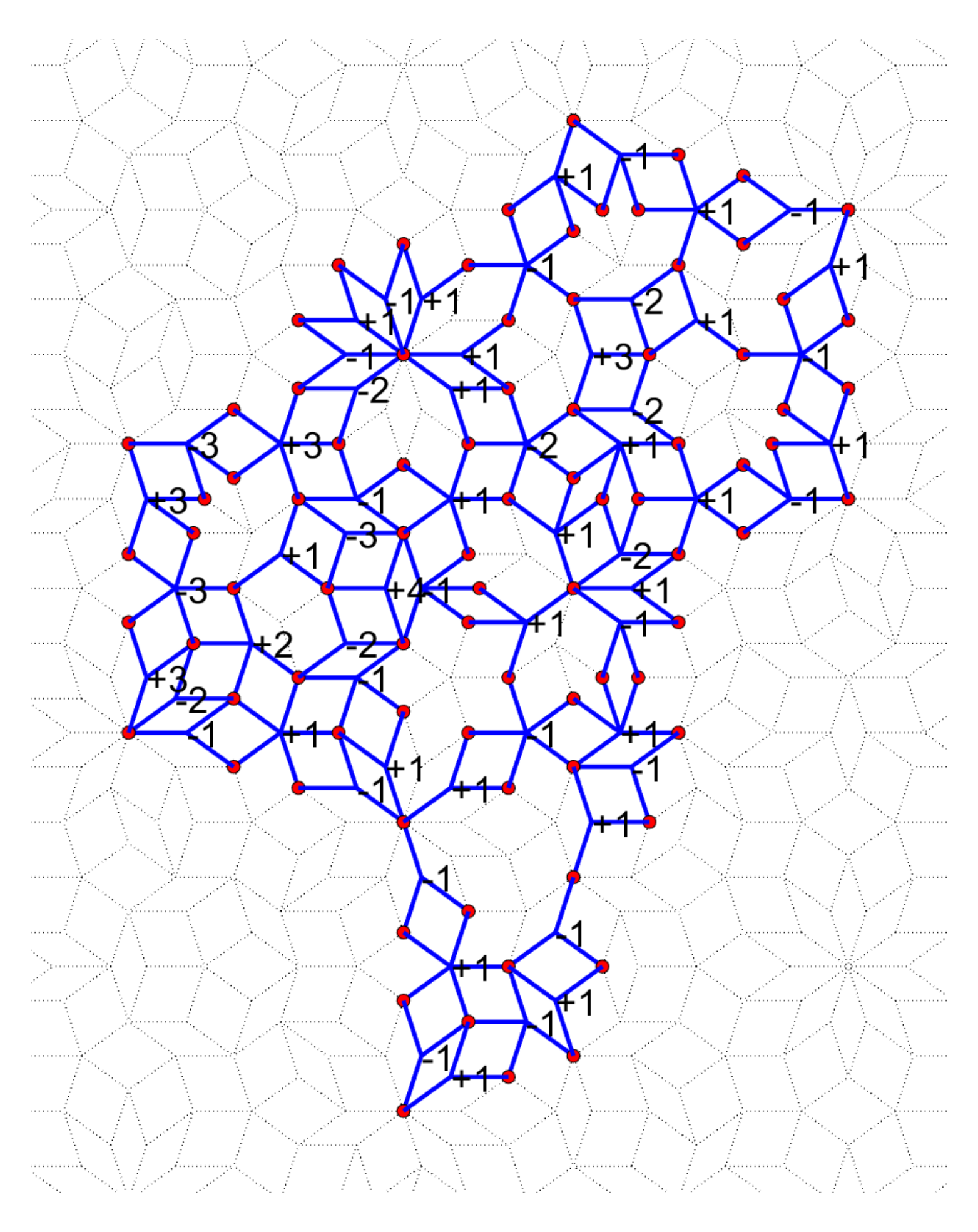}
    \includegraphics[clip,width=0.31\textwidth]{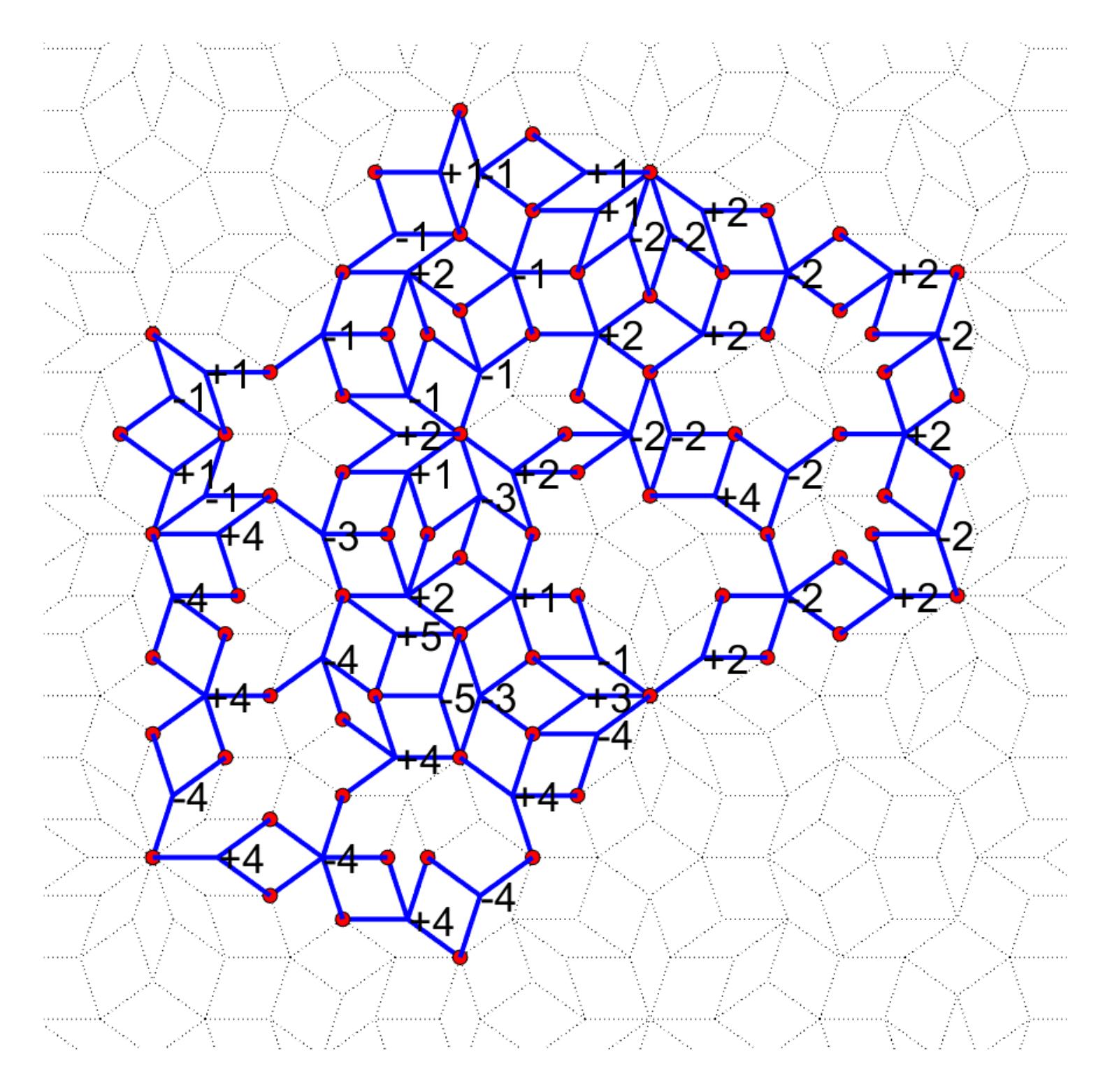}
    \includegraphics[clip,width=0.31\textwidth]{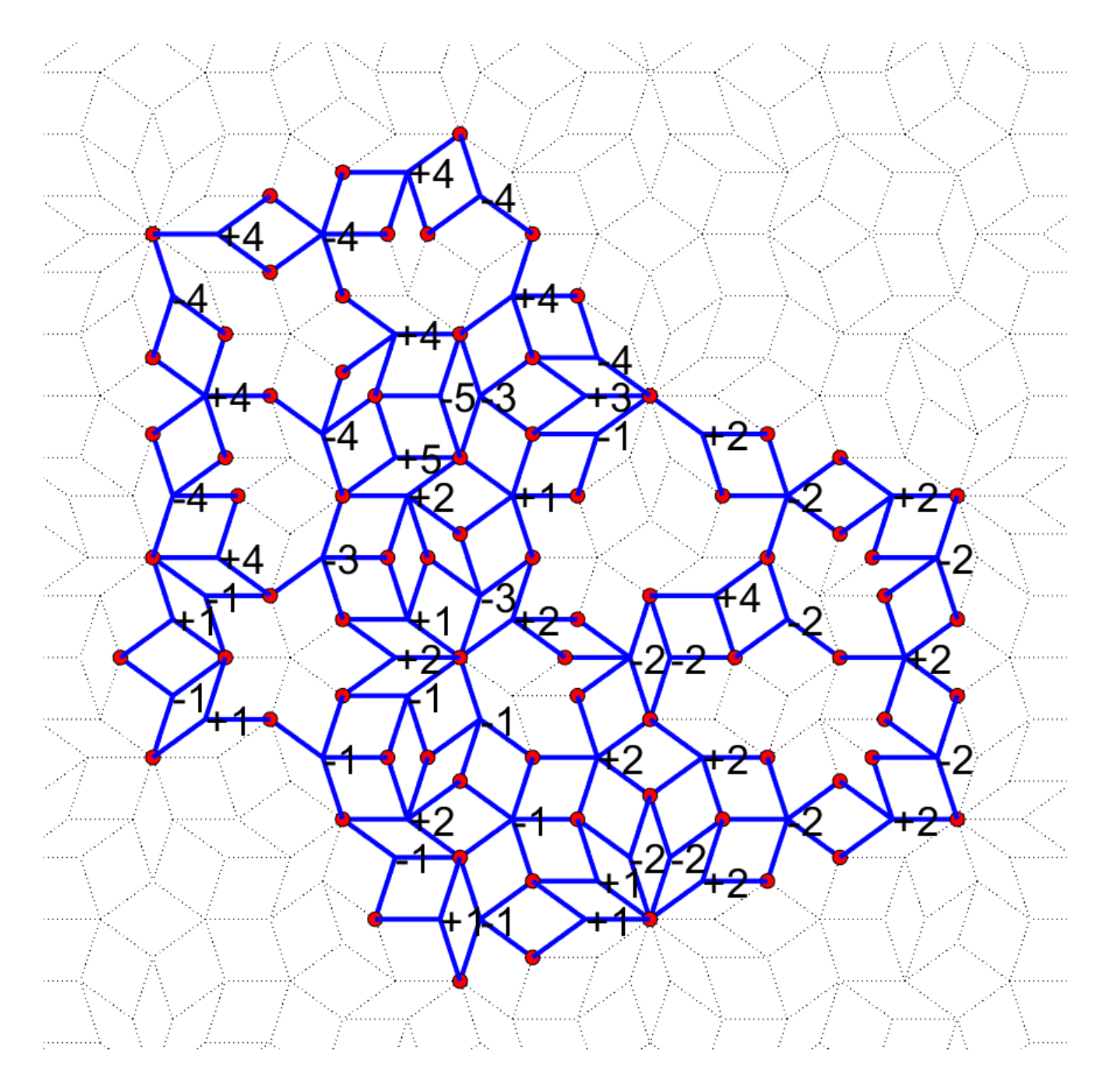}
    \includegraphics[clip,width=0.31\textwidth]{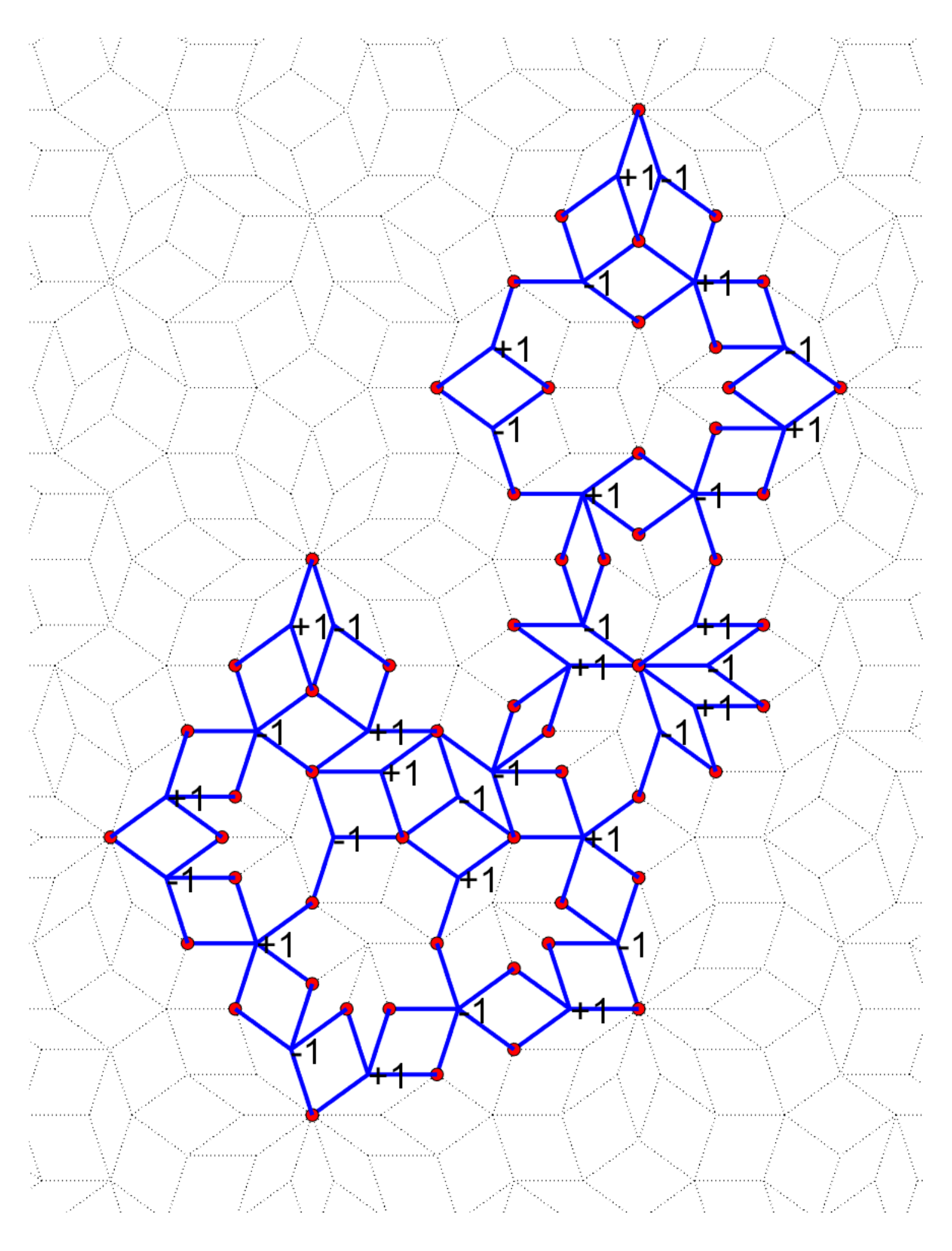}
    \caption{Types O31 to O39 on the odd sublattice. }
    \label{fig:LS3139OddRealSpace}
\end{figure}

\begin{figure}[!htb]
    \centering
    \includegraphics[clip,width=0.31\textwidth]{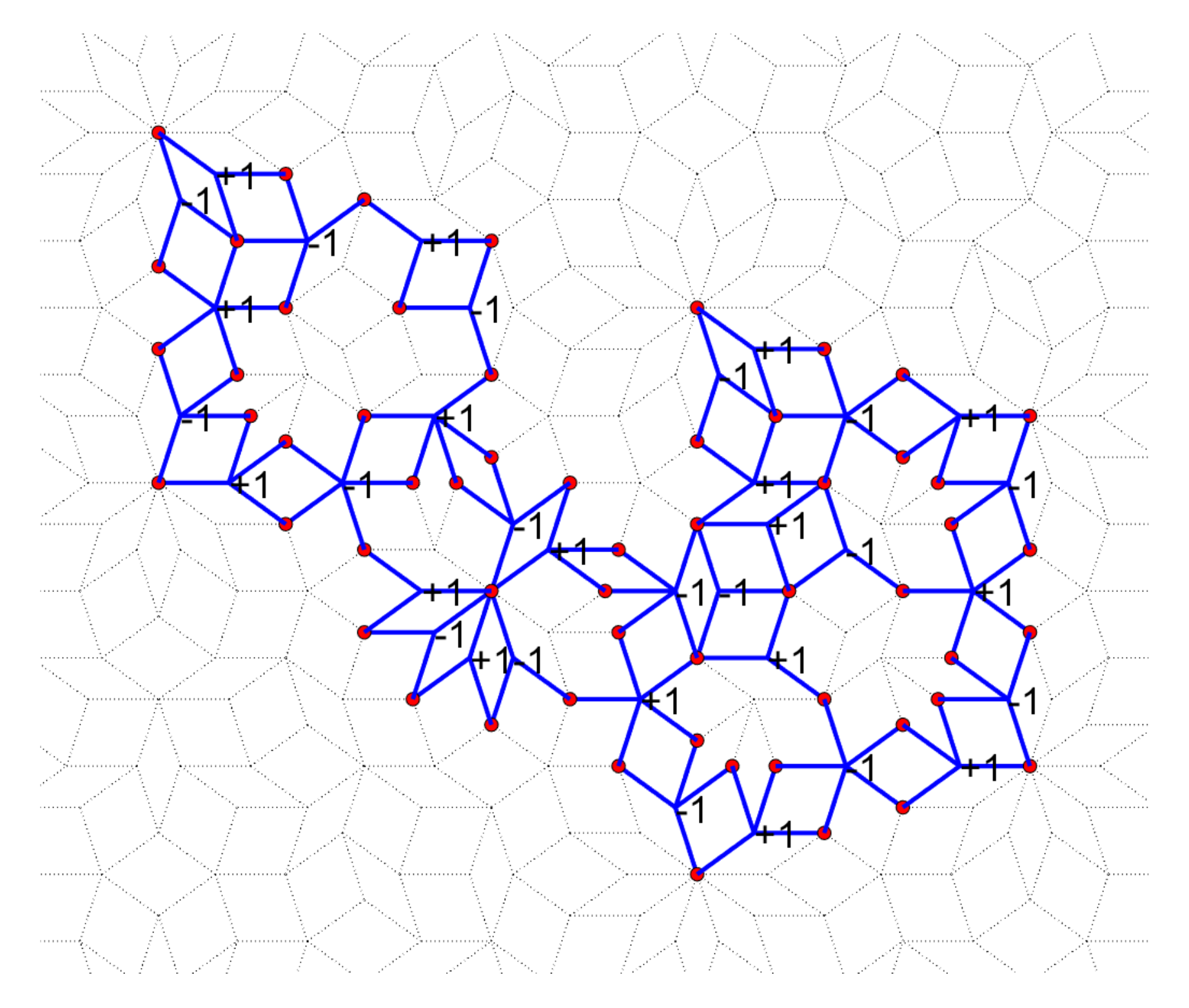}
    \includegraphics[clip,width=0.31\textwidth]{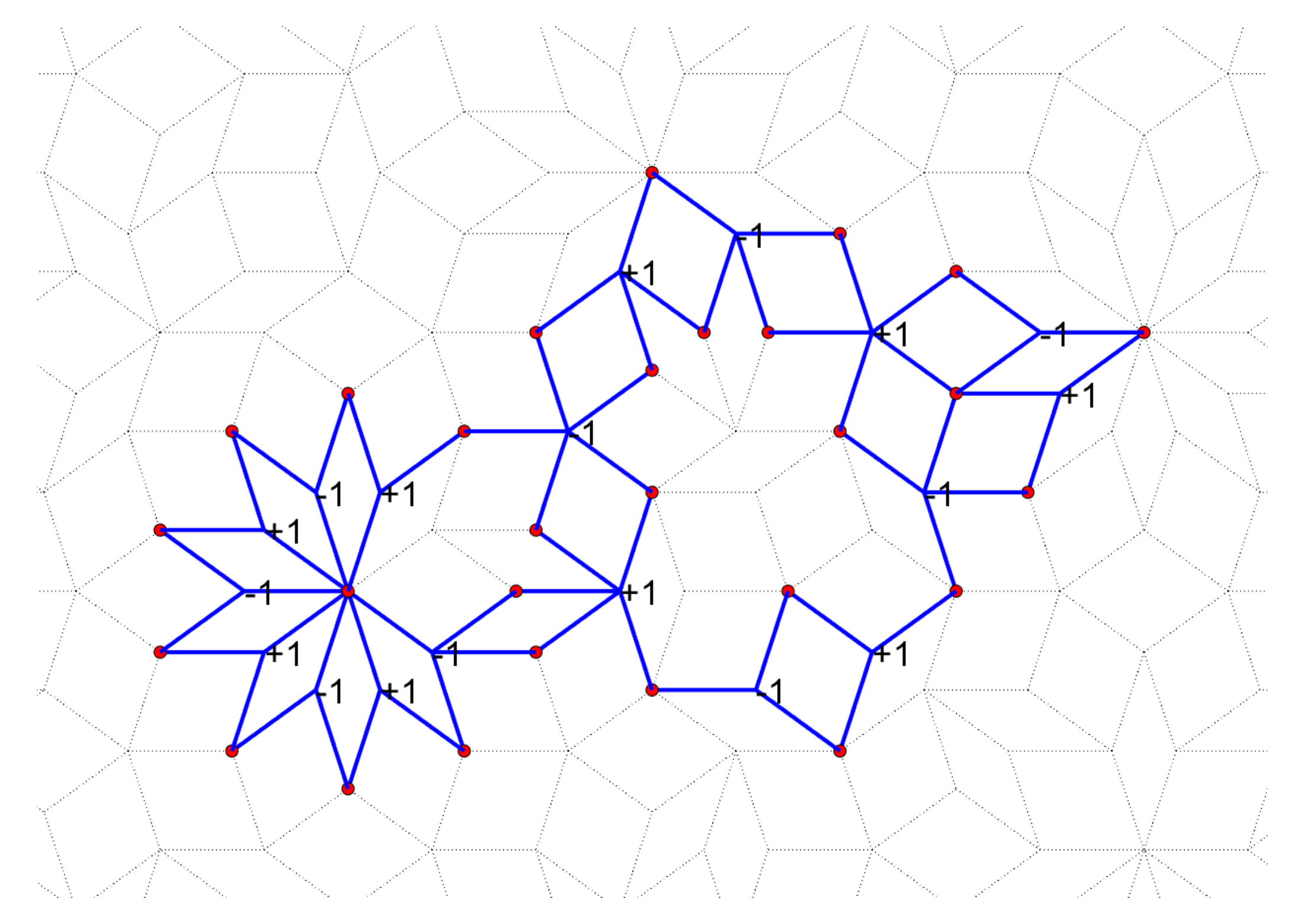}
    \includegraphics[clip,width=0.31\textwidth]{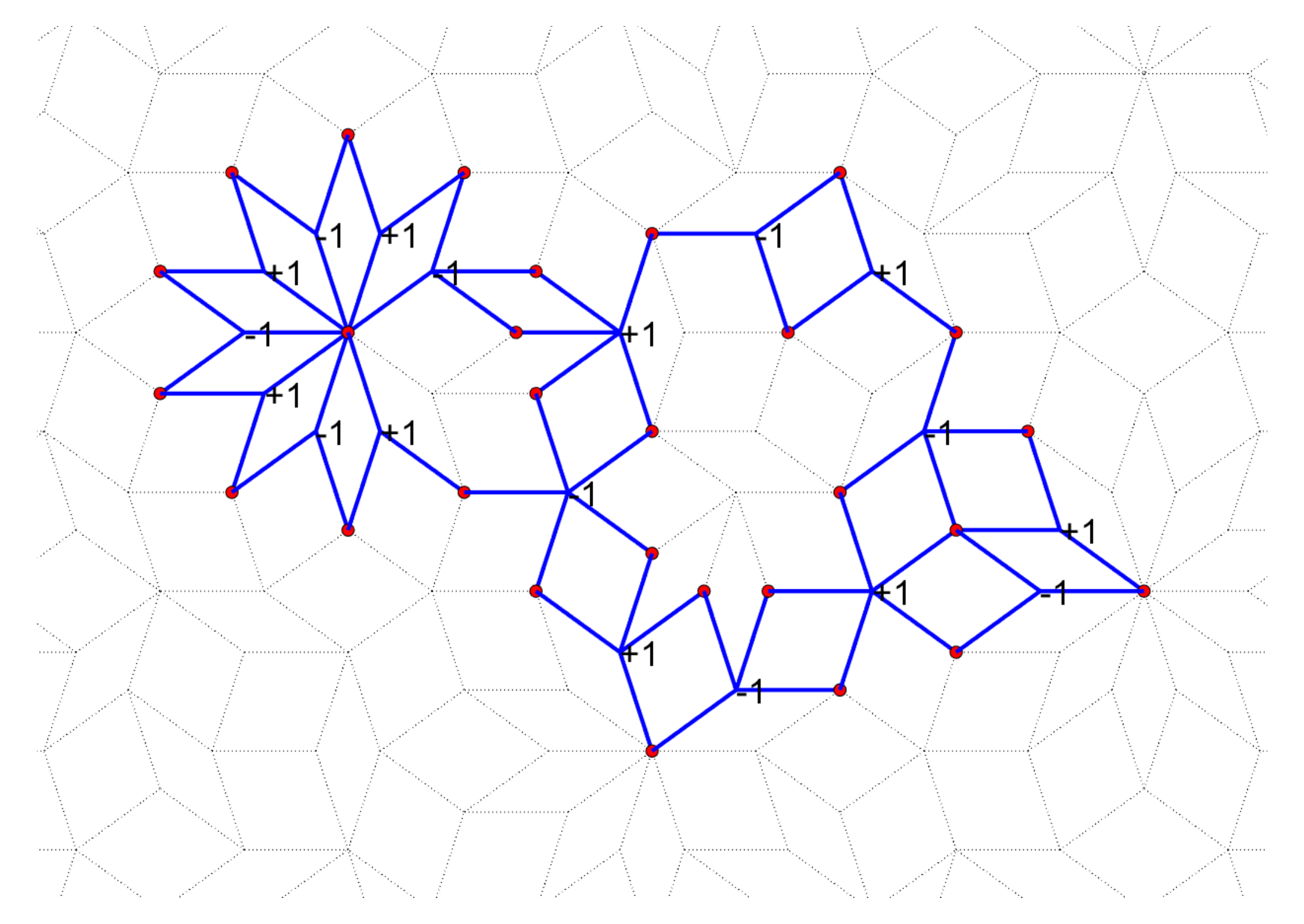}    
    \includegraphics[clip,width=0.31\textwidth]{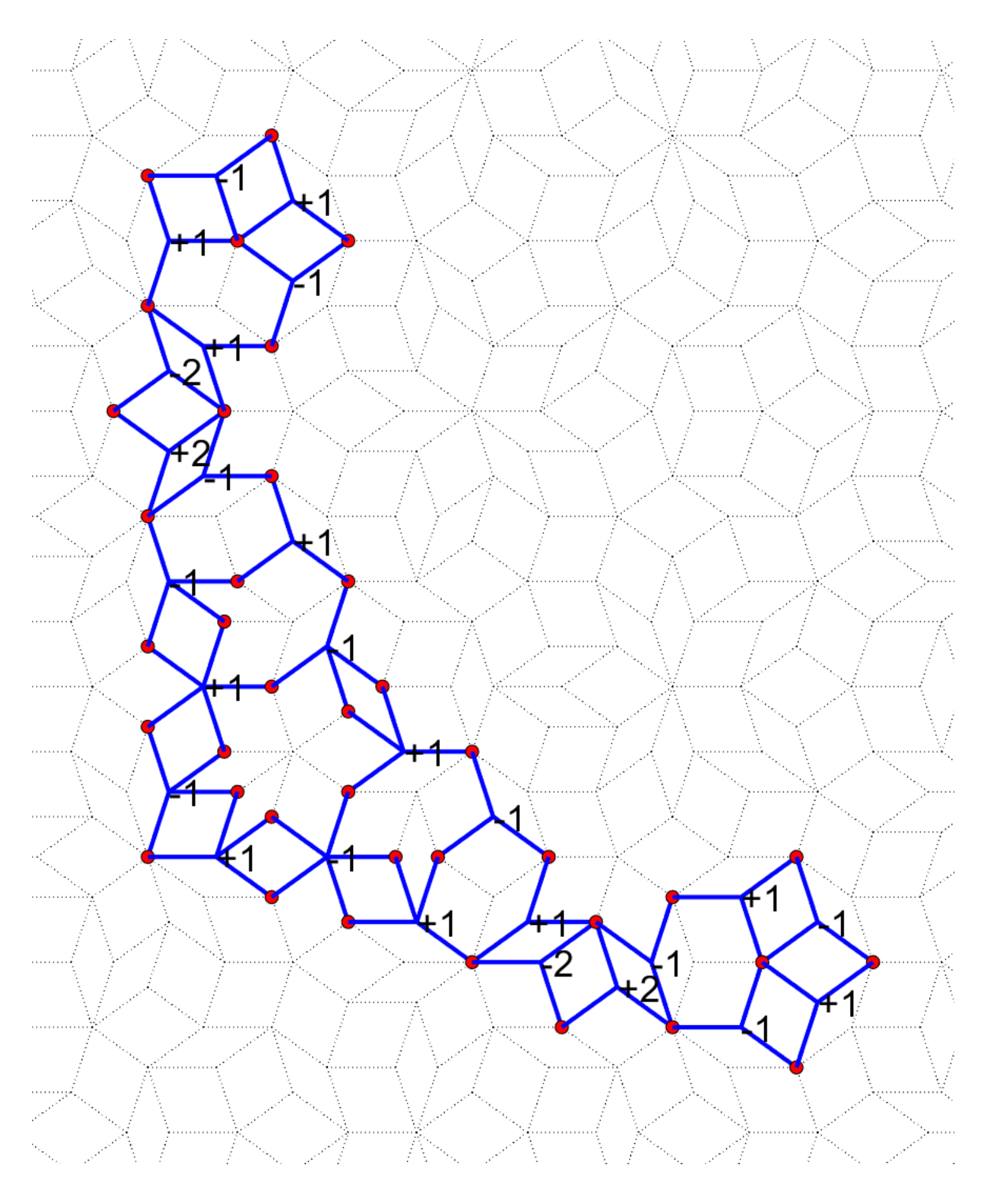}
    \includegraphics[clip,width=0.31\textwidth]{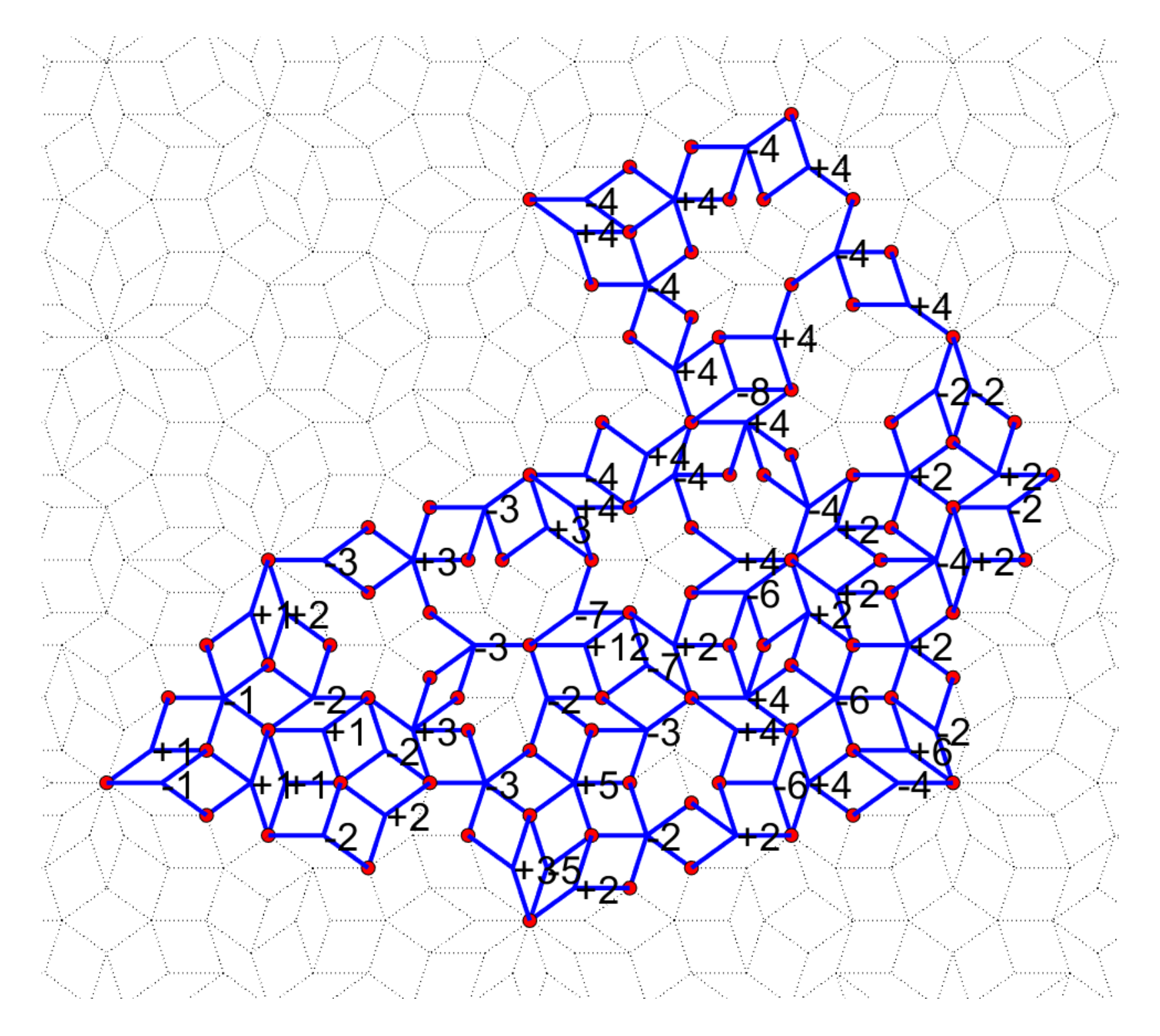}
    \includegraphics[clip,width=0.31\textwidth]{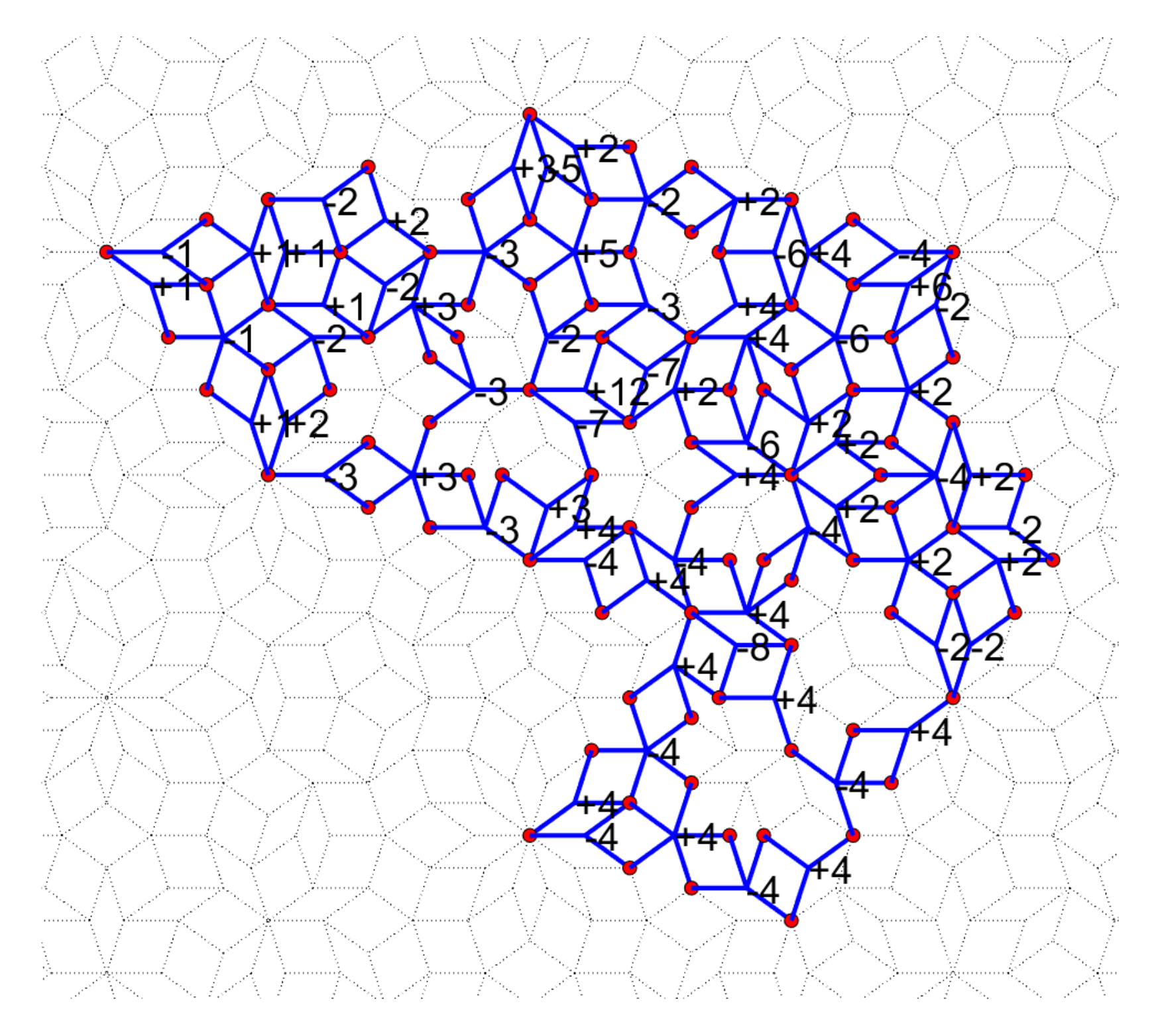}
\caption{Types O40 to O45 on the odd sublattice.}
    \label{fig:LS4045OddRealSpace}
\end{figure}

\begin{figure}[!htb]
    \centering
    \includegraphics[clip,width=0.31\textwidth]{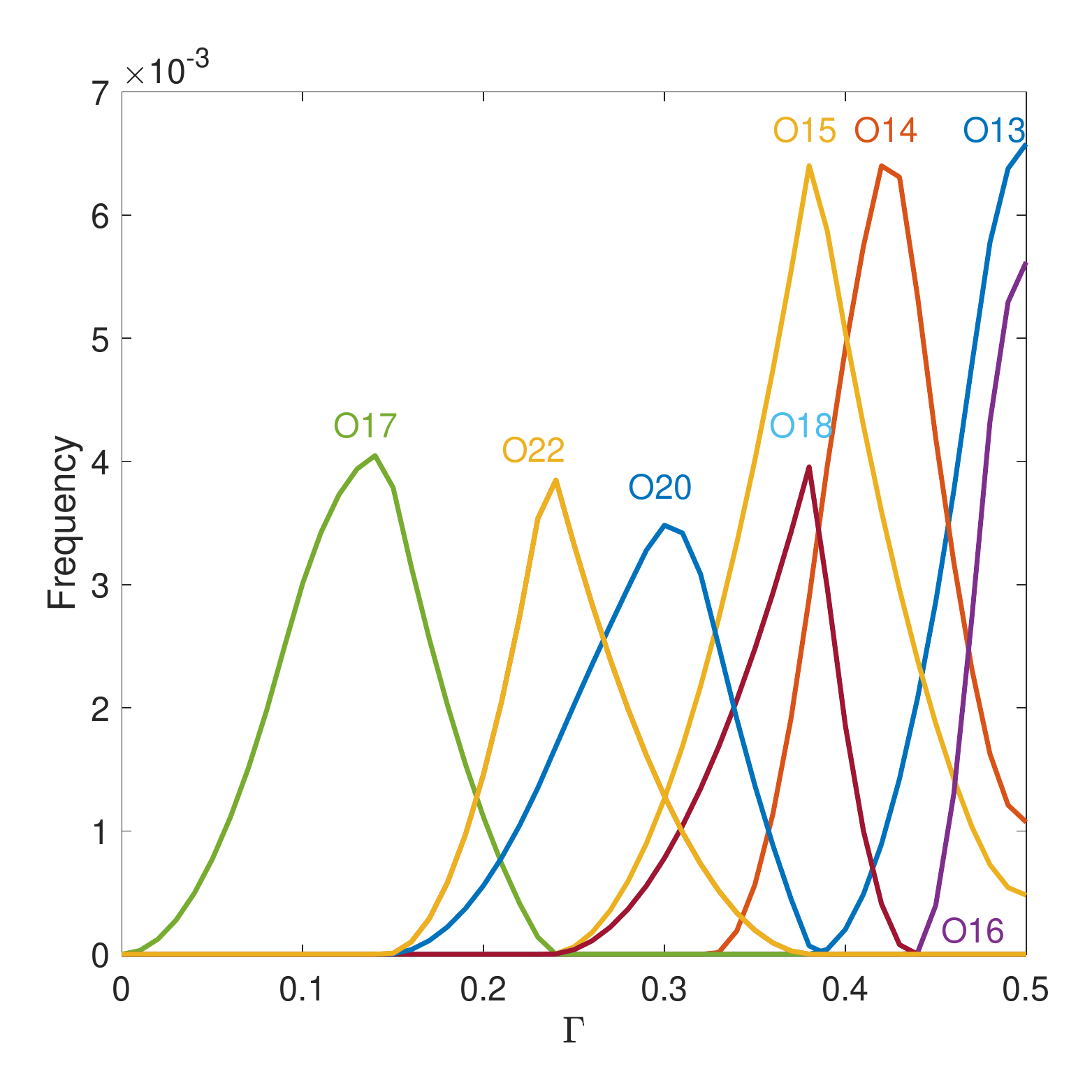}
    \includegraphics[clip,width=0.31\textwidth]{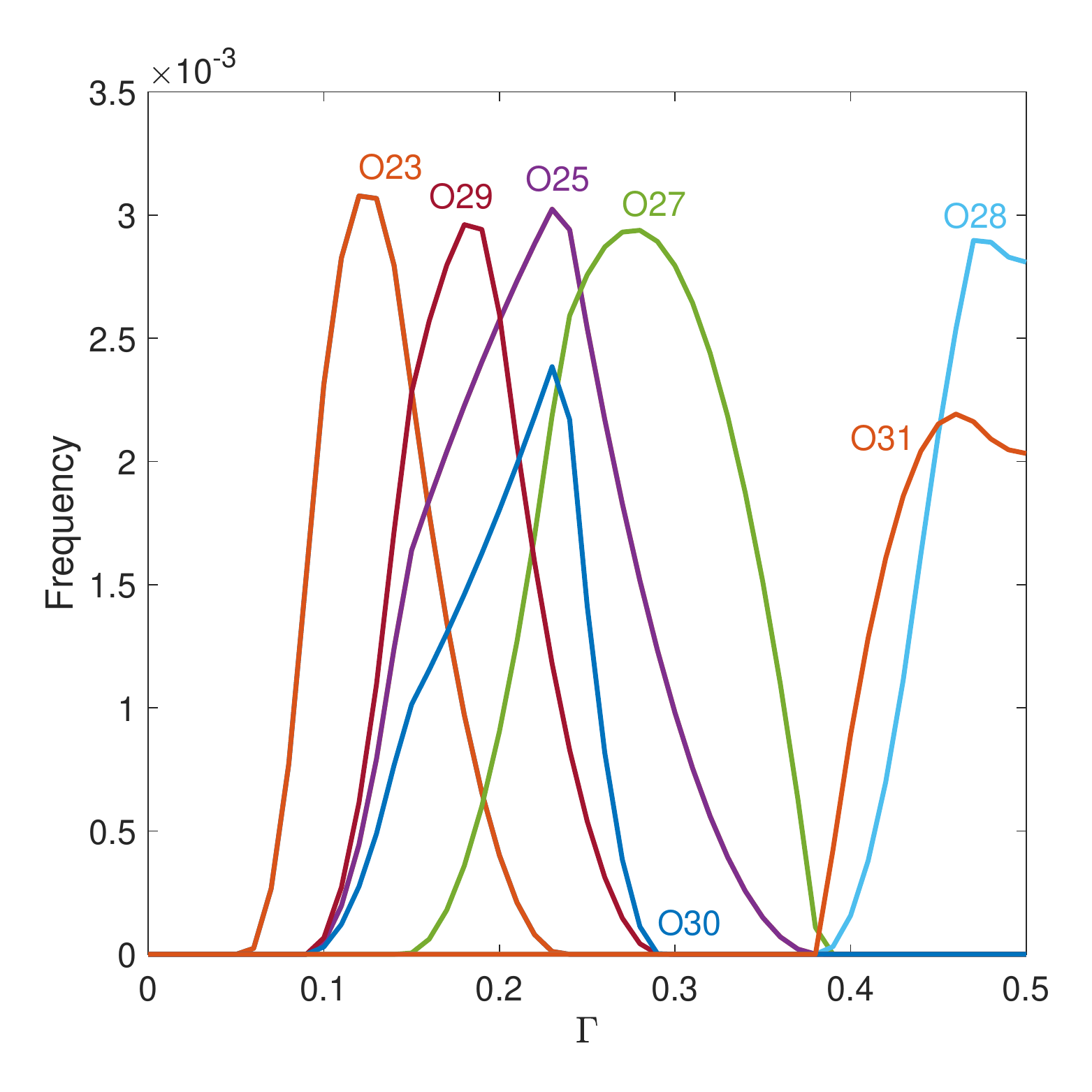}
    \includegraphics[clip,width=0.31\textwidth]{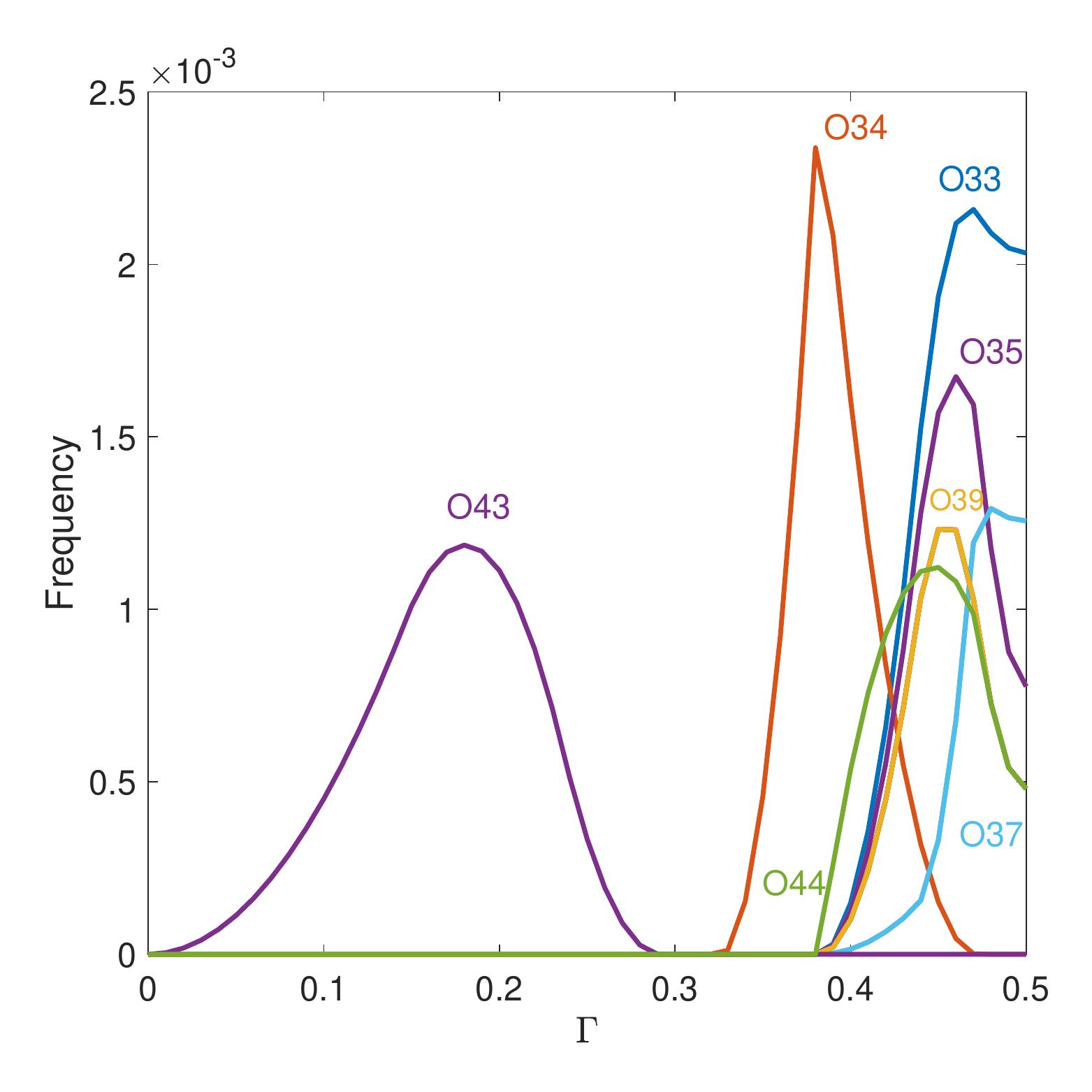}
    \caption{LS frequencies of the all the odd sublattice LS types given in the appendix. Type-O20 has the same frequency as Type-O19, hence is not shown separately. Similarly Types O21-O22,O23-O24,O25-O26,O31-O32,O35-O36,O37-O38, and O44-O45 form pairs with the same frequencies at all $\Gamma$. Finally all four types O39-O40-O41-O42 have the same frequency.}
    \label{fig:LastOddLSFrequency}
\end{figure}

We give the real space structure for the LS types which are not presented in the main text in Figs.\ref{fig:Last8EvenRealSpace},\ref{fig:LS1321OddRealSpace},\ref{fig:LS2230OddRealSpace},\ref{fig:LS3139OddRealSpace},\ref{fig:LS4045OddRealSpace}. The frequencies for these LS types are plotted in Figs.\ref{fig:Last8EvenLSFrequency},\ref{fig:LastOddLSFrequency}.

%
%
%



%

\end{document}